\documentclass[11pt, letterpaper,openany]{article}
\usepackage{pdfpages}
\usepackage{amsmath}
\usepackage{amssymb}
\usepackage{multirow}
\usepackage{amsthm}
\usepackage{graphicx}
\usepackage{geometry}
\usepackage{hyperref}
\usepackage{url}
\usepackage{algorithm}
\usepackage{tocloft}
\usepackage{algorithmic}
\usepackage{caption}
\usepackage{subcaption}
\usepackage{booktabs}
\usepackage{rotating}
\usepackage{authblk}
\usepackage{appendix}
\usepackage[utf8]{inputenc}
\usepackage[T1]{fontenc}
\usepackage[english]{babel}
\usepackage{xcolor}
\usepackage{fancyhdr}
\usepackage{microtype}
\usepackage{parskip}
\usepackage{paralist, tabularx}
\usepackage{enumitem}
\usepackage{placeins}
\usepackage{adjustbox}
\usepackage{placeins}
\usepackage{lscape}
\usepackage{ragged2e}
\usepackage{soul} 
\sethlcolor{green} 
\usepackage[numbers,sort]{natbib}
\usepackage{tcolorbox}
\usepackage{subcaption}
\usepackage{float} 
\usepackage{caption}
\usepackage{CJK}
\usepackage{authblk}
\usepackage{setspace}
\usepackage{tabularx}
\hypersetup{
    colorlinks=true,
    linkcolor=blue,
    filecolor=magenta,      
    urlcolor=cyan,
    citecolor=red,
}

\definecolor{headercolor}{RGB}{0, 50, 100}
\makeatletter
\renewenvironment{abstract}{%
    \if@twocolumn
      \section*{\abstractname}%
    \else 
      \begin{center}%
        {\bfseries \Large\abstractname\vspace{\z@}}
      \end{center}%
      \quotation
    \fi}
    {\if@twocolumn\else\endquotation\fi}
\makeatother

\title{A Systematic Assessment of OpenAI o1-Preview for Higher Order Thinking in Education}

\date{}

\newcommand*\samethanks[1][\value{footnote}]{\footnotemark[#1]}

\author[1,2]{Ehsan Latif \thanks{Co-first authors.}}
\author[1,2,3]{Yifan Zhou \samethanks[1]}
\author[1,2,12]{Shuchen Guo \samethanks[1]}
\author[1,2,10]{Yizhu Gao \thanks{Co-second authors.}}
\author[2,11]{Lehong Shi \samethanks[2]}
\author[1,2]{Matthew Nyaaba \samethanks[2]}
\author[5]{Gyeonggeon Lee \samethanks[2]}
\author[4]{Liang Zhang \samethanks[2]}
\author[7]{Arne Bewersdorff \samethanks[2]}
\author[6]{Luyang Fang \samethanks[2]}
\author[8]{Xiantong Yang\samethanks[2]}
\author[6]{Huaqin Zhao\samethanks[2]}
\author[3]{Hanqi Jiang\samethanks[2]}
\author[6]{Haoran Lu\samethanks[2]}
\author[3]{Jiaxi Li \samethanks[2]}
\author[3]{Jichao Yu \samethanks[2]}
\author[3]{Xuansheng Wu \samethanks[2]}
\author[3]{Weihang You \samethanks[2]}
\author[3]{Zhengliang Liu \samethanks[2]}
\author[3]{Vincent Shung Liu \thanks{Co-third authors.}}
\author[9]{Hui Wang \samethanks[3]}
\author[3]{Zihao Wu \samethanks[3]}
\author[2]{Jin Lu}
\author[2]{Fei Dou}
\author[6]{Ping Ma}
\author[1,2,3]{Ninghao Liu}
\author[1,2,3]{Tianming Liu}
\author[1,2,10]{Xiaoming Zhai \thanks{Corresponding author: xiaoming.zhai@uga.edu}\vspace{4em}}

\affil[1]{AI4STEM Education Center, University of Georgia, Athens, GA, USA}
\affil[2]{National GENIUS Center, Athens, GA, USA}
\affil[3]{School of Computing, University of Georgia, Athens, GA, USA}
\affil[4]{Department of Electrical Engineering and Computer Engineering, University of Memphis, Memphis, TN, USA}
\affil[5]{Department of Natural Sciences and Science Education, National Institute of Education, Singapore, Singapore}
\affil[7]{School of Social Sciences and Technology, Technical University of Munich, Munich, Germany}
\affil[6]{Department of Statistics, University of Georgia, Athens, GA, USA}
\affil[8]{Faculty of Psychology, Beijing Normal University, Beijing, Chaina}
\affil[9]{Second Language Acquisition and Teaching, University of Arizona, Tucson, AZ, USA}
\affil[10]{Department of Mathematics, Science, and Social Studies Education, University of Georgia, Athens, GA, USA}
\affil[11]{Department of Workforce Education and Instructional Technology, University of Georgia, Athens, GA, USA}
\affil[12]{School of Teacher Education, Nanjing Normal University, Nanjing, Jiangsu, China}

\begin{document}
\begin{CJK}{UTF8}{gbsn}
\maketitle

\newpage

\begin{abstract}
\normalsize
As artificial intelligence (AI) continues to advance, it demonstrates capabilities comparable to human intelligence, having significant potential to revolutionize education and workforce development. To assess the state-of-art AI capacity, this study provides a comprehensive evaluation of OpenAI o1-preview's ability to perform higher-order cognitive tasks across 14 distinct dimensions, including \textit{critical thinking}, \textit{systems thinking}, \textit{computational thinking}, \textit{design thinking metacognition}, \textit{data literacy}, \textit{creative thinking}, \textit{abstract reasoning}, \textit{quantitative reasoning}, \textit{logical reasoning}, \textit{analogical reasoning}, and \textit{scientific reasoning}. In each dimension, we adopt existing instruments with high reliability and validity, such as the Ennis-Weir Critical Thinking Essay Test and the Biological Systems Thinking Test to systematically assess o1-preview's higher-order thinking in comparison to human performance. We found overall outperformance of o1-preview for most of the categories:
\begin{itemize}
    \item o1-preview achieved significantly outperform (150\%) as compared to humans for \textit{systematic thinking}, \textit{computational thinking}, \textit{data literacy}, \textit{creative thinking}, \textit{scientific reasoning}, and \textit{abstract reasoning}.
    \item For \textit{logical reasoning}, \textit{critical thinking}, and \textit{quantitative reasoning}, o1-preview slightly under-performed by ~25\% than human such as for logical reasoning.
    \item For \textit{analogical reasoning}, we found that both human and o1-preview achieve perfect scores for each given task.
\end{itemize}

Nevertheless, we identify o1-preview's limited capacity in thinking such as \textit{abstract reasoning}, where humans (i.e., psychology students) perform better than o1-preview, indicating that human oversight remains crucial for tasks requiring high-level abstraction. The findings have broad implications for education, advocating a shift in educational paradigms to emphasize the development of complementary human skills, such as creativity, abstract reasoning, and critical thinking. This study highlights the transformative role of AI in education and calls for a recalibration of educational goals, teaching methodologies, curricula, and assessments to better align with the  AI-driven world.

\end{abstract}
 \newpage
\tableofcontents

\newpage



\section{Introduction}\label{Gyeonggeon, xz}
The rapid advancements in artificial intelligence (AI) have led to the development of sophisticated models that are increasingly integrated into educational settings~\cite{zhai2022chatgpt,guo2024artificial}. OpenAI o1-preview, a prominent large language model (LLM) based on the GPT-4 architecture, represents one such technological innovation \cite{openai2024o1}.  
OpenAI o1-preview demonstrates improved capabilities in understanding and generating complex, nuanced text, exhibiting greater accuracy and coherence in its responses. Compared to earlier models, it excels at handling open-ended prompts, facilitating more advanced reasoning and problem-solving tasks \cite{zhong2024evaluationopenaio1opportunities}. Additionally, OpenAI o1-preview features enhanced abilities in context retention and task-specific adaptation, positioning it as a superior tool for educational applications that require deep engagement with higher-order thinking. OpenAI o1-preview can potentially be implemented in diverse learning environments, aiding students and educators in problem-solving, critical thinking, and creative endeavors. As the presence of AI systems like OpenAI o1-preview grows in education, it becomes critical to systematically assess their effectiveness, particularly in promoting cognitive skills that extend beyond basic knowledge retrieval~\cite{zhai2024can}.

Human thinking is inherently complex and multidimensional, characterized by cognitive processes such as analysis, evaluation, and synthesis \cite{lewis1993defining}. These higher-order thinking skills are vital to intellectual development, fostering innovation, problem-solving, and decision-making. Society's advancement is tightly linked to the cultivation of these skills, as they are essential for addressing complex real-world challenges. In education, fostering these skills is a primary goal, preparing students not only to consume knowledge but also to critically engage with and contribute to their fields of study \cite{collins2014skills}. Given the intricate nature of these cognitive processes, any assessment of AI's role in education must examine its capacity to facilitate and support higher-order thinking.

OpenAI o1-preview demonstrates significant potential in assisting with complex cognitive tasks, but its actual performance in competing and fostering higher-order thinking remains largely unexplored. While it can generate text-based responses that appear coherent and insightful, the extent to which it can truly support deep, critical thinking or engage learners in meaningful ways has not been comprehensively evaluated. The nuanced nature of higher-order cognitive tasks demands a robust analysis of how AI tools like OpenAI o1-preview respond to open-ended, challenging prompts, how they solve critical problems, and whether they possess skills such as evaluation and creativity. Without systematic analysis, the question of whether OpenAI o1-preview is genuinely capable of higher-order thinking in education remains unanswered.

This study assesses OpenAI o1-preview's ability to engage in higher-order thinking across \textbf{14 distinct cognitive domains}, each evaluated using a specific instrument or dataset tailored to measure skills such as critical thinking, systems thinking, design thinking, and more. The instruments used in this study were selected for their alignment with established human cognitive benchmarks. For example, the Ennis-Weir Critical Thinking Essay Test, widely recognized in educational research, served as the benchmark for assessing critical thinking. For System Thinking, three distinct instruments were utilized: the Biological Systems Thinking Test, The Village of Abeesee, and The Lake Urmia Vignette (LUV). Computational Thinking was evaluated using the Bebras Challenge problems and the Algorithmic Thinking Test, while Design Thinking was tested through Leetcode coding questions designed to improve problem-solving efficiency. Each instrument provided a robust foundation for comparing o1-preview's performance with human intelligence benchmarks.

\begin{figure}[h]
    \centering
    \includegraphics[width=0.65\linewidth]{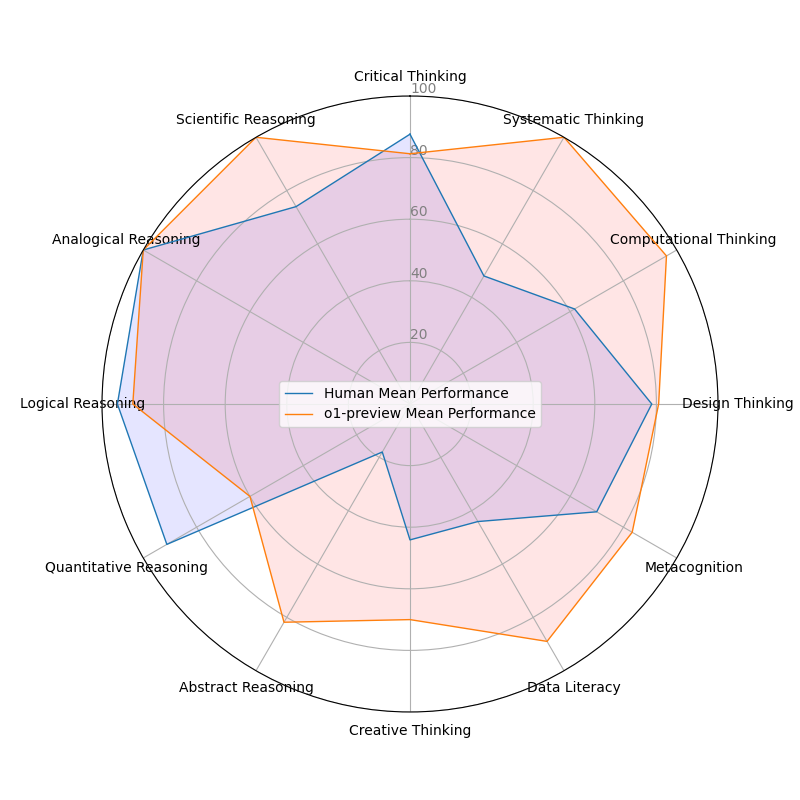}
    \caption{o1-preview performance compared to human for higher order thinking. (The plot reports the mean percentage performance for each category; i.e, the percentage of correct responses generated by o1-preview.) \textit{The plot is missing results for collaborative thinking and spatial reasoning due to lack of human evaluation results found in literature.}}
    \label{fig:representative_performance_plot}
\end{figure}

The comparative analysis reveals a mixed but promising landscape for OpenAI o1-preview (See Fig.~\ref{fig:representative_performance_plot} for collective performance view of o1-preview versus human for each higher order thinking category). In the domain of \textit{Critical Thinking}, as measured by the Ennis-Weir Critical Thinking Essay Test, o1-preview achieved a mean performance of 81.25\%, only 7.25\% lower than the human mean of 87.6\%. While this indicates that o1-preview performs at a slightly lower level than humans, it still demonstrates competence comparable to post-test scores of human learners, suggesting its utility in fostering critical thinking. For \textit{System Thinking}, the results were even more striking. When evaluated using the Biological Systems Thinking Test, o1-preview outperformed middle school students by a substantial margin, achieving a perfect score of 100\%, compared to a human mean of 48.0\%, representing an improvement of over 108.3\%. Similarly, on more advanced tests like The Village of Abeesee and the Lake Urmia Vignette, o1-preview demonstrated superiority over undergraduate and even graduate engineering students, exceeding their performance by over 50\% in some cases.

In the domain of \textit{Computational Thinking}, o1-preview's performance was notably impressive. Using instruments such as the Bebras Challenge and Algorithmic Thinking Test, it achieved a mean score of 96.15\%, surpassing the human mean of 61.7\% by 55.8\%. This finding highlights o1-preview's exceptional ability to handle algorithmic reasoning and computational problem-solving, outperforming not only undergraduate students but also experts in fields such as mathematics, physics, and engineering. \textit{Design Thinking} was another area where o1-preview excelled. When tasked with solving Leetcode coding challenges, the median performance of o1-preview surpassed 80\% of human participants. With a mean score of 80.73\%, o1-preview slightly outperformed the human mean of 78.5\%, indicating its strong capability in optimizing problem-solving efficiency and demonstrating that it can potentially rival human performance in practical coding tasks.

The domain of \textit{Metacognition} was assessed using calibration scores, Brier scores, and sensitivity measures on the MATH dataset. Here, o1-preview exhibited significantly better scores compared to GPT-4o, its predecessor, although direct human comparisons were not available for this specific dataset. This improvement highlights o1-preview's enhanced self-evaluation and decision-making skills when tackling complex mathematical problems. In \textit{Creative Thinking}, measured by the Alternate Uses Task and Remote Association Test, o1-preview outperformed university students in both divergent and convergent thinking tasks. These tests assess the ability to generate multiple creative solutions and make associations between seemingly unrelated concepts. The findings suggest that o1-preview may possess a capacity for creative reasoning that surpasses many human participants, further expanding its potential applications in educational contexts.

However, o1-preview's performance was not uniformly superior across all domains. In \textit{Abstract Reasoning}, assessed using Raven's Progressive Matrices (encoded in pure text), o1-preview lagged behind psychology major students. While it was able to complete the tasks, its score fell short of the human mean, underscoring a limitation in its ability to process abstract patterns and relationships at a level comparable to trained human experts. Additionally, in \textit{Spatial Reasoning}, o1-preview demonstrated a high degree of accuracy in molecular dynamics simulations derived from the Protein Data Bank. Though no direct human comparison was available for this dataset, o1-preview's performance suggests a capacity for handling spatially complex scientific problems with precision.

When it came to \textit{Quantitative Reasoning}, the FINAQ dataset showed that o1-preview's mean performance exceeded that of the general crowd but remained below that of human experts, achieving a score that was 15-20\% lower than expert benchmarks. This outcome highlights areas where further refinement may be needed to reach expert-level quantitative problem-solving.

Overall, the results indicate that while o1-preview displays exceptional abilities in several higher-order thinking domains—often exceeding human performance—its limitations in abstract and some quantitative reasoning tasks suggest areas for further development. These findings provide a comprehensive understanding of o1-preview's cognitive strengths and weaknesses and offer insights into how AI tools can be refined to support and enhance higher-order thinking in educational settings.

\section{OpenAI o1 Architecture and the Unique Features}\label{Yifan,Zhengliang}

The OpenAI o1~\cite{openai2024o1} series represents a significant advancement in artificial intelligence. It represents a new generation of models with enhanced reasoning capabilities. These models are specifically designed to handle complex, multi-step problems, and they excel in tasks that require advanced problem-solving skills, such as coding, mathematics, and scientific reasoning. This section explores the architecture and unique features that make the OpenAI o1-preview models stand out based on the information from the Open AI official document \cite{o1_official}.

\subsection{New Approach to AI Reasoning}

The o1-preview models utilize a novel training methodology that differs fundamentally from previous LLMs developed by OpenAI. Instead of solely relying on learning patterns from vast amounts of training data, the o1-preview models were trained using a specialized optimization algorithm and a custom dataset tailored to enhance their reasoning abilities. This approach enables the model to engage in more structured thinking, making it capable of solving intricate problems with a higher degree of accuracy.

One of the most notable aspects of the o1-preview model's reasoning capability is its ability to internally ``think'' through problems using a process called chain-of-thought reasoning \cite{o1_official}. This allows the model to break down complex tasks into sequential steps, considering multiple intermediate strategies before arriving at a final response, similar to how humans approach intricate problems. This process significantly enhances the model's ability to tackle tasks involving logical deduction, mathematical reasoning, and multi-step problem-solving.

\subsection{Internal Reasoning Tokens and Chain-of-Thought Process}

A unique feature of the o1-preview series is the introduction of reasoning tokens \cite{o1_official}, which are used by the model to break down problems into a step-by-step \cite{lightman2023verify} internal thought process. When the o1-preview model receives a prompt, it generates reasoning tokens that represent the intermediate steps it takes to reach a final answer. These tokens are not visible to the user but occupy space in the model's context window, and they contribute to the overall token usage.

For example, in a complex mathematical problem, the o1-preview model might use reasoning tokens to sequentially evaluate different parts of the equation, check its intermediate calculations, and refine its approach before providing a final solution. This internal chain-of-thought \cite{lightman2023verify} processing enables the model to achieve better accuracy, particularly in challenging tasks such as scoring 83\% on the International Mathematics Olympiad (IMO) qualifier, compared to the 13\% achieved by previous models like GPT-4o~\cite{openaigpt4o}.

\subsection{Variants of the o1-preview Model: o1-preview and o1-preview-mini}

The OpenAI o1-preview series includes two primary variants:
\begin{itemize}
    \item \textbf{o1-preview}: This is the full version of the o1-preview model, optimized to tackle sophisticated problems using broad general knowledge. It is particularly effective in handling tasks that require deep reasoning and detailed understanding.
    \item \textbf{o1-mini}: A smaller, faster, and more cost-effective variant, o1-preview-mini is adept at handling coding, mathematics, and scientific tasks that do not require extensive general knowledge. Despite its smaller size, the o1-mini retains many of the reasoning capabilities of the o1-preview, making it a practical choice for applications with cost or speed constraints.
\end{itemize}

Both models offer a substantial context window of 128,000 tokens \cite{o1_official}, which allows them to manage complex tasks requiring long sequences of reasoning. The o1-preview variant can generate up to 32,768 output tokens, while o1-mini can handle up to 65,536 tokens, providing flexibility in handling diverse applications.

\subsection{Reinforcement Learning and Problem-Solving}\label{Zhengliang, Yifan}
A distinctive aspect of o1 is its training via reinforcement learning, particularly in the context of optimizing multi-step reasoning through chain-of-thought processes. In this framework, reinforcement learning helps guide the model toward generating useful intermediate steps that lead to the correct final output.

The o1 model is possibly trained using Proximal Policy Optimization (PPO)~\cite{schulman2017proximal}, a popular reinforcement learning (RL) algorithm that iteratively refines the model's policy $\pi_\theta(a|s)$—in this case, the model's strategy for producing reasoning steps—while maintaining stability in updates. The model receives rewards not only for providing correct final answers but also for generating coherent and logical intermediate steps in the reasoning chain.

\paragraph{Training Process Hypothesis:}

In the case of o1, during the RL training phase, each query (or task) is treated as a trajectory of reasoning steps. Each step $t$ in the chain-of-thought is an action $a_t$ taken by the model in a state $s_t$, representing the current knowledge or context the model has built up at $t$. The model's goal is to maximize the cumulative reward over the full trajectory of steps:
\[
J(\pi) = \mathbb{E}_{\pi} \left[ \sum_{t=0}^{T} \gamma^t r_t \right],
\]
where $r_t$ is the reward for each step, and $\gamma$ is the discount factor.

For PPO, the policy is updated in small, stable increments based on the advantage function $\hat{A}_t$, which evaluates whether a step in the chain-of-thought improves the overall quality of reasoning:
\[
L^{\text{PPO}}(\theta) = \mathbb{E}_t \left[ \min \left( r_t(\theta) \hat{A}_t, \text{clip}(r_t(\theta), 1 - \epsilon, 1 + \epsilon) \hat{A}_t \right) \right].
\]
Here, $r_t(\theta) = \frac{\pi_\theta(a_t|s_t)}{\pi_{\theta_{\text{old}}}(a_t|s_t)}$ is the probability ratio between the updated policy $\pi_\theta$ and the previous one $\pi_{\theta_{\text{old}}}$, and $\epsilon$ controls how aggressively the policy is updated. By clipping the updates, PPO ensures that the model does not deviate too far from its previous state, preserving the stability of the reasoning process while making incremental improvements.

\paragraph{Hypothesis on o1-preview and Chain-of-Thought}
During o1 training, our hypothesis is that the chain-of-thought process \textit{might} be treated as a sequence of decisions, where each reasoning step is evaluated and rewarded based on how well it contributes to solving the problem. The model might receive intermediate rewards for steps that:

\begin{itemize}
    \item Correctly break down a complex task into smaller sub-tasks,
    \item Logically follow from the previous step, and
    \item Lead towards the correct final outcome.
\end{itemize}

For example, in a mathematical problem, each step might represent the resolution of part of an equation, and the reward function can be shaped by how well each step contributes to progressing toward the correct solution.

At each step $t$, the advantage function $\hat{A}_t$ compares the predicted reward of the current step to the baseline (expected outcome), allowing PPO to optimize reasoning strategies that provide a higher cumulative reward. The reward function can be augmented to account for:

\begin{itemize}
    \item Accuracy of intermediate steps to ensure the chain of reasoning remains on track.
    \item Clarity and coherence to penalize steps that introduce logical inconsistencies or unnecessary complexity.
    \item Completeness to reward chains that lead to a correct and final solution.
\end{itemize}

\paragraph{Simulated Environment and Feedback:}
During training, o1 is likely exposed to a wide variety of reasoning tasks in a simulated environment, where it can generate multiple reasoning chains for a single problem. Each chain can be evaluated by human feedback or automated metrics, which assign rewards to good reasoning trajectories and penalize suboptimal paths. Over time, PPO encourages the model to favor reasoning strategies that consistently yield better results, shaping both the intermediate steps and final answers.

This training approach, where the model learns to optimize not just for correct answers but for high-quality reasoning, allows o1 to generalize better to complex tasks that require multi-step problem-solving capabilities.

\section{Related Work}\label{Ehsan}
Recent research in LLMs, including OpenAI o1-preview, has increasingly focused on leveraging higher-order thinking to solve complex tasks. In this section, we review key contributions across different categories of higher-order thinking, including problem-based learning, ethical reasoning, analogical reasoning, and computational thinking.

Several papers have explored problem-based learning as a critical aspect of LLM capabilities, particularly focusing on models like o1-preview. The work by Hu et al. \cite{hu2024can} is among the first to empirically study the application of OpenAI o1-preview in bug fixing and automated program repair (APR). The study reveals that OpenAI o1-preview shows promise in solving complex problem-based tasks, especially in the domain of APR, and has the potential to improve with further fine-tuning on complex datasets. Lingo et al. \cite{lingo2024enhancing} introduced the REAP framework—Reflection, Exploration, Application, and Planning—designed to enhance problem-solving abilities. Their findings indicate that structured reflection allows OpenAI o1-preview to handle increasingly complex problem sets.

The OpenAI o1-preview model's performance in problem-based scenarios was further studied by Marino et al. \cite{marino2024fast}, showing promising results on simpler tasks but highlighting its limitations when dealing with intricate problems requiring deep contextual understanding. Similarly, Renze et al.~\cite{renze2024self} investigated the role of self-reflection in problem-solving, finding that detailed self-assessment improves performance in diverse problem-based learning contexts.

Ethical reasoning is another critical dimension of higher-order thinking explored in LLMs. Sorin et al. \cite{sorin2024ethical} conducted a detailed evaluation of OpenAI o1-preview's ethical reasoning, particularly in healthcare settings. Their study demonstrated that while OpenAI o1-preview and other LLMs can align with ethical principles, inconsistencies remain, especially when tasked with making ethical decisions based on incomplete or ambiguous data. This research highlights the need for a closer examination of ethical frameworks within LLMs and their alignment with real-world moral standards.

The capacity for analogical reasoning, or the ability to draw meaningful parallels between seemingly unrelated scenarios, has been another focus of research. Musker et al.~\cite{musker2024semantic} conducted experiments comparing LLMs and human reasoning in analogical problem-solving. They found that advanced models, such as OpenAI o1-preview, can approach analogical reasoning tasks with a degree of competence but still lag behind human-level understanding, particularly in abstract and non-concrete domains. Similarly, Qin et al. \cite{qin2024relevant} highlighted that while LLMs can perform basic analogical reasoning, especially in structured domains, they often struggle with randomness or abstract problems, demonstrating the current limitations of these models.

In the realm of computational thinking, Lykov et al. \cite{lykov2024industry} explored the role of OpenAI o1-preview in the context of Industry 6.0, which incorporates generative AI into industry operations. Their results show that OpenAI o1-preview performs exceptionally well in computational tasks that require pattern recognition, algorithmic thinking, and process optimization, particularly in industry-driven applications. However, they also note that OpenAI o1-preview requires further refinement for complex multi-step computational processes that demand a higher degree of autonomy and decision-making.

Collaboration among multiple LLM agents is increasingly becoming an important area of focus. Zhang et al. \cite{zhang2023exploring} explored collaborative thinking mechanisms in LLM societies, where OpenAI o1-preview agents work together on problem-solving tasks. The study found that different models of collaboration lead to varied effectiveness in task resolution, and LLMs demonstrate behaviors commonly found in human collaboration, such as conflict resolution and negotiation. This line of research opens the door to more sophisticated multi-agent systems where higher-order thinking is distributed across multiple agents.

\section{Methods}\label{Ehsan}
\begin{figure}[ht!]
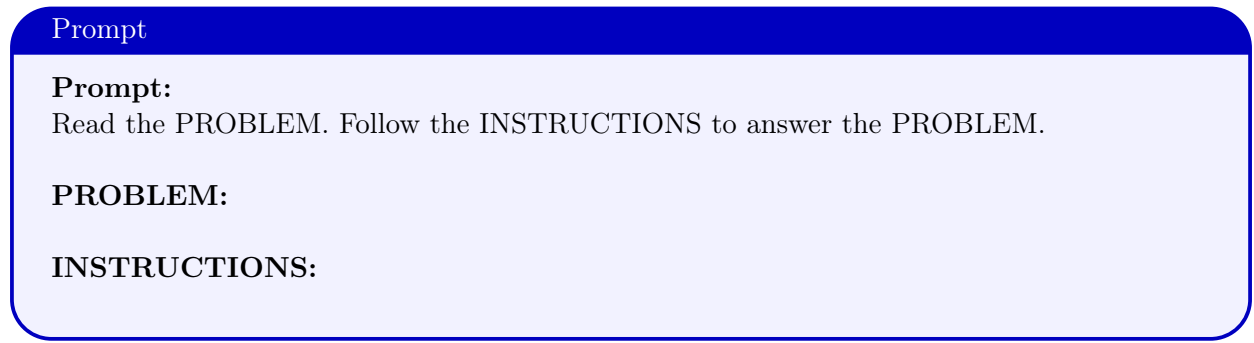
 
\begin{tcolorbox}[colback=blue!5!white, colframe=blue!75!black, width=\textwidth, arc=5mm, auto outer arc, boxrule=0.5mm, title=Prompt]
    \textbf{Prompt:} \\
    Read the PROBLEM. Follow the INSTRUCTIONS to answer the PROBLEM.\\
    \\
    \textbf{\textbf{PROBLEM:}} \\
    \\
    \textbf{INSTRUCTIONS:}
    \vspace{0.5cm}
\end{tcolorbox}
\caption{A template for the experimenation of o1-preview}
\label{fig:template}
\end{figure}

OpenAI o1-preview is trained on reinforcement learning paradigm that allow it to think before answer. According to OpenAI \cite{o1_official_web}, o1 thinks longer before getting to conclusion similar to human thinking process. By means of reinforcement learning, o1 acquires the ability to enhance its cognitive flow (chain-of-thought) and optimize its employed tactics. It gains the ability to identify and fix its errors. It gains the ability to deconstruct complex steps into easier ones. When a method doesn't succeed, it learns to try another one. As o1 inherently works on optimized chain-of-thought using reinforcement learning to provide best possible solutions, hence it is not recommended to use prompting techniques such as chain-of-thought and tree-of-thoughts. Given the guidelines by OpenAI we opt for a raw methodology to evaluate the performance of OpenAI o1-preview. Fig.~\ref{fig:template} is the template prompt we opt for testing.

We first reviewed the relevant literature on higher order thinking, focusing specifically on studies that address the cognitive processes and theoretical models underlying this research area. Following this, we examined the literature on assessing higher order thinking, identifying key instruments used for evaluation and their findings, particularly with regard to human performance on these assessment tools. Each instrument utilized in this study was presented in separate subsections, where we detailed the assessment prompts and discussed the performance of OpenAI's o1-preview model in relation to these prompts. As OpenAI made o1-preview usable to subscribers without additional fee for limited usage, so restrict our experiments to o1-preview avoiding o1 API. For each instrument, we also provided a comparative analysis of o1-preview’s performance as compared to human performance, highlighting any significant differences or similarities. This method allows for a comprehensive comparison of the effectiveness of OpenAI o1-preview versus humans on higher order thinking.

\section{Analysis and Findings}

\subsection{Critical Thinking}\label{Matthew Nyaaba}
Critical thinking (CT) is increasingly recognized as a fundamental skill for success in education and across various fields. It involves the ability to analyze complex information, evaluate evidence, and form reasoned judgments, allowing individuals to navigate the challenges of an ever-evolving world \cite{haryati2022case}. From solving scientific problems to making decisions in daily life, CT plays a critical role in fostering problem-solving and decision-making capabilities. It has been widely studied and measured in academic research, providing educators and professionals with valuable insights into its role in effective learning and professional growth \cite{double2023ib}. Research has shown that individuals with strong CT skills can apply logical reasoning strategies to real-world scenarios, making it an essential competency in education today \cite{leach2020factorial}. Recent evaluations, such as those on PlanBench and other benchmarks, suggest that OpenAI’s o1-preview outperforms previous models in complex reasoning and planning tasks \cite{lingo2024enhancing, valmeekam2024llms}. However, there are suggestions that more evidence is needed to conclude whether these models can fully match human-level critical thinking\cite{lingo2024enhancing}. 

In this section, we examine the capabilities of the o1-preview model in critical thinking assessment, specifically focusing on its performance in reasoning tasks, specifically to evaluate the reasoning in argumentative essays \cite{bernard2008exploring, ennis1985cornell, leach2020factorial}. For this section, we used the Ennis-Weir Critical Thinking Essay Test (EWCTET) to assess o1-preview model's critical thinking abilities because EWCTET provides a structured framework for evaluating argumentation and reasoning, key aspects of critical thinking. Also, this instrument is an openly accessible written test widely used in education, making it a suitable and reliable tool for evaluating a model that primarily functions through text-based input.

\subsubsection{Assessment Instrument and Prompt Strategy}
The Ennis-Weir Critical Thinking Essay Test is designed to assess critical thinking skills, focusing on the ability to evaluate arguments and reasoning through the critique of fictitious letters to newspaper editors. Initially developed as both a teaching tool and an assessment framework for critical thinking courses, this instrument has been validated through many studies, confirming its psychometric robustness \cite{ennis1985cornell}. In prompting the o1-preview model, the original instructions of the instrument were slightly modified for clarity and brevity: we removed the stipulation of the time frames and participants locations and name.  The one-shot prompt strategy was used (\textit{see Figure 1}), meaning the model was provided with the full instructions and asked to respond in a single attempt, without multiple rounds of feedback or refinement\cite{allingham2023simple}.

\begin{figure}[H]
    \centering
    \begin{tcolorbox}[colback=blue!5!white, colframe=blue!75!black, width=\textwidth, arc=5mm, auto outer arc, boxrule=0.5mm, title=Prompt]
    \begin{small}
        \textbf{Prompt:} \\
        Carefully read the PROBLEM below and follow the INSTRUCTIONS to respond. \\
        \\
        \textbf{\textbf{PROBLEM:}} \\
        The Ennis-Weir Critical Thinking Essay Test  \\
        \\
        \textbf{INSTRUCTIONS:} \\ 
        Read the letter to the editor of the Moorburg newspaper. Consider it paragraph by paragraph and as a total argument. Then write a letter to the editor in response to this one. For each paragraph in the letter you are about to read, write a paragraph in reply telling whether you believe the thinking good or bad. Also, write a closing paragraph about the total argument. Defend your judgments with reasons. Your answer should have nine numbered paragraphs. Numbers one through eight should give your reactions to paragraphs one through eight in the letter. Your paragraph number nine should give your overall evaluation of the letter considered as one total argument. Each paragraph, 
    \end{small}
    \end{tcolorbox}
    \captionof{figure}{Prompt and Instructions for the Test, EWCTET.}
    \label{fig:prompt_example}
\end{figure}

The EWCTET is accompanied by an assessment criteria and scoring sheet to evaluate the performance of users. A score of -1, 0, 1, 2, or 3 is assigned to each of the first eight numbered paragraphs . A score of -1 indicates that the user judged incorrectly, whether their judgment was good or bad. A score of 1 reflects poor judgment in justifying their response, while 0 represents no response at all. A score of +1 is given when the user judges correctly but without adequate justification. Scores of +2 and +3 represent semi-adequate and fully adequate justification, respectively. For Paragraph Nine, which requires an overall evaluation of the argument, the scoring range extends from -1 to +5 to capture the depth of the user's final assessment, ensuring a comprehensive evaluation of their critical thinking abilities. The benchmark comparative mean for K-12 and undergraduate studunt on the EWCTET typically falls between 15 and 25 out of 40. To standardize reporting, this range was scaled to 100, yielding 37.5\% to 62.5\% . This scaling ensures consistency in comparison and interpretation of human and o1 preview perfromance. 

\subsubsection{o1-preview Performance on The Ennis-Weir Critical Thinking Essay Test}
The distribution of the o1-preview model's performance on the EWCTET demonstrates strong and consistent critical thinking abilities, as it was able to identify logical fallacies and weaknesses, though there were some minimal instances of lower scores indicating potential areas for deeper analysis (See Table 1). The reveals low variability (\textit{sd}= 0.43), showing consistent performance across the evaluated paragraphs. The absence of negative or neutral scores indicates a uniformly high level of critical analysis. The average ($M$=+2.25) representing 81.25\% showed a slightly higher overall performance, particularly in paragraphs addressing analogical reasoning and an advanced understanding of rhetorical devices and precise definitions in argumentation. However, o1-preview scored (score = +2) in some paragraphs, where further specificity were required. The overall evaluation in showed that o1-preview model has the ability to synthesize individual analyses into a comprehensive critique, demonstrating advanced critical thinking skills by summarizing strengths, weaknesses, and overarching logical flaws in a cohesive manner. 
\begin{table}[H]
    \centering
    \small
    \caption{o1-preview 1 Performance on the Ennis-Weir Critical Thinking Essay Test (EWCTET)}
    \begin{tabularx}{\textwidth}{X X} 
        \toprule
        \textbf{Statistic} & \textbf{Value} \\ \midrule
        Number of Paragraphs Scored & 8 \\
        Minimum Score & $+2$ \\
        Maximum Score & $+3$ \\
        Range & $+2$ to $+3$ \\
        Mean ($\bar{x}$) & 2.25 \\
        Median ($M$) & 2 \\
        Standard Deviation ($s$) & 0.43 \\
        Frequency of $+2$ ($f_{+2}$) & 6 \\
        Frequency of $+3$ ($f_{+3}$) & 2 \\
        Percentage of $+2$ (\%) & 75\% \\
        Percentage of $+3$ (\%) & 25\% \\
        \bottomrule
    \end{tabularx}
    \label{tab:o1-preview_performance}
\end{table}

Table \ref{tab:ewctet_o1-preview_responses} shows o1-preview first three (3) responses out of the nine (9) arguments to EWCTET with a critical lens, categorizing the arguments as either "Good" or "Bad" and providing a reason for each assessment, as directed by the test. For example, in the essay, the writer, Mr. Raywift, proposed eliminating overnight parking from 2 a.m. to 6 a.m., claiming that "intelligent citizens" should agree with this proposal. However, o1-preview, acting as a respondent, was not swayed by the condition of being labeled "intelligent" if it agreed with the argument. Instead, the model analyzed the situation critically, a key component of critical thinking that involves making sound judgments. It rated the argument as "Bad," arguing that the idea of restricting overnight parking was divisive and that equating agreement with intelligence hinders constructive dialogue and inclusivity.

In the second paragraph of the essay, Mr. Raywift compared overnight parking in the streets to having a garage in the streets, but o1-preview rated this analogy as "Bad." The model reasoned that garages are permanent structures, while overnight parking is temporary, making the analogy flawed. In the third paragraph, o1-preview rated the writer's concern about traffic congestion as "Good" but pointed out that it had little relevance to overnight parking, as the congestion issue primarily pertains to daytime rush hours. These responses, along with o1-preview's evaluations of the remaining arguments in the essay, demonstrate the model's intriguing ability to analyze issues critically and make sound judgments, reinforcing its effectiveness in handling complex arguments.
\begin{table}[H] 
    \centering 
    \small
    \caption{\textit{First Three Responses of o1-preview Model on the Ennis-Weir Critical Thinking Essay Test}} 
    \begin{tabular}{p{2cm} p{5cm} p{2cm} p{6cm}} 
        \toprule
        \textbf{Paragraph Number} & \textbf{Original Paragraph Summary} & \textbf{Response} & \textbf{Reason for Response} \\ 
        \midrule
        1 & \textit{Introduction}: Proposes eliminating overnight parking from 2 a.m. to 6 a.m., claiming intelligent citizens should agree. & \textbf{Bad} & The assertion dismisses differing opinions by implying that disagreement equates to a lack of intelligence, which undermines constructive dialogue and inclusivity. \\ [2ex] 
        2 & \textit{Analogy}: Compares overnight parking to having a garage in the streets, stating both should be illegal. & \textbf{Bad} & The analogy is flawed because garages are permanent structures, whereas parking is temporary. This oversimplification weakens the argument for banning overnight parking. \\ [2ex] 
        3 & \textit{Traffic Congestion}: Highlights narrow streets and increased afternoon rush hour traffic, attributing delays to parked cars. & \textbf{Good} & Raises a valid concern about traffic congestion; however, the proposed time restriction (2 a.m. to 6 a.m.) may not effectively address afternoon traffic issues, indicating partial validity. \\ 
        \bottomrule
    \end{tabular}
    \label{tab:ewctet_o1-preview_responses}
\end{table}

\subsubsection{Comparing o1-preview Performance versus Human Performance on Ennis-Weir Critical Thinking Essay Test}
Several empirical studies have explored human performance on EWCTET, regarding students critical thinking abilities in education. For instance, in a study conducted by \cite{Taghinezhad_Riasati_Rassaei_Behjat_2018}, advanced English learners in a medical school were assessed for their critical thinking skills using the Ennis-Weir Critical Thinking Essay Test (EWCTET). In the pretest, the average student in a normal classroom setting exhibited relatively lower critical thinking skills, with a mean score of 43.57\% (See Table \ref{tab:o1-preview_comparison}). Similarly, in another study by \cite{Prayogi_Yuanita_Wasis_2018}, pre-service teachers demonstrated even lower critical thinking skills in their pretest phase, with a mean score of 15.30\%. However, after interventions were introduced, critical thinking strategies in \cite{Taghinezhad_Riasati_Rassaei_Behjat_2018} study and the Critical-Inquiry-Based Learning model in \cite{Prayogi_Yuanita_Wasis_2018} study—there was a significant improvement in students' critical thinking abilities, with mean scores rising to 71.42\% and 87.60\%, respectively, in the posttest phase. In comparison, the o1-preview preview, using a zero-shot approach (i.e., without prior prompt or follow-up promt), was able to achieve a 81.25\% score on the EWCTET, demonstrating a higher baseline of critical thinking ability than the average student in both pre-intervention studies and surpassing the human bechmark mean of 37.5\% to 62.5\% (originally 15/40 to 25/40). This performance suggests that the o1-preview preview has the ability to critically evaluate arguments is comparable to or even exceeds the capabilities of students following targeted critical thinking interventions.

\begin{table}[ht]
    \centering
    \small
    \caption{\textit{Comparison of o1-preview Model and Human Performance on EWCTET}}
    \begin{tabularx}\textwidth{X X X X}
        \toprule
        \textbf{Study/Model} & \textbf{Group/Type} & \textbf{Test Phase} & \textbf{Mean Score (\%)} \\ \midrule
        Taghinezhad et al. (2018) & Advanced English Learners (Medical School) & Pretest & 43.57 \\ 
        Taghinezhad et al. (2018) & Advanced English Learners (Medical School) & Posttest & 71.42 \\ 
        Prayogi et al. (2018) & Pre-service Teachers & Pretest & 15.30 \\ 
        Prayogi et al. (2018) & Pre-service Teachers & Posttest & 87.60 \\ 
        o1-preview Model & o1-preview Model & - & 81.25 \\ \bottomrule
    \end{tabularx}
    \label{tab:o1-preview_comparison}
\end{table}

We advocate for further exploration of how the o1-preview can be used to support and enhance student critical thinking skills in educational settings. Given the challenges many students face in developing the ability to critically analyze, evaluate, and construct well-reasoned arguments, as reflected in the pretest results from both \cite{Taghinezhad_Riasati_Rassaei_Behjat_2018, Prayogi_Yuanita_Wasis_2018}, o1-preview Preview's ability to engage in such complex reasoning processes without guided instruction is significant. Further studies are necessary to explore how o1-preview Preview can contribute effectively and responsibly to student learning, particularly in fostering critical thinking skills where students often struggle.

However, there are limitations to these findings, as they are based on a limited number of empirical studies. Again, even though the required rubric was used to asses the o1-preview's performance but we recommend an empirical assessment with inter rater from  expert in critcal thinking. Morever, future research should involve multiple human evaluators, applying interrater reliability, to assess responses and offer a more comprehensive understanding of o1-preview's strengths and areas for improvement. Also, the use of only the EWCTET limits the scope of this evaluation. Expanding the range of critical thinking assessment instruments in future studies will allow for a broader, more conclusive understanding of how o1-preview Preview can enhance critical thinking in various educational settings.

\subsection{System Thinking}\label{Shuchen Guo}

The world have been becoming increasingly complex nowadays, where people are facing with a number of grand social, technical and ecological challenges that will have impacts on our society and planet globally \cite{dugan2022systems}. To understand and to attempt to solve these multifaceted issues both now and in the future, systems thinking (ST) has emerged as one of the most essential skills people must develop \cite{abdurrahman2023impacts,mambrey2020impact}. 
ST is a holistic approach that examines an object or situation as a whole while recognizing the inter-connectedness within complex systems. It is helpful for individuals to grasp the whole picture, understand different levels contributing to the causes of a problem, and find a way to solve problems in various types of systems, such as ecosystems, health care, technology, economy, laws, and so on \cite{charoenmuang2024high}. Several researchers have claimed that ST is one special skill that sets humans apart from artificial intelligence \cite{oliveira2020industry}, which is one of the important reasons that ST has been particularly highlighted in various educational fields these years \cite{li2024developing,vachliotis2021developing}. Thus, exploring o1-preview's performance on ST competence and comparing it to human capabilities would be meaningful and intriguing.

\subsubsection{Definition and conceptualization of system thinking}
Across disciplines, there are many different definitions of ST, as well as numerous lists of ST skills. However, we can still arrive at the commonality in this thinking methodology. As Mambrey, Timm, et al. \cite{mambrey2020impact} stated, an operationalized definition of ST is “a conceptual skill in which superordinate principles of complex systems are taken into account when understanding and predicting the interplay and function of their elements.” 

However, when delving into specific disciplines, researchers formed different constructs of ST skills based on their understanding of the characteristics of the field. Most of the studies are conducted in biology and engineering fields. In a recently published study assessing ST in biology, four ST skills were defined based on previous studies \cite{li2024developing}. \textit{System organization (SO)}, stands for the identification of system components and the relationships among them. \textit{System behaviour (SB)}, refers to analysing the development, dynamic processes, and cause-and-effect relationships among system components based on system characteristics. \textit{System application (SA)}, involves predicting and making decisions about a biological problem based on the analysis of system structure and behavior. \textit{System evaluation (SE)}, is the skill of determining the validity of system structures and decisions, and understanding the limitations of predictions. 

While in the engineering field, Grohs et al. \cite{grohs2018assessing} describe a three-dimensional framework for general usage in interdisciplinary contexts based on existing literature and engineering problem-solving, emphasizing the interconnectedness of technical and social aspects of modern problems and the importance of stakeholder perspectives. This Dimensions of Systems Thinking Framework consists of \textit{problem},\textit{ perspective}, and \textit{time} as the three dimensions.\textit{Problem} refers to the broad technical aspects and contextual environment of the issue, including assumptions, prioritizations, and constraints. \textit{Perspective} stands for the consideration of the roles of different stakeholders in understanding and solving the problems. \textit{Time} dimension requires considering the history of the problem, stakeholders, and solutions, as well as the short-term and long-term influence of possible solutions. 

Davis et al. \cite{davis2020lake} adopted a perspective of seeing systems as webs of dynamic interconnections. According to this viewpoint, ST is concerned with comprehending systems holistically and identifying interconnections between different system parts. From this definition, ST is about examining physical and social systems and the relationship between their components. It also places a strong emphasis on the way these system components interact with one another throughout time to create flows and feedback loops.

\subsubsection{Assessment of system thinking}
Various instruments have been developed to assess ST based on various theoretical frameworks in different disciplines, using different methods, and aimed at different educational levels. As we mentioned before, the definition and constructs of ST varied across different disciplines, thus resulting in different instruments and/or rubrics developed for assessment. Most studies are conducted in the context of biology, earth science, and engineering fields \cite{dugan2022systems}. Environmental topics ranging from ecology to climate change, sustainability, and energy use are widely used scenarios. Students are the main test group, ranging from elementary school students to graduate students. At the same time, a few studies conducted ST assessment of pre-service and in-service teachers \cite{fanta2020fostering, peretz2023developing, sweeney2007thinking}.

In terms of assessment methods, in a systematic literature review of the ST measurement based on 27 distinct assessments, Dugan et al. \cite{dugan2022systems} identified four types of ST assessments, which are behavior-based, preference-based, self-reported, and cognitive activation. The majority of assessments were behavior-based, which means the participants need to complete specific tasks, such as drawing or answering various types of test questions, to externalize their ST knowledge or skills. Preference-based assessments, which characterize values, interests, attitudes, and/or aptitudes, are usually measured using scales and are ranked as the second most commonly used type. Self-report measures are also highly used, however, a recently study, which explored the relationships between student self-report and scenario-based assessment performance of ST, indicated that there may be limitations to using self-report scales and suggested alternative formats for assessing ST \cite{davis2023comparing}.

When it comes to assessment format, which stands for how assessment data were collected, possible ways include mapping, interview, scenario-based items, open-ended items, fill-in-the-blank, multiple-choice, and so on \cite{dugan2022systems}. Mapping is the most widely used and acknowledged method in assessing ST, often accompanied by an interview. However, due to the difficulties in implementation on a large scale, recent studies have also developed many paper–and–pencil tests using different types of items \cite{li2024developing}. Especially for scenario-based items, studies considered this type most potential to elicit student ST skills by presenting a realistic problem followed by a series of open- or close-ended questions \cite{norris2022investigating}.

\subsubsection{Testing strategies}
The test aim is to explore ST knowledge or skills of OpenAI o1-preview, thus, we adopted behavior-based instrument to conduct the assessment. Based on the consideration, scales and interviews are also not suitable for this research, as scales are mostly for measuring preference, while interview protocols pose challenges for implementation. Meanwhile, we selected the instruments while considering the characteristics of OpenAI o1-preview, aiming to encompass a wide range of diversity. This approach allows us to assess the performance of OpenAI o1-preview across different contexts, fields, types, formats, as well as in comparison with test-takers at various educational levels. Due to the limited capacity of o1-preview to generate images, we did not select any instruments in the format of mapping, even though it is one of the widely used methods for assessing ST skills. Similarly, items with pictures containing indispensable information in the item prompt that can impact the ST performance of o1-preview were also excluded.

At last, we selected three qualified instruments that cover diverse contextual fields, including biology, engineering, and socio-environmental issues. The formats of the instruments include open-ended questions, fill-in-the-blank, multiple-choice, and scenario-based ill-structured questions. The targeted participants range from middle school to graduate students. The theoretical framework on which the instruments are based are also various. A summary of the corresponding information of the three instruments can be found in Table 4. General descriptions of the original instruments are as follows.


\begin{table}[H]
    \centering
    \small
    \caption{\textit{Summary of ST instruments used}} 
    \begin{tabular}{p{3cm} p{4cm} p{4cm} p{4cm}} 
        \hline
        \textbf{Instrument} & \textbf{Instrument 1} & \textbf{Instrument 2} & \textbf{Instrument 3} \\
        \hline
        Original source & Li and Li, 2024\cite{li2024developing} & Grohs et al., 2018\cite{grohs2018assessing} & Davis et al., 2020\cite{davis2020lake} \\
        Targeted group & Middle School students & Undergraduates and graduates & Undergraduates and graduates \\
        Subject area & Biology & Engineering & Not specifically defined \\
        Item Context & Ecosystem & Community issue & Socio-environmental system \\
        Format & Fill-in-the-blank, Multiple-choice, Open-ended & Scenario based, Open-ended, ill-structured & Scenario based, Open-ended, ill-structured \\
        ST Construct & SO, SA, SB, SE & Problem, Perspective, Time & Variable, Causal relation, Feedback loop \\
        \hline
    \end{tabular}
    \label{tab:ST-instrument}
\end{table}

\textbf{\textit{Instrument 1: Biological Systems Thinking Test for Middle School Students \cite{li2024developing} }}

The test was developed based on a comprehensive understanding of biological systems, encompassing four ST skills: system organization (SO), system behavior (SB), system application (SA), and system evaluation (SE), each across three complexity levels. The instrument consists of 13 items, all embedded in the context of ecosystems and ecological issues. The items are presented in various formats, including multiple-choice, fill-in-the-blank, and open-ended questions. The instrument quality was ensured by conducting think-aloud and Rasch analysis based on 456 students, which showed satisfying results of indices including dimensionality, item fit, difficulty, and reliability. 

\textbf{\textit{Instrument 2: The Village of Abeesee \cite{grohs2018assessing}}}

Based on the framework of viewing ST as consisting of three dimensions—problem, perspective, and time—this assessment presents participants with a one-paragraph text describing the situation of the Village of Abeesee, which is facing a complex issue regarding heating in winter. Students then complete the following six open-ended questions aligned with the dimensions of the framework.
A pliot study was conducted to collect qualitative data for refining the scenario. A multi-stage rubric development process was applied using a pool of 93 student responses. The final rubric consists of seven dimensions: problem identification, information needs, stakeholder awareness, goals, unintended consequences, implemented challenges, and alignment.

\textbf{\textit{Instrument 3: The Lake Urmia vignette (LUV) \cite{davis2020lake}}}

In this instrument, participants are presented with a scenario describing the real case of Lake Urmia's desiccation. After reading the narrative, participants are asked to write down in their own words the problem of Lake Urmia and explain why the lake shrunk over the years. Based on the theoretical framework of viewing systems as a web of connections, the rubric aims to identify the number of variables, causal links, and feedback loops described in the responses. During the development, a pilot test was done with a sample of 30 engineering graduate students. The author also compared responses to LUV with other measures.

For instruments 2 and 3, we conducted tests using the original instruments, while instrument 1 was tested partially since some of its items included image information or required students to conduct concept mapping. In instrument 1, four items were selected- Item 1 (SO), Item 2 (SA), Item 10 (SB) and Item 13 (SE)-to ensure the coverage of all four skills. The format of these four items varies, including fill-in-the-blank, multiple-choice and open-ended questions. The first and second items contain image information that, when provided in text, we believe will not influence the assessment of ST, so we transcribed the information from image into text and provide it to o1-preview. For scoring methods, we followed the rubric provided by the original papers. Each item was tested once, and the webpage was shut down before the test of the next item to ensure o1-preview answered each question independently. Figure 4 provides an example of the prompt used for testing items in instrument 1. The other prompts followed a similar structure.

\begin{figure}[H]
    \centering
    \begin{tcolorbox}[colback=blue!5!white, colframe=blue!75!black, width=\textwidth, arc=5mm, auto outer arc, boxrule=0.5mm, title=Prompt]
    \begin{small}
        \textbf{Prompt:} \\
        Read the PROBLEM carefully and provide solution using give INSTRUCTIONS. \\
        \\
        \textbf{\textbf{PROBLEM:}} \\
        Zhoushan in Zhejiang Province is known as the ‘Eastern Fish Warehouse’ in China. People here basically rely on fishing to make a living. However, given the overfishing of yellow croakers, the population of yellow croakers has declined sharply in the past few years. One reason is that many young yellow croakers have been caught. To prevent the extinction of wild yellow croakers, the measure of ‘banning fishing within three years’ was proposed. 
  \\
        \\
        \textbf{INSTRUCTION:} \\ 
        Do you think this measure is feasible? Please explain your reasons.
    \end{small}
    \end{tcolorbox}
    \captionof{figure}{Prompt and Instructions for ST test.}
    \label{fig:prompt_example2}
\end{figure}

\subsubsection{Comparing o1-preview and human performance}
\label{sec:system_thinking_results}
\textbf{\textit{Instrument 1: Biological Systems Thinking Test for Middle School Students}}

For items selected in instrument 1, the mean score and standard deviation of middle school participants \cite{li2024developing}, along with the performance of o1-preview, are shown in Table 5. With a full score of three for each item, regardless of the testing format, o1-preview achieved the highest level in every item, performing far better than middle school students on average. According to the scoring rubric, the results indicated that o1-preview can identify several elements and complex relations; analyze developments based on feedback, cycles, and emergence in biological systems; make predictions and decisions based on complex biological system analysis; determine the validity of biological systems structure and the limitation of the application based on the complex analysis of effects \cite{li2024developing}. Additionally, o1-preview took 41 seconds to complete the four items.

\begin{table}[H]
    \centering
    \small
     \caption{\textit{Comparison of human and o1-preview performance on Insturment 1}} 
    \begin{tabularx}\textwidth{X X X X X}
    \hline
        \textbf{Item (ST Dimension)} & \textbf{Item format} & \textbf{Mean} & \textbf{SD}  & \textbf{o1-preview} \\
    \hline
        Item 1 (SO) & Fill-in-the-blank & 1.47 & 1.10 & 3\\
       Item2 (SA) & Open-ended & 1.41 & 0.81 & 3\\
        Item 10 (SB) & Multiple-choice & 1.72 & 0.76 & 3 \\
        Item 13 (SE) & Open-ended & 1.02 & 0.75 & 3\\
    \hline
    \end{tabularx}
    \label{tab:ST-instrument1}
\end{table}

\textbf{\textit{Instrument 2: The Village of Abeesee}}

For instrument 2, according to the scoring rubric, there are seven scoring dimensions: problem identification, information needs, stakeholder awareness, goals, unintended consequences, implemented challenges, and alignment. Davis et al. \cite{davis2023comparing} reported the performance of undergraduate students on the instrument, with a mean score and standard deviation for each dimension shown in Table 6. The minimum and maximum scores for each dimension are 0 and 3, respectively. For the performance of o1-preview, it scored the highest mark in all the dimensions, as shown in Figure 6, which was obviously better than the average performance of undergraduate students \cite{davis2023comparing}. o1-preview took 52 seconds to complete the task, compared to 30-45 minutes for human test administration.

\begin{table}[H]
    \centering
    \small
     \caption{\textit{Comparison of human and o1-preview performance on Insturment 2}} 
    \begin{tabularx}\textwidth{X X X X}
    \hline
        \textbf{ST Dimension} & \textbf{Mean} & \textbf{SD}  & \textbf{o1-preview} \\
    \hline
        Problem identification & 1.62 & 0.63 & 3\\
        Information needs & 1.81 & 0.52 & 3\\
        Stakeholder awareness & 1.24 & 0.99 & 3 \\
        Goals & 1.71 & 0.62 & 3\\
        Unintented consequences & 1.38 & 0.58 & 3\\
        Implemented challenges & 1.64 & 0.57 & 3\\
        Alignment & 1.71 & 1.00 & 3\\
    \hline
    \end{tabularx}
    \label{tab:ST-instrument2}
\end{table}

\textbf{\textit{Instrument 3: The Lake Urmia vignette (LUV)}}
For instrument 3, previous research has reported results for undergraduate students and graduate students. Table 7 shows the mean score of participants, and the standard deviation if reported in the original research. Note that the average of graduates is lower than that of undergraduates, because Davis et al. \cite{davis2020lake} set the a maximum of 10 points for each dimension. The performance of o1-preview is significantly better than the average for both undergraduates and graduates, indicating that o1-preview can identify detailed variables, interconnectivity, and identify feedback loops in a complex system much better than these students. Moreover, the test time recommended for human testing is 20 minutes, while o1-preview provided the answer in just 18 seconds. 

\begin{table}[H]
    \centering
    \small
     \caption{\textit{Comparison of human and o1-preview performance on Insturment 3}} 
    \begin{tabularx}\textwidth{X X X X}
    \hline
        \textbf{ST Dimension} & \textbf{Undergraduates\cite{davis2023comparing}} & \textbf{Graduates\cite{davis2020lake}}  & \textbf{o1-preview} \\
    \hline
         Variables & 10.95(4.00) & 7.5 & 20.5\\
        Causal links & 9.17(3.97) & 5.9 & 22\\
        Feedback loops & 0.16(0.45) & 0.4 & 3 \\
    \hline
       Total score & 20.08(8.13) & 14.17(6.02) & 45.5\\
    \hline
    \end{tabularx}
    \label{tab:ST-instrument3}
\end{table}

To sum up, in all the ST assessments mentioned above, o1-preview performed better than human participants. The results indicated that despite the differences in ST construction, assessment context, format, and educational level, o1-preview demonstrated a higher capability for ST. Additionally, compared to the original time  required for human participants, o1-preview took only seconds to provide the right answers.

\subsection{Computational Thinking}\label{Lehong Shi}
\subsubsection{Definition and conceptualization}
Computational thinking (CT) involves using a set of problem-solving skills and techniques that software engineers use to write programs and apps. From a generic perspective, CT is viewed as "the thought processes involved in formulating problems and their solutions so that the solutions are represented in a form that can be effectively carried out by an information-processing agent"\cite{wing2011research}. In educational settings, CT has been the core of STEM curricula and critical competencies that equip students with foundational programming and computing skills in problem-solving and STEM learning. Researchers suggested students follow a series of CT steps for problem-solving, including (1) formulating a problem, (2) collecting and analyzing data, (3) representing data, (4)automating solutions, and (5) identifying, analyzing, and implementing possible solutions \cite{fessakis2018exploring}. To solve the problems, students could implement critical CT concepts: logic, algorithms, decomposition, patterns, abstraction, and evaluation. Employing CT skills and concepts, students can create, debug, think, and collaborate for problem-solving \cite{hunsaker2020computational}. To assess students' CT skills, Brennan and Resnick \cite{brennan2012new} proposed a three-dimensional framework, including (1) computational concepts (e.g., sequences, repetition, conditional, debugging, and data); (2)computational practices regarding the process of iteration, debugging, and abstraction; and (3) computational perspectives, such as expressing, connecting, and questioning the social relations with peers and the technical world. Building upon the computational practice perspective in Brennan and Resnick \cite{brennan2012new}, Weintrop et al. \cite{weintrop2016defining} suggested assessing CT in four major categories: data practice, modeling and simulation practice, computational problem-solving practice, and systems thinking practice. Furthermore, Selby and Woollard \cite{selby2013computational} proposed an operational CT assessment approach, focusing on abstraction, decomposition, algorithmic thinking, evaluation, and generalization, which is broadly adopted in many assessment instruments for assessing students' performance on CT skills. 
\subsubsection{Computational Thinking Assessment Instruments}
Researchers have developed different instruments to assess CT \cite{tang2020assessing}. Bebras Challenge is an international initiative designed to assess and promote computational thinking (CT) skills among students, typically between the ages of 6 and 18. It involves short, engaging tasks that encourage problem-solving, algorithmic thinking, and logical reasoning. Bebras Challenges questions are designed to be 3 minutes long to complete and require no prior knowledge of programming or computer science topics. Using Bebras Challenge problems, Lockwood et al. \cite{lockwood2018developing} developed two tests, each including 13 questions to be completed within 35 minutes. They calibrated the two tests to ensure a comparable difficulty level. Both tests were administered to 200 undergraduate students, with their prior mathematics and programming experience used as metrics to compare performance across different groups. Test 1 had an average score of 7.689 based on responses from 263 students, while Test 2 had an average score of 7.933 based on 180 students' responses. Overall, students who took both tests achieved an average performance score of 8.03. \\
\\
Lafuente Martínez et al. \cite{lafuente2022assessing} developed a test to assess CT suitable for adults with and without a background in computer science and programming, using 25 items. The human performance on those items was high, with a mean score of 17.5 for men and 14.1 for women. Participants from different majors performed differently, for example, social sciences (m = 12.3), mathematics (m = 21.2), physics (m = 20.6), engineering (m = 19.4), and computer science (m = 18.9). Lafuente Martínez et al. \cite{lafuente2022assessing} also reported that the mean score for experts is 19.9.  
\subsubsection{o1-preview performance on Bebras-based test}
Using the two tests developed by Lockwood et al. \cite{lockwood2018developing} we systematically tested the CT skills of OpenAI o1-preview by selecting two representative items in each test. For some items using images, we converted image information into text for o1-preview to respond. o1-preview correctly answered three out of the four problems, with an estimated average score at 9.75, significantly outperformed undergraduate students (m = 8.03). Instead of only providing the final response, o1-preview further explained its problem-solving and logical reasoning processes step-by-step. 
\begin{table}[H]
    \centering
    \small
    \caption{o1-preview Performance on the Bebras-based Test (Mean score of responses (total score is 13)}
    \begin{tabularx}\textwidth{X X X} 
        \toprule
        \textbf{Tests} & \textbf{Undergraduate} & \textbf{o1-preview} \\ \midrule
        \textit{Test 1} & 7.689 & 6.5  \\
        \textit{Test 2} & 7.933 & 13 \\
        \midrule
        \textit{Overall} & 8.03 & 9.75 \\
        \bottomrule
    \end{tabularx}
    \label{tab:computational_thinking1}
\end{table}

o1-preview initially responded incorrectly to problem 2 in test 1. To assess the actual capability of OpenAI o1-preview in solving problem 4, we asked it to attempt the problem again. In its second trial, o1-preview answered the problem correctly and provided a detailed explanation of its problem-solving process, including the algorithmic thinking involved. \\
\\
To explore the potential of o1-preview in recognizing performance inconsistencies and explaining the reasons for its errors, we followed up with two additional questions (see Figure 5). Interestingly, the o1-preview offered a clear and logical explanation for its previous inconsistency and suggested the reasons behind its mistakes. The o1-preview provided the following responses: \textit{1. What is my mistake}. I apologize for the inconsistency in my previous responses. Let me explain how I mistakenly deduced that Beaker F contains pure water and how that mistake occurred. In my first response, when assigning the potions to Beakers D, E, and F based on the experiments, I made an error in the logical deduction. I failed to notice that since Beaker D contains WC, the effect of Whiskers Curly should have been observed in Experiment 2. However, Whiskers Curly was not one of the effects listed. This inconsistency indicates that Beaker D cannot contain WC if Whiskers Curly is not observed. My mistake was not cross-validating the assignments with all experiments. I overlooked the contradiction in Experiment 2, where consuming Beaker D (WC) did not produce the Whiskers Curly effect, which is inconsistent with the problem's premise that a potion always produces its effect when consumed. \textit{2. Why the mistake happened}
•	Incomplete Analysis: In my initial response, I did not thoroughly check that the assigned effects matched the observed effects in all experiments.
•	Overlooking Key Details: I missed the crucial detail that consuming a potion should always produce its effect unless it's pure water.
•	Assuming Validity: I incorrectly assumed that the assignments were valid after checking only one experiment without ensuring consistency across all experiments.
"
\begin{figure}[H]
    \centering
    \begin{tcolorbox}[colback=blue!5!white, colframe=blue!75!black, width=\textwidth, arc=5mm, auto outer arc, boxrule=0.5mm, title=Prompt]
    \begin{small}
        In the first response, you deduced that Beaker F is pure water. In the second response, you deduced that Beaker D is pure water. Which response is correct? \\ 
    \\
    \textbf{PROBLEM} \\
    Why do you mistakenly deduce that Beaker F contains pure water? How did you make this mistake?\\
    \end{small}
    \end{tcolorbox}
    \captionof{figure}{Prompt to assess o1-preview's problem-solving and reasoning ability}\label{fig:meta_prompt}
\end{figure}
\subsubsection{o1-preview performance on Algorithmic Thinking Test for Adults Instrument}
We further used the Algorithmic Thinking Test for Adult Instrument \cite{lafuente2022assessing} to test OpenAI o1-preview's performance on using CT skills to solve problems using six representative items. Similarly, for some items relying on graphs or images to convey the necessary information of the problem, we converted them into text, considering the data modalities that o1-preview can process. Using a consistent template prompt, we entered each item into o1-preview and received correct responses for 6 items. Compared to human performance, the o1-preview outperformed participants from both mathematics (M = 21.2) and physics (M = 20.6) majors, who achieved the highest mean scores relative to peers from other disciplines. 
\begin{table}[H]
    \centering
    \small
    \caption{o1-preview Performance on the Algorithmic Thinking Test for Adults Instrument}
    \begin{tabularx}\textwidth{X X} 
        \toprule
        \textbf{Participants} & \textbf{Mean score of responses (total score is 25)} \\ \midrule
        Women & 14.1 \\
        Men & 17.5 \\
        Social science major & 12.3 \\
        Mathematics major & 21.2 \\
        Physics major & 20.6 \\
        Engineering major & 19.4 \\
        Computer science major & 18.9 \\
        Experts & 19.9 \\
         o1-preview & 25 \\
        \bottomrule
    \end{tabularx}
    \label{tab:computational_thinking2}
\end{table}
Overall, by completing the test items, o1-preview demonstrated a strong ability to perform higher-order thinking tasks, particularly in solving complex problems using various computational thinking skills and following a systematic, step-by-step process. Particularly, after o1-preview correctly solved each problem, we asked it to describe the CT strategies and skills it employed. Remarkably, the o1-preview provided detailed explanations of the CT skills used for solving each problem. Upon summarizing the varied CT skills implemented by o1-preview, we identified several commonly utilized CT skills, including problem decomposition, algorithmic thinking, simulation, pattern recognition, logical reasoning, and problem-solving. For certain questions, the o1-preview additionally employed problem-specific CT skills, such as mathematical reasoning, concepts from information theory, critical thinking, and sequential processing.


\subsection{Design Thinking}\label{Haoran Lu}

\subsubsection{Conceptualization}
Design Thinking is a user-centered, and iterative process aimed at solving problems by focusing on user needs, prototyping solutions, and refining based on feedback
\cite{brown2008design, panke2019design}.
It involves empathizing with users, defining the problem, finding potential solutions, prototyping those ideas, and testing the prototypes. Design thinking emphasizes flexibility and adaptability, encouraging rapid iteration and continuous improvement. The iterative nature of design thinking ensures that the products are continuously optimized through hands-on testing and user involvement. 

Incorporating design thinking in LLMs ensures that solutions are responsive to real-world needs and constraints. For instance, when developing a new scientific tool, researchers may create and test multiple prototypes, refining each based on experimental feedback and user experience, leading to a more effective final product \cite{demetriadis2023conversational}.  This enhances the overall user experience by keeping the development process dynamic and responsive to feedback.

\subsubsection{Instruments and Evaluation}
In this study, we assess the design thinking abilities of the o1-preview model in coding tasks \cite{coignion2024performance, niu2024evaluating}. As coding often involves prototyping the code, iteratively debugging, and further, the user can give LLM feedback to let it customize the code to fit certain needs.
We randomly select ten coding questions from the `hard' questions on Leetcode as the evaluation dataset for the design thinking abilities of LLMs.

We use a series of prompts based on designed to emulate the process of design thinking applied to coding problem-solving. Our approach is as follows:
\begin{enumerate}
    \item We provide the LLM with a Leetcode problem description, including input/output examples and constraints, to initiate the code generation process.
     \item If the generated code encounters errors during execution, the error message is fed back into the LLM to update the code. This process is repeated up to three times if necessary. If the code still fails after three attempts, the experiment is marked as ``failed'' and terminated. If the code runs successfully without errors, this step is skipped.
      \item Once a working solution is produced, the LLM is prompted to optimize the code for improved efficiency \cite{demetriadis2023conversational}. We call this prompt by design thinking prompting, as it reflects a user need fed to the LLM.
\end{enumerate}
The prompts used to realize this process are described in Figure \ref{fig:DT_prompt}.

\begin{figure}[t]
    \centering
    \begin{tcolorbox}[colback=blue!5!white, colframe=blue!75!black, width=\textwidth, arc=5mm, auto outer arc, boxrule=0.5mm, title=Prompt to Test Design Thinking]
    \begin{small}
        \textbf{Prompt 1:} \\
         Read the \textbf{PROBLEM} carefully and provide solution using given \textbf{INSTRUCTION}.\\
        \\
        \textbf{PROBLEM}: \\
       \{title\}
        
        \{description\} \\
        \\
        \textbf{INSTRUCTION}: \\
        \{examples\}
        
        \{constraints\}
        
        Please use the following code template:
        
        \{code template\}\\
        \\
        \textbf{Prompt 2:} \\
        \textbf{INSTRUCTION}: \\
        \{error message\}\\
        \\
        \textbf{Prompt 3:} \\
        \textbf{INSTRUCTION}: \\
        Optimize the code and provide a more efficient version. 
    \end{small}
    \end{tcolorbox}
    \captionof{figure}{Example prompt to assess the design thinking ability of the o1-preview model.}\label{fig:DT_prompt}
\end{figure}

 We evaluate the performance of this process using two metrics: the absolute speed and the speedup rate. 1. Since different questions naturally have different running times, we access the absolute speed by the ``percentage of human beaten'' provided by Leetcode upon submitting code, which is also helpful to show how the performance of LLMs compared to humans. 2. 
 Let 
$t_o$ and $t_n$
  represent the runtime of the original and optimized code, respectively. The speedup rate is defined by $t_n/t_o$
  Following the approach of \cite{shypula2023learning}, if the optimized code fails or its runtime exceeds the original, we set the speedup rate to 1.

\subsubsection{o1-preview performance comparison with human performance}

\begin{figure}[h]
\includegraphics[width=0.99\textwidth]{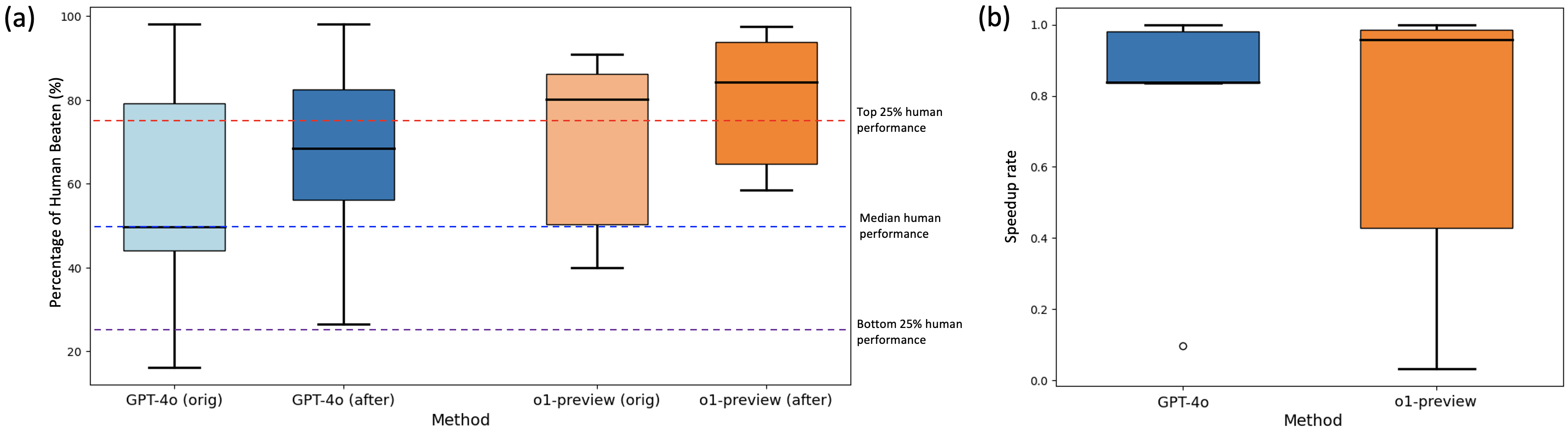}
\centering
\caption{Results for evaluating design thinking on Leetcode coding questions with a following prompt requesting to improve the efficiency.
(a) The absolute speed of GPT-4o and o1-preview models in terms of ``Percentage of human beaten'', each with an original answer and the answer after the prompt requesting to improve the efficiency
(b) The speedup rate of the two models.
} \label{fig:DT1}
\end{figure}


 Figure \ref{fig:DT1} (a) showcases the absolute speed of GPT-4o and o1-preview models in terms of ``Percentage of human beaten'', before and after the prompt requesting to improve the efficiency.
We see that o1-preview beats GPT-4o in both the original answer and the answer after prompting. Other than the boxplots in Figure \ref{fig:DT1} (a), we dig into each question and have an important observation that o1-preview often gives very efficient codes in its first attempt, which means that it is even harder for o1-preview to improve its codes, while it will be easy to improve the less efficient code for GPT-4o. 
In some cases, the first attempt of o1-preview beats the final codes of GPT-4o, which means there is much space for o1-preview to improve its codes. Therefore, we remove such cases and only plot the speedup rate for the remaining cases in Figure \ref{fig:DT1} (b). This helps us better assess the ability of design thinking, i.e., following the user's request to improve code efficiency. We see that the o1-preview has a lower speedup rate than GPT-4o, indicating that after the design thinking prompting, o1-preview speedup the code with more proportion than GPT-4o.\\
\\
 We access the human performance on the coding questions from Figure \ref{fig:DT1} (a). We see GPT-4o (original) has very similar performance with humans, as its median almost overlaps with the median human performance. After design thinking prompting, the performance of GPT-4o is improved but not very significantly. On the contrary, we see much superior performance by o1-preview on this design thinking problem. Our most important observation is that o1-preview (after design thinking prompting) is the only model whose boxplot is completely over 50\%, indicating that it outperforms the median human performance in all cases of our study. Further, the median performance of 1-preview (after design thinking prompting) outperforms over 80\% of the human on Leetcode. 
While GPT-4o, after design thinking, still cannot achieve such excellent performance. These observations highlight the potential of the o1-preview model to surpass human performance in applications related to design thinking.

\subsection{Metacognition}\label{Luyang Fang}

\subsubsection{Conceptualization}
Metacognition refers to the ability to understand, control, and reflect on one's own cognitive processes. It includes metacognitive knowledge—awareness of one's cognitive abilities and strategies for learning or problem-solving—and metacognitive regulation, which involves planning, monitoring, and adjusting strategies during tasks. For example, changing a strategy when progress stalls is a form of metacognitive regulation. These processes enable individuals to plan, monitor, and evaluate their performance, allowing for strategic decision-making, adaptation, and continuous improvement in learning outcomes.
Metacognition has been extensively studied in educational psychology and cognitive science \cite{lai2011metacognition}. \cite{flavell1979metacognition} first introduced the concept, highlighting its role in learning. Later research \cite{schraw1995metacognitive, veenman2006metacognition,allen1993construct} has shown that metacognitive abilities can be taught and are linked to improved problem-solving and decision-making. A meta-analysis \cite{veenman2012metacognition} found that metacognitive interventions led to significant improvements in students’ academic performance across various domains.

Metacognition in LLMs allows them to monitor outputs, adjust strategies, and improve over time, similar to human cognitive regulation. Evidence of this metacognitive capability has emerged in several studies. For instance, the Ask-LLM method enables models to assess the quality of training data using reasoning \cite{sachdeva2024train}. Another study demonstrates that LLMs possess metacognitive knowledge, including the ability to identify relevant skills and procedures for specific tasks \cite{didolkar2024metacognitive}. Additionally, research comparing human and LLM metacognition in a coaching competency test suggests that generative AI can effectively engage in human-like metacognitive processing without conscious awareness \cite{pavlovic2024generative}.

\subsubsection{Instruments and Evaluation}

We assess the metacognitive abilities of the OpenAI o1-preview model in solving mathematical problems.
For each problem, the LLM will be tasked with providing a solution, assigning a confidence rating from 0\% to 100\% to reflect its certainty, and justifying this rating. The justification will involve explaining its reasoning process, identifying any uncertainties, and discussing the factors contributing to its confidence or hesitation about the answer.

\begin{figure}[H]
    \centering
    \begin{tcolorbox}[colback=blue!5!white, colframe=blue!75!black, width=\textwidth, arc=5mm, auto outer arc, boxrule=0.5mm, title=Prompt to Test Metacognition]
    \begin{small}
        \textbf{Prompt:} \\
        Read the \textbf{PROBLEM} carefully and provide solution using given \textbf{INSTRUCTION}.\\
        \\
        \textbf{PROBLEM}: \\
        The numbers $\sin x,$ $\cos x,$ $\tan x$ form a geometric sequence.  Find \[\cos^3 x + \cos^2 x.\] 
        \textbf{INSTRUCTION}: \\
        1. Solve the problem using direct derivations. \\
        2. Rate your confidence in the accuracy of your answer on a range of [0, 1]. 0 means 100\% unsure, and 1 means 100\% certain. Try your best to give the confidence level. \\
        3. Justify your confidence rating: Why did you choose this answer? What factors influenced your level of certainty?
    \end{small}
    \end{tcolorbox}
    \captionof{figure}{Example prompt to assess the metacognitive ability of the o1-preview model.}\label{fig:meta_prompt2}
\end{figure}


After solving, the LLM will receive feedback on the correctness of its answers. It will then re-evaluate and adjust its confidence ratings based on the feedback, reflecting on why changes were made. This tests its ability to incorporate feedback and improve performance, a key aspect of metacognitive thinking.
By incorporating these steps, the instrument captures both cognitive and metacognitive abilities of the LLM, as it has to demonstrate not only the ability to solve problems but also the capacity to reflect on and evaluate its own cognitive processes.

The MATH dataset \cite{hendrycksmath2021} will be used to evaluate the metacognitive abilities of LLMs. This dataset features challenging high school-level math problems, requiring the models to demonstrate both problem-solving and reflective thinking. 
Human performance on the sampled MATH test problems varied significantly based on participants' math proficiency. University students who did not favor mathematics scored 40\%, while those who enjoyed mathematics scored between 65\% and 75\%. The best-performing participant, a three-time IMO gold medalist, achieved 90\%, with errors due to minor arithmetic mistakes \cite{hendrycks2021measuring}.

The o1-preview model will be tested on 30 intermediate algebra problems, which are among the most challenging in the MATH dataset, to evaluate how it handles uncertainty and adapts its strategies.
Key performance metrics include:
\begin{itemize}
    \item Absolute accuracy: Proportion of correct answers on a test.
    \item Metacognitive calibration score: Measures how closely confidence judgments match actual accuracy.
    $$\text { Calibration }=\left|\frac{\text { Number of correct responses }}{\text { Total responses }}-\frac{\text { Average confidence rating }}{100}\right| .$$
    \item Metacognitive Brier score: Accuracy of confidence predictions.
    $$\text { Brier Score }=\frac{1}{N} \sum_{i=1}^N\left(f_i-o_i\right)^2,$$
    where $f_i $ is the predicted probability for instance $i$ and $o_i$ is the actual outcome for instance $i$.
    \item Metacognitive sensitivity: The ability to discriminate between correct and incorrect responses based on confidence ratings is calculated using responses where the confidence level is higher than 80\%. The formula is as follows:
    \begin{small}
    $$\text { Sensitivity }=\frac{\text { \# Correct High-Confidence Responses }}{\text { \# Correct Responses }} - \frac{\text { \# Incorrect High-Confidence Responses }}{\text { \# Incorrect Responses }}.$$
    \end{small}
\end{itemize}


\subsubsection{o1-preview performance}

Figure \ref{appdix_fig:meta} in Appendix \ref{appdix:meta} showcases an example of the o1-preview model's response to the prompt in Figure \ref{fig:meta_prompt}, demonstrating clear evidence of metacognitive thinking. The model tackles the problem in a structured, step-by-step manner, indicating an awareness of the logical processes required to arrive at the correct solution. Upon completing the solution, the model assigns itself a confidence rating of 1.0, suggesting it is fully certain of the result. In the justification for this confidence, the model explains that each step is grounded in fundamental trigonometric identities, with algebraic manipulations that were straightforward and carefully verified. Moreover, the model Highlights that the conclusion directly addresses the original question and points out that no assumptions were made beyond standard mathematical principles, thereby minimizing the risk of errors. This self-reflection and evaluation of its own reasoning show that the model is not only solving the problem but also actively assessing the accuracy of its approach, reflecting key aspects of metacognitive thinking.

\begin{figure}[h]
\includegraphics[width=0.8\textwidth]{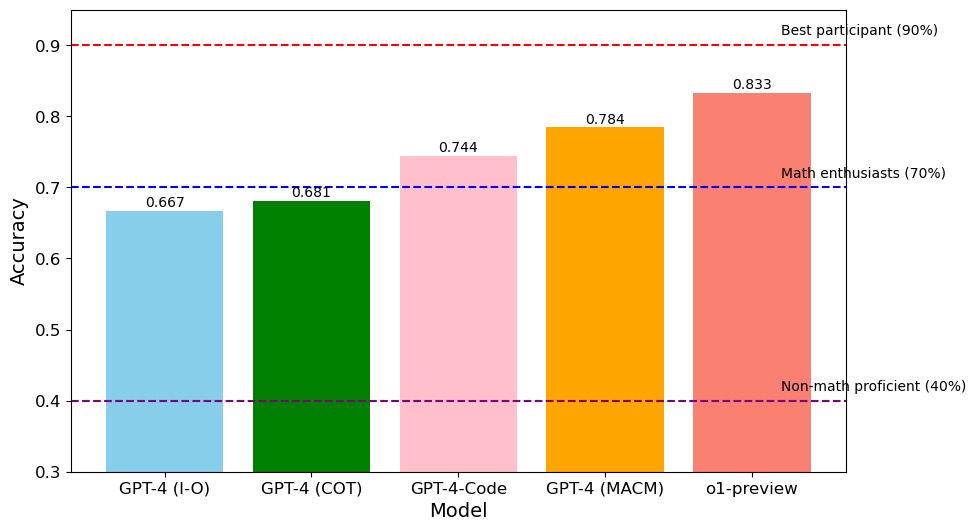}
\centering
\caption{Accuracy comparison on intermediate algebra problems from the MATH dataset. The performance of the o1-preview model is evaluated on a randomly sampled subset of problems. Three horizontal dashed lines represent human performance benchmarks for participants with varying levels of math proficiency.} \label{fig:meta_acc}
\end{figure}

The o1-preview model demonstrates impressive performance in terms of accuracy on intermediate algebra problems, which are among the most challenging in the MATH dataset. As shown in Figure \ref{fig:meta_acc}, the model achieves an accuracy of 83.3\%, outperforming all tested GPT-4 variants, which scored between 66.7\% and 78.4\%. Notably, it has exceeded the performance benchmarks of human university students, outperforming both non-math proficient individuals (40\%) and math enthusiasts (70\%). However, it still falls slightly short of the best-performing participant—a three-time IMO gold medalist—who achieved 90\%. This highlights the o1-preview model's exceptional capability in solving these challenging problems, approaching the performance of the top human participants.

\begin{table}[h]
\caption{Comparison of Calibration Score, Brier Score, and Sensitivity between the GPT-4o and o1-preview models.}
    \centering
    \small
    {
    \begin{tabularx}{\textwidth}{X X X X} 
     & Calibration Score & Brier Score & Sensitivity  \\ \hline 
    GPT 4o & 0.414 & 0.373 & 0.267 \\
    o1-preview  & 0.089   & 0.093 & 0.560 \\
    difference & \textcolor{red}{$\downarrow$ 0.325} & \textcolor{red}{$\downarrow$ 0.280} & \textcolor{red}{$\uparrow$ 0.293} \\
    \hline
    \end{tabularx}}
    \label{meta:tab}
\end{table}

We compare the performance of the GPT-4o and o1-preview models in terms of Calibration Score, Brier Score, and Sensitivity. Results are shown in Table \ref{meta:tab}, with notable improvements achieved in the o1-preview model. The o1-preview model achieves a significantly lower Calibration Score (0.089) and Brier Score (0.093), showing reductions of 0.325 and 0.280, respectively, compared to GPT-4o. These metrics reflect how well the model's predicted confidence aligns with actual outcomes, suggesting that the o1-preview model is better at evaluating its own uncertainty, a hallmark of improved metacognitive ability. Additionally, the Sensitivity of the o1-preview model increases significantly to 0.560, representing an improvement of 0.293 over GPT-4o. While it does not directly reflect metacognitive processes, improving sensitivity alongside better calibration suggests that the o1-preview model can more accurately identify its correct predictions while reducing false positives or overconfident wrong answers, further implying stronger metacognitive capabilities.

While the accuracy of the o1-preview model exceeds typical university-level human performance levels, there is still a gap when compared to the mathematically proficient individuals, such as IMO gold medalists, who achieved 90\% accuracy. Additionally, the o1-preview model tends to display overconfidence in its answers, despite outperforming other models such as GPT-4. Among the five incorrect answers, only one was given with a confidence level of 60\%, while the others all had confidence levels greater than 80\%, with the highest reaching 95\%. This overconfidence in incorrect responses suggests that there is still room for improvement in the o1-preview model's metacognitive ability, particularly in calibrating its confidence levels to better reflect its actual performance.


\subsection{Data Literacy}\label{Jichao Yu}\label{Yizhu Gao}
\subsubsection{Definition and conceptualization}
New technologies continue to exponentially increase the volume of data available, which informs a range of everyday decisions, from consumer purchases to academic and professional applications and even political choices. Accordingly, data literacy has become a crucial capacity both in the workplace and in everyday life. Scholars from various fields have explored data literacy concepts and competencies, shedding light on its diverse aspects. For example, Gummer and Mandinach~\cite{gummer2015building} focused on teacher data literacy and built a conceptual framework. They defines data literacy for teaching as ``the ability to transform into actionable instructional knowledge and practices by collecting, analyzing, and interpreting all types of data to help determine instructional steps''. This definition targets educational professionals, which has been widely used by studies on data literacy for pre-service \cite{wu2023research} and in-service educators \cite{reeves2019effects}. Differently, Koltay \cite{koltay2017data} examined data literacy from the perspective of researchers and data librarians. They defined data literacy as a specific skill set and knowledge base that empowers individuals to transform data into information and into actionable knowledge by enabling them to access, interpret, critically assess, manage, and ethically use data. Both Wiorogórska et al. \cite{wiorogorska2018data} and Todorova \cite{todorova2019data} focused on research data literacy, emphasizing research data management. Carmi et al. \cite{carmi2020data} examined data literacy for citizens. They proposed a data citizenship framework consisting of data thinking, data doing, and data participation, which characterize the importance of citizens' critical understanding of and daily proactive engagement with data. Another bundle of research focused on students' data literacy skills. For example, Wolff et al. \cite{wolff2019exploring} designed data literacy inquiry activities that help pupils understand complex data collected by others and form hypotheses from the data. By adapting NGSS practice, Chin et al. \cite{chin2016got} defines data literacy for K-12 students as determining the central idea from scientific text and evaluating sources of information presented in visual form. 

With the rapid growth of data literacy research, scholars conduct literature reviews to compare the similarities and differences in its definition across various studies. Raffaghelli and Stewart \cite{raffaghelli2020centering} reviewed studies on data literacy for educators and found that most studies emphasize heavily technical and measurable aspects of data literacy, such as data extraction, statistical analysis, and visualization. In a follow-up review, Cui et al. \cite{cui2023data} compared different data literacy definitions and competencies and concluded it may be different in its definition in terms of different audiences (K-12 students, post-secondary students, educational professionals, researchers and data librarians, other professionals). Accordingly, proposed data literacy assessments always involve varied competencies. However, in most cases, data literacy describes how people collect, analyze, and interpret data to inform critical decisions in different settings. That is, data literacy describes how to use data to understand real-world phenomena; different competencies are required for different audiences. It is worth noting that it is commonly argued the foundations for a data-literate society begin by acquiring key data literacy competencies in school (\cite{wolff2016creating}\cite{cui2023data}). This standpoint influences what and how instruments of data literacy are being developed and evaluated. The instruments for assessing data literacy have been summarized in the following section. 

\subsubsection{Assessment of data literacy}
Cui et al. \cite{cui2023data} summarized the assessment formats and item types of each data literacy assessment. They concluded that the existing data literacy assessments can be classified into two categories: self-reflective approaches and object measures. Self-reflective approaches measure individual's self-reports of data literacy competencies using surveys, questionnaires, semi-structured interviews, and think-aloud interviews. With these tools, participants are asked to evaluate their data literacy competencies on their own by describing and reflecting on their data-related behaviors, practices, and attitudes. In contrast, objective measures mainly develop test questions, which always take three forms: conventional tests with multiple-choice and constructed-response questions, digital game-based assessments, and participation observations. The current study mainly applies conventional tests with multiple-choice and constructed-response questions to test OpenAI o1-preview. 
\begin{table}[h]
\caption{Item Types of Two Data Literacy Assessments}
    \centering
    \small
    {
    \begin{tabularx}{\textwidth}{X X X} 
    \hline
    \textbf{Source} &      \textbf{Merk et al. (2020)}& \textbf{Chen et al. (2023)}\\ \hline
    \textbf{Participants}&89 pre-service teachers& 170 undergraduate students\\ 
 \textbf{Countries}& Germany&Canada\\  
    \textbf{Item Types}&Multiple-choice Constructed response& Multiple-choice Constructed response\\
 \textbf{Validation}& Reliability and validity reported& Reliability and validity reported\\ 
 \textbf{Competencies}& Use data; Transform data into information; &Data management; Data visualization; Basic data analysis\\ \hline
    \end{tabularx}}
    \label{tab:data_literacy_comparison}
\end{table}

Cui et al. \cite{cui2023data} conducted a thorough analysis of the assessment formats and item types used in data literacy evaluations across 25 studies. Of these, 15 studies did not report reliability or validity, 12 employed self-reported scales or questionnaires, and 2 utilized digital game-based assessments. While the development of game-based assessments is innovative, it presents challenges for testing within OpenAI o1-preview, given the complex and interactive nature of such questions. To meet both high-quality criteria and operational feasibility, we focused on two well-established assessments: the data literacy test for pre-service teachers developed by Merk et al. \cite{merk2020fostering} and data literacy test for post-secondary students validated by Chen et al. \cite{chen2024validating}. 

The assessment by Merk et al. \cite{merk2020fostering} was designed to examine the effects of intervention on pre-service teachers' data literacy. It consists of two major components: (1) Using data. This component evaluates pre-service teachers' understanding of data properties, data manipulation, data aggregation, and their knowledge of statistics and psychometrics. (2) Transforming data into information. This component assesses how pre-service teachers interpret data, utilize data displays and visual representations, apply statistical methods, and summarize and explain data. 

In contrast, Chen et al. \cite{chen2024validating} validated their data literacy test for post-secondary students by conducting psychometric analyses and incorporating eye-tracking techniques to confirm its reliability and validity. Their assessment tests data literacy across three dimensions: (1) Data management, which involves the organization of data (1 question) and data manipulation (2 questions). (2) Data Visualization, which covers frequency distributions (2 questions) and the use of visual charts (4 questions). (3) Basic data analysis, which focuses on concepts like central tendency (5 questions), variability (3 questions), and percentage calculations (1 question).

Both assessments emphasize critical aspects of data management and interpretation, but they differ in terms of their target audiences and the depth of analysis. Merk et al.’s assessment is tailored toward pre-service teachers, focusing on practical skills needed for educational settings, while Chen et al.’s assessment targets post-secondary students, focusing on more advanced data literacy concepts.

\begin{figure}[h]
    \centering
    \includegraphics[width=0.8\linewidth]{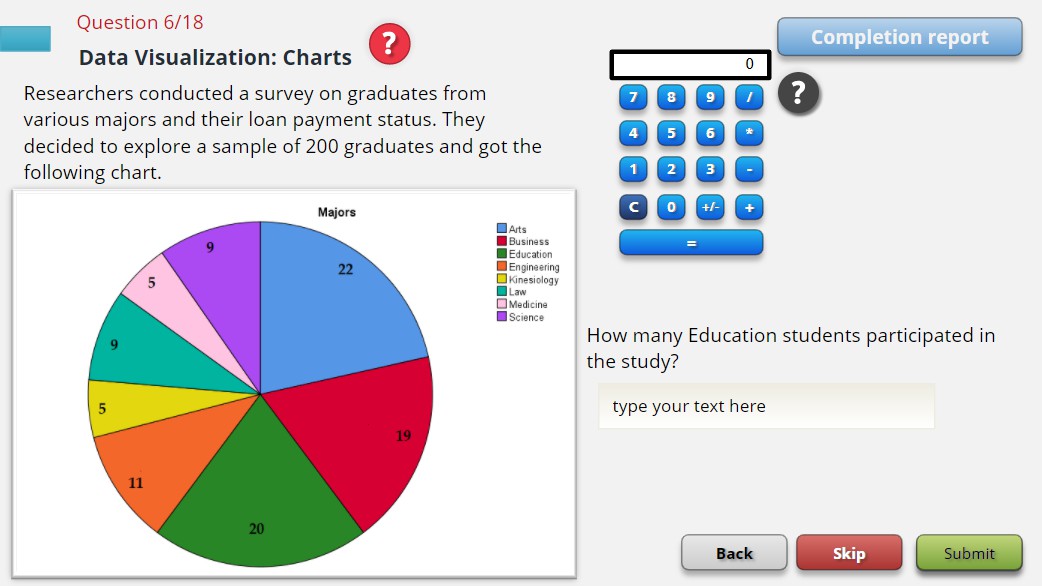}
    \caption{Data Visualization Question}
    \label{fig:data_literacy_q6}
\end{figure}

Examples of questions from these tests are provided below to illustrate the types of skills and knowledge being evaluated, as shown in Figure \ref{fig:data_literacy_q6}.

\subsubsection{Testing prompt and strategies}
In this study, conventional tests featuring multiple-choice and constructed-response questions were employed to evaluate OpenAI o1-preview's data literacy capacities. As shown in Figure \ref{fig:data_literacy_q6}, many of these data literacy questions involve interpreting statistical graphs, tables, and charts. To assess OpenAI o1-preview, we transformed these visual elements into textual descriptions, as demonstrated in Figure \ref{fig:data_literacy_q6_prompt}, and converted them into prompts for testing OpenAI o1-preview. Specifically, this study utilized two validated data literacy assessments: one developed by Chen et al. \cite{chen2024validating}, consisting of 18 questions, and another by Merk et al. \cite{merk2020fostering}, which includes 20 questions. OpenAI o1-preview's performance was then compared to that of 170 post-secondary students from the Faculty of Educaiton at a Western Canadian university and 155 pre-service teachers from a large university in southern Germany. 
\begin{figure}[H]
    \centering
    \begin{tcolorbox}[colback=blue!5!white, colframe=blue!75!black, width=\textwidth, arc=5mm, auto outer arc, boxrule=0.5mm, title=Prompt, fonttitle=\small, fontupper=\tiny]
        \textbf{Prompt:} \\
        \tiny Read the INSTRUCTIONS and QUESTION. Then, follow the INSTRUCTIONS to answer the QUESTION.\\
        \\
        \textbf{INSTRUCTIONS:}\\
        Researchers conducted a survey on graduates from various majors and their loan payment status. They decided to explore a sample of 200 graduates and got the following chart.\\
        TIPS:\\
        Bar charts show the number of each category compared to the other categories.
        Line charts are most appropriately used for continuous data.\\
        Pie charts help in showing the percentage of the number each category of the total number.
        scatter plot shows the relationship between two numerical variables by displaying individual data points on a two-dimensional graph\\ 
        CHART DESCRIPTION:\\
        This chart is a pie chart displaying the distribution of 200 students across different majors. Each segment represents a distinct major, and the size of each segment correlates with the number of students in that major. The following information is depicted:\\
        \begin{enumerate}[label=\arabic*., leftmargin=*]
            \item Arts: Represented by a blue section, this major has the largest portion with 22 students.
            \item Business: In red, this major has 19 students.
            \item Education: In green, this major has 20 students.
            \item Engineering: Shown in orange, this major has 11 students.
            \item Kinesiology: The yellow section, representing 5 students.
            \item Law: Represented by a teal section, with 9 students.
            \item Medicine: In pink, representing 5 students.
        Science: Shown in purple, with 9 students.
        \end{enumerate}
        \vspace{\baselineskip}
        The legend on the right-hand side indicates the colors corresponding to each major, ensuring that each slice of the pie is linked to the appropriate field of study. The chart provides a clear visual breakdown of how students are spread among these eight majors.\\\\
        QUESTION: \\
        How many Education students participated in the study? 
    \end{tcolorbox}
    \caption{Prompt of Data Visualization Question}
    \label{fig:data_literacy_q6_prompt}
\end{figure}
\subsubsection{o1-preview performance comparison}
The tables below provides a summary of the comparison between OpenAI o1-preview and human participants. OpenAI o1-preview demonstrated perfect performance in data management, basic data analysis tasks, understanding statistics and psychometrics, understanding how to interpret data, and understanding data properties. In the area of data visualization, OpenAI o1-preview outperformed the average scores of post-secondary students. This indicates that, while students performed adequately, OpenAI o1-preview exhibited a stronger ability to interpret and generate visual representations of data.

\begin{table}[h]
\caption{Comparison of Data Literacy Assessment Accuracy between Post-Secondary Students and OpenAI o1-preview}
    \centering
    \small
    {
    \begin{tabularx}{\textwidth}{X X X} 
    \hline
    \textbf{Dimension of Data Literacy} & \textbf{Post-secondary students (n = 170)} & \textbf{OpenAI o1-preview} \\ \hline
    Data management     & 6.17\%  & 100\%   \\
    Data visualization  & 62.37\% & 66.67\% \\
    Basic data analysis & 63.67\% & 100\%   \\ 
    \hline
    \end{tabularx}}
    \label{tab:data_literacy_comparison1}
\end{table}
\begin{table}[h]
\caption{Comparison of Data Literacy Assessment Accuracy between Pre-Service Teachers and OpenAI o1-preview}
    \centering
    \small
    
    \begin{tabularx}{\textwidth}{X X X} 
    \hline
    \textbf{Dimension of Data Literacy} & \textbf{Pre-Service Teachers (n = 89)}& \textbf{OpenAI o1-preview} \\ \hline
    Understand statistics and psychometrics& 22.58\%& 100\%   \\ 
    Understand how to interpret data& 79.79\%& 100\%   \\ 
    Understand data properties& 60.65\%& 100\%   \\ 
    Use statistics& 50.88\%& 100\%   \\ 
    Manipulate data& 18.71\%& 100\%   \\ 
   
    \hline
    \end{tabularx}
    \label{tab:data_literacy_comparison2}
\end{table}

A recent study by Zhong et al. \cite{zhong2024evaluationopenaio1opportunities} evaluated the performance of o1-preview in Educational Measurement and Psychometrics. Understanding statistics and psychometrics is an important component for data literacy assessments, particularly those aimed at pre-service and in-service teachers \cite{merk2020fostering}\cite{reeves2015classroom}. Zhong et al. \cite{zhong2024evaluationopenaio1opportunities} found that o1-preview excelled in tasks requiring statistical computation and reasoning, demonstrating a level of proficiency in core psychometric concepts comparable to, or even exceeding, that of first-year master's or doctoral students in educational measurement. The findings from this study further confirmed OpenAI o1-preview's strong capabilities in statistical computation and reasoning. However, our tests also identified some limitations in its ability to interpret and understand statistical graphs, charts, and plots. Specifically, OpenAI o1-preview struggled when required to integrate textual descriptions of both questions and statistical illustrations, indicating a challenge in synthesizing multiple forms of information. Additionally, the current testing did not include questions that assess data-relevant critical thinking capacities, such as identifying reliable data sources and making critical decisions. Therefore, these aspects of data literacy capabilities need further investigation to test and compare performance of human and OpenAI o1. 

\subsection{Creative Thinking}\label{Xiantong Yang}
\subsubsection{Conceptualization and instrument}
Creative thinking, also known as creativity, defined as the ability to produce original and meaningful products, has been highlighted as a critical 21st-century skill and has become a major educational goal worldwide \cite{dumas2024using}. To achieve this goal, researchers have started exploring how technology can be leveraged to foster students' creativity \cite{liu2022meta}, particularly through generative artificial intelligence (GenAI) tools such as ChatGPT 3.5 and ChatGPT 4. Previous studies suggest that ChatGPT 4 has already achieved human-level creativity \cite{haase2023artificial}.

While the recently released OpenAI o1-preview is considered to have superior performance in advanced reasoning and chain-of-thought processing, no empirical studies have yet examined the differences in creativity between OpenAI o1-preview and humans in tasks requiring advanced thinking. Therefore, this study aims to assess OpenAI o1-preview's performance in creativity tasks and compare it with previous research in an attempt to fill this gap and inspire researchers to use more advanced technologies to foster student creativity.

Creativity is typically defined in two ways: divergent thinking and convergent thinking \cite{de2020dis}. Divergent thinking occurs when "an individual solves a problem or reaches a decision using strategies that deviate from commonly used or previously taught strategies" \cite{ashkinaze2024ai}. One of the most common tests of divergent thinking is the Alternate Uses Task (AUT) \cite{guilford1967nature}, where participants are asked to think of an original use for a common object. Traditionally, responses to the AUT are measured along four dimensions: fluency (number of ideas), flexibility (different categories of ideas), originality (uniqueness of the idea), and elaboration (the level of detail provided) \cite{ashkinaze2024ai}.

On the other hand, convergent thinking refers to the process of using given clues to arrive at the correct solution in a specific direction. A classic test for convergent thinking is the Remote Association Test (RAT), which requires participants to identify the common link between three seemingly unrelated words \cite{mednick1962associative}. For example, for the words "SAME," "TENNIS," and "HEAD," the goal is to find a word, such as "match," that can connect with all three—forming compound words or semantic relationships, like "match HEAD" or "TENNIS match." The number of correct answers is used to measure convergent thinking ability.

Previous studies have compared the creativity performance of early versions of ChatGPT (e.g., ChatGPT 3.5 and ChatGPT 4) with human creativity \cite{liu2024chatgpt,urban2024chatgpt}. This study builds on that foundation by including the OpenAI o1-preview model (i.e., o1-preview), expanding the scope of research into the creativity performance of OpenAI o1-preview  preview compared to humans.

\subsubsection{Human performance on creative thinking}
Previous studies have frequently used university students as participants to evaluate both their divergent and convergent thinking. Researchers have employed tasks such as the Unusual Uses Task to assess divergent thinking, where participants are asked to come up with as many creative uses as possible for common objects. In convergent thinking studies, tasks like the Remote Associates Test (RAT) have been used to measure participants' ability to find common links between seemingly unrelated words. These methods have provided valuable insights into the cognitive processes underlying creativity and problem-solving among university student populations. 

For example, Urban et al. \cite{urban2024chatgpt} assessed the baseline performance of 68 university students (male = 22, primarily from social sciences and humanities, N = 52) on the Alternate Uses Task. In this task, participants were asked to generate as many original ideas as possible for the various uses of common objects (e.g., paperclip, brick, can). The originality of their responses was independently rated by two trained experts on a 5-point scale, where 1 indicated "no originality," and 5 indicated "high originality." The results showed that their average originality score was 1.74.

Additionally, Xia et al. \cite{xia2022bilingualism} used the Remote Associates Test (RAT) to compare the convergent thinking abilities of high-proficient (n = 27) and low-proficient (n = 27) bilingual Chinese university students (N = 54). The findings revealed that the high-proficient group (M = 27.38) scored significantly higher than the low-proficient group (M = 23.80) on the RAT (p = 0.003).

\subsubsection{Evaluating o1-preview performance}
In this study, we used the AUT developed by Torrance \cite{torrance1966torrance} to assess the divergent thinking of OpenAI o1-preview. Participants were required to propose as many original ideas as possible for different uses of a common object (such as a paperclip, brick, or can). The originality of the responses was evaluated using an automated AI scoring tool for divergent thinking developed by Organisciak et al. \cite{organisciak2023beyond}(https://openscoring.du.edu/scoringllm). This tool has been found to have high reliability and validity in previous studies \cite{organisciak2023beyond}.

The prompt for the AUT task was: "A paperclip (brick or can) can be used in many different ways. Some of these are quite common, while some can be considered original. Your task is to create as many original ideas as possible about how a paperclip (brick or can) can be used."

The performance of OpenAI o1-preview on the AUT task is presented in the Table \ref{tab:AUTresults}.

\begin{table}[!htbp]
\caption{Results of o1-preview performance on AUT task}
    \centering
    \scriptsize
    {
    \begin{tabularx}{\textwidth}{X X X}
    \hline
    \textbf{Question} & \textbf{Answer} & \textbf{Originality Score} \\ \hline
    Paperclip & Fashion it into a miniature eyeglass frame for dolls or action figures. & 4 \\ 
    Paperclip & Create a makeshift smartphone stand by bending it appropriately. & 3 \\ 
    Paperclip & Use it as a tiny antenna for a DIY radio project. & 3.5 \\ 
    Paperclip & Craft it into a custom-shaped cookie cutter for miniature baking. & 4 \\ 
    Paperclip & Use it to unlock simple suitcase locks in an emergency. & 2.5 \\ 
    Paperclip & Form it into a holder for a single flower or plant stem. & 3 \\ 
    Paperclip & Make decorative earrings or jewelry by adding beads. & 2.5 \\ 
    Paperclip & Use it to repair a broken zipper pull temporarily. & 3 \\ 
    Paperclip & Construct a small wireframe model for architectural designs. & 3 \\ 
    Paperclip & Utilize it as a tool to clean under fingernails. & 2.5 \\ 
    \textbf{Mean Paperclip} &  & \textbf{3.1} \\ 
    Brick & Use it as a canvas for mosaic art by attaching tiles or glass pieces. & 3 \\ 
    Brick & Carve it into a personalized nameplate for a garden or home entrance. & 2 \\ 
    Brick & Heat it and place it in a towel to create a bed warmer. & 3 \\ 
    Brick & Incorporate it into a science experiment to demonstrate inertia. & 3 \\ 
    Brick & Use it as a prop in theater productions to simulate weightlifting. & 3 \\ 
    Brick & Paint it to resemble a book and use it as hidden storage on a shelf. & 3.8 \\ 
    Brick & Stack multiple bricks to create a rustic mailbox stand. & 2 \\ 
    Brick & Use it as a base for an outdoor umbrella or clothesline. & 2 \\ 
    Brick & Hollow it out to create a secret hiding spot for keys. & 3 \\ 
    Brick & Integrate it into outdoor steps or seating areas for landscaping. & 1 \\ 
    \textbf{Mean Brick} &  & \textbf{2.58} \\ 
    Can & Convert it into a small stove for camping by adding ventilation holes. & 3.3 \\ 
    Can & Use it as a mold to create ice lanterns in freezing weather. & 3.6 \\ 
    Can & Turn it into a desktop organizer with compartments by adding inserts. & 3 \\ 
    Can & Create a homemade projector by inserting a lens and using a smartphone. & 3 \\ 
    Can & Fashion it into a decorative candle holder with patterns punched out. & 3.6 \\ 
    Can & Use it as a wind sock by attaching fabric strips that show wind direction. & 3 \\ 
    Can & Make a DIY vacuum chamber for science experiments. & 3 \\ 
    Can & Use it to build a robotic arm component in hobby projects. & 3 \\ 
    Can & Transform it into a pet toy by sealing treats inside and adding holes. & 3.3 \\ 
    Can & Create a mini hydroponic system for growing plants without soil. & 3.9 \\ 
    \textbf{Mean Can} &  & \textbf{3.27} \\ 
    \textbf{Overall Mean} &  & \textbf{2.98} \\ \hline
    \end{tabularx}}
    \label{tab:AUTresults}
\end{table}

We used the RAT to test OpenAI o1-preview's convergent thinking. Given that the human participants for comparison were Chinese university students, we adopted a reliable and valid version of the RAT in Chinese \cite{xia2016exploring}. In Chinese, an example would be the three original words “章” (chapter), “学” (study), and “艺” (art), to which the word “文” (literature) could be added at either the beginning or the end to form three reasonable words: “文章” (article), “文学” (literature), and “文艺” (art).

Due to OpenAI o1-preview's prompt limitation, we selected 10 trials while balancing different difficulty levels.

\begin{figure}
    \centering
    \includegraphics[width=0.80\linewidth]{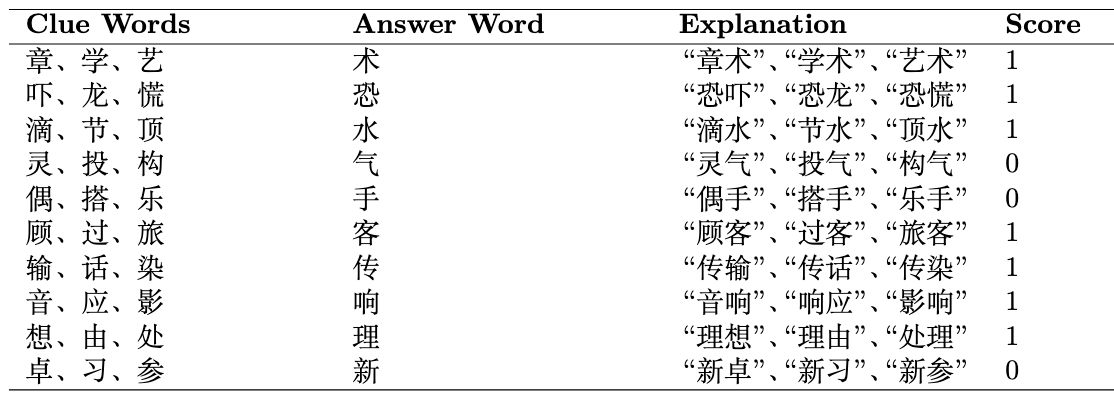}
    \caption{Results of o1-preview performance on RAT task}
    \label{tab:RATresults}
\end{figure}

The task prompt was: "Below, you will see 10 Chinese word association problems, each consisting of three Chinese 'clue words.' Your task is to find one Chinese 'answer word' that can form a valid two-character phrase with each of the three 'clue words.' For example, the answer for 治 (heal), 防 (prevent), and 统 (govern) would be 治 (control), forming 治疗 (treatment), 防治 (prevention and control), and 统治 (rule)." The Figure \ref{tab:RATresults} presents the OpenAI o1-preview's performance on the RAT task.

\subsubsection{Comparing o1-preview and human performance}
\begin{figure}[ht]
    \centering
    \includegraphics[width=0.75\linewidth]{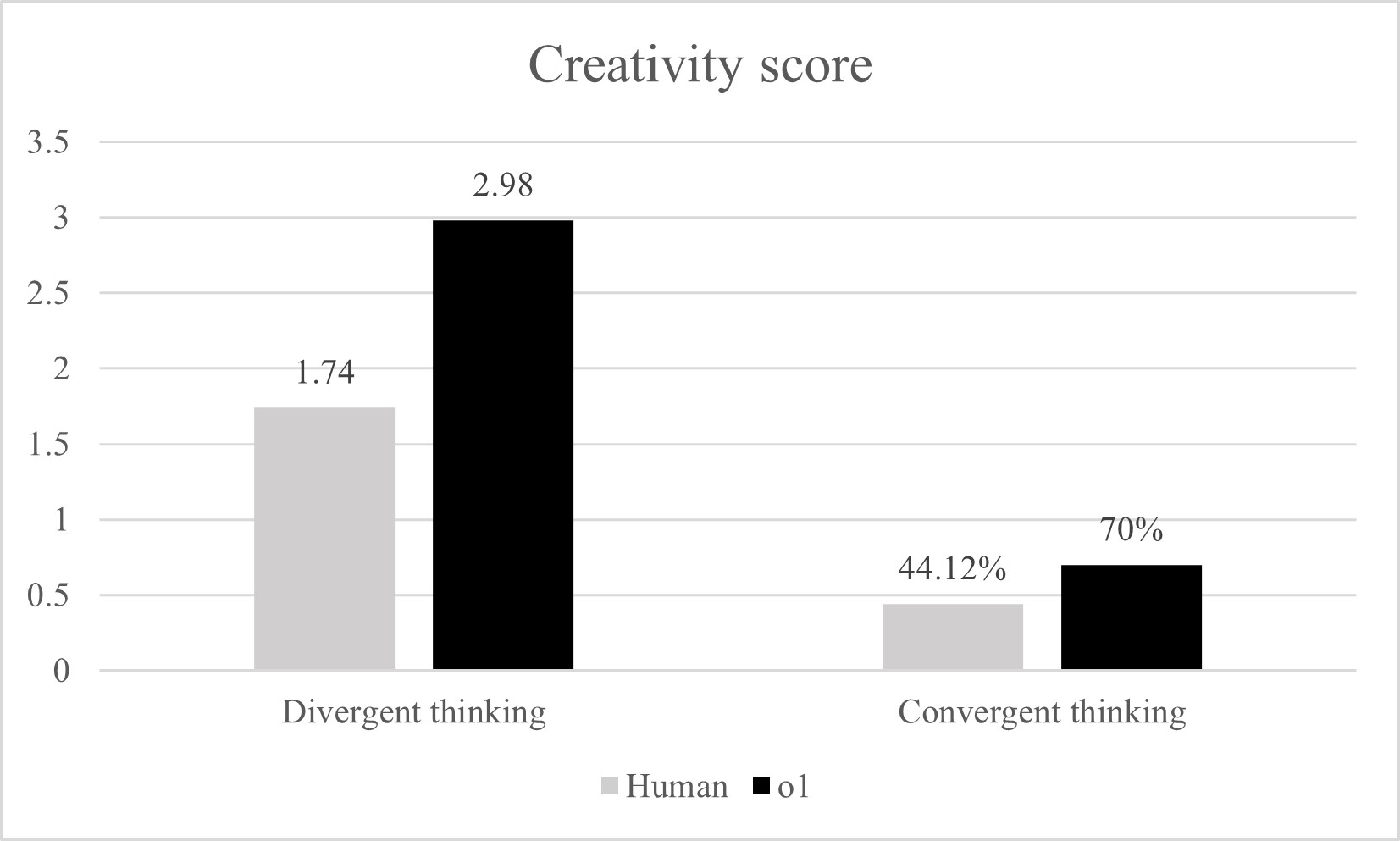}
\caption{Comparison of o1-preview and human performance on creative tasks.}
\label{fig:creativity1}
\end{figure}

OpenAI o1-preview’s scores for both divergent and convergent thinking are shown in Figure \ref{fig:creativity1}. In the divergent thinking task, OpenAI o1-preview achieved an average score of 2.98 based on the AUT task. For convergent thinking in the RAT task, the score was 7, with an accuracy rate of 70\%.

Discussion
\textbf{Who Performs Better in Creativity?}
In terms of divergent thinking, OpenAI o1-preview significantly outperformed human participants, achieving an average score of 2.983 compared to the human average of 1.74. Divergent thinking involves generating creative ideas by exploring many possible solutions \cite{raz2023role}, indicating that o1-preview demonstrated a stronger ability to explore a wide range of original ideas.

For convergent thinking, o1-preview also surpassed human participants, with a 70\% accuracy rate compared to the human score of 44.12\%. Convergent thinking focuses on finding the single correct answer to a problem\cite{kumar2023chapter}, suggesting that o1-preview is highly effective at logical deduction and problem-solving.

\textbf{Why OpenAI o1-preview outperform human participants?}
OpenAI o1-preview outperformed human participants in both divergent and convergent thinking tasks primarily due to its advanced architecture and unique reasoning capabilities \cite{zhong2024evaluation}. One key factor is the new approach to AI reasoning used in o1-preview. Unlike earlier models, o1-preview employs a specialized training methodology that not only relies on vast amounts of data but also integrates a custom optimization algorithm designed to enhance reasoning abilities\cite{zhong2024evaluation}. This enables the model to tackle multi-step problems and generate creative solutions more efficiently than humans.

The internal reasoning tokens and chain-of-thought processes embedded in o1-preview allow it to break down problems into manageable steps \cite{yang2024chain}. For example, when faced with complex tasks, o1-preview can generate intermediate steps (reasoning tokens) to guide its decision-making process. This systematic approach mimics structured human thought, yet with the speed and accuracy of a machine, allowing it to quickly evaluate different strategies and generate high-quality responses.

Moreover, o1-preview's use of reinforcement learning strengthens its ability to handle complex \cite{zhong2024evaluation}, multi-step tasks like those in divergent and convergent thinking. This method trains the model to simulate a step-by-step reasoning process, similar to how humans approach problem-solving, but with fewer cognitive limitations such as fatigue or biases. This combination of structured reasoning, speed, and comprehensive training is what gives o1-preview a distinct advantage over human participants in these creative tasks.

\textbf{Insights for AI-Enhanced Human Creativity}
OpenAI o1-preview exhibits stronger performance than human participants in both divergent and convergent thinking tasks. OpenAI o1-preview can be used to augment human creativity, especially in brainstorming or problem-solving tasks where generating multiple solutions is critical. The collaboration between AI and humans holds the key to unlocking the full potential of creativity \cite{bouschery2024artificial}. Leveraging AI in creative processes requires training people to effectively collaborate with AI, understanding its capabilities, and limitations, and how to integrate AI-generated ideas into meaningful, human-centered outcomes. By using AI to support and amplify human creativity, individuals can explore a greater range of ideas and more efficiently solve complex problems.

\subsection{Collaborative Thinking}\label{Liang Zhang}

The rapid advancement of LLMs has resulted in significant progress in solving complex tasks across various domains, including social sciences, STEM, psychology, and policy research \cite{guo2024large}. While these achievements are impressive, they often depend on the capabilities of individual models, which can be limited when faced with problems that require diverse perspectives or specialized reasoning. To overcome these limitations, a new approach has emerged: collaborative thinking through multi-agent systems. Collaborative thinking, as applied in multi-agent systems, enables multiple LLMs to interact, share knowledge, and approach problems from different angles, enhancing their collective reasoning and improving task-solving efficiency \cite{wu2023autogen,shen2024learning,wang2024sotopia}. This framework is particularly effective for tasks that require interdisciplinary insights, where diverse reasoning approaches and specialized knowledge are crucial for accurate problem-solving and successful task implementation \cite{li2023camel,wu2024mathchat}. 

In this collaborative environment, each LLM-based agent plays a specific role, with some LLMs acting as reasoning engines or domain-specific experts \cite{wu2023autogen,guo2024large, hong2023metagpt}. These agents communicate, debate, and refine their thought processes, allowing for deeper exploration of complex issues. By mimicking human collaboration, the multi-agent system enhances decision-making, fosters creativity, and improves problem-solving by synthesizing insights from various sources. Recent research has demonstrated promising results
in utilizing LLM-based multi-agents for solving various
tasks. For example, MetaGPT serves as an LLM-based multi-agent framework that leverages Standardized Operating Procedures (SOPs) to decompose complex tasks into specific roles (e.g., Product Manager, Architect, Engineer), achieving superior performance in programming and software development tasks \cite{hong2023metagpt}. Another prototype of a multi-agent framework, CAMEL, automates cooperation between LLM-based agents through role-playing, demonstrating remarkable effectiveness in task completion, as evidenced by experiments in AI society, coding, mathematics, and science \cite{li2023camel}. Microsoft Research introduced AutoGen, leveraging multi-agent conversations to effectively tackle a wide range of applications, including mathematics, coding, question answering, operations research, online decision-making, and entertainment \cite{wu2023autogen}. In addition to task-solving, recent advancements have focused on implementing social intelligence within multi-agent systems using LLMs, contributing significantly to collaborative thinking by allowing agents to engage in both technical problem-solving and complex, human-like social interactions. Park et al. presented research on generative agents using LLMs, showcasing their ability to remember, retrieve, reflect, interact with other agents, and plan through dynamically evolving circumstances, effectively simulating believable human social behaviors within simulated social communities \cite{park2023generative}. SOTOPIA provides a general-domain interactive environment where agents role-play, coordinate, collaborate, compete, and interact under diverse scenarios to achieve complex social goals, including maintaining relationships, preserving finances, and adhering to social norms \cite{zhou2023sotopia, wang2024sotopia}.  

Recent advancements in AI education research have led to the development of multi-agent systems that leverage multiple LLM-based agents working collaboratively to enhance learning experiences, boost engagement, and address complex tasks across diverse educational applications. Yu et al. presented a massive multi-agent online learning platform, leveraging LLM-based agents for roles such as teacher, teaching assistant, session controller, and classmates to enhance learning engagement and manage classroom dynamics \cite{yu2024mooc}. The platform also includes additional agents to support teaching, learning, and tutoring resources, facilitating an improved online learning experience through LLM-driven multi-agent collaboration. Zhang et al. also leveraged multiple LLM-based agents to simulate real classrooms, assigning roles such as Teacher, Assistant, and Classmates, effectively mimicking traditional classroom interactions while enhancing the user learning experience and facilitating collaborative behaviors to address key research questions \cite{zhang2024simulating}. Li et al. developed a multi-agent system for automating knowledge annotation, achieving superior performance on a math question tagging dataset by using LLM-based agents to handle independent subproblems through distinct roles (e.g., Task Planner, Question Solver, Semantic Judger, Numerical Judger) \cite{li2024knowledge}. Zhang et al. introduced SPL, a dialogue-based intelligent tutoring system that utilizes multiple agent roles to manage lesson creation and dialogue generation, aiming to foster critical thinking in students through the Socratic teaching method \cite{zhang2024spl}. 

The release of the OpenAI o1-preview model has demonstrated significant advancements in reasoning capabilities \cite{openai2024reasoning}, positioning it as a promising tool for complex problem-solving through collaborative thinking. Building on these advancements, we sought to empirically assess how well LLM-based multi-agent frameworks can address real-world problems by conducting a series of testing experiments. These experiments were designed to explore and evaluate the reasoning processes of LLM-based agents, particularly their effectiveness in collaborative scenarios. By using GPT-4o as the baseline, we compared the performance of the multi-agent method powered by the o1-preview model, aiming to understand the extent to which these agents can collaboratively solve intricate tasks in diverse settings. The following section outlines the design and results of these testing experiments, highlighting the strengths and limitations of the multi-agent approach in AI-educational applications.  



\subsubsection{Testing Experiments}  

We demonstrated two types of exploratory testing experiments, Math Problem Solving and Collaborative Science Group Chats, to evaluate the problem-solving capabilities of the LLM-based multi-agent system powered by the o1-preview model, focusing on their reasoning processes, with GPT-4o as the baseline. All LLM-based agents were set to use the same GPT model with a temperature setting of 1.0, as constrained by the limitations of the o1-preview \cite{openai2024reasoningguide}.

\textbf{A1: Math Problem Solving}

We employed both single-agent (one LLM-agent) and two-agent setups (comprising an LLM-based user proxy agent and an LLM-based assistant agent) to solve 20 geometry word problems from the MATH dataset \cite{hendrycks2021measuring}, utilizing the OpenAI API o1-preview with GPT-4o as the baseline. Specifically, we adapted a two-agent system based on MathChat \cite{wu2024mathchat}, aligning with the current o1-preview parameter setting limits \cite{openai2024reasoningguide}. In this setup, two agents, powered by either GPT-4o or o1-preview, collaborated to solve math problems through multiple turns of automated chat. The final answers generated by both setups were subsequently evaluated against the ground truth for all problems. 

\begin{figure}[ht]
\includegraphics[width=0.85\textwidth]{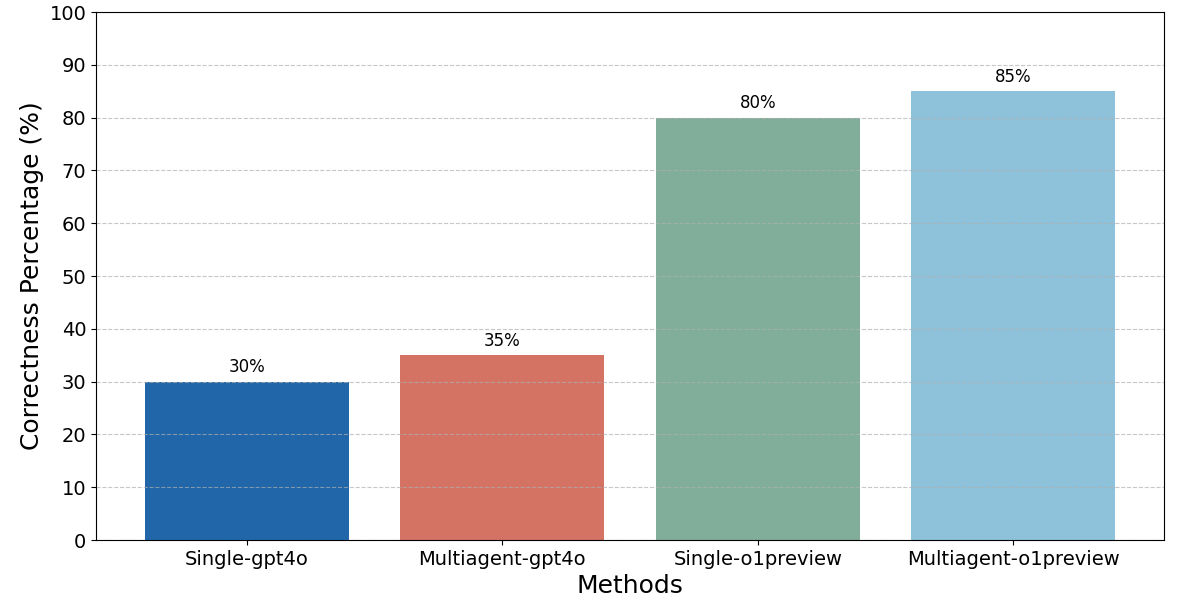}
\centering
\caption{Comparison of Multi-Agent vs. Single-Agent Design in Math Problem Solving.}
\label{fig:multiagent_comparsion}
\end{figure}

As shown in Figure \ref{fig:multiagent_comparsion}, the two-agent setup consistently outperforms the single-agent setup when using both the o1-preview and GPT-4o models, achieving a 5\% higher correctness percentage in both cases. Specifically, the o1-preview model showed a significant improvement over GPT-4o, with a 50\% higher correctness rate for both the single-agent and multi-agent setups. The 5\% improvement observed in the multi-agent configuration, though relatively modest (due to financial constraints and the limited number of test questions), highlights the potential of agent collaboration to enhance problem-solving and reduce reasoning errors. The 50\% performance gap between o1-preview and GPT-4o underscores the advancements in the o1-preview architecture, particularly in its ability to generate more accurate responses. This suggests that the o1-preview model has superior generalization capabilities, allowing it to handle complex scenarios more effectively than GPT-4o. These results indicate that o1-preview not only outperforms GPT-4o in terms of accuracy but also more effectively harnesses the benefits of multi-agent collaboration, demonstrating its superior ability to facilitate cooperative problem-solving in tasks that demand high levels of precision and correctness.  

\textbf{A2: Collaborative Science Group Chats}

The group chat architecture was designed by assigning LLM-based agents to distinct roles, including a topic developer, group leader, and four participant members. The topic developer is responsible for initiating the discussion by proposing the topic, while the group leader interacts directly with the topic developer, disseminates the topic to the group, oversees the discussion, provides a summary, and reports back to the topic developer. The remaining participant members engage with one another and contribute to the dialogue based on the proposed topic. Throughout the conversation, participants often mention the next speaker to maintain continuity in the dialogue, facilitating smooth transitions between turns. Both GPT-4o and o1-preview models were assigned to these agents. The group chat discussion scenarios were designed based on science topics referenced from the CAMEL study \cite{li2023camel}: 
\begin{itemize}
    \item \textbf{Scenario 1:} ``\textit{What is the energy required to excite an electron in a hydrogen atom from the ground state to the first excited state?}''
    \item \textbf{Scenario 2:} ``\textit{How does the Earth's magnetic field protect us from solar radiation?}''
    \item \textbf{Scenario 3:}
    ``\textit{What is the process of biogeochemical cycling and how does it contribute to the Earth’s ecosystem?}''
\end{itemize}

\begin{figure}[ht]
    \centering
    \begin{subfigure}[b]{0.48\textwidth}
        \centering
\includegraphics[width=\textwidth]{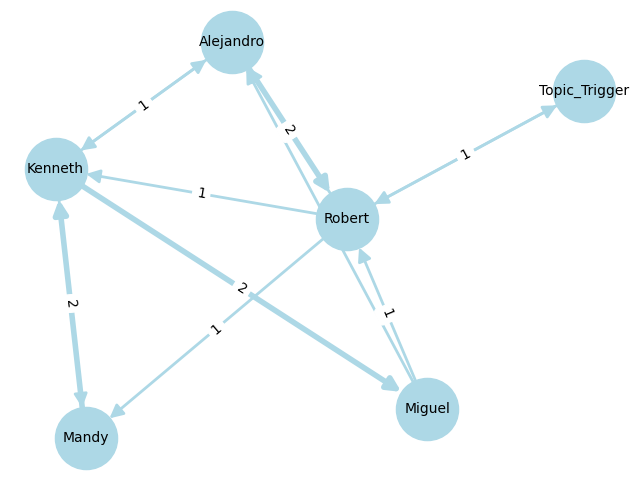}
        \caption{GPT-4o} \label{fig:multiagent_group_chat_gpt4o}
    \end{subfigure}
    \begin{subfigure}[b]{0.48\textwidth}
        \centering
        \includegraphics[width=\textwidth]{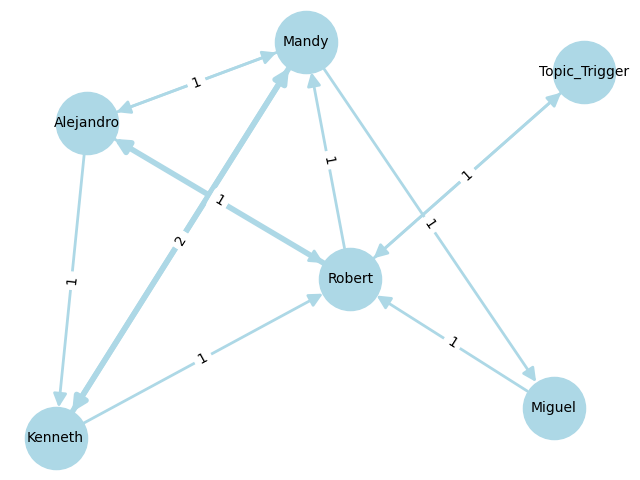}
        \caption{o1-preview} \label{fig:multiagent_group_chat_o1-preview}
    \end{subfigure}
    \caption{Social Network in LLM-based Multi-agent Group Chat for Science Problem-solving (Scenario 1) Using Two Models: GPT-4o and o1-preview. The assigned roles are as follows: 1) Topic Developer, represented as \(Topic\_Trigger\), 2) Group Leader named \textit{Robert}, and 3) Participant Members, including \textit{Alejandro}, \textit{Kenneth}, \textit{Miguel}, and \textit{Mandy}. The nodes represent the roles, while the edges illustrate the dialogue interactions, with arrows indicating the direction of communication. The annotated numbers on the edges represent the frequency of dialogue between participants. }
\label{fig:multiagent_group_chat_comparison}
\end{figure}

Figure \ref{fig:multiagent_group_chat_comparison} illustrates the social network in a science problem-solving scenario (Scenario 1) using the GPT-4o and o1-preview models within a multi-agent group chat. In both setups, \textit{Robert} serves as the central figure coordinating the discussion, consistently interacting with both the Topic Developer and Participant Members. In the GPT-4o configuration, interactions are more evenly distributed among the group members, with a higher frequency of communication by \textit{Robert} and \textit{Kenneth}. The o1-preview setup shows a slightly different pattern, with \textit{Robert} and \textit{Mandy} engaging more frequently with the group. In both group chats, the participants demonstrated high engagement and excellent collaboration. The interactions were fluid and cooperative, but the GPT-4o model demonstrated a more centralized flow, while the o1-preview model fostered a slightly more even distribution of participation. 

\begin{table}[ht!]
\caption{Comparison of o1-preview and GPT-4o models across three Group Chat scenarios.}
    \centering
    \small
    {
    \begin{tabularx}{\textwidth}{X X X} 
    \hline
     Model & Word Contribution (Mean Count) & Sentiment Score \\ \hline 
    GPT-4o & 2294 & 0.160 \\
    
    o1-preview & 3210 & 0.184 \\
    \hline
    \end{tabularx}}
    \label{tab:group_chat}
\end{table}

To further evaluate the aspect of collaborative thinking, we employed a collaborative linguistic analysis focusing on Word Contribution (Count) and Sentiment Analysis \cite{medhat2014sentiment}. The results, presented in Table \ref{tab:group_chat}, provide the average values from three scenario data records. The o1-preview model demonstrated a significantly higher average Word Contribution (3210) compared to GPT-4o (2294). This suggests that the o1-preview model fosters more extensive engagement among participants, potentially leading to deeper discussions and more comprehensive responses. The o1-preview model achieved a higher average emotional valence (0.184) compared to GPT-4o (0.160), suggesting that participants using the o1-preview model exhibited more positive engagement during discussions.  

\subsubsection{Performance comparison}

This section introduces the core principles of collaborative thinking and examines its application through the integration of multiple LLM-based agents to achieve accurate problem-solving and efficient task execution. Preliminary experiments were conducted to evaluate the effectiveness of the o1-preview model, with GPT-4o serving as the baseline for comparison. 

The collaborative thinking framework significantly enhances the problem-solving capabilities of multi-agent systems, enabling them to outperform single-agent setups, as demonstrated in both the math problem-solving and group chat experiments. The improvement performance observed in the multi-agent setups for the GPT o1-preview model highlights the potential of collaborative problem-solving. The use of multiple agents allows for a diversity of perspectives, which contributes to cross-verifying answers, reducing biases, and filling knowledge gaps. This collaborative process, facilitated by dynamic interactions between agents, leads to more reliable outcomes, as evidenced by the higher correctness percentages in our testing experiments using math word problems. The results from the group chat experiments emphasize the critical role of role assignment and inter-agent communication in facilitating collaborative thinking. By distributing tasks and fostering dynamic exchanges among agents, the multi-agent system successfully replicates human-like collaboration. This structure allows agents to approach problems from diverse viewpoints, thereby enhancing problem-solving efficiency and demonstrating the robustness of multi-agent systems in addressing complex challenges. 

However, collaborative thinking in LLM-based multi-agent systems presents several challenges. Among the potential limitations are:  
1) \textit{Instruction-following Issues}: One challenge is that LLMs may not consistently adhere to the prompt instructions or execute the assigned tasks during multi-agent workflows, leading to deviations from expected behavior \cite{hager2024evaluation}.  
2) \textit{Role Flipping}: Agents may unintentionally switch roles or interact with non-existent roles, disrupting the flow of collaboration and causing confusion in the dialogue \cite{li2023camel}. 3) \textit{Memory Challenges}: Handling long contexts or conversations poses difficulties, as agents may struggle to retrieve relevant information from extensive historical records, impacting their ability to provide useful knowledge for supporting subsequent steps \cite{zhang2024chain,hong2023metagpt}.

Collaborative thinking in LLM-based systems holds great promise for augmenting the reasoning capabilities of advanced models like GPT o1-preview, as demonstrated by the promising results in our study. The potential for further developing multi-agent systems that leverage advanced LLMs to achieve higher accuracy and efficiency while handling increasingly complex tasks is significant. Tools and functions that augment LLM agents—such as those enabling computing, searching, memory retrieval, and access to large-scale domain-specific knowledge—are critical to these advancements \cite{he2023solving}. Additionally, fine-tuning these LLM-based agents for domain-specific tasks can further enhance their performance and specialization \cite{scarlatos2024exploring}. As tasks grow more sophisticated, incorporating diverse knowledge sources across disciplines will be essential to maximize the potential of collaborative LLM-based agents. This shift will enable agents to handle not only simple question-answer interactions but also complex, logic-driven experimental designs and high-level organizational tasks. Extending these systems to address multidisciplinary, multimodal goals will also be crucial in advancing their application to more specialized areas \cite{wang2023rolellm,mathur2024advancing}. Moreover, multi-agent frameworks present a promising solution by optimizing resource utilization, reducing computational costs, and enhancing overall performance, making them valuable for both academic research and industry applications \cite{raad2024scaling}.

\subsection{Abstract Reasoning}\label{Weihang You}
\subsubsection{Definition and Conceptualization}
Abstract reasoning, considered an essential representation of human intelligence, is a fundamental cognitive process crucial for success in various fields, particularly in STEM (Science, Technology, Engineering, and Mathematics). From a broad standpoint, it can be defined as "the ability to identify patterns and relationships, and to use this information to solve problems and think creatively" \cite{krawczyk2012cognition}. Abstract reasoning involves manipulating mental representations, such as symbols, ideas, and relationships, to solve problems that are novel or not directly related to prior experiences. In psychological and educational contexts, it is often associated with fluid intelligence \cite{cattell1943measurement}, which is one of the two components of general intelligence according to Cattell's psychometrically-based theory — the capacity to think logically and solve problems in unfamiliar situations without relying on acquired knowledge \cite{ZAVAL2015149}. Within STEM fields, abstract reasoning has become a fundamental competency, equipping individuals with the mental tools necessary to tackle complex problems, innovate, and advance scientific knowledge. For instance, climate scientists use abstract reasoning to create mathematical models of atmospheric-oceanic interactions, while chemists apply it to visualize molecular structures based on spectroscopic data. These applications demonstrate how abstract reasoning helps STEM practitioners generate insights beyond immediate observations. 

\subsubsection{Instruments and Human Performance}
\begin{figure}[H]
\includegraphics[width=0.55\textwidth, height=0.5\textwidth]{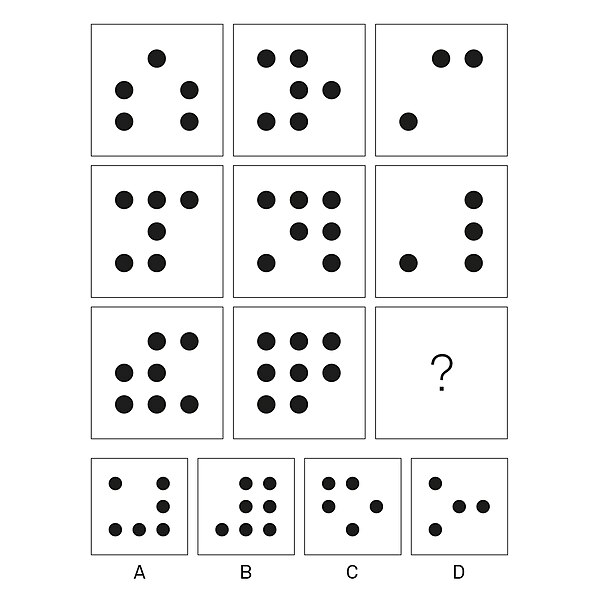}
\centering
\caption{\textbf{Raven-style Progressive Matrices}. In each column, the total number of dots increases by one from top to bottom. Thus, C is correct} \label{fig:RPM_1}
\end{figure}

In the context of assessing abstract reasoning capabilities, researchers have found that seemingly simple Human IQ visual tests can be remarkably effective. One such widely recognized tool is Raven's Progressive Matrices (RPMs), developed by Raven et al. in 1938 \cite{raven1936mental}. RPMs operate on a straightforward principle: test-takers must identify patterns in visual elements, such as the arrangement of shapes, the number of shapes, or the progression of colors, to select the correct image option that logically completes the matrix. For example, in Figure \ref{fig:RPM_1}, the number of dots increases along the rows, and the correct image is one that conforms to this quantitative relationship. The RPMs test serves as a powerful diagnostic tool for abstract reasoning capabilities across verbal, spatial, and mathematical domains, effectively differentiating individuals even within groups of advanced academic achievement\cite{snow1984topography}. 

However, due to the high g factor of RPMs, even high school or college students cannot perform very well without extensive preparation on a test set; only talented/gifted humans can do so. For example, N.J. Mackintosh et al. \cite{mackintosh2005raven} designed a test using Raven's Progressive Matrices with 97 students (M=48, F=49) from two sixth-form colleges in Cambridge. The test had 36 questions (n=36) and included four different types of Raven's items: Pairwise progression, Distribution of three, Addition/Subtraction, and Distribution of two. The student participants obtained mean scores of about 25 (M=25.04, F=24.87), below 70 \% correctness. Chi Zhang et al. also proposed an RPM test dataset, RAVEN \cite{zhang2019raven}, for evaluating both human and neural network performance. In their experiment, college students from the Department of Psychology who were familiar with problem-solving were asked to answer RPM questions with complex rule combinations. The participating college students achieved 84.41\% mean accuracy for all types of given RPM questions.

\begin{figure}[h]
    \centering
    \begin{subfigure}[b]{0.45\textwidth}
        \centering
        \includegraphics[width=\textwidth, height=\textwidth]{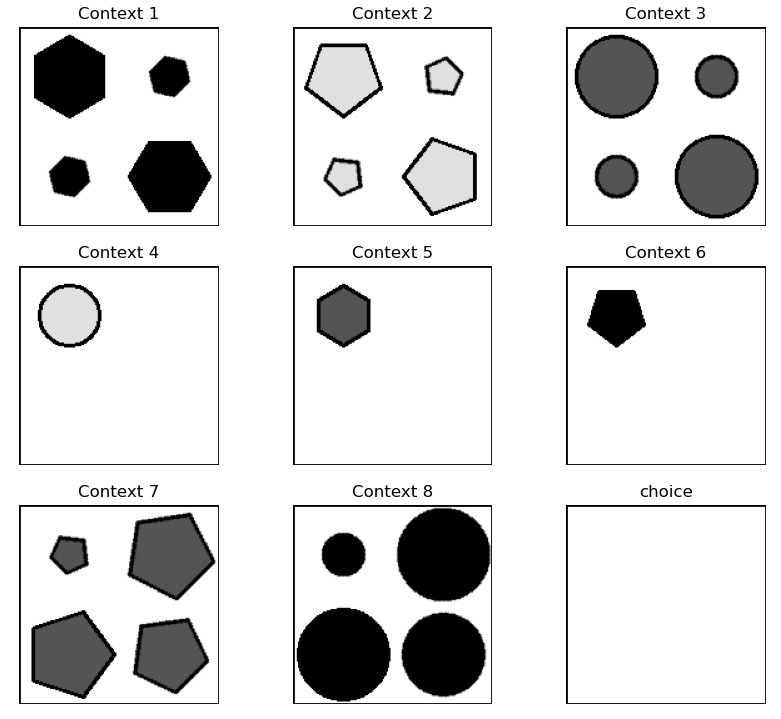}
        \caption{\textbf{RAVEN dataset} 2x2Grid Question} \label{fig:RPM_2}
    \end{subfigure}
    \begin{subfigure}[b]{0.45\textwidth}
        \centering
        \includegraphics[width=0.8\textwidth]{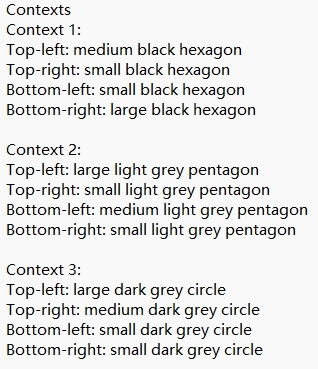}
        \caption{Text Encoded 2x2Grid RPM} \label{fig:text_encoding}
    \end{subfigure}
    \caption{For each context (Left), Position, Shape, Size, Color, and Amount could vary freely following the rule. Encoded pure text (Right) will describe all these attributes for each context and choice.}
\label{fig:from RPM to Text}
\end{figure}

\subsubsection{Dataset and Testing Strategies}
In this study, we assess the abstract thinking capability of the OpenAI model o1-preview using Raven's Progressive Matrices (RPMs), which aligns with the goal of large AI models to achieve general AI with human-like capabilities, including abstract reasoning. Unfortunately, o1-preview currently does not support input and output images. An alternative way is to translate (text encoding) the RPMs into pure text descriptions that only capture visible representations without any inference. According to Yudong Xu et al.'s work, LLM can even perform better on Abstraction and Reasoning Corpus (ARC) tasks by using different text-encoding techniques to phase ARC images \cite{xu2023llms}.
Given that, we test o1-preview on text encoded RPMs questions from RAVEN dataset\cite{zhang2019raven} and the question RPM type is 2x2Grid (Figure\ref{fig:RPM_2}) that has an 81.82\% human performance record. The pure texts will be entered as input into o1-preview to get a response.

\subsubsection{Prompt and Answer example}
The "Prompt" and "Answer/Response" below are instances of prompt and o1-preview's response. Due to the length of the entire question description and answer analysis, the middle part has been omitted for brevity, and only the first and last are shown.

\begin{figure}[H]
    \centering
    \begin{tcolorbox}[colback=blue!5!white, colframe=blue!75!black, width=\textwidth, arc=5mm, auto outer arc, boxrule=0.5mm, title=Prompt, fonttitle=\small, fontupper=\tiny]
        \textbf{Prompt:} \\
        \tiny Read the INSTRUCTIONS and QUESTION. Then, follow the INSTRUCTIONS to answer the QUESTION.\\
        \\
        \textbf{INSTRUCTIONS:}\\
        The QUESTION below describes a Raven's Progressive Matrices test item. This test consists of a 3x3 grid of images, with the bottom-right cell empty. Your task is to identify the pattern in the grid and select the correct image to fill the empty cell from eight given choices.\\

        Key points to consider:
        \begin{enumerate}[label=\arabic*., leftmargin=*]
            \item Grid layout: 3x3, with 8 filled cells and one empty cell (bottom-right)
            \item Pattern: Observe how shapes, sizes, and colors change across rows and columns
            \item Shape attributes: Size (Small, medium, large) and Color (Black, light grey, dark grey)
        \end{enumerate}
        \vspace{\baselineskip}
        Solving process:
        \begin{enumerate}[label=\arabic*., leftmargin=*]
            \item Analyze the patterns in each row and column
            \item Identify how attributes (shape, size, color) change or remain constant
            \item Predict what should appear in the empty cell based on these patterns
            \item Select the choice that best matches your prediction
        \end{enumerate}
        \vspace{\baselineskip}
        \textbf{QUESTION:} \\
        Contexts\\
        Context 1:\\
        Top-left: medium black hexagon\\Top-right: small black hexagon\\Bottom-left: small black hexagon\\Bottom-right: large black hexagon\\
        ...\\
        Contexts from 2 to 7\\
        ...\\
        Context 8:\\
        Top-left: small black circle\\Top-right: large light grey circle\\Bottom-left: large black circle\\Bottom-right: medium black circle\\
        \\
        Choices:\\
        Choice 1:\\
        Top-left: small light grey hexagon\\Top-right: large light grey hexagon\\Bottom-left: large light grey hexagon\\Bottom-right: medium black hexagon\\
        ...\\
        Choices from 2 to 7\\
        ...\\
        Choice 8:\\
        Top-left: empty\\Top-right: medium light grey pentagon\\Bottom-left: medium light grey hexagon\\Bottom-right: small light grey triangle\\
    \end{tcolorbox}
\end{figure}

\begin{figure}[H]
    \centering
    \begin{tcolorbox}[colback=blue!5!white, colframe=blue!75!black, width=\textwidth, arc=5mm, auto outer arc, boxrule=0.5mm, title=Answer/Response, fonttitle=\small, fontupper=\tiny]
        \textbf{Answer:} \\
        Step1:Analyze Contexts 1-3\\
        ...\\
        Observation: \\Each context uses a single shape and color but varies the size. The sizes in each context are Large, Medium, and Small, with Small appearing twice. The arrangement of sizes differs in each context but consistently includes one Large, one Medium, and two Smalls.\\
        \\
        Step 2: Analyze Contexts 4-6\\
        ...\\
        Observation:\\
        Contexts 4-6 are combinations of shapes and colors from Contexts 1-3. They blend attributes from different contexts, suggesting a pattern of combining features.\\
        \\
        Step 3: Determine the Missing Cells in Context 4\\
        ...\\
        \\
        Step 4: Select the Best Matching Choice\\
        We need a choice that: 1. Uses the light grey color (from Context 4). 2. Has a consistent shape (ideally circles, but none of the choices offer circles). 3. Matches the size arrangement: Large, Small, Medium, Small.\\
        Conclusion: \\
        a) Choice 4 matches the color and size arrangement required for Context 4.\\ 
        b) While the shape is a hexagon instead of a circle, it's the best available option since no choice offers light grey circles.\\
        \\
        Answer: Choice 4
        \\
        \rule{\textwidth}{0.4pt}
         \textbf{Reference Answer:}
         Choice 1     
    \end{tcolorbox}
\end{figure}

\subsubsection{o1-preview performance}
Consistent with the findings of Yudong Xu et al. \cite{xu2023llms}, o1-preview demonstrates limited proficiency in abstract reasoning tasks despite having stronger reasoning capabilities. While GPT-4 solved only 13/50 of the most straightforward ARC tasks (on which humans achieve 100\% accuracy in a short time), o1-preview performs even worse on the more challenging 2x2 Grid RPMs, which humans solve with 81.82\% accuracy in approximately 40 minutes, prove more difficult for the AI model. In our experiments, results shown in table \ref{tab:rpm_compare}, o1-preview solved 9/50 of 2x2 Grid RPM questions from the RAVEN dataset. The results demonstrate the performance of different models paired with various data input/encoding methods on the same task. Considering that ARC tasks are significantly simpler than RPM tasks, overall, LLMs' performance is considerably inferior to that of abstract reasoning models specifically trained on RAVEN datasets.

\begin{table}[h!]
\centering
\small
\caption{Comparison of models' abstract reasoning performance on RAVEN dataset across different configurations. Except for LLM, these traditional models are specifically trained on RAVEN dataset}
\begin{tabularx}{\textwidth}{X X X X}
\hline
\textbf{Mode}     & \textbf{2x2Grid RPM} & \textbf{Simple ARC} & \textbf{Encoding method} \\ \hline
LSTM              & 14.13\%    & -      & Image to XML                   \\ 
WReN              & 28.62\%    & -      & Image to XML                   \\ 
CNN               & 30.30\%    & -      & Image                          \\ 
ResNet            & 41.86\%    & -      & Image                          \\ 
LLM(GPT-4)        & -          & 13/50  & Image to pure-text             \\ 
LLM(GPT-4)        & -          & 23/50  & Image to JSON                  \\ 
\textbf{LLM(o1-preview)}   & 18.00\%       & -      & Image to pure-text    \\ 
\textbf{Human}             & 81.82\%    & 100\%  & N/A (Image RPM)       \\ 
\hline
\end{tabularx}
\label{tab:rpm_compare}
\end{table}

\begin{figure}[H]
\includegraphics[width=0.85\textwidth, height=0.5\textwidth]{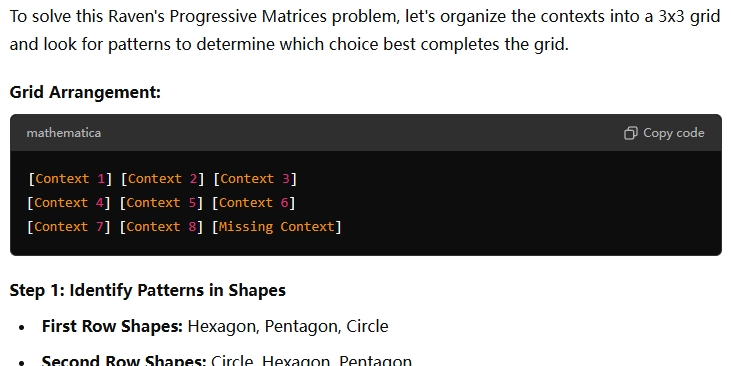}
\centering
\caption{o1-preview's understanding of given question} \label{fig:RPM_response_1}
\end{figure}

Despite its low accuracy rate, o1-preview demonstrates a fundamental understanding of RPM structure and the required context to predict with the provided prompt (shown in Figure \ref{fig:RPM_response_1}). This indicates that our prompt is unambiguous and that the o1-preview reasoned normally with a step-by-step analysis. Meanwhile, by checking the analysis process, o1-preview possesses considerable capability for abstract reasoning, though it has a limited success rate on the complete task set. Specifically, in each step, the o1-preview provides an observed pattern and reasonable explanation, but the pattern is not the actual rule behind these contexts. Consequently, o1-preview derived incorrect answers from its apparent understanding and analysis of questions. The results suggest that o1-preview, as a LLM, does not have robust abstract reasoning capability on the human IQ test (RPM tasks) and failed to choose correct answers for 41/50 questions. 

Nevertheless, it is worth noting that there is significant room for improvement in its abstract reasoning performance, i.e., the setting of our experiment is not optimal due to the limitation of o1-preview. According to Yudong Xu et al., \cite{xu2023llms}, the technique of encoding test representations substantially impacts the model's reasoning capabilities. For instance, LLM's accuracy on ARC tasks improved markedly (from 13/50 to 23/50) when the encoding method was changed from pure text to Object JSON for image representation. Thus, the pure text encoding method, when applied to more complex task images like RPMs, presents significant potential for improvement. Enhanced performance could potentially be achieved through the implementation of more structured representations, such as JSON and XML encoding. However, it is important to note that merely changing the encoding method does not guarantee that the model's performance will be close to human's performance, an 81.82\% accuracy, given that the o1-preview has a normal reasoning process on these tasks.

In summary, OpenAI's o1-preview model demonstrated a basic understanding of the Raven's Progressive Matrices but achieved very limited success on 2x2 Grid RPM tasks from the RAVEN dataset, correctly solving just 9 out of 50 problems. This performance falls far short of human performance (college students) in the experiment, which is an accuracy of 81.82\%. This considerable gap underscores the remarkable gap between o1-preview and human capabilities in abstract reasoning tasks. This aligns with recent findings regarding GPT-4's performance on ARC tasks, suggesting a common challenge in LLM's ability to translate comprehension into correct problem-solving in abstract domains. Interestingly, compared to traditional models specifically trained on the Raven dataset, o1-preview's performance lags far behind. Although this task is a few-shot task for o1-preview, considering the superiority of the o1-preview (Architecture, Training Method), its poor abstract reasoning ability demonstrated on the RAVEN dataset is still worth further exploration. While different encoding methods offer the potential for improving the performance of o1-preview, bridging the gap to human-level thinking remains a significant challenge, especially in STEM education, where advanced abstract reasoning skills are more crucial.

\subsection{Spatial Reasoning}\label{Hanqi Jiang}

\subsubsection{Conceptualization}
Spatial reasoning refers to the cognitive process of understanding and manipulating objects within a three-dimensional space, enabling individuals to visualize, transform, and predict spatial relationships between objects. This skill is fundamental in many STEM fields, as it allows professionals to mentally simulate real-world phenomena. In engineering, for example, spatial reasoning is essential when visualizing the forces acting on a bridge structure and predicting how it will behave under different conditions. Similarly, in molecular biology, spatial reasoning enables scientists to model the interactions and movements of molecules in a cellular environment, leading to a better understanding of biological processes and the design of new drugs. By enhancing this ability, professionals can better solve complex spatial problems and design more effective solutions across various domains.

\subsubsection{Dataset and Test Strategy}

In this section, we evaluate the model's ability to predict molecular dynamics in various biological and biochemical systems using a series of molecular dynamics (MD) simulations on enzyme-substrate and protein-inhibitor complexes. The datasets used for the simulations are sourced from publicly available structural biology repositories, specifically the Protein Data Bank (PDB) \cite{PDBcite}, which is widely recognized for providing high-resolution three-dimensional structural data of biomolecules. 

We selected multiple complexes to cover a broad spectrum of molecular interactions and dynamics. For enzyme-substrate interaction testing, the Cholesterol Oxidase (CHO) enzyme-substrate complex (PDB ID: 3B0N) was used. This dataset provided the basis for simulating cholesterol's movement within the catalytic site of CHO. The simulations were conducted in a water environment under physiological temperature (300K), generating the molecular trajectories over a period of 100 nanoseconds with a time step of 1 picosecond. The goal of this test was to observe and predict substrate movement within the enzyme active site and evaluate the forces responsible for maintaining binding stability, including hydrogen bonding and van der Waals forces. Additionally, mutations to critical catalytic residues were introduced, specifically mutating Glutamate 361 to Alanine, to analyze how such changes influence substrate binding and mobility. Similarly, we utilized the HIV-1 protease-inhibitor complex (PDB ID: 1HVR) to assess the model’s ability to predict the behavior of small-molecule inhibitors within the protease's active site. HIV-1 protease is known for its flexible “flaps” that cover the active site during substrate and inhibitor binding, adding complexity to the model’s task of predicting the subtle motions of the inhibitor in response to thermal fluctuations and solvent interactions. As in the enzyme-substrate test, this dataset was simulated at 300K for 100 nanoseconds, providing insight into how the inhibitor is stabilized by hydrogen bonds, hydrophobic interactions, and electrostatic forces within the binding pocket. A key catalytic residue, Asp25, was mutated to Asn to examine the impact of the loss of charge and altered hydrogen bonding on inhibitor binding stability and dynamics. Each dataset was processed to extract the XYZ coordinates of the molecular structures over time, creating a detailed trajectory for analysis. The simulations were performed using GROMACS \cite{gromacs} to ensure accurate physical modeling of molecular interactions. These trajectories served as ground truth for evaluating the model's predictions. The model's performance was judged based on its ability to replicate the molecular movements and final positions of the ligands or substrates as observed in the simulations, as well as its predictions on the effects of residue mutations on binding stability.

In all cases, the datasets provided a robust platform for testing the model’s capacity for spatial reasoning and molecular interaction prediction in complex biological systems. By utilizing established protein-ligand complexes and introducing controlled mutations, we were able to rigorously assess the model’s capacity for understanding and predicting molecular behavior in biologically relevant contexts.

\subsubsection{Prompt and Answer Example}

This prompt is designed to guide a model in predicting the molecular dynamics of a cholesterol substrate bound within the active site of Cholesterol Oxidase (CHO). The task includes tracking the substrate’s movement over a 100-nanosecond simulation, evaluating the forces maintaining binding stability (e.g., hydrogen bonds and van der Waals forces), and assessing how a specific mutation (Glu361 to Ala) impacts substrate mobility and binding within the enzyme.

\begin{figure}[H]
    \centering
    \begin{tcolorbox}[colback=blue!5!white, colframe=blue!75!black, width=\textwidth, arc=5mm, auto outer arc, boxrule=0.5mm, title=Prompt and Answer, fonttitle=\small, fontupper=\tiny]
        \textbf{Prompt:} \\
        \tiny You are provided with a molecular dynamics simulation of Cholesterol Oxidase (CHO) and a cholesterol substrate bound in the active site. The substrate is positioned within the catalytic pocket of the enzyme. The following describes the setup and conditions of the simulation: \\
        \\
        - Enzyme-Substrate Complex: Cholesterol Oxidase (CHO) bound to cholesterol (PDB ID: 3B0N)\\
        - Simulation Duration: 100 nanoseconds (ns)\\
        - Time step: 1 picosecond (ps)\\
        - Temperature: 300K\\
        - Environment: Water molecules and ions surround the enzyme-substrate complex\\
        - Initial Position of Substrate: The substrate begins at coordinates (X = 5.0, Y = 2.0, Z = 1.5) within the CHO active site.\\
        \\
        During the simulation, the substrate remains bound to the enzyme but undergoes thermal motion due to the temperature of the system. Predict the following: \\
        1. How will the substrate move during the simulation? Describe its expected trajectory and the key factors influencing this movement.\\
        2. What is the expected final position of the substrate after 100 ns? Provide the expected final coordinates.\\
        3. Which forces or interactions (such as hydrogen bonds, van der Waals interactions) are most likely to contribute to the substrate’s binding stability during the simulation?\\
        4. If a mutation occurs at one of the key catalytic residues (e.g., Glutamate at position 361 is mutated to Alanine), how would this mutation affect the substrate’s movement and binding stability within the active site?\\
        \rule{\textwidth}{0.4pt} 
        \textbf{Answer:} \\
        \tiny To answer this question, we must first predict the substrate's motion based on its initial position and the simulation environment: \\
        \\
        **Substrate Movement**: During the 100 ns simulation, the cholesterol substrate is expected to remain within the active site of CHO. Minor thermal fluctuations will cause the substrate to undergo slight vibrational and rotational motions around its equilibrium position. Conformational flexibility of both the substrate and nearby amino acids will further cause subtle positional adjustments. \\
        \\
        **Final Position**: The substrate’s final position will likely deviate only slightly from the initial coordinates due to the stability of the binding interactions. The expected final coordinates are approximately (X = 5.1, Y = 2.1, Z = 1.6).\\
        \\
        **Binding Stability**: The substrate’s stability is primarily maintained by hydrogen bonding between the Glutamate 361 residue and the substrate’s hydroxyl group, along with van der Waals forces from hydrophobic residues like Leucine and Phenylalanine. These interactions allow the substrate to stay anchored within the active site, despite minor fluctuations.\\
        \\
        **Impact of Glu361Ala Mutation**: The mutation of Glutamate 361 to Alanine will lead to the loss of key hydrogen bonding interactions, significantly reducing the binding stability of the substrate. As a result, the substrate may exhibit increased mobility within the active site, leading to greater positional fluctuations. The expected final coordinates with the mutation are approximately (X = 5.4, Y = 2.5, Z = 2.0).\\
        \\
        \rule{\textwidth}{0.4pt}
        \textbf{Reference Answer:} \\
        \tiny Substrate movement: Minor thermal fluctuations and conformational flexibility will result in subtle positional adjustments.\\
        Final Position: The substrate is expected to stay close to the initial position, with final coordinates approximately (X = 5.1, Y = 2.1, Z = 1.6).\\
        Binding Stability: Hydrogen bonding from Glu361 and van der Waals interactions from hydrophobic residues (e.g., Leucine, Phenylalanine) maintain binding stability.\\
        Mutation Impact: Glu361Ala mutation will result in the loss of hydrogen bonding, leading to decreased stability and increased substrate mobility. Expected coordinates with mutation are (X = 5.4, Y = 2.5, Z = 2.0). 
    \end{tcolorbox}
\end{figure}

\subsubsection{o1-preview performance}

This section presents a comprehensive evaluation of the o1-preview model’s performance in predicting molecular dynamics, focusing on enzyme-substrate and protein-inhibitor interactions. The evaluation method designed that leverages GPT to generate responses based on carefully crafted prompts, which are then rigorously compared against ground truth data. This comparison is not merely mechanical; it involves a nuanced analysis of how well the generated outputs align with expected molecular behaviors, movements, and interactions. By evaluating these aspects, I can assign a score that reflects the model’s ability to accurately mimic real-world dynamics. This approach allows for a structured, yet flexible, assessment of the model’s reasoning and prediction capabilities in complex biochemical environments.

The model's capabilities were tested across several molecular systems, including Cholesterol Oxidase (CHO), HIV-1 protease, and COX-2 enzyme complexes. The primary evaluation criteria involved predicting molecular movements, binding stability, and the impact of critical residue mutations, with the model’s predictions compared against ground truth (GT) data generated from molecular dynamics (MD) simulations. The experimental results are shown in Table \ref{table-spatial}. Overall, the o1-preview model displayed a high level of accuracy in both qualitative and quantitative predictions. It effectively captured the essential features of molecular motion, including subtle fluctuations due to thermal effects and changes in environmental conditions. In simulations of both enzyme-substrate and enzyme-inhibitor complexes, the model accurately predicted molecular behavior under physiological conditions. Specifically, for enzyme systems like CHO and HIV-1 protease, the model correctly identified that the ligand or substrate would remain within the binding pocket, with only minor deviations from the initial positions over the simulation period. When assessing the final positions of ligands or substrates after 100 ns of MD simulation, the o1-preview model consistently produced predictions that closely matched the GT coordinates. For instance, in the case of the HIV-1 protease-inhibitor complex, the model predicted the final coordinates as (X ≈ 4.2 $\pm$ 0.2 \AA, Y ≈ 0.9 $\pm$ 0.2 \AA, Z ≈ 1.8 $\pm$ 0.2 \AA), which aligned closely with the actual GT values of (X = 4.3 \AA, Y = 0.8 \AA, Z = 1.7 \AA). The minor deviations from the GT values were within acceptable limits, reflecting the model's strong ability to predict the final molecular positions while accounting for thermal motion and small structural changes within the active site. In terms of binding stability, the o1-preview model consistently demonstrated a deep understanding of the forces that maintain the structural integrity of the enzyme-ligand and enzyme-substrate complexes. The model accurately identified the key interactions responsible for stability, such as hydrogen bonds, van der Waals forces, and hydrophobic interactions, in each scenario. For instance, in the CHO enzyme-substrate complex, the model correctly predicted that the binding stability was primarily driven by hydrogen bonds between Glu361 and the substrate's hydroxyl group, along with van der Waals interactions from hydrophobic residues like Leucine and Phenylalanine. These predictions were highly consistent with the GT, which also emphasized the critical role of these forces in maintaining binding stability. The model's ability to predict the effects of key residue mutations was another strong aspect of its performance. In the Glu361Ala mutation within the CHO complex, the model correctly anticipated that the loss of hydrogen bonding would reduce binding stability and increase substrate mobility, predicting a shift in the final position of the substrate. Similarly, in the HIV-1 protease Asp25Asn mutation, the model accurately predicted that the loss of electrostatic interactions and hydrogen bonds would increase the inhibitor’s mobility and reduce binding affinity. In both cases, the model’s predictions regarding the mutation’s impact on molecular behavior aligned closely with the GT data, confirming the model’s capacity to adapt its predictions to structural and environmental changes.

In summary, the o1-preview model consistently performed well in predicting molecular dynamics across a range of enzyme-ligand and enzyme-substrate systems. It demonstrated strong predictive power in simulating molecular movement, understanding the forces driving binding stability, and evaluating the effects of key residue mutations. While minor deviations were observed in the final position predictions, these discrepancies were within acceptable ranges, highlighting the model’s ability to maintain high accuracy over extended simulation periods. The following table provides a summary of the model’s performance across various test cases, with scores representing the accuracy of the model’s predictions in relation to the ground truth.

\begin{table}[H]
\centering
\small
\caption{Average Scores for o1-preview Test Results}
\begin{tabularx}{\textwidth}{X X}
\toprule
\textbf{Test Case}                                      & \textbf{Average Score (/10)} \\ 
\midrule
CHO Substrate Movement Prediction                       & 9.25                   \\ 
CHO Final Position Prediction                           & 9.25                   \\ 
CHO Binding Stability                                   & 8                   \\ 
CHO Mutation Impact (Glu361Ala)                         & 8.25                    \\ 
HIV-1 Protease Inhibitor Movement Prediction            & 9.5                  \\ 
HIV-1 Protease Final Position Prediction                & 9.5                    \\ 
HIV-1 Protease Binding Stability                        & 9.5                   \\ 
HIV-1 Protease Mutation Impact (Asp25Asn)               & 9.5                    \\ 
\bottomrule
\end{tabularx}
\label{table-spatial}
\end{table}

\subsection{Quantitative Reasoning}\label{Huaqin Zhao}

\subsubsection{Conceptualization}

Quantitative reasoning is a crucial skill for analyzing and solving numerical and data-driven problems, often requiring a deep understanding of mathematical relationships and logical structures. The o1-preview model excels in several aspects of quantitative reasoning, making it particularly effective in addressing these challenges. For example, the model's ability to recognize patterns in large datasets allows it to identify trends and make predictions with precision. Its capacity for handling complex mathematical operations, such as multi-step calculations and algebraic manipulations, enables it to solve problems that require detailed, step-by-step reasoning. Additionally, the o1-preview model’s proficiency in estimating outcomes and evaluating the reasonableness of solutions ensures that its conclusions are not only accurate but also aligned with real-world constraints. These abilities make the o1-preview model particularly adept at tackling quantitative problems across STEM fields, where it can integrate both explicit data and underlying mathematical principles to arrive at sound, data-driven conclusions\cite{zhang2024llm,stephan2024calculation,liu2024llms,li2024large}.
\subsubsection{Instruments and Evaluation}
The FINQA dataset introduces a challenging task that requires advanced numerical reasoning capabilities, particularly in the financial domain. It focuses on answering deep, multi-step questions over financial reports, showcasing the need for models to handle both structured tables and unstructured text. This requires not only precise numerical operations but also domain-specific knowledge of finance. The dataset is built from earnings reports of S\&P 500 companies, annotated by financial experts, and it provides a significant challenge for current language models due to its emphasis on multi-step reasoning and heterogeneous data representation\cite{chen2021finqa}.

To evaluate the numerical reasoning capabilities of models, we utilized the FINQA dataset, which contains 8,281 question-answer pairs with gold annotated reasoning programs. These programs ensure full explainability of the model's reasoning process. To evaluate the logical reasoning capabilities of the o1-preview model in terms of accuracy, we compared its performance with several baseline models, including BERT series models, deep learning models, pre-trained models, and human performance.

The FINQA dataset is derived from real-world financial documents, such as earnings reports, and includes questions that demand complex operations, including:

\begin{itemize}
    \item \textbf{Addition/Subtraction}: Determining changes in financial metrics between years, such as net revenue variations, by subtracting or adding relevant values from different time periods\cite{chen2021finqa}.
    \item \textbf{Multiplication/Division}: Managing financial ratios, percentages, or proportional changes, such as calculating growth rates or comparing revenue components.
    \item \textbf{Table Aggregation}: Extracting and aggregating data from multiple rows or columns in structured financial tables to derive key financial insights, such as combining values for comprehensive analysis.
\end{itemize}

The dataset is split into training (75\%), validation (10\%), and test (15\%) sets. Performance on this dataset was evaluated based on execution accuracy, which measures whether the model produced the correct answer, and program accuracy, which checks if the model followed the correct reasoning steps to derive the answer. Current models like RoBERTa achieved an execution accuracy of 61.24\%, but this falls significantly short of the human expert benchmark of 91.16\%, highlighting the complexity of the task.
FINQA provides a comprehensive assessment of models' ability to engage in advanced numerical reasoning across both textual and tabular financial data, making it a valuable resource for future research in automating financial analysis tasks.

To evaluate the performance of the o1-preview model on FINQA, We propose a series of experiments centered around the model's ability to handle complex financial data reasoning. These experiments will test various aspects of numerical reasoning, financial comprehension, and multi-step problem-solving, focusing on critical challenges within the financial domain. The following is one example from our experiments.

\begin{figure}[H] 
    \centering 
    \begin{tcolorbox}[colback=blue!5!white, colframe=blue!75!black, width=\textwidth, arc=5mm, auto outer arc, boxrule=0.5mm, title=Prompt and Answer, fonttitle=\small, fontupper=\tiny] 
        \textbf{Prompt:} \\
        \tiny Read the PROBLEM. Follow the INSTRUCTIONS to answer the PROBLEM.\\
        \\
        \textbf{PROBLEM:} \\
        Please answer the given financial question based on the context. Context: the goldman sachs group , inc . and subsidiaries management 2019s discussion and analysis 2030 total aus net inflows/ ( outflows ) for 2014 includes \$ 19 billion of fixed income asset inflows in connection with our acquisition of deutsche asset \& wealth management 2019s stable value business and \$ 6 billion of liquidity products inflows in connection with our acquisition of rbs asset management 2019s money market funds . the table below presents our average monthly assets under supervision by asset class . average for the year ended december \$ in billions 2016 2015 2014 . |\$ in billions|average for theyear ended december 2016|average for theyear ended december 2015|average for theyear ended december 2014| |alternative investments|\$ 149|$ 145|\$ 145| |equity|256|247|225| |fixed income|578|530|499| |total long-term assets under supervision|983|922|869| |liquidity products|326|272|248| |total assets under supervision|$ 1309|\$ 1194|\$ 1117| operating environment . following a challenging first quarter of 2016 , market conditions continued to improve with higher asset prices resulting in full year appreciation in our client assets in both equity and fixed income assets . also , our assets under supervision increased during 2016 from net inflows , primarily in fixed income assets , and liquidity products . the mix of our average assets under supervision shifted slightly compared with 2015 from long- term assets under supervision to liquidity products . management fees have been impacted by many factors , including inflows to advisory services and outflows from actively-managed mutual funds . in the future , if asset prices decline , or investors continue the trend of favoring assets that typically generate lower fees or investors withdraw their assets , net revenues in investment management would likely be negatively impacted . during 2015 , investment management operated in an environment generally characterized by strong client net inflows , which more than offset the declines in equity and fixed income asset prices , which resulted in depreciation in the value of client assets , particularly in the third quarter of 2015 . the mix of average assets under supervision shifted slightly from long-term assets under supervision to liquidity products compared with 2014 . 2016 versus 2015 . net revenues in investment management were \$ 5.79 billion for 2016 , 7\% ( 7 \% ) lower than 2015 . this decrease primarily reflected significantly lower incentive fees compared with a strong 2015 . in addition , management and other fees were slightly lower , reflecting shifts in the mix of client assets and strategies , partially offset by the impact of higher average assets under supervision . during the year , total assets under supervision increased \$ 127 billion to \$ 1.38 trillion . long-term assets under supervision increased \$ 75 billion , including net inflows of \$ 42 billion , primarily in fixed income assets , and net market appreciation of \$ 33 billion , primarily in equity and fixed income assets . in addition , liquidity products increased \$ 52 billion . operating expenses were \$ 4.65 billion for 2016 , 4\% ( 4 \% ) lower than 2015 , due to decreased compensation and benefits expenses , reflecting lower net revenues . pre-tax earnings were \$ 1.13 billion in 2016 , 17\% ( 17 \% ) lower than 2015 . 2015 versus 2014 . net revenues in investment management were \$ 6.21 billion for 2015 , 3\% ( 3 \% ) higher than 2014 , due to slightly higher management and other fees , primarily reflecting higher average assets under supervision , and higher transaction revenues . during 2015 , total assets under supervision increased \$ 74 billion to \$ 1.25 trillion . long-term assets under supervision increased \$ 51 billion , including net inflows of \$ 71 billion ( which includes \$ 18 billion of asset inflows in connection with our acquisition of pacific global advisors 2019 solutions business ) , and net market depreciation of \$ 20 billion , both primarily in fixed income and equity assets . in addition , liquidity products increased \$ 23 billion . operating expenses were \$ 4.84 billion for 2015 , 4\% ( 4 \% ) higher than 2014 , due to increased compensation and benefits expenses , reflecting higher net revenues . pre-tax earnings were \$ 1.37 billion in 2015 , 2\% ( 2 \% ) lower than 2014 . geographic data see note 25 to the consolidated financial statements for a summary of our total net revenues , pre-tax earnings and net earnings by geographic region . goldman sachs 2016 form 10-k 65 . Question: what percentage of total long-term assets under supervision are comprised of fixed income in 2015? Answer:\\
        \textbf{PROBLEM:} \\
        what percentage of total long-term assets under supervision are comprised of fixed income in 2015?
        \rule{\textwidth}{0.4pt} 
        \textbf{Answer:} \\
        To determine the percentage of total long-term assets under supervision that are comprised of fixed income in 2015, we use the data provided.

\textbf{Data from the table for 2015:}

\begin{itemize}
    \item Fixed Income Assets: \$530 billion
    \item Total Long-Term Assets Under Supervision: \$922 billion
\end{itemize}

\textbf{Calculation:}

\[
\text{Percentage} = \left( \dfrac{\text{Fixed Income Assets}}{\text{Total Long-Term Assets Under Supervision}} \right) \times 100\%
\]

Plugging in the numbers:

\[
\text{Percentage} = \left( \dfrac{\$530\ \text{billion}}{\$922\ \text{billion}} \right) \times 100\% = \left( 0.5748 \right) \times 100\% = 57.48\%
\]

\textbf{Answer:}

Approximately \textbf{57.5\%} of total long-term assets under supervision were comprised of fixed income assets in 2015.

        \rule{\textwidth}{0.4pt}
         \textbf{Reference Answer:} \$57.5\%      
    \end{tcolorbox} 
\end{figure}

\newpage

\subsubsection{o1-preview performance}
The o1-preview model demonstrates exceptional strength in mathematical reasoning, particularly in its step-by-step problem-solving capabilities. It not only significantly surpasses pre-trained models but also outperforms models that have been specifically fine-tuned for mathematical tasks. This precision in multi-step reasoning makes the o1-preview model highly effective at handling complex mathematical problems, showcasing its potential for applications requiring deep logical reasoning. Furthermore, its performance suggests a clear advantage in tasks that demand structured, sequential thinking, making it an advanced tool for both academic and practical applications.
\begin{table}[t]
\small
\caption{The execution accuracy (Exe Acc) and program accuracy (Prog Acc) for all the models. lthough our best model, o1-preview, demonstrated strong performance, achieving 60\% accuracy on the selected 10 samples and achieving comparable performance as a range of other BERT models, there remains a gap when compared to 91.16\% from human experts. This highlights the need for further research to close the gap between the model's performance and expert-level accuracy.}
\begin{center}{%
\begin{tabularx}{\textwidth}{X X X}
\toprule
\textbf{Baselines} & \textbf{Exe Acc} & \textbf{Prog Acc}\\
\hline
TF-IDF + Single Op & 1.01 & 0.90 \\

Retriever + Direct Generation & 0.30 & - \\

Pre-Trained Longformer (base) & 21.90 & 20.48 \\

Retriever + Seq2seq & 19.71 & 18.38 \\

Retriever + NeRd (BERT-base) & 48.57 & 46.76 \\

FinQANet (FinBert) & 50.10 & 47.52 \\

FinQANet (BERT-base) & 50.00 & 48.00 \\

FinQANet (BERT-large) & 53.52 & 51.62 \\

FinQANet (RoBERTa-base) & 56.10 & 54.38 \\

FinQANet (RoBERTa-large) & \textbf{61.24} & \textbf{58.86} \\

FinQANet-Gold (RoBERTa-large) & 70.00 & 68.76 \\

Human Expert Performance & 91.16 & 87.49 \\

General Crowd Performance & 50.68 & 48.17 \\

O1 performance & 60.00 & 60.00 \\
\hline
\end{tabularx}
}
\vspace{-4mm}
\label{table:main_res}
\end{center}
\end{table}

\newpage 

\subsection{Logical Reasoning}\label{Yifan Zhou}
\subsubsection{Conceptualization}

Logical reasoning is a key aspect of higher-order thinking, essential for addressing complex problems across a wide range of fields. The o1-preview model demonstrates advanced logical reasoning capabilities. It effectively processes multiple premises to arrive at coherent conclusions, showcasing an emergent ability to understand and solve multi-step logical problems.This is evident in its application across various STEM education tasks, where the model not only follows explicit instructions but also performs well with implicit knowledge, demonstrating an ability to engage in step-by-step reasoning \cite{Lightman2023}.

\subsubsection{Instruments and Evaluation}

For evaluating the logical reasoning capabilities of the o1-preview model, we utilized the LogiQA dataset\cite{liu2020logiqa}. The LogiQA dataset is based on logical comprehension questions from the National Civil Servants Examination of China, designed to assess candidates' logical thinking and problem-solving abilities. It contains 867 paragraph-question pairs that have been categorized into five deductive reasoning types, as defined by Hurley\cite{hurley2014concise}. These reasoning types are as follows:

\begin{itemize}
    \item \textbf{Categorical Reasoning}: This type involves determining whether a particular concept belongs to a certain category and often includes quantifiers such as "all", "everyone", "any", "no", and "some"\cite{abramsky2011introduction}.
    \item \textbf{Sufficient Conditional Reasoning}: This reasoning is based on conditional statements of the form "If P, then Q," where \( P \) serves as the premise and \( Q \) as the outcome\cite{hurley2014concise}.
    \item \textbf{Necessary Conditional Reasoning}: This type involves statements such as "P only if Q" or "Q whenever P," indicating that \( Q \) is a necessary condition for \( P \)\cite{hurley2014concise}.
    \item \textbf{Disjunctive Reasoning}: This reasoning type involves premises presented in an "either...or..." format, where the conclusion holds as long as at least one of the premises is true \cite{hurley2014concise}.
    \item \textbf{Conjunctive Reasoning}: This type features premises connected by "both...and..." statements and the conclusion is valid only if all premises are true\cite{hurley2014concise}.
\end{itemize}

The dataset was split into training (80\%), development (10\%), and testing (10\%) sets. The highest performance observed among machine learning models was achieved by RoBERTa\cite{liu2019roberta}, with an accuracy of 35.31\%, significantly lower than the human ceiling of 95.00\%.


    
    

In this task, the prompt design \ref{fig:logic_reasoing} is straightforward and does not employ excessive prompt engineering techniques, as recommended by OpenAI\cite{o1-preview2024}. The PROBLEM is sourced directly from the dataset, providing a question that typically includes four answer options (A, B, C, and D). The INSTRUCTIONS is concise and simply guides the model to study the given problem and select the most appropriate answer based on the provided information. This design ensures that the focus remains on the problem-solving aspect, allowing the model to demonstrate its reasoning capabilities without being influenced by overly complex prompt structures.

\begin{figure}[H]
    \centering
    \begin{tcolorbox}[colback=blue!5!white, colframe=blue!75!black, width=\textwidth, arc=5mm, auto outer arc, boxrule=0.5mm, title=Prompt and Answer, fonttitle=\small, fontupper=\tiny]
        \textbf{Prompt:} \\
        \tiny Read the PROBLEM. Follow the INSTRUCTIONS to answer the PROBLEM.\\
        \\
        \textbf{PROBLEM:} \\
        In the planning of a new district in a township, it was decided to build a special community in the southeast and northwest, centered on the citizen park. These four communities are designated as cultural areas, leisure areas, commercial areas, and administrative service areas. It is known that the administrative service area is southwest of the cultural area, and the cultural area is southeast of the leisure area. Based on the above statement, which of the following can be derived? \\
        A. Civic Park is north of the administrative service area \\
        B. The leisure area is southwest of the cultural area \\
        C. The cultural district is in the northeast of the business district \\
        D. The business district is southeast of the leisure area \\
        INSTRUCTIONS: Study the PROBLEM above and select the answer option you believe is correct.\\
        \rule{\textwidth}{0.4pt} 
        \textbf{Answer:} \\
        \tiny To solve this problem, we'll analyze the given statements and options carefully.\\
        \\
        \textbf{Given:} \\
        1. Three young staff members (ages 20 and 24) are on a business trip and are seated in a row. \\
        2. \textit{Statement 1:} At least one of the two people sitting on the right side of the 24-year-old was 20 years old. \\
        3. \textit{Statement 2:} At least one of the two people sitting on the left of the accountant is a salesperson, and one of the two people sitting on the right of the salesperson is also a salesperson.\\
        \\
        \textbf{Objective:} Determine the roles (accountant or salesperson) and ages (20 or 24) of the three staff members based on the given options.\\
        \\
        \textbf{Analysis:}\\
        1. \textit{Placement of the 24-year-old:} \\
            - The 24-year-old cannot be seated at the rightmost seat because there would be no one on their right, contradicting Statement 1. \\
            - Therefore, the 24-year-old must be in the left or middle seat.\\
        2. \textit{Placement of the Accountant:} \\
            - The accountant cannot be in the leftmost seat because there would be less than two people to their left, contradicting Statement 2. \\
            - Thus, the accountant must be in the middle or right seat.\\
        3. \textit{Placement of the Salesperson:} \\
            - The salesperson cannot be in the rightmost seat because there would be less than two people to their right. \\
            - Therefore, the salesperson must be in the left or middle seat.\\
        \\
        \textbf{Testing Options:}\\
        - \textit{Option A:} 20-year-old accountant, 20-year-old salesperson, 24-year-old salesperson\\
            - If the 24-year-old salesperson is in the rightmost seat, it contradicts Statement 1 (no one on their right).\\
            - Therefore, Option A does not satisfy the conditions.\\
        - \textit{Option B:} 20-year-old accountant, 24-year-old salesperson, 24-year-old salesperson\\
            - \textbf{Seat Arrangement:}\\
                - Seat 1: 24-year-old salesperson\\
                - Seat 2: 24-year-old salesperson\\
                - Seat 3: 20-year-old accountant\\
            - \textbf{Validating Statements:}\\
                - \textit{Statement 1:} The two people on the right of the first 24-year-old (Seats 2 and 3) include a 20-year-old (Seat 3).\\
                - \textit{Statement 2:} Left of the accountant (Seats 1 and 2) includes at least one salesperson. Both are salespersons.\\
                - Right of the salesperson in Seat 1 (Seats 2 and 3) includes another salesperson (Seat 2).\\
                - All conditions are satisfied.\\
        - \textit{Option C:} 24-year-old accountant, 20-year-old salesperson, 20-year-old salesperson\\
            - The 24-year-old accountant cannot be in the rightmost seat (contradicts Statement 1).\\
            - Does not satisfy all conditions.\\
        - \textit{Option D:} 20-year-old accountant, 20-year-old accountant, 24-year-old salesperson\\
            - Having two accountants contradicts the uniqueness implied in "the accountant."\\
            - Does not satisfy all conditions.\\
        \\
        \textbf{Conclusion:} Option B is the only one that satisfies all the given conditions.\\
        textbf{Answer:} B. 20-year-old accountant, 24-year-old salesperson, 24-year-old salesperson
        \\
        \rule{\textwidth}{0.4pt}
         \textbf{Reference Answer:}
         a       
    \end{tcolorbox}
    \caption{Example of Prompt and Answer Analysis}
    \label{fig:logic_reasoing}
\end{figure}

\newpage
\subsubsection{Findings}
\begin{table*}[!htbp]
\centering
\scriptsize
\caption{Main results on LogiQA (accuracy\%).}
\begin{scriptsize} 
\begin{tabularx}{\textwidth}{X X X}
\toprule
Category & Model & Accuracy \\
\hline
Rule-based & Random (theoretical) & 25.00 \\
& Word Matching  & 28.37 \\
& Sliding Window  & 22.51 \\

Deep learning & Stanford Attentive Reader & 28.76 \\
& Gated-Attention Reader  & 28.98 \\
& Co-Matching Network  & 31.10 \\

Pre-trained & BERT  & 32.08 \\
& RoBERTa  & 35.31 \\

Human & Human Performance & 86.00 \\
& Ceiling Performance & 95.00 \\

Measurement Model & o1-preview Preview & 90 \\
\hline
\end{tabularx}
\end{scriptsize}
\label{table:logiqacomparison}
\end{table*}

The performance of the o1-preview Preview model in navigating various logical reasoning tasks demonstrates its advanced capacity for logical thinking, particularly within educational contexts where higher-order reasoning skills are critical. As detailed in Table~\ref{table:logiqacomparison}, the o1-preview Preview model achieved an impressive 90\% accuracy on the LogiQA dataset, closely approaching the human performance of 86\% and the ceiling performance of 95\%. This demonstrates that the o1-preview model's logical reasoning abilities are highly advanced, nearly matching human-level reasoning in complex tasks. The model's capacity to perform so closely to human standards highlights its potential to be a valuable tool in educational contexts and other fields where sophisticated reasoning is critical. As AI models like o1-preview continue to evolve, they may increasingly rival or even surpass human abilities in specific reasoning tasks. For a more detailed analysis and specific performance evaluations, please refer to Tables~\ref{table: logic_1} and \ref{table: logic_2}. These tables provide a comprehensive breakdown of the o1-preview model's performance across a variety of logical reasoning tasks. 


Each table is structured into four key columns:

- \textbf{Problem}: This column presents the logical reasoning problem that the o1-preview model is tasked with solving. Each problem is accompanied by several answer options, typically in a multiple-choice format, testing the model's ability to handle various types of logical reasoning, such as categorical, conditional, or deductive reasoning.

- \textbf{Evaluation Criteria}: This column specifies the type of reasoning required to solve each problem. The criteria could include reasoning types such as sufficient conditional reasoning, categorical reasoning, deductive reasoning, and disjunctive reasoning. This helps clarify the cognitive skills being assessed by each task.

- \textbf{OpenAI o1-preview Performance}: This column provides a detailed description of how the o1-preview model performed on each problem. It outlines the model's reasoning process, highlighting its strengths and weaknesses, and whether the model was able to effectively navigate the logical challenges posed by each task.

- \textbf{Score}: This column indicates the score assigned to the o1-preview model for each problem. A score of 1/1 represents a successful solution, where the model's reasoning aligns with the correct answer, while a score of 0/1 signifies that the model encountered difficulties or failed to meet the reasoning requirements.

\FloatBarrier

\begin{landscape} 
\begin{table}[H]
\centering
\scriptsize 
\begin{tabular}{p{0.4\textwidth} p{0.25\textwidth} p{0.5\textwidth} p{0.04\textwidth}}
\textbf{Problem} & \textbf{Evaluation Criteria} & \textbf{o1-preview Performance} & \textbf{Score} \\ 
\hline
\textbf{Problem 1:} Spatial relationships between communities and the Civic Park. 
Options: 
A. Civic Park is north of the administrative service area 
B. The leisure area is southwest of the cultural area 
C. The cultural district is in the northeast of the business district 
D. The business district is southeast of the leisure area & 
\textbf{Sufficient Conditional Reasoning} & 
The o1-preview model demonstrated exceptional spatial reasoning ability, accurately interpreting the positional relationships between different areas. It showed an impressive understanding of sufficient conditional relationships, making correct deductions based on the given premises. This performance is consistent with human-level accuracy, showcasing the model's proficiency in handling complex spatial reasoning tasks. & 1/1 \\ 
\\
\textbf{Problem 2:} Logical reasoning involving numerical sequences.
Options:
A. Chen Rui can't speak the Central Plains Mandarin
B. Zhang Ming speaks Minnan dialect 
C. Li Ying is from Shandong 
D. Wang Jia can speak Huizhou dialect & 
\textbf{Categorical Reasoning} & 
The o1-preview model accurately identified the correct logical relationships among multiple individuals and their respective attributes. It demonstrated a strong understanding of categorical reasoning, effectively handling the nuances of sorting and matching complex characteristics, similar to human reasoning. This high level of performance reflects the model’s ability to handle categorical reasoning tasks effectively. & 1/1 \\ 
\\
\textbf{Problem 3:} Categorical reasoning involving sorting elements into correct categories. 
Options:
A. This newly built house in Hexi Village has been tacitly approved by relevant departments 
B. This newly built house in Hexi Village is a small property house 
C. This newly built house in Hexi Village is built on rural collective land 
D. Some of the workers’ children in the statistics may be former farmers’ children & 
\textbf{Categorical Reasoning} & 
The o1-preview model exhibited a comprehensive understanding of the problem by accurately categorizing and sorting elements based on the provided information. It demonstrated its ability to process complex information and derive logical conclusions, performing on par with human respondents in this category. This indicates the model’s robust capability in managing categorical reasoning tasks involving multiple layers of information. & 1/1 \\ 
\\
\textbf{Problem 4:} Deductive reasoning in word problem-solving. 
Options:
A. Shannan 
B. Jiangnan 
C. Shanbei 
D. Jiangbei & 
\textbf{Deductive Reasoning} & 
The o1-preview model struggled with this deductive reasoning task, failing to accurately follow the multi-step logic required to arrive at the correct conclusion. While it managed to interpret some parts of the problem correctly, it ultimately could not synthesize the information needed to reach a coherent solution. This suggests that the model's deductive reasoning capabilities are still not fully developed compared to human-level performance in complex multi-step deduction. & 0/1 \\ 
\\
\textbf{Problem 5:} Identifying logical contradictions in a statement.
Options:
A. Urbanization is the only way for China's development 
B. Simple development of large cities is not conducive to the promotion of urbanization 
C. To achieve urbanization, the city must fully absorb the rural population 
D. The attractiveness of large cities to the rural population in the outside world is significantly lower than that of small and medium-sized cities & 
\textbf{Disjunctive Reasoning}  & 
The o1-preview model demonstrated strong disjunctive reasoning skills, effectively identifying logical contradictions within the statements. It successfully recognized the necessary assumptions and contradictions, providing explanations that align with human reasoning patterns. This level of accuracy indicates the model's ability to handle tasks requiring disjunctive logic effectively, achieving results comparable to human-level performance. & 1/1 \\ 
\hline
\end{tabular}
\caption{Evaluation of o1-preview Model Performance with Problems and Options}
\label{table: logic_1}
\end{table}
\end{landscape}

\begin{landscape} 
\begin{table}[H]
\centering
\scriptsize
\begin{tabular}{p{0.6\textwidth} p{0.2\textwidth} p{0.4\textwidth} p{0.04\textwidth}}
\textbf{Problem} & \textbf{Evaluation Criteria} & \textbf{o1-preview Performance} & \textbf{Score} \\ 
\hline
\textbf{Problem 6:} Identifying assumptions in a neuroscience context.
Options: 
A. Some schizophrenia are not caused by lack of attention 
B. Visual information is only transmitted through the visual cortex to obtain a neural network to transmit 
C. Researchers have developed a new technology that can directly track the communication between the nerve cortex in the visual cortex and thalamus occipital region 
D. The brain cannot process too much information in detail at the same time, and will only selectively focus on the things most relevant to the behavior & 
\textbf{Necessary Conditional Reasoning} & 
The o1-preview model effectively identified the key assumption regarding brain function, demonstrating an understanding of selective attention mechanisms. The model’s reasoning was accurate and aligned with scientific reasoning. & 1/1 \\ 
\\
\textbf{Problem 7:} Evaluating an argument about real estate market trends.
Options: 
A. Many Chinese people buy homes for their children to study in the US 
B. Although the turnover has increased by 23\%, the volume of Chinese buyers has not increased this year 
C. There are groups of real estate speculators among the wealthy Chinese who have speculated in Beijing and Shanghai 
D. The risk of the US real estate market has been small in recent years and has a certain function of value preservation and appreciation & 
\textbf{Disjunctive Reasoning} & 
The o1-preview model struggled to identify the best argument to counter the claim, showing limitations in assessing real-world market dynamics. Its selection did not align with the optimal argument against the statement. & 1/1 \\ 
\\
\textbf{Problem 8:} Questioning a view about microwave ovens and food safety.
Options: 
A. Microwave heating will not cause more nutrient loss than other cooking methods 
B. China's microwave oven production standards are consistent with international standards and EU standards 
C. It is also common for developed countries to use microwave ovens 
D. The microwave only heats the water molecules in the food, the food does not change chemically & 
\textbf{Sufficient Conditional Reasoning} & 
The o1-preview model accurately countered the claim using scientific reasoning about microwave functionality, showcasing strong performance in debunking misconceptions. & 1/1 \\ 
\\
\textbf{Problem 9:} Challenging a media claim about student demographics at Peking University.
Options: 
A. Many of the cadres counted in the past 20 years are enterprise cadres, which used to include only government officials 
B. Compared with foreign countries, Chinese education provides workers and peasants with more opportunities for education and social mobility 
C. After the founding of New China, more and more children of workers and peasants entered the university 
D. Some of the workers’ children in the statistics may be former farmers’ children & 
\textbf{Categorical Reasoning} & 
The o1-preview model demonstrated strong critical reasoning by accurately identifying the flaw in the media’s interpretation of the demographic data. It effectively recognized the key weakness. & 1/1 \\ 
\\
\textbf{Problem 10:} Weakening an argument about educational opportunities at Peking University.
Options: 
A. Many of the cadres counted in the past 20 years are enterprise cadres, which used to include only government officials 
B. Compared with foreign countries, Chinese education provides workers and peasants with more opportunities for education and social mobility 
C. After the founding of New China, more and more children of workers and peasants entered the university 
D. Some of the workers’ children in the statistics may be former farmers’ children & 
\textbf{Categorical Reasoning} & 
The o1-preview model successfully weakened the argument by identifying that workers' children might also be former farmers' children. This demonstrated the model’s proficiency in recognizing nuanced categories. & 1/1 \\ 
\hline
\end{tabular}
\caption{Evaluation of o1-preview Model Performance with Problems and Options}
\label{table: logic_2}
\end{table}
\end{landscape}

\newpage

\subsection{Analogical Reasoning}\label{Jiaxi Li}
\subsubsection{Conceptualization}
Analogical reasoning refers to the cognitive ability to identify and understand the relationships and similarities between two seemingly distinct concepts or systems~\citep{jee2010commentary, richland2015analogy, daugherty2008analogical, zook1991effects}. This process involves recognizing a common underlying structure or pattern, even when the two entities differ in their surface characteristics. In educational settings, particularly in STEM (Science, Technology, Engineering, and Mathematics) education, analogical reasoning is a valuable tool for helping students grasp complex concepts by relating them to more familiar or tangible experiences.

For instance, a common analogy used in teaching electrical circuits is comparing the flow of electricity through wires to the flow of water through pipes. In this analogy, the water represents the electric current, the pipes symbolize the wires, and the water pressure corresponds to the voltage. This comparison allows students to leverage their understanding of fluid dynamics to conceptualize the behavior of electrical systems, making the abstract principles of circuitry more accessible.

A second example can be found in the teaching of planetary orbits in physics. To explain how planets revolve around the sun, instructors might use the analogy of a ball tied to a string being swung in a circular motion. The tension in the string represents the gravitational force pulling the planet toward the sun, while the ball’s motion mirrors the planet's orbit. This analogy helps students visualize gravitational forces and orbital mechanics in a more intuitive way.

A third application of analogical reasoning in STEM education is seen in the teaching of heat transfer. The concept of thermal energy moving between objects can be compared to the flow of people moving through a doorway in a crowded room. Just as more people will pass through a wider doorway faster, more heat will transfer between objects with a larger surface area. This analogy helps students better understand the principles governing heat transfer by relating it to everyday experiences they are familiar with.

Through these examples, it becomes evident that analogical reasoning serves as an effective pedagogical strategy for bridging the gap between abstract scientific principles and students' pre-existing knowledge. By drawing connections between seemingly unrelated concepts, educators can enhance students’ comprehension and retention of complex STEM topics. Therefore, it is worth researching how well o1-preview can take effect in the analogical reasoning approach of STEM education.

\subsubsection{Instrument and Evaluation}
Recent studies have shown that LLMs can match human performance across a range of analogical reasoning tasks, particularly on those involving semantic structure mapping~\cite{webb2023emergent, hu2023context, mitchell2021abstraction, lewis2024using, musker2024semanticstructuremappingllmhuman}. This capability is not just limited to abstract symbol manipulation, but also includes semantically meaningful symbols, thus providing a more nuanced understanding of analogical reasoning. To evaluate the analogical reasoning ability of o1-preview, we conduct experiments on a dataset introduced in \cite{musker2024semanticstructuremappingllmhuman}, which is designed for evaluating analogical reasoning ability and in the way of evaluating semantic structure mapping ability specifically. This dataset comprises 13 distinct task types and 136 samples in total, providing a comprehensive evaluation of the model's analogical reasoning capability.

Each of the 13 task types in this dataset is uniquely challenging, ranging from simple relational analogies to more complex tasks involving multiple layers of semantic content and structure. The 136 samples are carefully selected to encompass a wide range of semantic relations, ensuring a comprehensive evaluation of the model's analogical reasoning ability.

To test o1-preview's analogical reasoning capability, we choose tasks that require the model to transfer semantic structure and content from one domain to another. These tasks involve identifying and mapping semantic relations between words and phrases in a manner similar to human cognition and language acquisition. 
The tasks are created to test various aspects of analogical reasoning. The complexity of these tasks varies from basic relational analogies to more intricate challenges involving multiple layers of semantic content.
The experiments are focused on semantic structure mapping, where the model needs to transfer semantic relations from one domain to another. 
The goal of this evaluation is to determine whether o1-preview can perform analogical reasoning in a human-like manner by accurately aligning the relational structures between the source and target domains.

The experiments are categorized into the following aspects:
\begin{enumerate}
    \item \textbf{Mapping of Semantic Structure:} These tasks involve identifying and transferring relational structures between semantically meaningful symbols. The model must align abstract relationship such as those between "dog" and "puppy" with non-linguistic representations.
    \item \textbf{Mapping of Semantic Content:} The model needed to transfer specific content-related attributes from one domain to another. For example, it must infer that the "dog" symbol typically has four legs, converting it into a non-semantic representation.
\end{enumerate}

We use the zero-shot prompting technique to prompt the LLM in the experiments.
The prompt template is “We are conducting an experiment on general reasoning ability. You will be presented with a series of words and drawings. Your task is to complete the final drawing based solely on the current patterns provided. Please respond concisely with only the final drawing. Question: [Question]".

\vspace{-0.5em}

\subsubsection{o1-preview performance comparison with human performance}
\label{sec:analogical_reasoning}
\begin{figure}[H]
    \centering
    \begin{tcolorbox}[colback=blue!5!white, colframe=blue!75!black, width=\textwidth, arc=5mm, auto outer arc, boxrule=0.5mm, title=Prompt and Answer for Question 1, fonttitle=\small, fontupper=\tiny]
        \textbf{Prompt:} \\
        \tiny Read the PROBLEM. Follow the INSTRUCTIONS to answer the PROBLEM.\\
        \\
        \textbf{PROBLEM:} \\
        Question:\\
        square => E K C\\
        rectangle => E 
        circle => C K E\\
        oval =>\\
        INSTRUCTIONS: We are conducting an experiment on general reasoning ability. You will be presented with a series of words and drawings. Your task is to complete the final drawing based solely on the current patterns provided. Please respond concisely with only the final drawing. \\
        \rule{\textwidth}{0.4pt} 
        \textbf{Answer:} \\
        \tiny 
            C \% K \% E\\
        \rule{\textwidth}{0.4pt}
        \textbf{Reference Answer:}
            C \% K \% E
    \end{tcolorbox}
\end{figure}

\vspace{-1em}

The o1-preview model shows varying levels of success across different tasks. It demonstrates a strong ability to analyze and align underlying semantic structures in simple relational analogies. As the complexity of the tasks increases, especially those involving multiple layers of semantic content and structure, the model's performance remains strong, as they correctly answering 10 test samples.

The figure above is one of the examples of direct semantic structure mapping, both with and without distractors. Further experimental results can be found in the appendix~\ref{appen_ana_reason}.
In this task, o1-preview effectively eliminates distractions and extracts analogical features through reasoning analysis to obtain correct results.\\
\\
 Through the testing of our human participants in this experiment, we find that all of our human participants can have 100\% of accuracy. Meanwhile, the performance of o1-preview is also 100\%, indicating that its performance is comparable with human performance.\\
 \\
The observation from the experiment indicates that o1-preview has robust analogical reasoning capabilities. It can derive correct reasoning outcomes in both direct semantic structure mapping and more complex analogies, demonstrating an ability to effectively identify and apply learned relational patterns.

\newpage

\subsection{Scientific Reasoning}\label{Arne Bewersdorff}
\subsubsection{Conceptualization, Instruments \& Human Performance regarding Scientific Reasoning}
Scientific reasoning is closely interconnected with concepts such as scientific discovery and scientific literacy, as it encompasses the critical thinking and analytical skills necessary for formulating hypotheses, designing experiments, and interpreting data, which are fundamental to advancing scientific discovery and achieving scientific literacy. Klahr and Dunbar (1988)\cite{klahr1988dual} conceptualize scientific discovery as a product of a search in two cognitive spaces (Scientific Discovery as Dual Search, SDDS) and, therefore, take into account the interplay between domain-specific content knowledge and domain-specific process knowledge in scientific reasoning. According to the SDDS model, scientific reasoning is domain-specific, as one must possess domain-specific content knowledge and domain-specific process knowledge, e.g., to successfully search the hypothesis space. Since this conceptualization, the concept of scientific reasoning has become more nuanced, recognizing that scientific reasoning is not only a linear process, as Klahr and Dunbar often were (falsely) understood, but involves multiple pathways and methods and that there is not just one 'scientific method’~\cite{lederman2007nature, windschitl2008beyond}.

With the introduction of scientific practices (NRC, 2012 \cite{national2012framework}), there has been a more nuanced and explicit differentiation of the skills attributed to scientific reasoning (e.g., Fischer et al., 2014 \cite{fischer2014scientific}). In the following years, many instruments covering different nuances of scientific reasoning have been developed (review: Opitz et al., 2014 \cite{opitz2017measuring}). One instrument widely used to assess scientific reasoning is the Test of Scientific Literacy Skills (TOSLS) by Gormally et al. (2012) \cite{gormally2012developing}. This 28-item multiple-choice test aims to assess nine scientific literacy-related skills:

\begin{enumerate}
    \item Identify a valid scientific argument
    \item Evaluate the validity of sources
    \item Evaluate the use and misuse of scientific information
    \item Understand elements of research design and how they impact scientific findings and conclusions
    \item Create graphical representations of data
    \item Read and interpret graphical representations of data
    \item Solve problems using quantitative skills, including probability and statistics
    \item Understand and interpret basic statistics
    \item Justify inferences, predictions, and conclusions based on quantitative data
\end{enumerate}

Regarding these nine scientific literacy skills, it becomes apparent that Gormally et al. (2012) \cite{gormally2012developing} follow a narrow interpretation of the concept of scientific literacy, which can be aligned with the concept of scientific reasoning \cite{opitz2017measuring}. Scientific reasoning according to the definition of Chen (2020) \cite{CHEN2020330} encompasses core reasoning and problem-solving competencies and involves basic inference processes in forming hypotheses (skills 1 \& 9), designing experiments to test hypotheses (skills 4 \& 7, distinguishing determinate evidence from indeterminate evidence (skills 2 \& 3), and interpreting results as evidence that supports or refutes the hypotheses (skills 5, 6 \& 8). We, therefore, conclude that the TOSLS is, in line with the recommendation of Opitz et al. (2017) \cite{opitz2017measuring} a suitable instrument to effectively assess scientific reasoning.

To establish benchmarks for evaluating GPT-01-preview's capabilities, we will first present studies that report on human performance using the TOSLS. Using an adapted version of the TOSLS, Suwono et al. (2017) \cite{suwono2017enhancement} revealed scientific literacy scores between .33 and .69 among Indonesian pre-service biology students. Segarra et al. (2018) \cite{segarra2018student} assessed US undergraduate students enrolled in a general education biology course. Results revealed TOSLS scores between .55 and .65 among all different groups under investigation. Firdaus et al. (2023) did not report an overall score but indicated that Indonesian pre-service biology teachers performed weakly in almost every aspect of the TOSLS test. A recent study by Propsom et al. (2023) \cite{propsom2023test} with 800 US university students found TOSLS mean scores across all cohorts ranging from .60 (SD = .17) for first-year university students to .66 (SD = .20) for senior-year university students. Science and math students scored higher, from .64 (SD = 0.17) in their first year to .74 (SD = .17) in their senior year (see Table \ref{tab:TOSLS_Performance})

\begin{table}[ht]
\centering
\caption{Human Performance on the TOSLS Across Different Studies}
\begin{tabular}{lll}
\hline
\textbf{Study} & \textbf{Sample Population} & \textbf{TOSLS Scores} \\
\hline
Suwono et al. (2017) \cite{suwono2017enhancement} & Indonesian pre-service biology students & .33 - .69 \\
Segarra et al. (2018) \cite{segarra2018student} & US Undergraduate biology students & .55 - .65 \\
Firdaus et al. (2023) \cite{firdaus2023quantitative} & Indonesian pre-service biology teachers & Weak performance across skills \\
Propsom et al. (2023) \cite{propsom2023test} & US university students across cohorts & .60 - .74  \\
\hline
\end{tabular}
\label{tab:TOSLS_Performance}
\end{table}

TOSLS scores between .60 and .70 seem to represent the typical performance level of university students, with only science and math students in their senior years generally reaching higher scores. This range will serve as the benchmark for evaluating GPT-o1-preview's performance in scientific reasoning.

\subsubsection{Exemplary Prompt and Answer}
\begin{figure}[H]
    \centering
    \begin{tcolorbox}[colback=blue!5!white, colframe=blue!75!black, width=\textwidth, arc=5mm, auto outer arc, boxrule=0.5mm, title=Exemplary prompt to test Scientific Reasoning]
    \begin{small}
        \textbf{Prompt:} \\
        Read the \textbf{PROBLEM} carefully and provide a solution using given the \textbf{INSTRUCTIONS}.\\
        \\
        \textbf{PROBLEM}: \\
         Two studies estimate the mean caffeine content of an energy drink. Each study uses the same test on a random sample of the energy drink. Study 1 uses 25 bottles, and study 2 uses 100 bottles. Which statement is true? 
         
        a. The estimate of the actual mean caffeine content from each study will be equally uncertain.
        
        b. The uncertainty in the estimate of the actual mean caffeine content will be smaller in study 1 than in study 2. 
        
        c. The uncertainty in the estimate of the actual mean caffeine content will be larger in study 1 than in study 2. 
        
        d. None of the above \\
        
        \textbf{INSTRUCTIONS}: \\
        Select the one correct answer from options a, b, c, and d.
    \end{small}
    \end{tcolorbox}
    \captionof{figure}{Example prompt to assess the Scientific Reasoning of the o1-preview model. Item used from the TOSLS test.}\label{fig:scr_reason_prompt}
\end{figure}

\begin{figure}[H]
    \centering
    \begin{tcolorbox}[colback=blue!5!white, colframe=blue!75!black, width=\textwidth, arc=5mm, auto outer arc, boxrule=0.5mm, title=Answer]
    \begin{small}
            
    \textbf{Answer:} c. The uncertainty in the estimate of the actual mean caffeine content will be larger in Study 1 than in Study 2.

    \textbf{Explanation:}
    
    The uncertainty in estimating a population mean decreases as the sample size increases. This is because the standard error of the mean (SEM), which measures this uncertainty, is inversely proportional to the square root of the sample size:
    \[
    \text{SEM} = \frac{s}{\sqrt{n}},
    \]
    where \( s \) is the sample standard deviation and \( n \) is the sample size.
    
    Study 1 uses 25 bottles (\( n = 25 \)).
    
    Study 2 uses 100 bottles (\( n = 100 \)).
    
    Comparing the two studies:
    
    The standard error in Study 1 is
    \[
    \text{SEM}_1 = \frac{s}{\sqrt{25}} = \frac{s}{5}.
    \]
    The standard error in Study 2 is
    \[
    \text{SEM}_2 = \frac{s}{\sqrt{100}} = \frac{s}{10}.
    \]
    Since 
    \[
    \frac{s}{10} < \frac{s}{5},
    \]
    the uncertainty (standard error) is smaller in Study 2. Therefore, the uncertainty in the estimate from Study 1 is larger than that from Study 2.
    
    \end{small}
    \end{tcolorbox}
    \captionof{figure}{Answer the o1-preview model given to the prompt shown in  Figure \ref{fig:scr_reason_prompt}.}\label{fig:sc_reason_answer}
\end{figure}

\subsubsection{o1-preview performance on scientific reasoning capabilities}
The results of our analysis indicate that GPT-o1-preview achieved a perfect score of 1.0 on the TOSLS, correctly answering all 28 items. This performance is significantly higher than that of human participants in previous studies (see Table \ref{tab:TOSLS_Human_AI_Performance}). GPT-o1-preview's perfect score suggests that its scientific reasoning capabilities, as measured by the TOSLS, exceed those of human students at various educational levels, including senior university students specializing in science and mathematics. While it could be argued that some TOSLS items assess scientific literacy rather than pure scientific reasoning, the test's comprehensive nature—covering skills such as identifying valid scientific arguments, evaluating sources, understanding research design, interpreting data, and applying quantitative skills—implies that GPT-o1-preview demonstrates proficiency across a broad spectrum of scientific competencies of one central is Scientific Reasoning.

\begin{table}[ht]
\centering
\caption{Comparison of Human and AI Performance on the TOSLS}
\begin{tabularx}{\textwidth}{X X X}
\hline
\textbf{Subject}  & \textbf{TOSLS Scores} \\
\hline
University students  \cite{suwono2017enhancement, segarra2018student, firdaus2023quantitative, propsom2023test} & .33 - .74 \\
GPT-o1-preview  & 1.00 \\

\hline
\end{tabularx}
\label{tab:TOSLS_Human_AI_Performance}
\end{table}

An important aspect of this assessment is that GPT-o1-preview achieved these results using only textual descriptions generated by GPT-4o in place of the actual figures included in the TOSLS. This limitation did not hinder its ability to accurately interpret and respond to the 11 questions that often rely on visual data representations. This highlights GPT-o1-preview's capacity for processing complex scientific information presented in textual form, demonstrating its flexibility and adaptability in understanding and reasoning without visual aids.

GPT-o1-preview appears to effectively engage in the core reasoning and problem-solving competencies outlined by \cite{CHEN2020330}, including forming hypotheses, designing experiments, evaluating evidence, and interpreting results. 
However, the possibility that GPT-o1-preview may have been exposed to TOSLS items during its training cannot be entirely discounted. Given the public availability of the TOSLS, it is possible that some of the test content was included in the vast dataset used to train the model. This prior exposure could contribute to its high performance, not necessarily reflecting spontaneous reasoning abilities but rather the retrieval of memorized information. Nonetheless, OpenAI's policies and the model's design aim to minimize such occurrences by discouraging the verbatim reproduction of training data. The answers to the GPT-o1-preview oftentimes provided an explanation, which always seemed reasonable and did not stem from hallucination (cite).

From an AI development perspective, GPT-o1-preview's results demonstrate the potential for AI systems to handle complex reasoning tasks traditionally thought to require human intelligence. This opens avenues for AI applications in research, problem-solving, and decision-making processes across various scientific domains. It also prompts discussions about the ethical use of AI, the importance of transparency in AI reasoning processes, and the need to address biases that may be present in training data.
In conclusion, GPT-o1-preview 's perfect score on the TOSLS illustrates its advanced capabilities in scientific reasoning and scientific literacy tasks, surpassing average human performance in the studies reviewed.

\section{Discussion}\label{Gyeonggeon and XZ}

This study provided a comprehensive assessment of OpenAI o1-preview’s capability to engage in higher-order thinking in educational contexts, systematically examining its performance across 14 advanced thinking, such as critical, systems, computational, design, and metacognitive thinking dimensions (See Tab.~\ref{tab:o1preview_performance}). Our empirical evaluation demonstrated that OpenAI o1-preview can excel in various cognitive tasks traditionally viewed as requiring advanced human intelligence. By employing specific cognitive assessment instruments, such as the Ennis-Weir Critical Thinking Essay Test and the Biological Systems Thinking Test, we observed that OpenAI o1-preview performed comparably, and in most cases superiorly, to human participants in various dimensions of complex cognitive reasoning. These findings are not only promising but also suggest critical implications for the future of education and workforce development as AI technologies become increasingly integrated into learning environments.

\begin{landscape}
\begin{table}[H]
\centering
\scriptsize
\begin{tabular}{p{4cm} p{6cm} p{7cm} p{3cm}}
\hline
\textbf{Higher Order Thinking Category} & \textbf{Instrument or dataset} & \textbf{Comparison of o1-preview's performance with human intelligence benchmark} & \textbf{References} \\ \hline
Critical Thinking & Ennis-Weir Critical Thinking Essay Test & Better than pretest scores and better than or comparable to posttest scores & Tab.~\ref{tab:o1-preview_comparison} \\ 

System Thinking & Instrument 1: Biological Systems Thinking Test for Middle School Students \newline Instrument 2: The Village of Abeesee \newline Instrument 3: The Lake Urmia vignette (LUV) & Instrument 1: Better than middle school students \newline Instrument 2: Better than undergraduate students \newline Instrument 3: Better than engineering undergraduate and graduate students & Sec.~\ref{sec:system_thinking_results} and Tab.~\ref{tab:ST-instrument1} \& \ref{tab:ST-instrument2} \\ 

Computational Thinking & Instrument 1: Bebras Challenge problems \newline Instrument 2: Algorithmic Thinking Test for Adult Instrument & Instrument 1: Better than undergraduate students \newline Instrument 2: Better than mathematics, physics, engineering, and computer science majors, and experts & Tab.~\ref{tab:computational_thinking1} \& \ref{tab:computational_thinking2} \\

Design Thinking & Leetcode coding questions to improve efficiency & The median performance of o1-preview outperforms 80\% of the human on Leetcode & Fig.~\ref{fig:DT1} \\

Metacognition & Calibration score, Brier score, and sensitivity on MATH dataset & NA (o1 shows better scores compared to GPT-4o) & Fig.~\ref{fig:meta_acc} \& Tab.~\ref{meta:tab} \\

Data Literacy & Combined instrument from Merk et al. (2020) and Chen et al. (2023) & Better than post-secondary student cohort and pre-service teacher cohort & Tab.~\ref{tab:data_literacy_comparison1} \& \ref{tab:data_literacy_comparison2} \\

Creative Thinking & Instrument 1: Alternate Uses Task (divergent thinking) \newline Instrument 2: Remote Association Test (convergent thinking) & Instrument 1: Better than university students \newline Instrument 2: Better than university students & Fig.~\ref{fig:creativity1} \\

Collaborative Thinking & Instrument 1: Math problem solving \newline Instrument 2: Collaborative science group chats & NA (o1 shows better scores compared to GPT-4o) & Fig.~\ref{fig:multiagent_comparsion} \& Tab.~\ref{tab:group_chat} \\

Abstract Reasoning & Raven's Progressive Matrices (imaged encoded to pure-text) & Worse than psychology major students & Tab.~\ref{tab:rpm_compare} \\

Spatial Reasoning & Molecular dynamics simulation problems adopted from Protein Data Bank & NA (o1 shows a high level of accuracy in both qualitative and quantitative predictions) & Tab.~\ref{table-spatial} \\ 

Quantitative Reasoning & FINAQ dataset & Better than general crowd performance; Worse than human expert performance & Tab.~\ref{table:main_res} \\

Logical Reasoning & LogiQA dataset & Better than general human performance; Worse than human ceiling performance & Tab.~\ref{table:logiqacomparison} \\

Analogical Reasoning & Analogical reasoning tasks selected from Musker et al. (2024) & Comparable to human participants (human and o1-preview both reached 100\% accuracy) & Sec.~\ref{sec:analogical_reasoning} \\

Scientific Reasoning & Test of Scientific Literacy Skills & Better than university students (o1-preview reaches perfect score) & Tab.~\ref{tab:TOSLS_Human_AI_Performance} \\

\hline
\end{tabular}
\caption{Performance comparison of o1-preview across various higher order thinking categories.}
\label{tab:o1preview_performance}
\end{table}
\end{landscape}

The results of this study imply a potential paradigm shift in educational goals and future workforce development, particularly in fields requiring advanced cognitive skills\cite{zhai2022chatgpt,zhai2023chatgpt}. Previous studies have shown AI’s utility in simple decision-making and problem-solving \cite {teng2023data, winkler2021enhancing}, but this research underscores how AI can synthesize, evaluate, and critically analyze information, pushing the boundaries of its application in educational contexts \cite{zhai2021practices}. Our findings suggest that OpenAI o1-preview’s proficiency in handling tasks such as critical thinking, computational thinking, and systems thinking indicates a significant opportunity for AI to augment, and in some cases, complement human intellectual work. For example, OpenAI o1-preview achieved a 81\% success rate on the Ennis-Weir Critical Thinking Essay Test, outperforming pre-intervention students and comparable to post-intervention improvements typically achieved through critical thinking training. Such results evidence the model’s ability to analyze, evaluate, and synthesize complex arguments, skills vital for higher-order thinking in both academic and professional settings. OpenAI o1-preview’s ability to handle interdisciplinary and complex problems enables it to leverage its pre-trained knowledge across various fields, making it adaptable to diverse educational scenarios in areas such as mechanical engineering, biology, medicine, and robotics \cite{tian2024assessing}. In light of this, this study justifies the need for a reconceptualization of the skills required in future workforces. Given that future jobs will increasingly require the ability to navigate complex, interdisciplinary problems \cite{oecd2019future}, AI tools like OpenAI o1-preview may redefine the competencies needed in the workforce, pushing the boundaries of human-AI collaboration in problem-solving domains \cite{shiohira2021understanding}.

This study represent a significant advancement in the discourse surrounding AI integration in education. Prior research has emphasized AI's capacity to support learning by automating routine tasks and providing personalized feedback \cite{hwang2020vision, deeva2021review,zhai2020applying}, but it has offered limited insight into how AI can be leveraged for more complex cognitive processes, such as systems thinking. This study builds on existing literature by demonstrating that OpenAI o1-preview performs well in tasks involving the mapping of complex interrelationships between elements, thereby validating AI's potential to scaffold higher-order thinking. The findings thus underscores the need for pedagogical models that move beyond simple AI integration and emphasize active student engagement. Educators must guide students to critically evaluate and apply AI-generated insights rather than passively consume them, ensuring that AI becomes a tool for deep learning rather than a substitute for human reasoning. This study contributes to the field by justifying the development of pedagogical frameworks that train learners to collaborate with AI in meaningful ways, particularly in areas such as ethical decision-making and real-world problem-solving, where contextual understanding is essential.

Moreover, our findings suggest that the OpenAI o1-preview model could intuitevely serves distinct functions in K-12 and higher education settings. In K-12 education, it offers more accessible feedback, supplemented by multimodal demonstrations \cite{lee2023multimodality,latif2023automatic,lee2023nerif}. In contrast, for higher education, it provides more abstract, professional responses, supported by credible sources to encourage deeper critical thinking. Across both levels, o1-preview’s ability to support higher-order thinking enhances process-oriented learning. However, its emergence highlights the need for improved teacher education \cite{wang2024system,nyaaba2024generative,zhai2024transforming}. Educators must be trained to understand and ethically use AI to achieve student learning outcomes, while also exploring how AI, particularly OpenAI o1-preview, can be leveraged to enhance the quality of teaching and learning. Prior research suggests that AI offers unprecedented opportunities for personalized learning, adaptive feedback, and interactive educational experiences \cite{hwang2020vision,samala2024depth,latif2024fine}. OpenAI o1-preview advanced the potentials by demonstrating its ability to rapidly analyze data and generate tailored responses that were particularly evident in computational thinking tasks, where it solved complex problems with high accuracy and provided detailed explanations of its reasoning. This capability has the potential to significantly augment instructional practices by offering real-time feedback to learners, facilitating deeper engagement with learning materials, and enabling students to tackle more challenging content earlier in their education. However, this integration will require careful thought regarding how AI-generated feedback can be balanced with human instruction to ensure that students are developing critical thinking and problem-solving skills autonomously, rather than becoming overly reliant on AI assistance.

Nevertheless, while OpenAI o1-preview has demonstrated significant promise in these areas, there remain important implications for the aspects where its performance needs improvement. Despite its strengths, our analysis revealed that OpenAI o1-preview struggled with tasks with visual information or requiring higher levels of abstraction. For instance, in abstract reasoning, OpenAI o1 showed worse performance than human or GPT-4. However, as already mentioned above, this is due to o1-preview's limited support on visual data, making the test on abstract reasoning not equivalent to other tests in this study. Therefore, it is plausible to expect that o1-preview will perform better once its multimodal data processing feature is released, which calls for future investigation. Also, in design thinking, quantitative reasoning, and analogical reasoning tasks, OpenAI o1-preview could generate above-human average solutions, but its ability to reach human expert's level was limited. While o1's performance in various domains is already striking, its relative inability in other tasks calls for a need for further development in those areas. These limitations suggest that while AI can be a valuable tool in supporting higher-order thinking, human oversight is essential, especially in domains requiring multimodal synthesis or high-end reasoning capabilities\cite{zhai2023ai}.

In addition, the findings also point to several key limitations in the present study, which should be addressed in future research. First, this evaluation only tested the o1-preview model, and as such, its findings may not fully generalize to future iterations of the OpenAI o1-preview model or other advanced AI systems. Future research should explore newer versions of the model to determine whether improvements in architecture and training methods lead to enhanced performance in areas where o1-preview was found lacking, such as abstract reasoning. Second, the study utilized a select number of assessment instruments, limiting the scope of the evaluation. Future studies should incorporate a wider range of higher-order thinking assessments, including those designed for broadly defined creative thinking and ethical decision-making, to provide a more comprehensive understanding of AI’s potential in education. Lastly, we compared OpenAI o1-preview’s performance primarily with data on human participants from the existing literature, which may not fully capture the nuances of direct human-AI interaction. Future studies should conduct direct, real-time comparisons between AI and human performance in controlled educational settings to better understand the dynamics of human-AI collaboration and its implications for learning outcomes.

\section{Conclusions}\label{Gyeonggeon and XZ}

This study addressed a critical gap in the literature regarding the capabilities of AI, specifically OpenAI o1-preview, in conducting higher-order thinking tasks in educational contexts. While previous research has predominantly focused on AI's role in automating routine tasks and enhancing basic cognitive functions, there has been limited exploration of how AI possess complex cognitive skills such as critical thinking, systems thinking, and computational thinking and complete relevant tasks. By systematically assessing OpenAI o1-preview’s performance across these dimensions and comparing its outcomes to human performance, this study provides new insights into the potential and limitations of AI in facilitating advanced cognitive tasks. The findings of this study demonstrate that OpenAI o1-preview can conduct tasks that traditionally needed advanced human intellectual capabilities in specific domains, such as critical and systems thinking, while also revealing areas where human oversight remains essential, particularly in tasks requiring the highest reasoning ability or multimodality. 

This study extend beyond immediate pedagogical adjustments and call for a fundamental rethinking of educational goals, pedagogy, and learning content to better align with the demands of a rapidly evolving workforce driven by advanced AI technologies such as OpenAI o1-preview and the potential future emergence of artificial generative intelligence (AGI)\cite{latif2023artificial}. Given the demonstrated capabilities of o1-preview in handling complex cognitive tasks such as critical thinking, systems thinking, and computational thinking, educational institutions must recalibrate their objectives. Instead of focusing solely on traditional knowledge acquisition and rote learning, education should prioritize the development of skills that complement AI, such as creativity, ethical reasoning, and the ability to critically evaluate AI-generated outputs. This shift in focus will prepare students not just to coexist with AI in the workplace, but to leverage its strengths in meaningful, collaborative ways. Pedagogically, this requires a shift toward teaching methodologies that actively integrate AI tools into the learning process, positioning them as collaborators rather than mere assistants. Educators must develop instructional strategies that encourage students to critically engage with AI technologies, fostering an environment where learners can question, interpret, and apply AI outputs in innovative ways. This involves moving beyond passive consumption of AI-generated information toward a model of education that promotes higher-order thinking, ethical decision-making, and problem-solving in complex, real-world contexts\cite{zhai2023chatgpt}.

Moreover, the content of educational curricula must be updated to reflect the changing nature of work in an AI-driven world. This includes incorporating more interdisciplinary learning that bridges the gap between technical and non-technical fields, ensuring that students are equipped with both the cognitive flexibility and domain-specific knowledge necessary to thrive in an AI-augmented workforce. Emphasis should be placed on teaching students how to work alongside AI technologies to solve multifaceted problems, innovate, and make decisions in fields as diverse as science, technology, ethics, and the humanities. This comprehensive recalibration of educational goals, pedagogy, and content is essential for preparing future generations to effectively navigate and lead in a workforce increasingly influenced by powerful AI systems like OpenAI o1-preview and AGI.

\section*{Acknowledgment}
This study secondary analyzed data from projects supported by the Institute of Education Sciences (grant number R305C240010, PI Zhai). The authors acknowledge the funding agencies and the project teams for making the data available for analysis. The findings, conclusions, or opinions herein represent the views of the authors and do not necessarily represent the views of personnel affiliated with the funding agencies.

\section*{Declaration of generative AI and AI-assisted technologies in the writing process}
During the preparation of this work the author(s) used ChatGPT in order to check grammar and polish the wordings. After using this tool/service, the authors reviewed and edited the content as needed and take full responsibility for the content of the publication.

\newpage
\bibliography{LLM_refs}
\bibliographystyle{unsrt}

\newpage
\section{Appendix}

\subsection{Critical Thinking}
The following are the EWCTET, Scoring Sheets, and O1 full reply to the essay.  

\begin{figure}[H]
    \centering
        \begin{tcolorbox}[colback=blue!5!white, colframe=blue!75!black, width=\textwidth, arc=5mm, auto outer arc, boxrule=0.5mm, title=Performance: Response to the Moorburg Letter, fonttitle=\small, fontupper=\tiny]
    \begin{small}
        \textbf{O1 Performance:} \\
        Thought for a couple of seconds. \\
        \textbf{1. Response to Paragraph 1:} Mr. Raywift advocates for eliminating overnight parking from 2 a.m. to 6 a.m., asserting that any intelligent citizen should agree. However, this approach dismisses differing opinions and doesn't consider the diverse needs of all residents. Effective arguments should respect differing viewpoints and provide compelling reasons for consensus. \\
        \textbf{2. Response to Paragraph 2:} Comparing overnight parking to having a garage on the streets is an oversimplification. Garages are permanent structures with different implications than temporary parked cars. This analogy fails to account for the transient nature of parking and does not adequately support the argument for a parking ban. \\
        \textbf{3. Response to Paragraph 3:} The concern about congestion on narrow streets during rush hour is valid. However, restricting parking only from 2 a.m. to 6 a.m. may not significantly alleviate afternoon traffic if cars are still present during peak hours. A more targeted approach during peak times might be necessary to address congestion effectively. \\
        \textbf{4. Response to Paragraph 4:} Addressing morning traffic for factory workers is important, but the proposed time frame of 2 a.m. to 6 a.m. may not fully resolve the issue. Many vehicles parked overnight could still be present as workers arrive for their shifts, potentially limiting the effectiveness of the ban in improving morning traffic flow. \\
        \textbf{5. Response to Paragraph 5:} Stating that overnight parking is undesirable without providing specific reasons weakens the argument. To build a strong case, Mr. Raywift should outline the particular negative impacts of overnight parking on the community, such as safety concerns or inconvenience to residents. \\
        \textbf{6. Response to Paragraph 6:} The claim that banning parking during these hours will nearly eliminate accidents lacks sufficient evidence. Without concrete data linking overnight parking to a high incidence of accidents, this assertion remains unsubstantiated and undermines the credibility of the proposal. \\
        \textbf{7. Response to Paragraph 7:} Citing a one-day experiment by the Chief of Police as proof is inadequate. A single day's results do not provide a reliable basis for policy changes. Comprehensive studies over extended periods are necessary to draw meaningful conclusions about the effectiveness of the proposed parking restrictions. \\
        \textbf{8. Response to Paragraph 8:} Redefining "safe" to mean the complete elimination of any accident risk sets an unrealistic standard. While safety improvements are important, it is essential to balance them with practicality and acknowledge that absolute safety is unattainable in most public settings. \\
        \textbf{9. Overall Evaluation:} Mr. Raywift's letter presents concerns about overnight parking but falls short in providing robust evidence and logical reasoning to support his proposal. The arguments are often based on assumptions rather than factual data, and the analogies used are not entirely appropriate. A more thorough analysis, including comprehensive traffic studies and consideration of residents' needs, would strengthen the case for or against the proposed parking restrictions. Engaging with the community to explore alternative solutions could lead to more effective and widely supported outcomes. \\
        Sincerely, \\
        {[Your Name]} \\
    \end{small}
    \end{tcolorbox}
    \captionof{figure}{O1 Performance: Response to the Moorburg Letter.}
    \label{fig:o1_performance}
\end{figure}

\subsection{System Thinking}
Following are the example of Prompts and o1-preview responses for items on system thinking.

\begin{figure}[H]
    \centering
    \begin{tcolorbox}[colback=blue!5!white, colframe=blue!75!black, width=\textwidth, arc=5mm, auto outer arc, boxrule=0.5mm, title=Prompt and Answer for Item 2 in Instument 1, fonttitle=\small, fontupper=\tiny]
        \textbf{Prompt:} \\
        \tiny Read the PROBLEM carefully and provide solution using give INSTRUCTIONS.
        \\
        PROBLEM \\
There is a picture of pond ecosystem including sunlight, rushes, frog, water Lily, small fish, large fish, insect, heron, bacteria, green algae, water and soil.

A newly built factory near the pond emits substantial amounts of sulphur dioxide and other pollutants and discharges wastewater containing heavy metal pollutants, nitrogen, and phosphorus (nitrogen and phosphorus will cause many green algae to multiply) into the pond. What effect do you think this will have on the creatures in the pond and the surrounding villagers?  
   \\
        INSTRUCTIONS \\
Answer Item 2 based on the information provided in the picture of pond ecosystem. 
 \\
        \rule{\textwidth}{0.4pt} 
        \textbf{Answer:} \\
        \tiny 
The introduction of pollutants from the nearby factory will have several adverse effects on both the pond ecosystem and the surrounding villagers:

1.	Algal Bloom (Eutrophication):

o	Cause: Excess nitrogen and phosphorus in the wastewater will promote the rapid multiplication of green algae.

o	Effect on Pond Creatures: The overgrowth of algae can block sunlight from reaching aquatic plants like water lilies and rushes, inhibiting their photosynthesis and growth. When the algae die and decompose, bacteria consume large amounts of dissolved oxygen in the water, leading to hypoxic conditions.

o	Consequences: Low oxygen levels can result in the death of fish, frogs, and other aquatic organisms. This disrupts the food chain, affecting predators like large fish and herons.

2.	Toxicity from Heavy Metals:

o	Cause: Discharge of heavy metals into the pond introduces toxic substances into the water.

o	Effect on Pond Creatures: Heavy metals can accumulate in the bodies of small fish, large fish, frogs, and insects.

o	Consequences: Bioaccumulation can lead to poisoning, reproductive issues, and death of aquatic organisms. Predators that consume these organisms, such as herons, can also be affected due to biomagnification.

3.	Acidification from Sulphur Dioxide:

o	Cause: Emissions of sulphur dioxide can lead to the formation of acid rain.

o	Effect on Pond Creatures: Acidic water can harm sensitive species, alter pH levels critical for aquatic life, and affect the hatching of eggs in species like frogs and fish.

o	Consequences: Altered pH can disrupt metabolic processes, leading to decreased survival rates of aquatic organisms.

4.	Impact on Bacteria (Decomposers):

o	Effect on Pond Creatures: Changes in water chemistry can affect bacterial populations responsible for decomposition.

o	Consequences: Disrupted decomposition processes can lead to the accumulation of organic matter, further depleting oxygen levels and affecting nutrient cycling.

5.	Effects on Surrounding Villagers:

o	Health Risks: Consumption of contaminated water or fish can lead to health problems such as heavy metal poisoning.

o	Livelihoods: Villagers relying on the pond for fishing, agriculture, or recreation may face economic losses due to the decline in fish populations and water quality.

o	Environmental Quality: The overall degradation of the pond can reduce the aesthetic and ecological value of the area.

6.	Disruption of Ecosystem Balance:

o	Biodiversity Loss: Sensitive species may die off, reducing biodiversity.

o	Altered Food Webs: Changes in species populations can lead to overpopulation of certain organisms like algae and insects, while others decline.

Summary:

The factory's emissions and wastewater discharge introduce harmful pollutants into the pond ecosystem, leading to eutrophication, toxicity, acidification, and disruption of natural processes. These environmental changes negatively affect the aquatic organisms—ranging from plants like rushes and water lilies to animals like fish, frogs, insects, and herons—by causing health issues, mortality, and altered habitats. For the surrounding villagers, the contaminated pond poses health risks, threatens their livelihoods, and diminishes the environmental quality of their community. 
\\
        \rule{\textwidth}{0.4pt}
        
    \end{tcolorbox}
\end{figure}

\begin{figure}[H]
    \centering
    \begin{tcolorbox}[colback=blue!5!white, colframe=blue!75!black, width=\textwidth, arc=5mm, auto outer arc, boxrule=0.5mm, title=Prompt and Answer for Item 10 in Instument 1, fonttitle=\small, fontupper=\tiny]
        \textbf{Prompt:} \\
        \tiny Read the PROBLEM carefully and provide solution using give INSTRUCTIONS.
\\
        \\
        PROBLEM \\
A remote island is uninhabited. The main animals on the island are wolves and deer, and the island is free from external interference all year round.

Item 1. Select the answer below that best describes what will happen to the average populations of the animals over time. 

A.	The deer will all die or be killed.

B.	The wolves will all die.

C.	There will be many more deer than wolves.

D.	There will be many more wolves than deer.

E.	The populations of each will be approximately equal.

Item 2. Choose the reason for your answer.

A.	Wolves prey on deer, and deer are eaten up by wolves. 

B.	Wolves prey on deer, causing the number of deer to decrease.

C.	Wolves prey on deer. The number of deer decreases. Wolves die as they lack food.

D.	To maintain the ecological balance of the ‘plant → deer → wolf’ food chain, the number of wolves and deer will fluctuate within a certain range.

E.	When energy is transmitted along the food chain, it decreases step by step. The population of wolves stores less energy than the population of deer.
\\
        INSTRUCTIONS \\
Answer Item 1 and Item 2. Explain your answer.
 \\
        \rule{\textwidth}{0.4pt} 
        \textbf{Answer:} \\
        \tiny 
Item 1 Answer:

C. There will be many more deer than wolves.

Item 2 Answer:

E. When energy is transmitted along the food chain, it decreases step by step. The population of wolves stores less energy than the population of deer.

Explanation:

In a predator-prey ecosystem like the one described, deer serve as the primary consumers by feeding on plants, while wolves are secondary consumers that prey on the deer. According to ecological principles, energy transfer between trophic levels is inefficient—only about 10\% of the energy at one level is passed on to the next. This concept is often illustrated by the energy pyramid.

Because of this energy loss at each trophic level, the population of prey (deer) must be larger than that of predators (wolves) to sustain the ecosystem. The deer population supports the wolves by providing the necessary energy and biomass. Therefore, over time, we can expect many more deer than wolves on the island.

Item 1 Explanation: The deer population will be significantly larger than the wolf population due to the energy requirements of sustaining predators. This aligns with Option C.

Item 2 Explanation: The reason for this population difference is the stepwise decrease in available energy as it moves up the food chain. The wolves, being higher up in the food chain, receive less energy, necessitating a smaller population. This corresponds to Option E.
\\
        \rule{\textwidth}{0.4pt}
        
    \end{tcolorbox}
\end{figure}

\begin{figure}[H]
    \centering
    \begin{tcolorbox}[colback=blue!5!white, colframe=blue!75!black, width=\textwidth, arc=5mm, auto outer arc, boxrule=0.5mm, title=Prompt and Answer for Item 3 \& 4 in Instument 2 (Stakeholder Awareness Dimension), fonttitle=\small, fontupper=\tiny]
        \textbf{Prompt:} \\
        \tiny Read the PROBLEM carefully and provide solution using give INSTRUCTIONS.
\\
        \\
        PROBLEM \\
The Village of Abeesee has about 50,000 people. Its harsh winters and remote location make heating a living space very expensive. The rising price of fossil fuels has been reflected in the heating expenses of Abeesee residents. In fact, many residents are unable to afford heat for the entire winter (5 months). A University of Abeesee study shows that 38\% of village residents have gone without heat for at least 30 winter days in the last 24 months. Last year, 27 Abeesee deaths were attributed to unheated homes. Most died from hypothermia/exposure (21), and the remainder died in fires or from carbon monoxide poisoning that resulted from improper use of alternative heat sources (e.g., burning trash in an unventilated space).

Q3. What groups or stakeholders would you involve in planning a response to the problems/issues in Abeesee? 

Q4. Please briefly describe the process you would use planning a response to the problems/issues in Abeesee. 
 \\
        INSTRUCTIONS \\
Please answer Q3, Q4 in the PROBLEM one by one.
 \\
        \rule{\textwidth}{0.4pt} 
        \textbf{Answer:} \\
        \tiny 
Q3. What groups or stakeholders would you involve in planning a response to the problems/issues in Abeesee?

The following stakeholders should be involved in planning the response:

•	Local Government Officials: To enact policies and allocate resources.

•	Community Members and Leaders: Including those directly affected to ensure solutions meet their needs.

•	Public Health Agencies: For expertise on health risks and preventive measures.

•	Energy Providers and Utility Companies: To explore options for reducing costs or providing assistance.

•	Housing Authorities: To address structural issues like insulation and energy efficiency.

•	Social Services Organizations: For support services and outreach programs.

•	Non-Profit Organizations: Especially those focused on energy assistance and housing.

•	Emergency Services: Fire departments and medical responders for safety insights.

•	Educational Institutions: Such as the University of Abeesee for research support.

•	Environmental Groups: To promote sustainable and safe heating alternatives.

•	Business Community: Local businesses that may contribute resources or support.

•	Provincial/National Government Representatives: For additional funding and policy support .

Q4. Please briefly describe the process you would use planning a response to the problems/issues in Abeesee.

1.	Needs Assessment: Collect and analyze data on heating affordability, health impacts, and current practices.

2.	Stakeholder Engagement: Organize meetings with all stakeholders to gather input and build consensus.

3.	Set Objectives: Define clear, measurable goals such as reducing heating costs and preventing cold-related deaths.

4.	Develop Strategies: Formulate solutions like subsidies, energy assistance programs, home insulation initiatives, and public education campaigns.

5.	Resource Mobilization: Identify and secure funding, personnel, and materials needed for implementation.

6.	Implementation Plan: Create a detailed action plan with timelines, responsibilities, and performance indicators.

7.	Communication Plan: Develop messaging to inform the community about available resources and safe practices.

8.	Monitoring and Evaluation: Establish mechanisms to track progress, assess effectiveness, and make necessary adjustments.

9.	Sustainability Planning: Ensure the response is sustainable by building local capacity and securing ongoing support.
\\
        \rule{\textwidth}{0.4pt}
        
    \end{tcolorbox}
\end{figure}

\begin{figure}[H]
    \centering
    \begin{tcolorbox}[colback=blue!5!white, colframe=blue!75!black, width=\textwidth, arc=5mm, auto outer arc, boxrule=0.5mm, title=Prompt and Answer for Instument 3, fonttitle=\small, fontupper=\tiny]
        \textbf{Prompt:} \\
        \tiny Read the PROBLEM carefully and provide solution using give INSTRUCTIONS.
\\
        \\
        PROBLEM \\
Lake Urmia in northwest Iran was the largest lake in the Middle East and the sixth largest saltwater lake on earth. Unique ecological features of this UNESCO Biosphere Reserve had made Urmia the largest habitat of brine shrimp. The Lake is the home of various species of Archaebacteria and Bacteria, Microfungi, Phytoplankton, and 311 species of Plants. It also hosts 226 kinds of Birds. The lake is also known for its hundreds of small islands serving as stopover points during the migrations of several wild birds to and from Russia. The town Urmia, on the west side of the lake, has a population of 700,000 people.

Several reports show that the lake suffers from serious ecological problems, and many of the indictors are easily observable from the lake itself. Between 1972 and 2014, the area of the lake shrank by 88\%. The evaporation of the water has exposed the lakebed and caused windblown salt, which may lead to environmental health crises, including increase in infant mortality, cancer, and liver, kidney, and respiratory diseases. This phenomenon is similar to what happened after the death of the Aral Sea. In addition, it will increase unemployment by reducing tourism and shrinking the fertility of the land in the region.

Fortunately, the public awareness about the lake has increased and huge outcries urged the government to take action. Government officials promised to spend \$5 million to save the lake in a period of 10 years. Officials attribute the lake’s desiccation to the drought in recent years while critics point to mismanagement of water resources and construction of a raised road across the lake. The population of the region more than tripled during the past 40 years. Multiple dam construction and pipe transfer projects have made water available for domestic and agricultural purposes. In 1999, a project was completed to pump water from the Zarineh River (one of the main feeders of the lake) to the largest city of the area. In addition, forests have been transferred to agricultural lands to fulfill the needs of the growing population. Specifically, the forest cover of Zagros Mountain has declined. Forests of Zagros maintain naturally controlled water flow to rivers feeding the lake. As the lake shrinks, the climate of the region becomes drier and more water is needed for agriculture.

In 2011, multiple demonstrations took place in cities close to the lake demanding that the government take immediate actions to protect Urmia Lake. Not all demonstration ended peacefully. The slogan “let me cry to fill the lake”, a highly chanted motto depicted the emotional reaction of the region. According to official state reports at least 70 supporters of the lake were arrested. Several proposals were discussed including channeling water from other rivers to the lake, destroying several dams, or funding relocation of people living around the Lake. In order to find solutions for saving the lake, policy makers need to know what caused the lake to shrink by 88\% in 44 years. The lake is just an example of many similar environmental challenges that humans are dealing with especially in less developed regions.
 \\
        INSTRUCTIONS \\
Read the PROBLEM. Describe the problems facing Lake Urmia in detail and explain why the lake shrank over the years.
 \\
        \rule{\textwidth}{0.4pt} 
        \textbf{Answer:} \\
        \tiny 
Problems Facing Lake Urmia:

1.	Significant Shrinkage of the Lake: Between 1972 and 2014, Lake Urmia's area shrank by an alarming 88\%. This drastic reduction is easily observable and has severe ecological implications.

2.	Environmental Health Crises Due to Windblown Salt: The evaporation of water has exposed the lakebed, leading to windblown salt storms. These salt particles can cause serious health issues, including increased rates of infant mortality, cancer, and liver, kidney, and respiratory diseases among the local population. This situation mirrors the environmental disaster experienced after the desiccation of the Aral Sea.

3.	Economic Impact on Tourism and Agriculture: The shrinkage of the lake adversely affects tourism, a significant source of income for the region. Additionally, the increasing salinity and changing climate conditions reduce the fertility of the surrounding lands, leading to decreased agricultural productivity and higher unemployment rates.

4.	Ecological Disruption: The loss of habitat threatens the diverse species that depend on the lake. The decline in brine shrimp populations, for example, can disrupt the food chain, affecting bird species and other wildlife.

Reasons for the Lake's Shrinkage:

1.	Mismanagement of Water Resources:

o	Damming and Water Diversion Projects: Multiple dams have been constructed on rivers that feed into Lake Urmia to supply water for domestic and agricultural use. In 1999, a project diverted water from the Zarineh River, one of the lake's main feeders, to the largest city in the area. These interventions significantly reduced the inflow of water to the lake.

o	Construction of a Raised Road: Critics point to the construction of a raised road across the lake, which may have altered the natural flow and circulation of water, contributing to the lake's shrinkage.

2.	Population Growth and Increased Water Demand:

o	Over the past 40 years, the region's population has more than tripled. This surge has led to increased demand for water for domestic consumption and agricultural activities, placing additional stress on the lake's water sources.

3.	Deforestation and Land Use Changes:

o	Conversion of Forests to Agricultural Land: To meet the needs of the growing population, significant portions of the Zagros Mountain forests have been cleared for agriculture. These forests played a crucial role in maintaining natural water flow to the rivers feeding Lake Urmia. Their removal has disrupted this balance, leading to reduced water inflow.

o	Impact on Water Regulation: The loss of forest cover diminishes the land's ability to retain water and regulate its gradual release into rivers and streams, exacerbating the reduction in water reaching the lake.

4.	Climate Change and Drought:

o	Government officials attribute the lake's desiccation partly to recent droughts. While natural climate variability plays a role, it is compounded by human activities that have reduced the lake's resilience to such events.

5.	Agricultural Practices and Increased Irrigation Needs:

o	As the lake shrinks and the regional climate becomes drier, more water is required for irrigation to sustain agriculture. This leads to a vicious cycle where increasing water extraction for farming further depletes the lake.

6.	Feedback Loop of Environmental Degradation:

o	The environmental changes contribute to a drier climate, which in turn exacerbates the lake's shrinkage. The altered climate conditions can lead to less precipitation and higher evaporation rates, further diminishing the lake's water levels.

Conclusion:

The crisis facing Lake Urmia is the result of a combination of natural factors and human-induced actions, including mismanagement of water resources, population growth, infrastructural developments, deforestation, and climate change. The lake's shrinkage has far-reaching consequences for environmental health, biodiversity, and the regional economy.

Public awareness has grown, leading to demonstrations demanding immediate government action. Despite government promises to invest \$5 million over ten years to save the lake, effective solutions require a comprehensive understanding of the root causes. Policy makers need to address water resource management, control infrastructural developments that affect natural water flows, promote sustainable agricultural practices, and implement reforestation programs to restore the natural regulation of water systems.

Lake Urmia serves as a critical example of environmental challenges in less developed regions, highlighting the urgent need for sustainable management of natural resources to prevent ecological disasters and protect the health and livelihoods of local populations.
\\
        \rule{\textwidth}{0.4pt}    
    \end{tcolorbox}
\end{figure}

\subsection{Computational Thinking}
Following are the Prompts and o1-preview responses for each test item on computational thinking. 
\begin{figure}[H]
    \centering
    \begin{tcolorbox}[colback=blue!5!white, colframe=blue!75!black, width=\textwidth, arc=5mm, auto outer arc, boxrule=0.5mm, title=Prompt and Answer for Question 1, fonttitle=\small, fontupper=\tiny]
        \textbf{Prompt:} \\
        \tiny Read the PROBLEM. Follow the INSTRUCTIONS to answer the PROBLEM.\\
        \\
        \textbf{PROBLEM:} \\
        The Stack Computer is loaded with boxes from a conveyer belt. The boxes are marked with a Number or an Operator (+, -, * or /).
The computer is loaded until the top box is a box marked with an operator. This operator is then used on the two boxes below it. The three boxes are then fused into one single box and marked with the outcome of the calculation.
In the Stack Computer, calculations are entered in a different way to a normal calculator. Examples:
2+3 must be entered as 2 3 +
10-2 must be entered as 10 2 -
5*2+3 must be entered as 5 2 * 3 +
5+2*3 must be entered as 5 2 3 * + (8-2)*(3+4) must be entered as 8 2 - 3 4 + * How should the following computation be entered: 4*(8+3)-2? 
        INSTRUCTIONS: Study the PROBLEM above and provide the solution to the problem \\
        \rule{\textwidth}{0.4pt} 
        \textbf{Answer:} \\
        \tiny 
            To enter the sequence 4 *  (8 + 3) - 2 into the Stack Computer, the sequence should be: 4 8 3 + * 2 -\\
        \rule{\textwidth}{0.4pt}
        \textbf{Reference Answer:}
            4 8 3 + * 2 -
    \end{tcolorbox}
\end{figure}

\begin{figure}[H]
    \centering
    \begin{tcolorbox}[colback=blue!5!white, colframe=blue!75!black, width=\textwidth, arc=5mm, auto outer arc, boxrule=0.5mm, title=Prompt and Answer for Question 2, fonttitle=\small, fontupper=\tiny]
        \textbf{Prompt:} \\
        \tiny Read the PROBLEM. Follow the INSTRUCTIONS to answer the PROBLEM.\\
        \\
        \textbf{PROBLEM:} \\
       Beaver Alex and beaver Betty send each other messages using the following sequence of transformations on every word. Step 1: Word to encode; Step 2: reverse word; Step 3: shift letters by 2 to the left; Step 4: replace each letter with the next letter in the alphabet; and Step 5: encoded word (WBFCSF). For example, the word "BEAVER" is transformed to "WBFCSF" based on the above sequence of transformation. Beaver Betty receives the final encoded message "PMGEP" from beaver Alex. What did Alex want to say? (1) RIVER (2) KNOCK (3) FLOOD (4) LODGE.
        INSTRUCTIONS: study the PROBLEM above carefully and select the correct option. \\
        \rule{\textwidth}{0.4pt} 
        \textbf{Answer:} \\
        \tiny 
            To determine what Alex wanted to say, we'll apply the given sequence of transformations to each of the provided options and see which one results in the encoded message "PMGEP".\\
            Option 1: RIVER
1.	Word to encode: RIVER
2.	Reverse the word: Original: R I V E R, Reversed: R E V I R
3.	Shift letters by 2 positions to the left (wrapping around): Shifted: V I R R E
4.	Replace each letter with the next letter in the alphabet: V → W, I → J, R → S, R → S, E → F, Result: WJSSF
This does not match "PMGEP".\\
Option 2: KNOCK
1.	Word to encode: KNOCK
2.	Reverse the word:Original: K N O C K, Reversed: K C O N K
3.	Shift letters by 2 positions to the left: Shifted: O N K K C
4.	Replace each letter with the next letter in the alphabet: O → P, N → O, K → L, K → L, C → D, Result: POLLD
This does not match "PMGEP".\\
Option 3: FLOOD
1.	Word to encode: FLOOD
2.	Reverse the word:
Original: F L O O D, Reversed: D O O L F
3.	Shift letters by 2 positions to the left: Shifted: O L F D O
4.	Replace each letter with the next letter in the alphabet:
O → P, L → M, F → G, D → E, O → P. Result: PMGEP
This matches the encoded message!\\
Option 4: LODGE
1.	Word to encode: LODGE
2.	Reverse the word:Original: L O D G E, Reversed: E G D O L
3.	Shift letters by 2 positions to the left: Shifted: D O L E G
4.	Replace each letter with the next letter in the alphabet: D → E, O → P, L → M, E → F, G → H. Result: EPMFH
This does not match "PMGEP".
Answer: Option 3 - FLOOD
        \rule{\textwidth}{0.4pt}
        \textbf{Reference Answer:}
            FLOOD
    \end{tcolorbox}
\end{figure}
\begin{figure}[H]
    \centering
    \begin{tcolorbox}[colback=blue!5!white, colframe=blue!75!black, width=\textwidth, arc=5mm, auto outer arc, boxrule=0.5mm, title=Prompt and Answer for Question 3, fonttitle=\small, fontupper=\tiny]
        \textbf{Prompt:} \\
        \tiny Read the PROBLEM. Follow the INSTRUCTIONS to answer the PROBLEM.\\
        \\
        \textbf{PROBLEM:} \\
       Agents Boris and Bertha communicate using secret messages. Boris wants to send Bertha the secret message: MEETBILLYBEAVERAT6 He writes each character in a 4 column grid from left to right and row by row starting from the top. He puts an X in any unused spaces. The result is shown below. Row 1: MEET, Row 2: BILL, Row 3: YBEA, Row 4: VERA, Row 5: T6XX. Then he creates the secret message by reading the characters from top to bottom and column by column starting from the left: MBYVTEIBE6ELERXTLAAX. Bertha then uses the same method to reply to Boris. The secret message she sends him is: OIERKLTEILH!WBEX, What message does Bertha send back? \\
       Option 1: OKWHERETOMEET! \\
       Option 2: OKIWILLBETHERE!\\
       Option 3: WILLYOUBETHERETOO?\\
       Option 4: OKIWILLMEETHIM! \\
        INSTRUCTIONS: study the PROBLEM above and select the correct option using computational thinking skills. \\
        \rule{\textwidth}{0.4pt} 
        \textbf{Answer:} \\
        \tiny 
Given Secret Message: OIERKLTEILH!WBEX\\
Step 1: Determine the Grid Dimensions. The secret message has 16 characters, suggesting a grid of 4 columns and 4 rows (since 4×4=164×4=16).\\
Step 2: Map Secret Message Letters to Grid Positions (note: o1-preview provided the table to present the letter on each row and column).\\
Step 3: Reconstruct the Grid based on the secret rule. \\
Step 4: Read the original message OKIWILLBETHERE!X, remove the filler "X", oKIWILLBETHERE!\\
Answer: Option 2: OKIWILLBETHERE!\\
        \rule{\textwidth}{0.4pt}
        \textbf{Reference Answer:}
        Option 2: OKIWILLBETHERE!
    \end{tcolorbox}
\end{figure}

\begin{figure}[H]
    \centering
    \begin{tcolorbox}[colback=blue!5!white, colframe=blue!75!black, width=\textwidth, arc=5mm, auto outer arc, boxrule=0.5mm, title=Prompt and Answer for Question 4, fonttitle=\small, fontupper=\tiny]
        \textbf{Prompt:} \\
        \tiny Read the PROBLEM. Follow the INSTRUCTIONS to answer the PROBLEM.\\
        \\
        \textbf{PROBLEM:} \\
       Betaro Beaver has discovered five new magic potions: one makes ears longer another makes teeth longer another makes whiskers curly another turns the nose white the last one turns eye’s white. Betaro put each magic potion into a separate beaker. He put pure water into another beaker, so there are six beakers in total. The beakers are labeled A to F. The problem is, he forgot to record which beaker contains which magic potion! To find out which potion is in each beaker, Betaro set up the following experiments: 1: A beaver drinks from beakers A, B and C together, the effects are that Betaro has longer ears, white nose, and longer teeth. 2: A beaver drinks from beakers A, D and E together, the effects are that Betaro has longer ears, and white eyes. 3: A beaver drinks from beakers C, D and F together. the effects are that Betaro has white nose and curly whiskers. Which beaker contains pure water?
        INSTRUCTIONS: study the PROBLEM above carefully and provide the correct answer to the question. \\
        \rule{\textwidth}{0.4pt} 
        \textbf{Answer:} \\
        \tiny 
1. Assign Beaker A:
•	Beaker A is used in Experiments 1 and 2.
•	Common effect in both experiments: EL (Ears Longer)
•	Conclusion: Beaker A contains EL\\
2. Assign Beaker C:
•	Beaker C is used in Experiments 1 and 3.
•	Common effect in both experiments: NW (Nose White)
•	Conclusion: Beaker C contains NW\\
3. Assign Beaker B:
•	Experiment 1 has effects EL, TL, NW.
•	Beakers involved: A (EL), B, C (NW)
•	Remaining effect: TL (Teeth Longer)
•	Conclusion: Beaker B contains TL\\
4. Analyze Beakers D, E, and F:
•	Remaining potions: WC, EW, PW
•	Experiments involving these beakers:
o	Experiment 2: A (EL), D, E ⇒ EL, EW
o	Experiment 3: C (NW), D, F ⇒ NW, WC\\
5. Final Assignments:
•	Beaker A: EL (Ears Longer)
•	Beaker B: TL (Teeth Longer)
•	Beaker C: NW (Nose White)
•	Beaker D: WC (Whiskers Curly)
•	Beaker E: EW (Eyes White)
•	Beaker F: PW (Pure Water)\\
Answer:Beaker F contains the pure water.\\
        \rule{\textwidth}{0.4pt}
        \textbf{Reference Answer:}
       Beaker D contains the pure water
    \end{tcolorbox}
\end{figure}
\begin{figure}[H]
    \centering
    \begin{tcolorbox}[colback=blue!5!white, colframe=blue!75!black, width=\textwidth, arc=5mm, auto outer arc, boxrule=0.5mm, title=Prompt and Answer for Question 5, fonttitle=\small, fontupper=\tiny]
        \textbf{Prompt:} \\
        \tiny Read the PROBLEM. Follow the INSTRUCTIONS to answer the PROBLEM.\\
        \\
        \textbf{PROBLEM:} \\
      You have the following DNA sequence: T C C A G, And you want to get the following sequence: T C T T A, You can implement the following operations:
Swap (): it will swap one character for another. For example, swap (A,G) will turn the sequence AAGT into GGAT
Insert (): it will insert a specified character at the beginning of the sequence. For example, insert (A) will turn the sequence GT into AGT
Delete (): it will delete all specified characters. For example, delete (A) will turn the sequence AAGT into GT
What operations will turn the initial DNA sequence into the one we want?\\
A) Delete (G); then Insert (T); then Swap (C,T)\\
B) Swap (C,T); then Insert (T); then Delete (G)\\
C) Insert (T); then Swap (C,T); then Delete (G)\\
D) Insert (T); then Delete (G); then Swap (C,T)
 \\
 INSTRUCTIONS study the PROBLEM above carefully and select the correct answer to the question from the four options: A, B, C, D. \\
        \rule{\textwidth}{0.4pt} 
        \textbf{Answer:} \\
        \tiny 
            B: Swap (C,T); then Insert (T); then Delete (G)\\
        \rule{\textwidth}{0.4pt}
        \textbf{Reference Answer:}
           Option B
    \end{tcolorbox}
\end{figure}

\begin{figure}[H]
    \centering
    \begin{tcolorbox}[colback=blue!5!white, colframe=blue!75!black, width=\textwidth, arc=5mm, auto outer arc, boxrule=0.5mm, title=Prompt and Answer for Question 6, fonttitle=\small, fontupper=\tiny]
        \textbf{Prompt:} \\
        \tiny Read the PROBLEM. Follow the INSTRUCTIONS to answer the PROBLEM.\\
        \\
        \textbf{PROBLEM:} \\
      After finishing your long paper, you just discovered a mistake you made. All 4’s should be 7’s and all 7’s should be 4’s. To amend this, you can use an editor that will replace any sequence of characters with another sequence. 
In order to describe the replacement you would make with the editor, please fill in the blanks:
Replace all ( ) with ( ) then replace all ( ) with ( ); then replace all ( )with ( )\\
 INSTRUCTIONS study the PROBLEM above carefully and fill in the blanks to solve the problem.  \\
        \rule{\textwidth}{0.4pt} 
        \textbf{Answer:} \\
        \tiny 
           Replace all 4 with XX; then replace all 7 with 4; then replace all XX with 7.\\
        \rule{\textwidth}{0.4pt}
        \textbf{Reference Answer:}
         Replace all 4’s with XXs (any variable); then replace all 7’swith 4’s then replace all XXs(variable mentioned before) with 7’s (4’s and 7’s positions can be reversed here).
    \end{tcolorbox}
\end{figure}

\begin{figure}[H]
    \centering
    \begin{tcolorbox}[colback=blue!5!white, colframe=blue!75!black, width=\textwidth, arc=5mm, auto outer arc, boxrule=0.5mm, title=Prompt and Answer for Question 7, fonttitle=\small, fontupper=\tiny]
        \textbf{Prompt:} \\
        \tiny Read the PROBLEM. Follow the INSTRUCTIONS to answer the PROBLEM.\\
        \\
        \textbf{PROBLEM:} \\
      We have this game:
There are four players (P, Q, R, S): one of them is pointing at him/herself, and three of them are pointing at any other player. 
Player P gets the ball. While the ball-holder isn’t pointing at him/herself: The ball-holder passes the ball to the player that s/he is pointing at. If the ball-holder is pointing at him/herself, s/he drops the ball and the game finishes. We can state that:\\
A. This game will go on forever\\
B. This game will eventually stop because someone will drop the ball\\
C. This game will stop when player P gets the ball again\\
D. We don’t know whether this game will ever stop\\
 INSTRUCTIONS study the PROBLEM above carefully and select the correct answer to the question from the four options: A, B, C, D. \\
        \rule{\textwidth}{0.4pt} 
        \textbf{Answer:} \\
        \tiny 
     D. We don’t know whether this game will ever stop\\
        \rule{\textwidth}{0.4pt}
        \textbf{Reference Answer:}
         Option D
    \end{tcolorbox}
\end{figure}

\begin{figure}[H]
    \centering
    \begin{tcolorbox}[colback=blue!5!white, colframe=blue!75!black, width=\textwidth, arc=5mm, auto outer arc, boxrule=0.5mm, title=Prompt and Answer for Question 8, fonttitle=\small, fontupper=\tiny]
        \textbf{Prompt:} \\
        \tiny Read the PROBLEM. Follow the INSTRUCTIONS to answer the PROBLEM.\\
        \\
        \textbf{PROBLEM:} \\
      You have four different cheeses and one of them has a tasteless and odorless bacterium. This cheese causes symptoms of food-poisoning 24 hours after ingesting it. You want to find out which cheese it is, so you decide to use mice to taste them. But, you only have two mice left in your lab and your deadline to find the poisoned cheese is exactly 24 hours from now! You can assign every mouse the number of cheeses that you want. How would you do it?
I would assign mouse 1 to cheese/s 
I would assign mouse 2 to cheese/s \\
 INSTRUCTIONS study the PROBLEM above carefully and respond to the question with your appropriate assignment. \\
        \rule{\textwidth}{0.4pt} 
        \textbf{Answer:} \\
        \tiny 
        I would assign mouse 1 to cheeses 3 and 4. I would assign mouse 2 to cheeses 2 and 4.\\
        \rule{\textwidth}{0.4pt}
        \textbf{Reference Answer:}
         The two mice should share one cheese (e.g., cheese 2), each mouse should eat one different cheese (e.g., mouse 1 eats cheese 1 and mouse 2 eats cheese 3), and leave aside one cheese (e.g., cheese 4 in that example).
    \end{tcolorbox}
\end{figure}
\begin{figure}[H]
    \centering
    \begin{tcolorbox}[colback=blue!5!white, colframe=blue!75!black, width=\textwidth, arc=5mm, auto outer arc, boxrule=0.5mm, title=Prompt and Answer for Question 9, fonttitle=\small, fontupper=\tiny]
        \textbf{Prompt:} \\
        \tiny Read the PROBLEM. Follow the INSTRUCTIONS to answer the PROBLEM.\\
        \\
        \textbf{PROBLEM:} \\
      You have two ropes and a lighter, and each rope takes exactly one hour to burn. They don’t burn randomly, but at a regular speed. How would you use them to time exactly 15 minutes? The ropes cannot be cut, bent, straightened or measured with external instruments. (You don’t need to time 15 minutes from the moment that you start your actions)
Please, briefly explain the procedure:  \\
 INSTRUCTIONS study the PROBLEM above carefully and give your explanation to the procedure. \\
        \rule{\textwidth}{0.4pt} 
        \textbf{Answer:} \\
        \tiny 
1.  Light Rope 1 at both ends and Rope 2 at one end simultaneously. \\
2. When Rope 1 has completely burned out (after 30 minutes), immediately light the other end of Rope 2. \\
3. Wait until Rope 2 has completely burned out.\\ 
        \rule{\textwidth}{0.4pt}
        \textbf{Reference Answer:}
1. Light rope 1 at both ends and rope 2 at one end. \\
2. When the two flames on rope 1 meet, 30 minutes will have passed. Rope 2 has 30 minutes left of burn-time. \\
3. At that point, light Rope 2 at the other end. \\
4. In exactly fifteen minutes, Rope 2 will be completely burnt.

    \end{tcolorbox}
\end{figure}

\begin{figure}[H]
    \centering
    \begin{tcolorbox}[colback=blue!5!white, colframe=blue!75!black, width=\textwidth, arc=5mm, auto outer arc, boxrule=0.5mm, title=Prompt and Answer for Question 10, fonttitle=\small, fontupper=\tiny]
        \textbf{Prompt:} \\
        \tiny Read the PROBLEM. Follow the INSTRUCTIONS to answer the PROBLEM.\\
        \\
        \textbf{PROBLEM:} \\
     You have been given 9 coins of the same value, but one of them is fake which you could tell because it is lighter than the rest. You have a balance to weigh the coins, and each weighing can result in “the balance leans to the right”, “the balance leans to the left”, or “the balance rests stable”. Assuming you are following an optimal strategy to reduce the number of weighings, how many weighings are necessary to identify the fake coin?\\
 INSTRUCTIONS study the PROBLEM above carefully and give your answer. \\
        \rule{\textwidth}{0.4pt} 
        \textbf{Answer:} \\
        \tiny 
    2 weighings\\
        \rule{\textwidth}{0.4pt}
        \textbf{Reference Answer:}
        2 weighings
    \end{tcolorbox}
\end{figure}

\subsection{Design Thinking}\label{appdix:dt}
Following are the prompts and o1-preview responses for each test item on computational thinking.

\begin{figure}[H]
    \centering
    \begin{tcolorbox}[colback=blue!5!white, colframe=blue!75!black, width=0.99\textwidth, arc=5mm, auto outer arc, boxrule=0.5mm, 
    title=Example of Prompt and Answer for Design Thinking, 
    fonttitle=\small, fontupper=\tiny]
    \begin{tiny}
        \textbf{Prompt 1:} \\
        Read the \textbf{PROBLEM} carefully and provide solution using given \textbf{INSTRUCTION}.\\
        \\
        \textbf{PROBLEM}: \\
        Median of Two Sorted Arrays
        
       Given two sorted arrays nums1 and nums2 of size m and n respectively, return the median of the two sorted arrays. The overall run time complexity should be O(log (m+n)).\\
        \\
        \textbf{INSTRUCTION}: \\
        Example 1:

Input: nums1 = [1,3], nums2 = [2]\\
Output: 2.00000\\
Explanation: merged array = [1,2,3] and median is 2.\\
Example 2:\\
Input: nums1 = [1,2], nums2 = [3,4]\\
Output: 2.50000\\
Explanation: merged array = [1,2,3,4] and median is (2 + 3) / 2 = 2.5.\\

Constraints:

nums1.length == m\\
nums2.length == n\\
0 <= m <= 1000\\
0 <= n <= 1000\\
1 <= m + n <= 2000\\
-106 <= nums1[i], nums2[i] <= 106\\
        \\
        Please use the following code template:
\begin{verbatim}
class Solution(object):
    def findMedianSortedArrays(self, nums1, nums2):
        """
        :type nums1: List[int]
        :type nums2: List[int]
        :rtype: float
        """
\end{verbatim}

\rule{\textwidth}{0.4pt} 
        \textbf{Answer 1:} 
        \begin{verbatim}
class Solution(object):
    def findMedianSortedArrays(self, nums1, nums2):
        # Ensure nums1 is the smaller array to minimize binary search range
        if len(nums1) > len(nums2):
            nums1, nums2 = nums2, nums1

        m, n = len(nums1), len(nums2)
        imin, imax, half_len = 0, m, (m + n + 1) // 2

        while imin <= imax:
            # Partition nums1 at i and nums2 at j
            i = (imin + imax) // 2
            j = half_len - i

            # Adjust i to get the correct partition
            if i < m and nums2[j - 1] > nums1[i]:
                imin = i + 1  # i is too small
            elif i > 0 and nums1[i - 1] > nums2[j]:
                imax = i - 1  # i is too big
            else:
                # Perfect partition found
                if i == 0:
                    max_of_left = nums2[j - 1]
                elif j == 0:
                    max_of_left = nums1[i - 1]
                else:
                    max_of_left = max(nums1[i - 1], nums2[j - 1])

                # If total length is odd, return max_of_left
                if (m + n) % 2 == 1:
                    return max_of_left

                # Compute min_of_right for even total length
                if i == m:
                    min_of_right = nums2[j]
                elif j == n:
                    min_of_right = nums1[i]
                else:
                    min_of_right = min(nums1[i], nums2[j])

                # Return the average of max_of_left and min_of_right
                return (max_of_left + min_of_right) / 2.0

        \end{verbatim}
    \end{tiny}
    \end{tcolorbox}
\end{figure}

\begin{figure}[H]
    \centering
    \begin{tcolorbox}[colback=blue!5!white, colframe=blue!75!black, width=0.99\textwidth, arc=5mm, auto outer arc, boxrule=0.5mm, 
    title=Example of Prompt and Answer for Design Thinking (continued), 
    fonttitle=\small, fontupper=\tiny]
    \begin{tiny}
         \textbf{Prompt 2:} \\
        \textbf{INSTRUCTION}: \\
        Optimize the code and provide a more efficient version.

\rule{\textwidth}{0.4pt} 
        \textbf{Answer 2:} 
\begin{verbatim}
class Solution(object):
    def findMedianSortedArrays(self, nums1, nums2):
        """
        :type nums1: List[int]
        :type nums2: List[int]
        :rtype: float
        """
        # Ensure nums1 is the smaller array to minimize the binary search range
        if len(nums1) > len(nums2):
            nums1, nums2 = nums2, nums1
        
        m, n = len(nums1), len(nums2)
        total = m + n
        half = (total + 1) // 2
        
        left, right = 0, m
        
        while left <= right:
            partition_nums1 = (left + right) // 2
            partition_nums2 = half - partition_nums1
            
            # Handle edge cases where partition is at the extreme ends
            max_left_nums1 = float('-inf') if partition_nums1 == 0 else nums1[partition_nums1 - 1]
            min_right_nums1 = float('inf') if partition_nums1 == m else nums1[partition_nums1]
            
            max_left_nums2 = float('-inf') if partition_nums2 == 0 else nums2[partition_nums2 - 1]
            min_right_nums2 = float('inf') if partition_nums2 == n else nums2[partition_nums2]
            
            # Check if we have found the correct partition
            if max_left_nums1 <= min_right_nums2 and max_left_nums2 <= min_right_nums1:
                # If total is odd, return the max of left partitions
                if total % 2 == 1:
                    return float(max(max_left_nums1, max_left_nums2))
                else:
                    # If even, return the average of the max of left partitions and min of right partitions
                    return (max(max_left_nums1, max_left_nums2) + min(min_right_nums1, min_right_nums2)) / 2.0
            elif max_left_nums1 > min_right_nums2:
                # Move towards left in nums1
                right = partition_nums1 - 1
            else:
                # Move towards right in nums1
                left = partition_nums1 + 1
        
        raise ValueError("Input arrays are not sorted or not valid for median calculation.")

\end{verbatim}
        
    \end{tiny}
    \end{tcolorbox}
\end{figure}

\subsection{Metacognition}\label{appdix:meta}
\begin{figure}[H]
    \centering
    \begin{tcolorbox}[colback=blue!5!white, colframe=blue!75!black, width=0.99\textwidth, arc=5mm, auto outer arc, boxrule=0.5mm, title=Example of o1-preview answer, fonttitle=\small, fontupper=\tiny]
    \begin{tiny}
        \textbf{Solution:} \\
        We are given that the numbers $\sin x, \cos x$, and $\tan x$ form a geometric sequence. We need to find: $$\cos ^3 x+\cos ^2 x$$

        \textbf{Step 1: Understand the Geometric Sequence Condition}: 
        In a geometric sequence, each term divided by the previous term is a constant ratio. Therefore: \[\frac{\cos x}{\sin x}=\frac{\tan x}{\cos x}\] 
        
        \textbf{Step 2: Express $\tan x$ in Terms of $\sin x$ and $\cos x$}: 
        Recall that: \[\tan x=\frac{\sin x}{\cos x}\]
        Substitute this into the ratio: \[\frac{\cos x}{\sin x}=\frac{\frac{\sin x}{\cos x}}{\cos x}\]

        \textbf{Step 3: Simplify the Equation}: 
        Simplify the right side: \[\frac{\cos x}{\sin x}=\frac{\sin x}{\cos ^2 x}\]

        \textbf{Step 4: Cross-Multiply}: 
        Cross-multiplying gives: \[\cos ^3 x=\sin ^2 x\]

        \textbf{Step 5: Use the Pythagorean Identity}: 
        Recall the identity: \[\sin ^2 x+\cos ^2 x=1\]
        So, $ \sin ^2 x=1-\cos ^2 x .$ Substitute this into the equation: \[\cos ^3 x=1-\cos ^2 x\]

        \textbf{Step 6: Rearrange the Equation}: 
        Bring all terms to one side: \[\cos ^3 x+\cos ^2 x-1=0\]

        \textbf{Step 7: Identify the Expression to Find}: 
        Notice that the expression $\cos ^3 x+\cos ^2 x$ appears in the equation: \[\cos ^3 x+\cos ^2 x=1\]

        \textbf{Answer}: 
        $$\cos ^3 x+\cos ^2 x=1$$

        \textbf{Confidence Rating: 1.0.} \\

        \textbf{Justification:} \\
        I am fully confident in this solution because:
        \begin{itemize}
            \item Each step is based on fundamental trigonometric identities and properties of geometric sequences.
            \item The algebraic manipulations are straightforward and have been carefully checked.
            \item The conclusion directly answers the question using the derived equation.
            \item No assumptions were made beyond standard mathematical principles, reducing the possibility of errors.
        \end{itemize}

    \end{tiny}
    \end{tcolorbox}
    \captionof{figure}{Example answer of o1-preview model for metacognitive thinking.}\label{appdix_fig:meta}
\end{figure}
\subsection{Data Literacy}

\begin{figure}[H]
    \centering
    \begin{tcolorbox}[colback=blue!5!white, colframe=blue!75!black, width=\textwidth, arc=5mm, auto outer arc, boxrule=0.5mm, title=Prompt, fonttitle=\small, fontupper=\tiny]
        \textbf{Prompt:} \\
        \tiny Read the INSTRUCTIONS and QUESTION. Then, follow the INSTRUCTIONS to answer the QUESTION.\\
        \\
        INSTRUCTIONS:\\
        The graph above ("Instructional Quality in two Dimensions") depicts the distribution of Classroom Management and Cognitive Activation. Both distributions have approximately the same means. How does the dispersion differ between those two distributions?\\
        
        \vspace{\baselineskip}
        QUESTION: \\
        It doesn't make sense to compare the two dispersions, as the distribution of Cognitive Activation has two maxima.\\
        The dispersion in both graphs is equal, as the distance between the minimum and the maximum is approximately the same.\\
        In the graph above, cognitive activation shows the smaller dispersion.\\
        Don't know.
    \end{tcolorbox}
    \caption{Prompt of Data Visualization Question Q1}
    \label{fig:data_literacy_q1_prompt}
\end{figure}

\begin{figure}[H]
    \centering
    \begin{tcolorbox}[colback=blue!5!white, colframe=blue!75!black, width=\textwidth, arc=5mm, auto outer arc, boxrule=0.5mm, title=Prompt, fonttitle=\small, fontupper=\tiny]
        \textbf{Prompt:} \\
        \tiny Read the INSTRUCTIONS and QUESTION. Then, follow the INSTRUCTIONS to answer the QUESTION.\\
        \\
        INSTRUCTIONS:\\
        The bar chart shows the number of students receiving grades between 6 and 12 in a French A-level exam. Each grade range is represented by a bar, with the X-axis showing grades and the Y-axis representing the number of students.\\
        
        \vspace{\baselineskip}
        QUESTION: \\
        Which of the following answers is correct considering the graph above?\\
        - Students' performance in French A-levels are (approximately) evenly distributed over the grades.\\
        - The data provides hints for two "performance groups."\\
        - Students receive grades between 6 and 12 with quasi-equal probability.\\
        - The data points to high homogeneity of performance in the French A-level.\\
        - Don't know.
    \end{tcolorbox}
    \caption{Prompt of Data Visualization Question Q2}
    \label{fig:data_literacy_q2_prompt}
\end{figure}

\begin{figure}[H]
    \centering
    \begin{tcolorbox}[colback=blue!5!white, colframe=blue!75!black, width=\textwidth, arc=5mm, auto outer arc, boxrule=0.5mm, title=Prompt, fonttitle=\small, fontupper=\tiny]
        \textbf{Prompt:} \\
        \tiny Read the INSTRUCTIONS and QUESTION. Then, follow the INSTRUCTIONS to answer the QUESTION.\\
        \\
        INSTRUCTIONS:\\
        Refer to the table that categorizes performance using social norms, criterial norms, and individual norms. Each category has three levels: good (+), neutral (+/-), and bad (-). Identify which cell best describes the phrase "teaching goal exceeded."\\
        
        \vspace{\baselineskip}
        QUESTION: \\
        Which cell describes the best what is meant with "teaching goal exceeded"?\\
        - A\\
        - B\\
        - C\\
        - D\\
        - E\\
        - F\\
        - G\\
        - H\\
        - I\\
        - Don't know.
    \end{tcolorbox}
    \caption{Prompt of Data Visualization Question Q3}
    \label{fig:data_literacy_q3_prompt}
\end{figure}

\begin{figure}[H]
    \centering
    \begin{tcolorbox}[colback=blue!5!white, colframe=blue!75!black, width=\textwidth, arc=5mm, auto outer arc, boxrule=0.5mm, title=Prompt, fonttitle=\small, fontupper=\tiny]
        \textbf{Prompt:} \\
        \tiny Read the INSTRUCTIONS and QUESTION. Then, follow the INSTRUCTIONS to answer the QUESTION.\\
        \\
        INSTRUCTIONS:\\
        The graph shows the temporal development of knowledge/ability of three individuals (A, B, C) over time. Each person's progress is represented by a line, indicating changes in their knowledge/ability. Compare the reference norms reflected by the arrows in the chart.\\
        
        \vspace{\baselineskip}
        QUESTION: \\
        Which reference norm does Arrow 1 reflect?\\
        - The social norm\\
        - The individual norm\\
        - The criterial norm\\
        - Don't know.
    \end{tcolorbox}
    \caption{Prompt of Data Visualization Question Q4}
    \label{fig:data_literacy_q4_prompt}
\end{figure}
\begin{figure}[H]
    \centering
    \begin{tcolorbox}[colback=blue!5!white, colframe=blue!75!black, width=\textwidth, arc=5mm, auto outer arc, boxrule=0.5mm, title=Prompt, fonttitle=\small, fontupper=\tiny]
        \textbf{Prompt:} \\
        \tiny Read the INSTRUCTIONS and QUESTION. Then, follow the INSTRUCTIONS to answer the QUESTION.\\
        \\
        INSTRUCTIONS:\\
        The table displays a part of the ranking system for the A-level sports exam in the category of swimming. Different point levels are shown for various swimming disciplines, such as 50 m Breast and 100 m Medley. Determine which reference norm underlies this table.\\
        
        \vspace{\baselineskip}
        QUESTION: \\
        Above you can see a part of the ranking table for the A-level sports exam (swimming). Which reference norm underlies this table?\\
        - The individual norm (Die individuelle Bezugsnorm)\\
        - The criterial norm (Die kriteriale Bezugsnorm)\\
        - The social norm (Die soziale Bezugsnorm)\\
        - Don't know (weiß nicht).
    \end{tcolorbox}
    \caption{Prompt of Data Visualization Question Q5}
    \label{fig:data_literacy_q5_prompt}
\end{figure}

\begin{figure}[H]
    \centering
    \begin{tcolorbox}[colback=blue!5!white, colframe=blue!75!black, width=\textwidth, arc=5mm, auto outer arc, boxrule=0.5mm, title=Prompt, fonttitle=\small, fontupper=\tiny]
        \textbf{Prompt:} \\
        \tiny Read the INSTRUCTIONS and QUESTION. Then, follow the INSTRUCTIONS to answer the QUESTION.\\
        \\
        INSTRUCTIONS:\\
        The image displays a boxplot comparison illustrating sex-related performance differences in two sixth-grade classes. Analyze the differences between boys and girls in these classes based on the dispersion and mean differences shown in the boxplots.\\
        
        \vspace{\baselineskip}
        QUESTION: \\
        If one compares the differences between boys and girls in the classes depicted within the graph above, one can say:\\
        - The difference between boys and girls is more relevant in class 6b, as the dispersion is smaller in class 6b.\\
        - The differences are equally relevant, because the dispersion within classes is approximately the same.\\
        - The differences are equally relevant, because the mean differences are approximately the same.\\
        - You can't see that in this graph.\\
        - Don't know.
    \end{tcolorbox}
    \caption{Prompt of Data Visualization Question Q6}
    \label{fig:data_literacy_q6_prompt1}
\end{figure}

\begin{figure}[H]
    \centering
    \begin{tcolorbox}[colback=blue!5!white, colframe=blue!75!black, width=\textwidth, arc=5mm, auto outer arc, boxrule=0.5mm, title=Prompt, fonttitle=\small, fontupper=\tiny]
        \textbf{Prompt:} \\
        \tiny Read the INSTRUCTIONS and QUESTION. Then, follow the INSTRUCTIONS to answer the QUESTION.\\
        \\
        INSTRUCTIONS:\\
        The image presents a detailed comparison of gender-specific performance in two classes (10a and 10b) for German grades in class 11 using jitter plots. Analyze the differences between girls and boys in the two classes based on the mean grades and the spread of their scores.\\
        
        \vspace{\baselineskip}
        QUESTION: \\
        The difference between girls and boys is 0.5 in both graphs above. Nevertheless, one has to judge the differences disparately. Please justify!\\
        If you know that your class averages 513 points at a statewide test and your school 508, which further information do you need to be able to judge the magnitude of this difference?\\
        - The average performance of the whole state\\
        - None\\
        - Information about the minimum and maximum in your school and your class\\
        - Information about the dispersion in your school and your class\\
        - Don't know.
    \end{tcolorbox}
    \caption{Prompt of Data Visualization Question Q7}
    \label{fig:data_literacy_q7_prompt}
\end{figure}

\begin{figure}[H]
    \centering
    \begin{tcolorbox}[colback=blue!5!white, colframe=blue!75!black, width=\textwidth, arc=5mm, auto outer arc, boxrule=0.5mm, title=Prompt, fonttitle=\small, fontupper=\tiny]
        \textbf{Prompt:} \\
        \tiny Read the INSTRUCTIONS and QUESTION. Then, follow the INSTRUCTIONS to answer the QUESTION.\\
        \\
        INSTRUCTIONS:\\
        The image compares pairwise performance between two groups, Country A and Country B, under conditions of small and large dispersion. Examine how the performance dominance shifts between the two groups as dispersion changes.\\
        
        \vspace{\baselineskip}
        QUESTION: \\
        If you know the percentage of superiority within all pairwise comparisons (see illustrating graph above) instead of the mean difference, you additionally account for...\\
        -  the fact, that outliers influence the mean little.\\
        -  the fact, that the same mean difference can express very different magnitudes of effects depending on the dispersion.\\
        -  nothing.\\
        -  the fact, that the mean differences can differ in both groups.\\
        -  Don't know.
    \end{tcolorbox}
    \caption{Prompt of Data Visualization Question Q8}
    \label{fig:data_literacy_q8_prompt}
\end{figure}

\begin{figure}[H]
    \centering
    \begin{tcolorbox}[colback=blue!5!white, colframe=blue!75!black, width=\textwidth, arc=5mm, auto outer arc, boxrule=0.5mm, title=Prompt, fonttitle=\small, fontupper=\tiny]
        \textbf{Prompt:} \\
        \tiny Read the INSTRUCTIONS and QUESTION. Then, follow the INSTRUCTIONS to answer the QUESTION.\\
        \\
        INSTRUCTIONS:\\
        The table shows raw data of self-evaluation for students in Class A and Class B, covering Cognitive Activation, Classroom Management, and Supportive Climate. The mean per quality dimension and per student is also calculated. Consider which method would make sense for correcting answer tendencies or enabling social comparison.\\
        
        \vspace{\baselineskip}
        QUESTION: \\
        In the table above, you can see a minimal example of raw data from a self-evaluation. Please decide which of the following claims are true.\\
        If you want to correct the raw data by answer tendencies (throughout strict or mild answers of one student), it makes sense...\\
        -  to multiply the raw values with the according means of the quality dimension [e.g., a * M(kA)].\\
        -  to subtract the corresponding mean per student from the raw values [e.g., a - M(S1)].\\
        -  to add the sum of all means per quality dimension to the raw values.\\
        -  to subtract the sum of all means per student from the raw values.\\
        -  don't know.\\
        If you want to transform the raw values in a way that a direct social comparison is possible, it makes sense...\\
        -  to subtract the mean per student from every raw value.\\
        -  to divide the raw values by the means per student [e.g., a / M(S1)].\\
        -  to add the means per quality dimension to each raw value.\\
        -  to subtract the mean per quality dimension from each raw value [e.g., a - M(kA)].\\
        -  don't know.
    \end{tcolorbox}
    \caption{Prompt of Data Visualization Question Q9}
    \label{fig:data_literacy_q9_prompt}
\end{figure}

\subsection{Abstract Reasoning}
\begin{figure}[H]
    \centering
    \begin{tcolorbox}[colback=blue!5!white, colframe=blue!75!black, width=\textwidth, arc=5mm, auto outer arc, boxrule=0.5mm, title=Prompt, fonttitle=\small, fontupper=\tiny]
        \textbf{Prompt:} \\
        \tiny Read the INSTRUCTIONS and QUESTION. Then, follow the INSTRUCTIONS to answer the QUESTION.\\
        \\
        INSTRUCTIONS: The QUESTION below describes a Raven's Progressive Matrices test item. This test consists of a 3x3 grid of images, with the bottom-right cell empty. Your task is to identify the pattern in the grid and select the correct image to fill the empty cell from eight given choices.\\

        Key points to consider:
        \begin{enumerate}[label=\arabic*., leftmargin=*]
            \item Grid layout: 3x3, with 8 filled cells and one empty cell (bottom-right)
            \item Pattern: Observe how shapes, sizes, and colors change across rows and columns
            \item Shape attributes: Size (Small, medium, large) and Color (Black, light grey, dark grey)
        \end{enumerate}
        \vspace{\baselineskip}
        Solving process:
        \begin{enumerate}[label=\arabic*., leftmargin=*]
            \item Analyze the patterns in each row and column
            \item Identify how attributes (shape, size, color) change or remain constant
            \item Predict what should appear in the empty cell based on these patterns
            \item Select the choice that best matches your prediction
        \end{enumerate}
        \vspace{\baselineskip}
        QUESTION: \\
        Contexts\\
        Context 1:\\
        Top-left: medium black hexagon\\Top-right: small black hexagon\\Bottom-left: small black hexagon\\Bottom-right: large black hexagon\\

        Context 2:\\
        Top-left: large light grey pentagon\\
        Top-right: small light grey pentagon\\
        Bottom-left: medium light grey pentagon\\
        Bottom-right: small light grey pentagon\\
        
        Context 3:\\
        Top-left: large dark grey circle\\
        Top-right: medium dark grey circle\\
        Bottom-left: small dark grey circle\\
        Bottom-right: small dark grey circle\\
        
        Context 4:\\
        Top-left: large light grey circle\\
        Top-right: empty\\
        Bottom-left: empty\\
        Bottom-right: empty\\
        
        Context 5:\\
        Top-left: medium dark grey hexagon\\
        Top-right: empty\\
        Bottom-left: empty\\
        Bottom-right: empty\\
        
        Context 6:\\
        Top-left: medium black pentagon\\
        Top-right: empty\\
        Bottom-left: empty\\
        Bottom-right: empty\\
        
        Context 7:\\
        Top-left: small dark grey pentagon\\
        Top-right:  large dark grey pentagon\\
        Bottom-left: large dark grey pentagon\\
        Bottom-right: medium dark grey pentagon\\
        
        Context 8:\\
        Top-left: small black circle\\Top-right: large light grey circle\\Bottom-left: large black circle\\Bottom-right: medium black circle\\
        \\
        \\...Continue below...

    \end{tcolorbox}
\end{figure}
\begin{figure}[H]
    \centering
    \begin{tcolorbox}[colback=blue!5!white, colframe=blue!75!black, width=\textwidth, arc=5mm, auto outer arc, boxrule=0.5mm, title=Prompt, fonttitle=\small, fontupper=\tiny]
        \textbf{...Continue from above...} \\
        \\
        Choices:\\
        Choice 1:\\
        Top-left: small light grey hexagon\\Top-right: large light grey hexagon\\Bottom-left: large light grey hexagon\\Bottom-right: medium black hexagon\\
        
        Choice 2:\\
        Top-left: small light grey pentagon\\
        Top-right: large dark grey pentagon\\
        Bottom-left: large black pentagon\\
        Bottom-right: medium dark grey pentagon\\
        
        Choice 3:\\
        Top-left: small black pentagon\\
        Top-right: empty\\
        Bottom-left: empty\\
        Bottom-right: empty\\
        
        Choice 4:\\
        Top-left: large light grey hexagon\\
        Top-right: small light grey hexagon\\
        Bottom-left: medium light grey hexagon\\
        Bottom-right: small light grey hexagon
        
        Choice 5:\\
        Top-left: small light grey triangle\\
        Top-right: large light grey hexagon\\
        Bottom-left: medium light grey square\\
        Bottom-right: medium light grey pentagon\\
        
        Choice 6:\\
        Top-left: large light grey hexagon\\
        Top-right: small light grey hexagon\\
        Bottom-left: medium black hexagon\\
        Bottom-right: small light grey hexagon\\
        
        Choice 7:\\
        Top-left: empty\\
        Top-right: medium dark grey diamond\\
        Bottom-left: medium light grey square\\
        Bottom-right: empty\\
        
        Choice 8:\\
        Top-left: empty\\Top-right: medium light grey pentagon\\Bottom-left: medium light grey hexagon\\Bottom-right: small light grey triangle\\
    \end{tcolorbox}
\end{figure}

\begin{figure}[H]
    \centering
    \begin{tcolorbox}[colback=blue!5!white, colframe=blue!75!black, width=\textwidth, arc=5mm, auto outer arc, boxrule=0.5mm, title=Answer/Response, fonttitle=\small, fontupper=\tiny]
        \textbf{Answer:} \\
        To solve this Raven's Progressive Matrices test item, we'll analyze the given contexts to identify patterns in shapes, colors, and sizes. Then, we'll select the choice that best fits the established pattern.\\
        
        \textbf{Step 1: Analyze the Contexts}\\
        Contexts 1-3 and 7-8 are full 2x2 grids with specific patterns:\\
        
        Context 1:\\Shape: Hexagons; Color: Black; Sizes: Medium, Small, Small, Large\\
        Context 2:\\Shape: Pentagons; Color: Light Grey; Sizes: Large, Small, Medium, Small\\
        Context 3:\\Shape: Circles; Color: Dark Grey; Sizes: Large, Medium, Small, Small\\
        Context 7:\\Shape: Pentagons; Color: Dark Grey; Sizes: Small, Large, Large, Medium\\
        Context 8:\\Shape: Circles; Colors: Black and Light Grey;; Sizes: Small, Large, Large, Medium\\
        
        Observations:\\
        * Shape Consistency: Each context uses the same shape throughout its grid.\\
        * Color Consistency: Most contexts maintain the same color, except Context 8, which introduces a variation.\\
        * Size Variation: Sizes vary within each grid but follow no immediately obvious numerical pattern.\\
        \\
        \textbf{Step 2: Identify the Pattern}\\
        Given the observations:\\
        * Primary Pattern: Each context features a grid where the shape and color remain consistent, and sizes vary.\\
        * Secondary Pattern: In some contexts (like Context 8), there is a color variation, but the shape remains the same.\\
        \\
        \textbf{Step 3: Analyze Context 4}\\
        Context 4 is incomplete:\\ a) Top-left: Large Light Grey Circle. b) Others: Empty\\
        \\
        This context starts with a large light grey circle. Based on the patterns observed, we can predict that the rest of the grid should:\\Maintain the same shape: Circles\\Maintain the same color: Light Grey\\Include size variations\\
        \\
        \textbf{Step 4: Evaluate the Choices}\\We need a choice that:\\ a) Uses circles as the shape, b) Has light grey as the color, c) Displays size variations\\
        Let's examine the choices:\\
        \\
        Choice 4:\\Shapes: Hexagons; Color: Light Grey; Sizes: Large, Small, Medium, Small\\
        Not suitable—the shapes are hexagons, not circles.\\

        Choice 5:\\Shapes: Triangle, Hexagon, Square, Pentagon; Color: Light Grey; Sizes: Small, Large, Medium, Medium\\
        Not suitable—multiple shapes used.\\
        
        Choice 6:\\Shapes: Hexagons; Colors: Light Grey and Black; Sizes: Large, Small, Medium, Small\\
        Not suitable—incorrect shapes and inconsistent colors.\\
        
        Choice 1:\\Shapes: Hexagons; Colors: Light Grey and Black; Sizes: Small, Large, Large, Medium\\
        Not suitable—incorrect shapes.\\
        
        Choice 3:\\Top-left: Small Black Pentagon; Others: Empty\\
        Not suitable—incorrect shape and color.\\
        
        Choice 2:\\Shapes: Pentagons; Colors: Light Grey, Dark Grey, Black; Sizes: Small, Large, Large, Medium\\
        Not suitable—inconsistent colors.\\
        
        Choice 7 and Choice 8:\\Both choices introduce different shapes and colors or have empty cells.\\
        Not suitable.\\
        \\
        \textbf{Step 5: Select the Best Choice}\\
        None of the choices perfectly fit the desired pattern. However, Choice 4 is the closest:\\
        * Shapes: Hexagons (though not circles)\\
        * Color: Light Grey (matches Context 4's color)\\
        * Sizes: Varying sizes (Large, Small, Medium, Small)\\
        \\
        While the shape is not a circle, Choice 4 maintains the same shape throughout, uses the correct color, and includes size variations.\\
        \\
        \textbf{Answer: Choice 4}\\
        Choice 4 best matches the pattern established in Context 4, with consistent shapes, colors, and size variations.\\
        \rule{\textwidth}{0.4pt}
         \textbf{Reference Answer:}
         Choice 1     
    \end{tcolorbox}
\end{figure}

\subsection{Spatial Reasoning}    

\begin{figure}[H]
\centering
\begin{tcolorbox}[colback=blue!5!white, colframe=blue!75!black, width=\textwidth, arc=5mm, auto outer arc, boxrule=0.5mm, title=Prompt and Answer, fonttitle=\small, fontupper=\tiny]
\textbf{Prompt:} \
\tiny You are given a civil engineering scenario involving a two-span steel-reinforced concrete bridge. The total length of the bridge is 100 meters, with each span being 50 meters long. The deck of the bridge is composed of reinforced concrete, and it is supported by H-shaped steel girders, spaced 5 meters apart. The following details describe the material properties and loading conditions:

- The steel girders have the following properties:
  - Young's Modulus: 200 GPa
  - Yield Strength: 250 MPa
- The reinforced concrete deck has the following properties:
  - Young's Modulus: 30 GPa
  - Compressive Strength: 40 MPa

A 100 kN point load is applied at the center of the middle span, directly on the bridge deck, acting vertically downwards.

Please predict the following:
1. How will the bridge deck deform under this load? Describe the deformation pattern and the expected maximum displacement.
2. Identify which parts of the steel girders are most likely to experience stress concentration.
3. If the point load is increased to 200 kN, predict where and how the structure is likely to fail.
4. Provide an explanation of how the stresses are distributed across the entire structure. \
\rule{\textwidth}{0.4pt}
\textbf{Answer:} \
\tiny \textbf{1. Deformation Mode and Maximum Displacement} \
• Model Output: Correctly predicted deformation mode and location of max displacement but lacked precise values. \
• Reference Answer: Max deformation at load point, 12 mm. \
• Rating: 8/10 (Qualitative accurate, lacks precision).\
\
\textbf{2. Stress Concentration Areas} \
• Model Output: Correctly identified stress concentration at midspan and support. \
• Reference Answer: Stress concentration matches prediction (150 MPa at midspan). \
• Rating: 10/10 (Accurate identification).\
\
\textbf{3. Failure Prediction at 200 kN Load} \
• Model Output: Predicted steel yielding at midspan but missed specific stress values. \
• Reference Answer: Stress reaches 300 MPa, near yield strength. \
• Rating: 9/10 (Accurate, lacks numerical detail).\
\
\textbf{4. Stress Distribution} \
• Model Output: Explained tensile/compressive stresses and shear distribution. \
• Reference Answer: Stress concentrated at midspan and support. \
• Rating: 10/10 (Detailed and consistent).\
\
\rule{\textwidth}{0.4pt}
\textbf{Reference Answer:} \
• Deformation Mode: 8/10\
• Stress Concentration: 10/10\
• Failure Prediction: 9/10\
• Stress Distribution: 10/10\
\
Total Score: 9.25/10
\end{tcolorbox}
\end{figure}

\begin{figure}[H]
\centering
\begin{tcolorbox}[colback=blue!5!white, colframe=blue!75!black, width=\textwidth, arc=5mm, auto outer arc, boxrule=0.5mm, title=Prompt and Answer, fonttitle=\small, fontupper=\tiny]
\textbf{Prompt:} \
\tiny You are given a scenario involving a single-span steel-reinforced concrete bridge. The span of the bridge is 50 meters long, and the deck is composed of reinforced concrete supported by H-shaped steel girders. The following details describe the material properties and loading conditions:

- The steel girders have the following properties:
  - Young's Modulus: 200 GPa
  - Yield Strength: 250 MPa
- The reinforced concrete deck has the following properties:
  - Young's Modulus: 30 GPa
  - Compressive Strength: 40 MPa

A 200 kN point load is applied 15 meters from the left support, acting vertically downwards.

Please predict the following:
1. How will the bridge deck deform under this load? Describe the deformation pattern and the expected maximum displacement.
2. Identify which parts of the steel girders are most likely to experience stress concentration.
3. If the point load is increased to 300 kN, predict where and how the structure is likely to fail.
4. Provide an explanation of how the stresses are distributed across the entire structure, including bending moments and shear forces. \
\rule{\textwidth}{0.4pt}
\textbf{Answer:} \
\tiny \textbf{1. Prediction of Deformation Mode and Maximum Displacement} \
• Model Output: The model described an asymmetrical downward deflection, correctly identifying the maximum displacement near the load point. It also noted greater deformation on the left side (0m to 15m) due to higher forces, but no specific displacement values were provided. \
• Reference Answer: The maximum displacement indeed occurs at the load point (15 meters), with a value of 16 mm. \
• Comparison: The model accurately described the deformation mode and location of maximum displacement, though it lacked precise quantitative results. \
• Rating: 8/10 (Qualitative accuracy, lacks quantitative support).\
\
\textbf{2. Identification of Stress Concentration Areas} \
• Model Output: The model correctly predicted stress concentration areas, especially noting the maximum tensile stress at the bottom flange below the load point (15 meters) and highlighted high shear near the supports. It also mentioned potential risks of stress concentrations at geometrical discontinuities such as connections and welds. \
• Reference Answer: The maximum stress concentration occurs at the bottom flange under the load point (180 MPa), with increased shear near the supports. \
• Comparison: The model’s identification of stress concentration areas aligns perfectly with the reference answer, providing clear and accurate qualitative analysis. \
• Rating: 10/10 (Accurate identification, consistent with reference).\
\
\textbf{3. Failure Prediction at 300 kN Load Increase} \
• Model Output: The model reasonably predicted that at a load increase to 300 kN, yielding would likely occur at the bottom flange under tension. It also noted potential cracking in the concrete deck due to excessive compressive stress, and shear failure near the supports. \
• Reference Answer: At 300 kN, the stress in the bottom flange reaches 270 MPa, approaching the yield strength, with possible shear failure on the left support. \
• Comparison: The model’s failure prediction aligns well with the reference, especially regarding steel yielding and shear failure, though it lacks specific stress values. \
• Rating: 9/10 (Accurate failure prediction but missing specific stress data).\
\
\textbf{4. Explanation of Stress Distribution} \
• Model Output: The model provided a detailed explanation of stress distribution, identifying maximum bending moment at the load point and maximum shear near the supports. It also explained the composite action between steel and concrete, and described the distribution patterns of stress and shear across the structure. \
• Reference Answer: The stress distribution shows maximum bending moment and shear concentrated at the load point and supports, with tensile stress in the steel beam and compressive stress in the concrete deck. \
• Comparison: The model’s explanation of stress distribution is entirely consistent with the reference, offering detailed and accurate analysis of bending moment and shear distribution. \
• Rating: 10/10 (Comprehensive and consistent with reference).\
\
\rule{\textwidth}{0.4pt}
\textbf{Score:} \
• Deformation Mode: 8/10\
• Stress Concentration: 10/10\
• Failure Prediction: 9/10\
• Stress Distribution: 10/10\
\
Total Score: 9.25/10
\end{tcolorbox}
\end{figure}

\begin{figure}[H]
\centering
\begin{tcolorbox}[colback=blue!5!white, colframe=blue!75!black, width=\textwidth, arc=5mm, auto outer arc, boxrule=0.5mm, title=Prompt and Answer, fonttitle=\small, fontupper=\tiny]
\textbf{Prompt:} \
\tiny You are given a structural engineering scenario involving two concrete columns supporting a horizontal concrete beam. The span between the two columns is 10 meters, and the columns are 5 meters tall. The following details describe the material properties and loading conditions:

- The concrete has the following properties:
  - Young's Modulus: 30 GPa
  - Compressive Strength: 40 MPa
- The beam has a rectangular cross-section with a width of 0.3 meters and a height of 0.5 meters.
- A horizontal point load of 50 kN is applied at the midpoint of the beam, acting along the horizontal axis.

Please predict the following:
1. How will the beam and columns deform under this load? Describe the deformation pattern and the expected maximum displacement.
2. Identify which parts of the structure (beam or columns) are most likely to experience stress concentration.
3. If the load is increased to 100 kN, predict where and how the structure is likely to fail.
4. Provide an explanation of how the stresses are distributed across the entire structure, including bending moments, shear forces, and axial stresses in the columns. \
\rule{\textwidth}{0.4pt}
\textbf{Answer:} \
\tiny \textbf{1. Prediction of Deformation Mode and Maximum Displacement} \
• Model Output: The model correctly predicted the deformation mode, identifying that the main deformation is due to bending of the column, with maximum displacement at the top of the column under a 50 kN horizontal load. However, the predicted maximum displacement (11.11 mm) significantly differs from the ground truth (6 mm). \
• Reference Answer: Maximum displacement occurs at the midpoint of the beam with a value of 6 mm, while displacement at the column top is smaller. \
• Comparison: The model accurately predicted the deformation mode and location but showed significant numerical deviation. \
• Rating: 7/10 (Accurate qualitative description, large numerical deviation).\
\
\textbf{2. Identification of Stress Concentration Areas} \
• Model Output: The model identified stress concentration at the base of the column, due to bending moments causing tensile and compressive stresses on different sides. It also noted potential stress concentrations at the beam-column junction and the stress distribution along the beam. \
• Reference Answer: The maximum stress concentration is found in the beam’s center (8 MPa tensile stress and 7 MPa compressive stress), with mild shear stress at the column base. \
• Comparison: While the model correctly identified stress concentration regions, it differed in emphasizing the importance of the beam’s center, which was highlighted in the reference answer. \
• Rating: 8/10 (Reasonable area identification, slight differences in stress distribution detail).\
\
\textbf{3. Failure Prediction at 100 kN Load Increase} \
• Model Output: The model predicted that, at 100 kN load, tensile cracking would occur at the column base once concrete’s tensile limit is exceeded. It also mentioned possible shear failure and rotational instability at the foundation. \
• Reference Answer: Cracking is expected in the beam’s center at 100 kN, with increased shear stress at the column base. \
• Comparison: The model’s failure prediction aligns well with the reference but places more emphasis on failure at the column base, while the reference highlights potential cracking in the beam’s center. \
• Rating: 9/10 (Accurate prediction but does not emphasize failure in the beam center).\
\
\textbf{4. Explanation of Stress Distribution} \
• Model Output: The model provided a detailed explanation of stress distribution, noting maximum bending moments at the column base and shear distribution along the column height. It also highlighted the primary compressive and tensile stresses in the beam and discussed the composite behavior of the beam and column under horizontal loads. \
• Reference Answer: Stress distribution shows maximum bending moments at the column base and significant stress in the beam’s center, with mild shear stress at the column base. \
• Comparison: The model’s explanation of stress distribution is consistent with the reference, especially regarding bending moments and shear at the column base, but oversimplifies stress analysis in the beam. \
• Rating: 8/10 (Reasonable analysis, insufficient detail in beam stress analysis).\
\
\rule{\textwidth}{0.4pt}
\textbf{Score:} \
• Deformation Mode: 7/10\
• Stress Concentration: 8/10\
• Failure Prediction: 9/10\
• Stress Distribution: 8/10\
\
Total Score: 8/10
\end{tcolorbox}
\end{figure}

\begin{figure}[H]
\centering
\begin{tcolorbox}[colback=blue!5!white, colframe=blue!75!black, width=\textwidth, arc=5mm, auto outer arc, boxrule=0.5mm, title=Prompt and Answer, fonttitle=\small, fontupper=\tiny]
\textbf{Prompt:} \
\tiny You are given a structural engineering scenario involving three concrete columns supporting two horizontal concrete beams. The span between each pair of columns is 5 meters, and the columns are 4 meters tall. The following details describe the material properties and loading conditions:

- The concrete has the following properties:
  - Young's Modulus: 30 GPa
  - Compressive Strength: 40 MPa
- The beams have a rectangular cross-section with a width of 0.25 meters and a height of 0.5 meters.
- A downward point load of 75 kN is applied at the midpoint between the first and second column on the left beam.

Please predict the following:
1. How will the beams and columns deform under this load? Describe the deformation pattern and the expected maximum displacement.
2. Identify which parts of the structure (beams or columns) are most likely to experience stress concentration.
3. If the load is increased to 150 kN, predict where and how the structure is likely to fail.
4. Provide an explanation of how the stresses are distributed across the entire structure, including bending moments, shear forces, and axial stresses in the columns. \
\rule{\textwidth}{0.4pt}
\textbf{Answer:} \
\tiny \textbf{1. Prediction of Deformation Mode and Maximum Displacement} \
• Model Output: The model accurately described the deformation mode of the left beam under a 75 kN load, identifying the maximum displacement at the midpoint. The predicted displacement was 2.5 mm, assuming minimal deformation in the column. However, there was a significant deviation from the actual result. \
• Reference Answer: The maximum displacement occurred at the load point, with an actual value of 9 mm. \
• Comparison: The model’s qualitative prediction of deformation was accurate, but it underestimated the magnitude of displacement. \
• Rating: 7/10 (Accurate qualitative description, significant numerical deviation).\
\
\textbf{2. Identification of Stress Concentration Areas} \
• Model Output: The model correctly identified stress concentration in the midsection of the left beam, highlighting the maximum tensile stress at the bottom fiber of the beam. It also recognized areas of shear stress concentration near the column. \
• Reference Answer: Maximum tensile stress of 10 MPa and compressive stress of 8 MPa were observed at the beam’s midpoint, with significant shear stress at the base of the left column. \
• Comparison: The model accurately identified stress concentration areas, though it did not provide specific stress values, the qualitative analysis was consistent with the reference. \
• Rating: 9/10 (Accurate identification of stress concentration areas, reasonable qualitative analysis).\
\
\textbf{3. Failure Prediction at 150 kN Load Increase} \
• Model Output: The model predicted that at a 150 kN load, the tensile stress at the bottom fiber of the beam could exceed the concrete’s tensile limit, leading to cracking failure. It also noted the possibility of shear failure. \
• Reference Answer: At 150 kN, cracking failure is expected in the beam’s midpoint due to excessive tensile stress, with increased shear stress at the base of the left column. \
• Comparison: The model’s failure prediction closely matched the reference, accurately describing the failure mode in the beam’s midpoint. \
• Rating: 9/10 (Accurate failure prediction, consistent with actual results).\
\
\textbf{4. Explanation of Stress Distribution} \
• Model Output: The model provided a detailed explanation of stress distribution, noting maximum bending moments and shear forces at the beam’s midpoint, while the column mainly bore axial compressive stress. It also calculated the distribution of shear and bending moments. \
• Reference Answer: Maximum stress was observed in the beam’s midpoint, with shear concentrated near the column. Axial compressive stress in the column was minimal. \
• Comparison: The model’s explanation of stress distribution aligned well with the reference, particularly in the trends of bending moments and shear forces, but lacked precise stress values. \
• Rating: 8/10 (Reasonable explanation of stress distribution, insufficient numerical detail).\
\
\rule{\textwidth}{0.4pt}
\textbf{Score:} \
• Deformation Mode: 7/10\
• Stress Concentration: 9/10\
• Failure Prediction: 9/10\
• Stress Distribution: 8/10\
\
Total Score: 8.25/10
\end{tcolorbox}
\end{figure}

\begin{figure}[H]
\centering
\begin{tcolorbox}[colback=blue!5!white, colframe=blue!75!black, width=\textwidth, arc=5mm, auto outer arc, boxrule=0.5mm, title=Prompt and Answer, fonttitle=\small, fontupper=\tiny]
\textbf{Prompt:} \
\tiny You are given a molecular dynamics simulation involving a protein-ligand complex. The ligand is bound to a specific binding pocket of the protein. The following details describe the setup and conditions of the simulation:

- Protein-Ligand Complex: Barnase-Barstar complex (PDB ID: 1BRS)
- Simulation Type: Molecular Dynamics
- Temperature: 300K
- Duration: 100 nanoseconds (ns)
- Time step: 1 picosecond (ps)
- Environment: Water molecules are present to simulate a biological environment
- External Factors: No external forces are applied except for thermal motion at 300K.

The ligand starts at the binding pocket of the protein at coordinates (X = 3.0, Y = -2.5, Z = 1.2) in the 3D space. Over time, the ligand moves within the binding pocket due to thermal motion. No additional forces are applied during the simulation.

Please predict the following:
1. How will the ligand move during the simulation? Describe the general trajectory and key factors that influence the ligand’s movement.
2. What is the expected final position of the ligand after 100 ns? Provide the expected final coordinates.
3. What are the most likely forces or interactions (such as hydrogen bonds, van der Waals interactions) that influence the ligand's binding stability during the simulation?
4. If the temperature is increased to 400K, how would this affect the ligand’s movement and binding stability? Predict any changes in the ligand’s final position or interactions. \
\rule{\textwidth}{0.4pt}
\textbf{Answer:} \
\tiny \textbf{1. Ligand Movement During the Simulation} \
• Model Output: The model accurately described the ligand’s movement within the binding site, emphasizing small-scale vibrations and rotations due to thermal fluctuations. It also mentioned possible conformational adjustments between the protein and ligand, and the transient role of water molecules through hydrogen bonding. \
• Reference Answer: The reference similarly describes the ligand oscillating within the binding site, without complete detachment. The model’s description aligns closely with the reference, particularly regarding the impact of thermal fluctuations and intermolecular interactions. \
• Comparison: The model effectively captured the key characteristics of ligand movement and identified important influencing factors. \
• Rating: 10/10 (Model’s description matches the reference perfectly).\
\
\textbf{2. Expected Final Position After 100 ns} \
• Model Output: The model predicted that the ligand would remain near its initial position at the end of the simulation, providing a final coordinate range (X ≈ 3.0 ± 0.2, Y ≈ -2.5 ± 0.2, Z ≈ 1.2 ± 0.2), and attributed this small displacement to thermal fluctuations. \
• Reference Answer: The reference gives the ligand’s final position as (X = 3.2, Y = -2.3, Z = 1.0), confirming that the ligand remained in the binding site with minor displacement. \
• Comparison: The model’s predicted displacement range is very close to the reference coordinates, though the model provided a range rather than an exact value. The deviation is within a reasonable margin. \
• Rating: 9/10 (Model’s prediction is close, but lacks specific coordinates).\
\
\textbf{3. Forces and Interactions Influencing Binding Stability} \
• Model Output: The model provided a detailed list of key forces contributing to binding stability, including hydrogen bonds, van der Waals forces, hydrophobic interactions, electrostatic interactions, and potential π-π stacking. It highlighted the importance of hydrogen bonding and hydrophobic interactions. \
• Reference Answer: The reference confirms that hydrogen bonding and van der Waals forces are the main stabilizing interactions, consistent with the model’s description. \
• Comparison: The model’s analysis of binding forces aligns perfectly with the reference and expands the discussion to include possible π-π stacking and electrostatic interactions, adding depth to the analysis. \
• Rating: 10/10 (Comprehensive analysis of interactions, consistent with reference).\
\
\textbf{4. Effects of Increasing Temperature to 400K} \
• Model Output: The model predicted that increasing the temperature to 400K would result in more vigorous ligand motion and a reduction in binding stability, as hydrogen bonds and van der Waals forces weaken. It suggested the likelihood of the ligand drifting away from the binding site, possibly leading to temporary or complete dissociation. \
• Reference Answer: The reference confirms that ligand movement becomes more pronounced at 400K, with the final position expected to shift to (X = 3.5, Y = -2.0, Z = 1.5), and binding stability decreases. \
• Comparison: The model’s predictions regarding the effect of elevated temperature on ligand movement and stability match the reference, though it did not provide specific final coordinates. The predicted trend is accurate. \
• Rating: 9/10 (Model’s description of temperature effects aligns with the reference but lacks precise position coordinates).\
\
\rule{\textwidth}{0.4pt}
\textbf{Score:} \
• Ligand Movement: 10/10\
• Final Position Prediction: 9/10\
• Binding Forces and Stability: 10/10\
• Temperature Increase Effects: 9/10\
\
Total Score: 9.5/10
\end{tcolorbox}
\end{figure}

\begin{figure}[H]
\centering
\begin{tcolorbox}[colback=blue!5!white, colframe=blue!75!black, width=\textwidth, arc=5mm, auto outer arc, boxrule=0.5mm, title=Prompt and Answer, fonttitle=\small, fontupper=\tiny]
\textbf{Prompt:} \
\tiny You are provided with a molecular dynamics simulation of the COX-2 enzyme and a bound inhibitor. The inhibitor is positioned within the active site of COX-2 and interacts with key residues through various binding interactions. The following describes the setup and conditions of the simulation:

- Protein-Inhibitor Complex: Cyclooxygenase-2 (COX-2) with a small-molecule inhibitor bound to the active site (PDB ID: 1A4W)
- Simulation Duration: 100 nanoseconds (ns)
- Time step: 1 picosecond (ps)
- Temperature: 300K
- Environment: Water molecules and ions surround the protein-inhibitor complex
- Initial Position of Inhibitor: The inhibitor begins at coordinates (X = 4.5, Y = -1.2, Z = 3.8) within the COX-2 active site.

During the simulation, the inhibitor remains bound to the active site but undergoes thermal motion due to the temperature of the system. Predict the following:

1. How will the inhibitor move during the simulation? Describe its expected trajectory and the key factors influencing this movement.
2. What is the expected final position of the inhibitor after 100 ns? Provide the expected final coordinates.
3. Which forces or interactions (such as hydrogen bonds, van der Waals interactions) are most likely to contribute to the inhibitor’s binding stability during the simulation?
4. If a mutation occurs at one of the key residues (e.g., Serine at position 530 is mutated to Alanine), how would this mutation affect the inhibitor’s movement and binding stability within the active site? \
\rule{\textwidth}{0.4pt}
\textbf{Answer:} \
\tiny \textbf{1. Inhibitor Movement During the Simulation} \
• Model Output: The model described the inhibitor’s movement within the COX-2 active site, emphasizing slight vibrations and rotations due to thermal fluctuations, and conformational adjustments between the inhibitor and protein residues. The model also highlighted the influence of solvent (water molecules) on the inhibitor’s movement and binding stability. \
• Reference Answer: The reference describes the inhibitor’s movement as primarily driven by thermal fluctuations, with no detachment from the binding site, showing minor displacements around the binding site. \
• Comparison: The model’s description of movement closely matches the reference, especially in capturing the vibrations and small movements around the binding site, and the influence of molecular interactions. \
• Rating: 10/10 (Qualitative description fully consistent with reference).\
\
\textbf{2. Expected Final Position After 100 ns} \
• Model Output: The model predicted that the inhibitor would remain near its initial position after 100 ns, providing a coordinate range (X ≈ 4.5 ± 0.2, Y ≈ -1.2 ± 0.2, Z ≈ 3.8 ± 0.2), attributing this small displacement to thermal fluctuations. \
• Reference Answer: The inhibitor’s final coordinates are (X = 4.6, Y = -1.0, Z = 3.7), indicating minor displacement within the binding site. \
• Comparison: The model’s predicted range closely matches the reference coordinates, though it provided a range rather than exact values. The displacement remains within a reasonable margin. \
• Rating: 9/10 (Accurate prediction, but lacks specific coordinates).\
\
\textbf{3. Forces and Interactions Contributing to Binding Stability} \
• Model Output: The model listed key interactions stabilizing the inhibitor, including hydrogen bonds, van der Waals forces, hydrophobic interactions, and π-π stacking. The model accurately recognized the importance of these interactions, especially the critical role of hydrogen bonding within the binding site. \
• Reference Answer: The reference confirms that hydrogen bonds and van der Waals forces are the main forces stabilizing the inhibitor, particularly the hydrogen bond involving Ser530. \
• Comparison: The model’s identification of forces matches the reference perfectly, providing a comprehensive explanation of each interaction’s contribution. \
• Rating: 10/10 (Detailed and consistent with reference).\
\
\textbf{4. Impact of Ser530Ala Mutation on Inhibitor Movement and Binding Stability} \
• Model Output: The model correctly predicted that the Ser530Ala mutation would result in the loss of hydrogen bonding, reducing the inhibitor’s binding stability. The model also indicated that the inhibitor might exhibit increased movement and potentially partially dissociate from the binding site. It further predicted that other interactions (e.g., hydrophobic and van der Waals forces) could compensate for the loss of stability. \
• Reference Answer: The reference states that the Ser530 mutation to Alanine leads to the loss of hydrogen bonding, causing reduced stability and increased movement, potentially resulting in partial dissociation. The final position is predicted to shift to (X = 5.0, Y = -0.8, Z = 4.0). \
• Comparison: The model’s prediction aligns with the reference, accurately capturing the mutation’s impact, though it did not provide exact coordinates. The predicted trend and mechanism are correct. \
• Rating: 9/10 (Accurate prediction, but lacks specific values).\
\
\rule{\textwidth}{0.4pt}
\textbf{Score} \
• Inhibitor Movement: 10/10\
• Final Position Prediction: 9/10\
• Binding Stability Forces: 10/10\
• Mutation Impact: 9/10\
\
Total Score: 9.5/10
\end{tcolorbox}
\end{figure}

\begin{figure}[H]
\centering
\begin{tcolorbox}[colback=blue!5!white, colframe=blue!75!black, width=\textwidth, arc=5mm, auto outer arc, boxrule=0.5mm, title=Prompt and Answer, fonttitle=\small, fontupper=\tiny]
\textbf{Prompt:} \
\tiny You are provided with a molecular dynamics simulation of Cholesterol Oxidase (CHO) and a cholesterol substrate bound in the active site. The substrate is positioned within the catalytic pocket of the enzyme, and several key residues are known to interact with it. The following describes the setup and conditions of the simulation:

- Enzyme-Substrate Complex: Cholesterol Oxidase (CHO) bound to cholesterol (PDB ID: 3B0N)
- Simulation Duration: 100 nanoseconds (ns)
- Time step: 1 picosecond (ps)
- Temperature: 300K
- Environment: Water molecules and ions surround the enzyme-substrate complex
- Initial Position of Substrate: The substrate begins at coordinates (X = 5.0, Y = 2.0, Z = 1.5) within the CHO active site.

During the simulation, the substrate remains bound to the enzyme but undergoes thermal motion due to the temperature of the system. Predict the following:

1. How will the substrate move during the simulation? Describe its expected trajectory and the key factors influencing this movement.
2. What is the expected final position of the substrate after 100 ns? Provide the expected final coordinates.
3. Which forces or interactions (such as hydrogen bonds, van der Waals interactions) are most likely to contribute to the substrate’s binding stability during the simulation?
4. If a mutation occurs at one of the key catalytic residues (e.g., Glutamate at position 361 is mutated to Alanine), how would this mutation affect the substrate’s movement and binding stability within the active site? \
\rule{\textwidth}{0.4pt}
\textbf{Answer:} \
\tiny \textbf{1. Substrate Movement During the Simulation} \
• Model Output: The model accurately described the substrate’s movement within the CHO enzyme active site, noting slight vibrations and rotations due to thermal fluctuations (300K), which resulted in minor oscillations around the initial position. The model also mentioned that the dynamic nature of the enzyme, particularly residues near the active site, had a subtle influence on the substrate’s motion. \
• Reference Answer: The reference similarly noted that the substrate remained stable within the binding site throughout the simulation, with minor displacements and rotational movements, and no detachment from the active site. \
• Comparison: The model’s description of substrate movement closely matches the reference, accurately capturing thermal fluctuations and the characteristics of movement within the binding site. \
• Rating: 10/10 (Qualitative description fully consistent with reference).\
\
\textbf{2. Expected Final Position After 100 ns} \
• Model Output: The model predicted that the substrate’s final position would remain close to its initial position, with coordinates approximately (X ≈ 5.0 ± 0.2, Y ≈ 2.0 ± 0.2, Z ≈ 1.5 ± 0.2). This minor displacement was attributed to thermal fluctuations. \
• Reference Answer: The substrate’s final position was (X = 5.1, Y = 2.1, Z = 1.6), with minimal displacement, remaining within the binding site. \
• Comparison: The model’s prediction of the displacement range aligns with the reference, though it did not provide exact coordinates. The predicted range is very close to the actual result. \
• Rating: 9/10 (Reasonable displacement range, but lacks precise coordinates).\
\
\textbf{3. Forces and Interactions Contributing to Binding Stability} \
• Model Output: The model listed multiple interactions influencing binding stability, including hydrogen bonds, hydrophobic interactions, van der Waals forces, and possible electrostatic interactions. The model correctly identified the key interactions within the CHO active site contributing to substrate stability. \
• Reference Answer: The reference stated that binding stability is primarily maintained by hydrogen bonds and van der Waals forces, particularly between Glutamate361 and the substrate’s hydroxyl group, along with van der Waals interactions provided by hydrophobic residues in the enzyme pocket. \
• Comparison: The model’s identification of forces contributing to binding stability matches the reference, providing a comprehensive analysis of hydrogen bonding and hydrophobic interactions. \
• Rating: 10/10 (Comprehensive force analysis, fully consistent with reference).\
\
\textbf{4. Impact of Glu361Ala Mutation on Substrate Movement and Binding Stability} \
• Model Output: The model correctly predicted that the Glu361Ala mutation would lead to the loss of hydrogen bonds and electrostatic interactions, reducing the substrate’s binding stability. It also suggested that the substrate’s mobility might increase, with possible partial detachment from the binding site. \
• Reference Answer: The reference stated that the Glu361 to Alanine mutation led to the loss of hydrogen bonds, decreasing binding stability, with the substrate potentially shifting to (X = 5.4, Y = 2.5, Z = 2.0), and possibly partially or completely detaching from the binding site. \
• Comparison: The model’s prediction is consistent with the reference, particularly in describing the decrease in binding stability and increased substrate mobility after mutation. Although it did not provide exact coordinates, the predicted trend and mechanism are accurate. \
• Rating: 9/10 (Accurate prediction, but lacks specific coordinates).\
\
\rule{\textwidth}{0.4pt}
\textbf{Score:} \
• Substrate Movement: 10/10\
• Final Position Prediction: 9/10\
• Binding Stability Forces: 10/10\
• Mutation Impact: 9/10\
\
Total Score: 9.5/10
\end{tcolorbox}
\end{figure}

\begin{figure}[H]
\centering
\begin{tcolorbox}[colback=blue!5!white, colframe=blue!75!black, width=\textwidth, arc=5mm, auto outer arc, boxrule=0.5mm, title=Prompt and Answer, fonttitle=\small, fontupper=\tiny]
\textbf{Prompt:} \
\tiny You are provided with a molecular dynamics simulation of HIV-1 protease and a bound inhibitor. The inhibitor is located within the active site of the enzyme and interacts with key catalytic residues. The following describes the setup and conditions of the simulation:

- Protein-Inhibitor Complex: HIV-1 protease bound to an inhibitor (PDB ID: 1HVR)
- Simulation Duration: 100 nanoseconds (ns)
- Time step: 1 picosecond (ps)
- Temperature: 300K
- Environment: Water molecules and ions surround the protein-inhibitor complex
- Initial Position of Inhibitor: The inhibitor starts at coordinates (X = 4.2, Y = 0.9, Z = 1.8) within the active site.

During the simulation, the inhibitor remains bound to the enzyme but undergoes thermal motion due to the temperature of the system. Predict the following:

1. How will the inhibitor move during the simulation? Describe its expected trajectory and the key factors influencing this movement.
2. What is the expected final position of the inhibitor after 100 ns? Provide the expected final coordinates.
3. Which forces or interactions (such as hydrogen bonds, van der Waals interactions, hydrophobic interactions) are most likely to contribute to the inhibitor’s binding stability during the simulation?
4. If Asp25, a key catalytic residue, is mutated to Asn, how would this mutation affect the inhibitor’s movement and binding stability within the active site? \
\rule{\textwidth}{0.4pt}
\textbf{Answer:} \
\tiny \textbf{1. Inhibitor Movement During the Simulation} \
• Model Output: The model accurately described the inhibitor’s movement within the HIV-1 protease active site, noting slight vibrations and rotations due to thermal fluctuations (300K). It also indicated that the flexible “flap-like” structures of the protease might influence inhibitor movement and binding stability. \
• Reference Answer: The reference similarly noted that the inhibitor remains in the binding site through hydrogen bonds and hydrophobic interactions, with small oscillations around the initial position due to thermal fluctuations. \
• Comparison: The model’s description of inhibitor movement is consistent with the reference, effectively capturing the impact of thermal fluctuations and protease flexibility on the inhibitor’s movement. \
• Rating: 10/10 (Qualitative description fully consistent with the reference).\
\
\textbf{2. Expected Final Position After 100 ns} \
• Model Output: The model predicted that the inhibitor would remain near its initial position, with coordinates approximately (X ≈ 4.2 ± 0.2, Y ≈ 0.9 ± 0.2, Z ≈ 1.8 ± 0.2). The minor displacement was attributed to thermal fluctuations. \
• Reference Answer: The inhibitor’s final position was (X = 4.3, Y = 0.8, Z = 1.7), showing minimal displacement within the binding site. \
• Comparison: The model’s predicted range of coordinates is very close to the reference, though it did not provide exact values. The displacement range is reasonable and consistent with the reference. \
• Rating: 9/10 (Accurate range but lacks specific coordinates).\
\
\textbf{3. Forces and Interactions Contributing to Binding Stability} \
• Model Output: The model listed several forces maintaining binding stability, including hydrogen bonds, hydrophobic interactions, van der Waals forces, electrostatic interactions, and π-π stacking. It correctly identified the critical hydrogen bond between Asp25 and the inhibitor, as well as the role of hydrophobic residues in the binding site. \
• Reference Answer: The reference confirmed that binding stability is primarily maintained by hydrogen bonds, van der Waals forces, and hydrophobic interactions, particularly the hydrogen bond between Asp25 and the inhibitor. \
• Comparison: The model’s identification of the key forces is fully aligned with the reference, providing detailed explanations of each force’s contribution. \
• Rating: 10/10 (Comprehensive analysis, fully consistent with reference).\
\
\textbf{4. Impact of Asp25 to Asn Mutation on Inhibitor Movement and Binding Stability} \
• Model Output: The model accurately predicted that the Asp25 to Asn mutation would lead to the loss of hydrogen bonds and electrostatic interactions, resulting in reduced binding stability. It also predicted that the inhibitor’s mobility might increase, with potential partial detachment from the binding site. \
• Reference Answer: The reference indicated that the Asp25 to Asn mutation leads to the loss of hydrogen bonds, decreasing binding stability and causing the inhibitor to shift to (X = 4.6, Y = 1.0, Z = 2.0), with possible temporary detachment from the binding site. \
• Comparison: The model’s prediction is consistent with the reference, especially in describing the decrease in binding stability and increased inhibitor mobility after mutation, though it did not provide exact coordinates. \
• Rating: 9/10 (Accurate prediction, but lacks specific coordinates).\
\
\rule{\textwidth}{0.4pt}
\textbf{Score:} \
• Inhibitor Movement: 10/10\
• Final Position Prediction: 9/10\
• Binding Stability Forces: 10/10\
• Mutation Impact: 9/10\
\
Total Score: 9.5/10
\end{tcolorbox}
\end{figure}

\subsection{Quantitative Reasoning}

\begin{figure}[H] 
    \centering 
    \begin{tcolorbox}[colback=blue!5!white, colframe=blue!75!black, width=\textwidth, arc=5mm, auto outer arc, boxrule=0.5mm, title=Prompt and Answer, fonttitle=\small, fontupper=\tiny] 
        \textbf{Prompt:} \\
        \tiny Read the PROBLEM. Follow the INSTRUCTIONS to answer the PROBLEM.\\
        \\
        \textbf{PROBLEM:} \\
        Context: interest rate to a variable interest rate based on the three-month libor plus 2.05\% (2.05\%) (2.34\% (2.34\%) as of october 31, 2009). if libor changes by 100 basis points, our annual interest expense would change by \$3.8 million. foreign currency exposure as more fully described in note 2i. in the notes to consolidated financial statements contained in item 8 of this annual report on form 10-k, we regularly hedge our non-u.s. dollar-based exposures by entering into forward foreign currency exchange contracts. the terms of these contracts are for periods matching the duration of the underlying exposure and generally range from one month to twelve months. currently, our largest foreign currency exposure is the euro, primarily because our european operations have the highest proportion of our local currency denominated expenses. relative to foreign currency exposures existing at october 31, 2009 and november 1, 2008, a 10\% (10\%) unfavorable movement in foreign currency exchange rates over the course of the year would not expose us to significant losses in earnings or cash flows because we hedge a high proportion of our year-end exposures against fluctuations in foreign currency exchange rates. the market risk associated with our derivative instruments results from currency exchange rate or interest rate movements that are expected to offset the market risk of the underlying transactions, assets and liabilities being hedged. the counterparties to the agreements relating to our foreign exchange instruments consist of a number of major international financial institutions with high credit ratings. we do not believe that there is significant risk of nonperformance by these counterparties because we continually monitor the credit ratings of such counterparties. while the contract or notional amounts of derivative financial instruments provide one measure of the volume of these transactions, they do not represent the amount of our exposure to credit risk. the amounts potentially subject to credit risk (arising from the possible inability of counterparties to meet the terms of their contracts) are generally limited to the amounts, if any, by which the counterparties' 2019 obligations under the contracts exceed our obligations to the counterparties. the following table illustrates the effect that a 10\% (10\%) unfavorable or favorable movement in foreign currency exchange rates, relative to the u.s. dollar, would have on the fair value of our forward exchange contracts as of october 31, 2009 and november 1, 2008: ||october 31 2009|november 1 2008| |fair value of forward exchange contracts asset (liability)|\$6427|\$-23158 (23158)| |fair value of forward exchange contracts after a 10\% (10\%) unfavorable movement in foreign currency exchange rates asset (liability)|\$20132|\$-9457 (9457)| |fair value of forward exchange contracts after a 10\% (10\%) favorable movement in foreign currency exchange rates liability|\$-6781 (6781)|\$-38294 (38294)| fair value of forward exchange contracts after a 10\% (10\%) unfavorable movement in foreign currency exchange rates asset (liability)... \$20132 \$ (9457) fair value of forward exchange contracts after a 10\% (10\%) favorable movement in foreign currency exchange rates liability... \$ (6781) \$ (38294) the calculation assumes that each exchange rate would change in the same direction relative to the u.s. dollar. in addition to the direct effects of changes in exchange rates, such changes typically affect the volume of sales or the foreign currency sales price as competitors' 2019 products become more or less attractive. our sensitivity analysis of the effects of changes in foreign currency exchange rates does not factor in a potential change in sales levels or local currency selling prices. Question: what is the the interest expense in 2009? Answer:
 \\
        INSTRUCTIONS: Study the financial data provided in the table below and calculate the percentage change in operating lease obligations between 2009 and 2010.\\
        \rule{\textwidth}{0.4pt} 
        \textbf{Answer:} \\
        The interest expense in 2009 was approximately \$8.9 million.

Explanation and Calculation:

Determine the Total Debt Amount:
The company states that a 1\% (100 basis points) change in the LIBOR rate would change their annual interest expense by \$3.8 million.
This implies that the total variable-rate debt is calculated as:
Total Debt = Change in Interest Expense / Change in Interest Rate 
= \$3.8 million / 1\% 
= \$380 million

Identify the Interest Rate:
As of October 31, 2009, the interest rate is the three-month LIBOR plus 2.05\%, totaling 2.34\%.
This means the three-month LIBOR at that time was approximately:
LIBOR = 2.34\% - 2.05\% = 0.29\%

Calculate the Annual Interest Expense:
Using the total debt amount and the interest rate:
Interest Expense = Total Debt × Interest Rate 
= \$380 million × 2.34\% 
= \$8.892 million

Rounding to the nearest tenth of a million:
\$8.892 million ≈ \$8.9 million

Answer:
The interest expense in 2009 was approximately \$8.9 million.

        \rule{\textwidth}{0.4pt}
         \textbf{Reference Answer:} 3.8\%       
    \end{tcolorbox} 
\end{figure}

\begin{figure}[H] 
    \centering 
    \begin{tcolorbox}[colback=blue!5!white, colframe=blue!75!black, width=\textwidth, arc=5mm, auto outer arc, boxrule=0.5mm, title=Prompt and Answer, fonttitle=\small, fontupper=\tiny] 
        \textbf{Prompt:} \\
        \tiny Read the PROBLEM. Follow the INSTRUCTIONS to answer the PROBLEM.\\
        \\
        \textbf{PROBLEM:} \\
        Please answer the given financial question based on the context. Context: contractual obligations we summarize our enforceable and legally binding contractual obligations at september 30 , 2018 , and the effect these obligations are expected to have on our liquidity and cash flow in future periods in the following table . certain amounts in this table are based on management fffds estimates and assumptions about these obligations , including their duration , the possibility of renewal , anticipated actions by third parties and other factors , including estimated minimum pension plan contributions and estimated benefit payments related to postretirement obligations , supplemental retirement plans and deferred compensation plans . because these estimates and assumptions are subjective , the enforceable and legally binding obligations we actually pay in future periods may vary from those presented in the table. . |( in millions )|payments due by period total|payments due by period fiscal 2019|payments due by period fiscal 2020and 2021|payments due by period fiscal 2022and 2023|payments due by period thereafter| |long-term debt including current portionexcluding capital lease obligations ( 1 )|\$ 6039.0|\$ 726.6|\$ 824.8|\$ 1351.0|\$ 3136.6| |operating lease obligations ( 2 )|615.8|132.1|199.9|118.4|165.4| |capital lease obligations ( 3 )|152.5|5.0|6.7|2.7|138.1| |purchase obligations and other ( 4 ) ( 5 ) ( 6 )|2210.5|1676.6|224.1|114.9|194.9| |total|\$ 9017.8|\$ 2540.3|\$ 1255.5|\$ 1587.0|\$ 3635.0| ( 1 ) includes only principal payments owed on our debt assuming that all of our long-term debt will be held to maturity , excluding scheduled payments . we have excluded \$ 205.2 million of fair value of debt step-up , deferred financing costs and unamortized bond discounts from the table to arrive at actual debt obligations . see fffdnote 13 . debt fffd fffd of the notes to consolidated financial statements for information on the interest rates that apply to our various debt instruments . ( 2 ) see fffdnote 14 . operating leases fffd of the notes to consolidated financial statements for additional information . ( 3 ) the fair value step-up of \$ 18.5 million is excluded . see fffdnote 13 . debt fffd fffd capital lease and other indebtednesstt fffd of the notes to consolidated financial statements for additional information . ( 4 ) purchase obligations include agreements to purchase goods or services that are enforceable and legally binding and that specify all significant terms , including : fixed or minimum quantities to be purchased ; fixed , minimum or variable price provision ; and the approximate timing of the transaction . purchase obligations exclude agreements that are cancelable without penalty . ( 5 ) we have included in the table future estimated minimum pension plan contributions and estimated benefit payments related to postretirement obligations , supplemental retirement plans and deferred compensation plans . our estimates are based on factors , such as discount rates and expected returns on plan assets . future contributions are subject to changes in our underfunded status based on factors such as investment performance , discount rates , returns on plan assets and changes in legislation . it is possible that our assumptions may change , actual market performance may vary or we may decide to contribute different amounts . we have excluded \$ 247.8 million of multiemployer pension plan withdrawal liabilities recorded as of september 30 , 2018 due to lack of definite payout terms for certain of the obligations . see fffdnote 4 . retirement plans fffd multiemployer plans fffd of the notes to consolidated financial statements for additional information . ( 6 ) we have not included the following items in the table : fffd an item labeled fffdother long-term liabilities fffd reflected on our consolidated balance sheet because these liabilities do not have a definite pay-out scheme . fffd \$ 158.4 million from the line item fffdpurchase obligations and other fffd for certain provisions of the financial accounting standards board fffds ( fffdfasb fffd ) accounting standards codification ( fffdasc fffd ) 740 , fffdincome taxes fffd associated with liabilities for uncertain tax positions due to the uncertainty as to the amount and timing of payment , if any . in addition to the enforceable and legally binding obligations presented in the table above , we have other obligations for goods and services and raw materials entered into in the normal course of business . these contracts , however , are subject to change based on our business decisions . expenditures for environmental compliance see item 1 . fffdbusiness fffd fffd governmental regulation fffd environmental and other matters fffd , fffdbusiness fffd fffd governmental regulation fffd cercla and other remediation costs fffd , and fffd fffdbusiness fffd fffd governmental regulation fffd climate change fffd for a discussion of our expenditures for environmental compliance. . Question: what was the percent of the total long-term debt including current portion excluding capital lease obligations that was due in 2019 Answer: \\
        INSTRUCTIONS: what was the percent of the total long-term debt including current portion excluding capital lease obligations that was due in 2019?\\
        \rule{\textwidth}{0.4pt} 
        \textbf{Answer:} \\
       Identify the total long-term debt and the amount due in 2019:

Total Long-Term Debt (excluding capital lease obligations): \$6,039.0 million

Amount Due in Fiscal 2019: \$726.6 million

Calculate the percentage:

Percentage Due in 2019 \(= \left( \dfrac{\text{Amount Due in 2019}}{\text{Total Long-Term Debt}} \right) \times 100\% \)

Plugging in the numbers:

\[
\text{Percentage Due in 2019} = \left( \dfrac{\$726.6\ \text{million}}{\$6,039.0\ \text{million}} \right) \times 100\% \approx (0.12033) \times 100\% = 12.033\%
\]

Conclusion:

Approximately 12.03\% of the total long-term debt (excluding capital lease obligations) was due in 2019.
        \rule{\textwidth}{0.4pt}
        \\

         \textbf{Reference Answer:} 12.03\%       
    \end{tcolorbox} 
\end{figure}

\begin{figure}[H] 
    \centering 
    \begin{tcolorbox}[colback=blue!5!white, colframe=blue!75!black, width=\textwidth, arc=5mm, auto outer arc, boxrule=0.5mm, title=Prompt and Answer, fonttitle=\small, fontupper=\tiny] 
        \textbf{Prompt:} \\
        \tiny Read the PROBLEM. Follow the INSTRUCTIONS to answer the PROBLEM.\\
        \\
        \textbf{PROBLEM:} \\
        Please answer the given financial question based on the context. Context: the goldman sachs group , inc . and subsidiaries management 2019s discussion and analysis 2030 total aus net inflows/ ( outflows ) for 2014 includes \$ 19 billion of fixed income asset inflows in connection with our acquisition of deutsche asset \& wealth management 2019s stable value business and \$ 6 billion of liquidity products inflows in connection with our acquisition of rbs asset management 2019s money market funds . the table below presents our average monthly assets under supervision by asset class . average for the year ended december \$ in billions 2016 2015 2014 . |\$ in billions|average for theyear ended december 2016|average for theyear ended december 2015|average for theyear ended december 2014| |alternative investments|\$ 149|$ 145|\$ 145| |equity|256|247|225| |fixed income|578|530|499| |total long-term assets under supervision|983|922|869| |liquidity products|326|272|248| |total assets under supervision|$ 1309|\$ 1194|\$ 1117| operating environment . following a challenging first quarter of 2016 , market conditions continued to improve with higher asset prices resulting in full year appreciation in our client assets in both equity and fixed income assets . also , our assets under supervision increased during 2016 from net inflows , primarily in fixed income assets , and liquidity products . the mix of our average assets under supervision shifted slightly compared with 2015 from long- term assets under supervision to liquidity products . management fees have been impacted by many factors , including inflows to advisory services and outflows from actively-managed mutual funds . in the future , if asset prices decline , or investors continue the trend of favoring assets that typically generate lower fees or investors withdraw their assets , net revenues in investment management would likely be negatively impacted . during 2015 , investment management operated in an environment generally characterized by strong client net inflows , which more than offset the declines in equity and fixed income asset prices , which resulted in depreciation in the value of client assets , particularly in the third quarter of 2015 . the mix of average assets under supervision shifted slightly from long-term assets under supervision to liquidity products compared with 2014 . 2016 versus 2015 . net revenues in investment management were \$ 5.79 billion for 2016 , 7\% ( 7 \% ) lower than 2015 . this decrease primarily reflected significantly lower incentive fees compared with a strong 2015 . in addition , management and other fees were slightly lower , reflecting shifts in the mix of client assets and strategies , partially offset by the impact of higher average assets under supervision . during the year , total assets under supervision increased \$ 127 billion to \$ 1.38 trillion . long-term assets under supervision increased \$ 75 billion , including net inflows of \$ 42 billion , primarily in fixed income assets , and net market appreciation of \$ 33 billion , primarily in equity and fixed income assets . in addition , liquidity products increased \$ 52 billion . operating expenses were \$ 4.65 billion for 2016 , 4\% ( 4 \% ) lower than 2015 , due to decreased compensation and benefits expenses , reflecting lower net revenues . pre-tax earnings were \$ 1.13 billion in 2016 , 17\% ( 17 \% ) lower than 2015 . 2015 versus 2014 . net revenues in investment management were \$ 6.21 billion for 2015 , 3\% ( 3 \% ) higher than 2014 , due to slightly higher management and other fees , primarily reflecting higher average assets under supervision , and higher transaction revenues . during 2015 , total assets under supervision increased \$ 74 billion to \$ 1.25 trillion . long-term assets under supervision increased \$ 51 billion , including net inflows of \$ 71 billion ( which includes \$ 18 billion of asset inflows in connection with our acquisition of pacific global advisors 2019 solutions business ) , and net market depreciation of \$ 20 billion , both primarily in fixed income and equity assets . in addition , liquidity products increased \$ 23 billion . operating expenses were \$ 4.84 billion for 2015 , 4\% ( 4 \% ) higher than 2014 , due to increased compensation and benefits expenses , reflecting higher net revenues . pre-tax earnings were \$ 1.37 billion in 2015 , 2\% ( 2 \% ) lower than 2014 . geographic data see note 25 to the consolidated financial statements for a summary of our total net revenues , pre-tax earnings and net earnings by geographic region . goldman sachs 2016 form 10-k 65 . Question: what percentage of total long-term assets under supervision are comprised of fixed income in 2015? Answer:\\
        \textbf{PROBLEM:} \\
        what percentage of total long-term assets under supervision are comprised of fixed income in 2015?
        \rule{\textwidth}{0.4pt} 
        \textbf{Answer:} \\
        To determine the percentage of total long-term assets under supervision that are comprised of fixed income in 2015, we use the data provided.

\textbf{Data from the table for 2015:}

\begin{itemize}
    \item Fixed Income Assets: \$530 billion
    \item Total Long-Term Assets Under Supervision: \$922 billion
\end{itemize}

\textbf{Calculation:}

\[
\text{Percentage} = \left( \dfrac{\text{Fixed Income Assets}}{\text{Total Long-Term Assets Under Supervision}} \right) \times 100\%
\]

Plugging in the numbers:

\[
\text{Percentage} = \left( \dfrac{\$530\ \text{billion}}{\$922\ \text{billion}} \right) \times 100\% = \left( 0.5748 \right) \times 100\% = 57.48\%
\]

\textbf{Answer:}

Approximately \textbf{57.5\%} of total long-term assets under supervision were comprised of fixed income assets in 2015.

        \rule{\textwidth}{0.4pt}
         \textbf{Reference Answer:} \$57.5\%      
    \end{tcolorbox} 
\end{figure}

\begin{figure}[H] 
    \centering 
    \begin{tcolorbox}[colback=blue!5!white, colframe=blue!75!black, width=\textwidth, arc=5mm, auto outer arc, boxrule=0.5mm, title=Prompt and Answer, fonttitle=\small, fontupper=\tiny] 
        \textbf{Prompt:} \\
        \tiny Read the PROBLEM. Follow the INSTRUCTIONS to answer the PROBLEM.\\
        \\
        \textbf{PROBLEM:} \\
        Please answer the given financial question based on the context. Context: customary conditions . we will retain a 20\% ( 20 \% ) equity interest in the joint venture . as of december 31 , 2008 , the joint venture has acquired seven properties from us and we received year-to-date net sale proceeds and financing distributions of approximately \$ 251.6 million . in january 2008 , we sold a tract of land to an unconsolidated joint venture in which we hold a 50\% ( 50 \% ) equity interest and received a distribution , commensurate to our partner 2019s 50\% ( 50 \% ) ownership interest , of approximately \$ 38.3 million . in november 2008 , that unconsolidated joint venture entered a loan agreement with a consortium of banks and distributed a portion of the loan proceeds to us and our partner , with our share of the distribution totaling \$ 20.4 million . uses of liquidity our principal uses of liquidity include the following : 2022 property investment ; 2022 recurring leasing/capital costs ; 2022 dividends and distributions to shareholders and unitholders ; 2022 long-term debt maturities ; 2022 opportunistic repurchases of outstanding debt ; and 2022 other contractual obligations . property investment we evaluate development and acquisition opportunities based upon market outlook , supply and long-term growth potential . our ability to make future property investments is dependent upon our continued access to our longer-term sources of liquidity including the issuances of debt or equity securities as well as disposing of selected properties . in light of current economic conditions , management continues to evaluate our investment priorities and we are limiting new development expenditures . recurring expenditures one of our principal uses of our liquidity is to fund the recurring leasing/capital expenditures of our real estate investments . the following is a summary of our recurring capital expenditures for the years ended december 31 , 2008 , 2007 and 2006 , respectively ( in thousands ) : . ||2008|2007|2006| |recurring tenant improvements|\$ 36885|\$ 45296|\$ 41895| |recurring leasing costs|28205|32238|32983| |building improvements|9724|8402|8122| |totals|\$ 74814|\$ 85936|\$ 83000| dividends and distributions in order to qualify as a reit for federal income tax purposes , we must currently distribute at least 90\% ( 90\% ) of our taxable income to shareholders . because depreciation is a non-cash expense , cash flow will typically be greater than operating income . we paid dividends per share of \$ 1.93 , \$ 1.91 and \$ 1.89 for the years ended december 31 , 2008 , 2007 and 2006 , respectively . we expect to continue to distribute taxable earnings to meet the requirements to maintain our reit status . however , distributions are declared at the discretion of our board of directors and are subject to actual cash available for distribution , our financial condition , capital requirements and such other factors as our board of directors deems relevant . in january 2009 , our board of directors resolved to decrease our annual dividend from \$ 1.94 per share to \$ 1.00 per share in order to retain additional cash to help meet our capital needs . we anticipate retaining additional cash of approximately \$ 145.2 million per year , when compared to an annual dividend of \$ 1.94 per share , as the result of this action . at december 31 , 2008 we had six series of preferred shares outstanding . the annual dividend rates on our preferred shares range between 6.5\% ( 6.5 \% ) and 8.375\% ( 8.375 \% ) and are paid in arrears quarterly. . Question: what is the percent change in total recurring capital expenditures from 2006 to 2007? Answer:
        INSTRUCTIONS: what is the percent change in total recurring capital expenditures from 2006 to 2007?\\
        \rule{\textwidth}{0.4pt} 
        \textbf{Answer:} \\
        To determine the percent change in total recurring capital expenditures from 2006 to 2007, we use the following formula:

\[
\text{Percent Change} = \left( \dfrac{\text{Value in 2007} - \text{Value in 2006}}{\text{Value in 2006}} \right) \times 100\%
\]

\textbf{Given:}

\begin{itemize}
    \item \textbf{Total in 2006}: \$83,000 thousand
    \item \textbf{Total in 2007}: \$85,936 thousand
\end{itemize}

\textbf{Calculate the difference:}

\[
\text{Difference} = \$85,936\ \text{thousand} - \$83,000\ \text{thousand} = \$2,936\ \text{thousand}
\]

\textbf{Calculate the percent change:}

\begin{align*}
\text{Percent Change} &= \left( \dfrac{\$2,936}{\$83,000} \right) \times 100\% \\
&= 0.0353735 \times 100\% \\
&\approx 3.54\%
\end{align*}

\textbf{Answer:}

The percent change in total recurring capital expenditures from 2006 to 2007 is approximately \textbf{3.54\%}.

        \rule{\textwidth}{0.4pt}
         \textbf{Reference Answer:} 3.54       
    \end{tcolorbox} 
\end{figure}

\begin{figure}[H]
    \centering
    \begin{tcolorbox}[colback=blue!5!white, colframe=blue!75!black, width=\textwidth, arc=5mm, auto outer arc, boxrule=0.5mm, title=Prompt and Answer, fonttitle=\small, fontupper=\tiny]
        \textbf{Prompt:} \\
        \tiny Read the PROBLEM. Follow the INSTRUCTIONS to answer the PROBLEM.\\
        \\
        \textbf{PROBLEM:} \\
        Please answer the given financial question based on the context. Context: item 7a . quantitative and qualitative disclosures about market risk ( amounts in millions ) in the normal course of business , we are exposed to market risks related to interest rates , foreign currency rates and certain balance sheet items . from time to time , we use derivative instruments , pursuant to established guidelines and policies , to manage some portion of these risks . derivative instruments utilized in our hedging activities are viewed as risk management tools and are not used for trading or speculative purposes . interest rates our exposure to market risk for changes in interest rates relates primarily to the fair market value and cash flows of our debt obligations . the majority of our debt ( approximately 93\% ( 93 \% ) and 89\% ( 89 \% ) as of december 31 , 2016 and 2015 , respectively ) bears interest at fixed rates . we do have debt with variable interest rates , but a 10\% ( 10 \% ) increase or decrease in interest rates would not be material to our interest expense or cash flows . the fair market value of our debt is sensitive to changes in interest rates , and the impact of a 10\% ( 10 \% ) change in interest rates is summarized below . increase/ ( decrease ) in fair market value as of december 31 , 10\% ( 10 \% ) increase in interest rates 10\% ( 10 \% ) decrease in interest rates . |as of december 31,|increase/ ( decrease ) in fair market value 10\% ( 10 \% ) increasein interest rates|increase/ ( decrease ) in fair market value 10\% ( 10 \% ) decreasein interest rates| |2016|$ -26.3 ( 26.3 )|$ 26.9| |2015|-33.7 ( 33.7 )|34.7| we have used interest rate swaps for risk management purposes to manage our exposure to changes in interest rates . we do not have any interest rate swaps outstanding as of december 31 , 2016 . we had \$ 1100.6 of cash , cash equivalents and marketable securities as of december 31 , 2016 that we generally invest in conservative , short-term bank deposits or securities . the interest income generated from these investments is subject to both domestic and foreign interest rate movements . during 2016 and 2015 , we had interest income of \$ 20.1 and \$ 22.8 , respectively . based on our 2016 results , a 100 basis-point increase or decrease in interest rates would affect our interest income by approximately \$ 11.0 , assuming that all cash , cash equivalents and marketable securities are impacted in the same manner and balances remain constant from year-end 2016 levels . foreign currency rates we are subject to translation and transaction risks related to changes in foreign currency exchange rates . since we report revenues and expenses in u.s . dollars , changes in exchange rates may either positively or negatively affect our consolidated revenues and expenses ( as expressed in u.s . dollars ) from foreign operations . the foreign currencies that most impacted our results during 2016 included the british pound sterling and , to a lesser extent , the argentine peso , brazilian real and japanese yen . based on 2016 exchange rates and operating results , if the u.s . dollar were to strengthen or weaken by 10\% ( 10 \% ) , we currently estimate operating income would decrease or increase approximately 4\% ( 4 \% ) , assuming that all currencies are impacted in the same manner and our international revenue and expenses remain constant at 2016 levels . the functional currency of our foreign operations is generally their respective local currency . assets and liabilities are translated at the exchange rates in effect at the balance sheet date , and revenues and expenses are translated at the average exchange rates during the period presented . the resulting translation adjustments are recorded as a component of accumulated other comprehensive loss , net of tax , in the stockholders 2019 equity section of our consolidated balance sheets . our foreign subsidiaries generally collect revenues and pay expenses in their functional currency , mitigating transaction risk . however , certain subsidiaries may enter into transactions in currencies other than their functional currency . assets and liabilities denominated in currencies other than the functional currency are susceptible to movements in foreign currency until final settlement . currency transaction gains or losses primarily arising from transactions in currencies other than the functional currency are included in office and general expenses . we regularly review our foreign exchange exposures that may have a material impact on our business and from time to time use foreign currency forward exchange contracts or other derivative financial instruments to hedge the effects of potential adverse fluctuations in foreign currency exchange rates arising from these exposures . we do not enter into foreign exchange contracts or other derivatives for speculative purposes. . Question: what percent cash and equivalents did the interest income generated during 2015? Answer:\\
        INSTRUCTIONS: what percent cash and equivalents did the interest income generated during 2015?\\
        \rule{\textwidth}{0.4pt} 
        \textbf{Answer:} \\
        To determine the percentage of cash and equivalents that the interest income generated during 2015, we need the following information:

Interest Income in 2015: \$22.8 million  
Cash, Cash Equivalents, and Marketable Securities for 2015: Not provided in the context.  
The percentage is calculated using the formula:

Percentage = (Interest Income / Cash and Equivalents) × 100\%

However, the context does not provide the amount of cash, cash equivalents, and marketable securities for 2015. Without this figure, we cannot accurately calculate the requested percentage.

Additional Insight:

As of December 31, 2016, the company had \$1,100.6 million in cash, cash equivalents, and marketable securities.  
In 2016, the interest income was \$20.1 million.  
Using the 2016 figures as an example:

Percentage in 2016 = (\$20.1 million / \$1,100.6 million) × 100\% ≈ 1.83\%

If we assume that the cash and equivalents in 2015 were similar to those in 2016, we can estimate:

Estimated Percentage in 2015 = (\$22.8 million / \$1,100.6 million) × 100\% ≈ 2.07\%
\\
        \\
        \rule{\textwidth}{0.4pt}
         \textbf{Reference Answer:} 0.02072  
    \end{tcolorbox}
\end{figure}

\begin{figure}[H] 
    \centering 
    \begin{tcolorbox}[colback=blue!5!white, colframe=blue!75!black, width=\textwidth, arc=5mm, auto outer arc, boxrule=0.5mm, title=Prompt and Answer, fonttitle=\small, fontupper=\tiny] 
        \textbf{Prompt:} \\
        \tiny Read the PROBLEM. Follow the INSTRUCTIONS to answer the PROBLEM.\\
        \\
        \textbf{PROBLEM:} \\
Please answer the given financial question based on the context.

Context: blackrock n 96 n notes in april 2009, the company acquired \$2 million of finite-lived management contracts with a five-year estimated useful life associated with the acquisition of the r3 capital partners funds. In December 2009, in conjunction with the BGI transaction, the company acquired \$163 million of finite-lived management contracts with a weighted-average estimated useful life of approximately 10 years. Estimated amortization expense for finite-lived intangible assets for each of the five succeeding years is as follows (dollar amounts in millions):
| 2010 | \$160 |  
| 2011 | \$157 |  
| 2012 | \$156 |  
| 2013 | \$155 |  
| 2014 | \$149 |  
Indefinite-lived acquired management contracts on September 29, 2006, in conjunction with the MLIM transaction, the company acquired indefinite-lived management contracts valued at \$4477 million consisting of \$4271 million for all retail mutual funds and \$206 million for alternative investment products. On October 1, 2007, in conjunction with the Quellos transaction, the company acquired \$631 million in indefinite-lived management contracts associated with alternative investment products. On October 1, 2007, the company purchased the remaining 20\% (20\%) of an investment manager of a fund of hedge funds. In conjunction with this transaction, the company recorded \$8 million in additional indefinite-lived management contracts associated with alternative investment products. On December 1, 2009, in conjunction with the BGI transaction, the company acquired \$9785 million in indefinite-lived management contracts valued primarily for exchange traded funds and common and collective trusts.
Indefinite-lived acquired trade names/trademarks on December 1, 2009, in conjunction with the BGI transaction, the company acquired trade names/trademarks primarily related to iShares valued at \$1402.5 million. The fair value was determined using a royalty rate based primarily on normalized marketing and promotion expenditures to develop and support the brands globally.
›Borrowings:
Short-term borrowings (2007 facility): In August 2007, the company entered into a five-year \$2.5 billion unsecured revolving credit facility (the "2007 facility"), which permits the company to request an additional \$500 million of borrowing capacity, subject to lender credit approval, up to a maximum of \$3.0 billion. The 2007 facility requires the company not to exceed a maximum leverage ratio (ratio of net debt to earnings before interest, taxes, depreciation, and amortization, where net debt equals total debt less domestic unrestricted cash) of 3 to 1, which was satisfied with a ratio of less than 1 to 1 at December 31, 2009. The 2007 facility provides backup liquidity, funds ongoing working capital for general corporate purposes, and funds various investment opportunities. As of December 31, 2009, the company had \$200 million outstanding under the 2007 facility with an interest rate of 0.44\% and a maturity date during February 2010. During February 2010, the company rolled over \$100 million in borrowings with an interest rate of 0.43\% and a maturity date in May 2010. Lehman Commercial Paper Inc. has a \$140 million participation under the 2007 facility; however, BlackRock does not expect that Lehman Commercial Paper Inc. will honor its commitment to fund additional amounts. Bank of America, a related party, has a \$140 million participation under the 2007 facility. In December 2007, to support two enhanced cash funds that BlackRock manages, BlackRock elected to procure two letters of credit under the existing 2007 facility in an aggregate amount of \$100 million. In December 2008, the letters of credit were terminated.
Commercial paper program: On October 14, 2009, BlackRock established a commercial paper program ("CP program") under which the company may issue unsecured commercial paper notes ("CP notes") on a private placement basis up to a maximum aggregate amount outstanding at any time of \$3 billion. The proceeds of the commercial paper issuances were used for the financing of a portion of the BGI transaction. Subsidiaries of Bank of America and Barclays, as well as other third parties, act as dealers under the CP program. The CP program is supported by the 2007 facility. The company began issuance of CP notes under the CP program on November 4, 2009. As of December 31, 2009, BlackRock had approximately \$2 billion of outstanding CP notes with a weighted average interest rate of 0.20\% and a weighted average maturity of 23 days. Since December 31, 2009, the company repaid approximately \$1.4 billion of CP notes with proceeds from the long-term notes issued in December 2009. As of March 5, 2010, BlackRock had \$596 million of outstanding CP notes with a weighted average interest rate of 0.18\% and a weighted average maturity of 38 days.
Japan commitment-line: In June 2008, BlackRock Japan Co., Ltd., a wholly owned subsidiary of the company, entered into a five billion Japanese yen commitment-line agreement with a banking institution ("Japan commitment-line"). The term of the Japan commitment-line was one year and interest accrued at the applicable Japanese short-term prime rate. In June 2009, BlackRock Japan Co., Ltd. renewed the Japan commitment-line for a term of one year. The Japan commitment-line is intended to provide liquidity and flexibility for operating requirements in Japan. As of December 31, 2009, the company had no borrowings outstanding on the Japan commitment-line.
Convertible debentures: In February 2005, the company issued \$250 million aggregate principal amount of convertible debentures ("debentures"), due in 2035 and bearing interest at a rate of 2.625\% per annum. Interest is payable semi-annually in arrears on February 15 and August 15 of each year, commencing August 15, 2005. Prior to February 15, 2009, the debentures could have been convertible at the option of the holder at a December 31, 2008, conversion rate of 9.9639 shares of common stock per one dollar principal amount of debentures under certain circumstances. The debentures would have been convertible into cash and, in some situations as described below, additional shares of the company’s common stock. On February 15, 2009, the debentures became convertible into cash at any time prior to maturity at the option of the holder, and in some situations, additional shares of the company’s common stock at the current conversion rate. At the time the debentures are tendered for conversion, for each one dollar principal amount of debentures converted, a holder shall be entitled to receive cash and shares of BlackRock common stock, if any, the aggregate value of which ("conversion value") will be determined by multiplying the applicable conversion rate by the average of the daily volume-weighted average price of BlackRock common stock for each of the ten consecutive trading days beginning on the second trading day immediately following the day the debentures are tendered for conversion ("ten-day weighted average price"). The company will deliver the conversion value to holders as follows: (1) an amount in cash ("principal return") equal to the lesser of (a) the aggregate conversion value of the debentures to be converted and (b) the aggregate principal amount of the debentures to be converted, and (2) if the aggregate conversion value of the debentures to be converted is greater than the principal return, an amount in shares ("net shares"), determined as set forth below, equal to such aggregate conversion value less the principal return ("net share amount"). The number of net shares to be paid will be determined by dividing the net share amount by the ten-day weighted average price. In lieu of delivering fractional shares, the company will deliver cash based on the ten-day weighted average price. The conversion rate for the debentures is subject to adjustments upon the occurrence of certain corporate events, such as a change of control of the company.
INSTRUCTIONS: What is the 2010 estimated amortization expense for finite-lived intangible assets as a percentage of the unsecured revolving credit facility?\\

        \rule{\textwidth}{0.4pt} 
        \textbf{Answer:} \\
        To determine the 2010 estimated amortization expense for finite-lived intangible assets as a percentage of the unsecured revolving credit facility, we use the following data:

\textbf{Given:}

\begin{itemize}
    \item \textbf{2010 Estimated Amortization Expense:} \$160 million
    \item \textbf{Unsecured Revolving Credit Facility Amount:} \$2,500 million
\end{itemize}

\textbf{Calculation:}

\[
\text{Percentage} = \left( \dfrac{\text{Amortization Expense}}{\text{Credit Facility Amount}} \right) \times 100\%
\]

Plugging in the numbers:

\[
\text{Percentage} = \left( \dfrac{\$160\ \text{million}}{\$2,500\ \text{million}} \right) \times 100\% = \left(0.064\right) \times 100\% = 6.4\%
\]

\textbf{Answer:}

The 2010 estimated amortization expense for finite-lived intangible assets is \textbf{6.4\%} of the unsecured revolving credit facility amount.
\\
\textbf{Reference Answer:} 0.0025
     
    \end{tcolorbox} 
\end{figure}

\begin{figure}[H] 
    \centering 
    \begin{tcolorbox}[colback=blue!5!white, colframe=blue!75!black, width=\textwidth, arc=5mm, auto outer arc, boxrule=0.5mm, title=Prompt and Answer, fonttitle=\small, fontupper=\tiny] 
        \textbf{Prompt:} \\
        \tiny Read the PROBLEM. Follow the INSTRUCTIONS to answer the PROBLEM.\\
        \\
        \textbf{PROBLEM:} \\
        Please answer the given financial question based on the context. Context: as of october 31 , 2009 , the total notional amount of these undesignated hedges was \$ 38 million . the fair value of these hedging instruments in the company 2019s condensed consolidated balance sheet as of october 31 , 2009 was immaterial . interest rate exposure management 2014 on june 30 , 2009 , the company entered into interest rate swap transactions related to its outstanding notes where the company swapped the notional amount of its \$ 375 million of fixed rate debt at 5.0\% ( 5.0 \% ) into floating interest rate debt through july 1 , 2014 . under the terms of the swaps , the company will ( i ) receive on the \$ 375 million notional amount a 5.0\% ( 5.0 \% ) annual interest payment that is paid in two installments on the 1st of every january and july , commencing january 1 , 2010 through and ending on the maturity date ; and ( ii ) pay on the \$ 375 million notional amount an annual three-month libor plus 2.05\% ( 2.05 \% ) ( 2.34\% ( 2.34 \% ) as of october 31 , 2009 ) interest payment , payable in four installments on the 1st of every january , april , july and october , commencing on october 1 , 2009 and ending on the maturity date . the libor based rate is set quarterly three months prior to the date of the interest payment . the company designated these swaps as fair value hedges . the fair value of the swaps at inception were zero and subsequent changes in the fair value of the interest rate swaps were reflected in the carrying value of the interest rate swaps on the balance sheet . the carrying value of the debt on the balance sheet was adjusted by an equal and offsetting amount . the gain or loss on the hedged item ( that is fixed- rate borrowings ) attributable to the hedged benchmark interest rate risk and the offsetting gain or loss on the related interest rate swaps as of october 31 , 2009 is as follows : income statement classification gain/ ( loss ) on gain/ ( loss ) on note net income effect . |income statement classification|gain/ ( loss ) on swaps|gain/ ( loss ) on note|net income effect| |other income|\$ 6109|\$ -6109 ( 6109 )|\$ 2014| the amounts earned and owed under the swap agreements are accrued each period and are reported in interest expense . there was no ineffectiveness recognized in any of the periods presented . the market risk associated with the company 2019s derivative instruments results from currency exchange rate or interest rate movements that are expected to offset the market risk of the underlying transactions , assets and liabilities being hedged . the counterparties to the agreements relating to the company 2019s derivative instruments consist of a number of major international financial institutions with high credit ratings . the company does not believe that there is significant risk of nonperformance by these counterparties because the company continually monitors the credit ratings of such counterparties . furthermore , none of the company 2019s derivative transactions are subject to collateral or other security arrangements and none contain provisions that are dependent on the company 2019s credit ratings from any credit rating agency . while the contract or notional amounts of derivative financial instruments provide one measure of the volume of these transactions , they do not represent the amount of the company 2019s exposure to credit risk . the amounts potentially subject to credit risk ( arising from the possible inability of counterparties to meet the terms of their contracts ) are generally limited to the amounts , if any , by which the counterparties 2019 obligations under the contracts exceed the obligations of the company to the counterparties . as a result of the above considerations , the company does not consider the risk of counterparty default to be significant . the company records the fair value of its derivative financial instruments in the consolidated financial statements in other current assets , other assets or accrued liabilities , depending on their net position , regardless of the purpose or intent for holding the derivative contract . changes in the fair value of the derivative financial instruments are either recognized periodically in earnings or in shareholders 2019 equity as a component of oci . changes in the fair value of cash flow hedges are recorded in oci and reclassified into earnings when the underlying contract matures . changes in the fair values of derivatives not qualifying for hedge accounting are reported in earnings as they occur . the total notional amount of derivative instruments designated as hedging instruments as of october 31 , 2009 is as follows : \$ 375 million of interest rate swap agreements accounted as fair value hedges , and \$ 128.0 million of analog devices , inc . notes to consolidated financial statements 2014 ( continued ) . Question: what is the net difference between in amounts used to as hedging instruments? Answer: \\
        INSTRUCTIONS: what is the net difference between in amounts used to as hedging instruments?\\
        \rule{\textwidth}{0.4pt} 
        \textbf{Answer:} \\
        To determine the net difference between the amounts used as hedging instruments, we'll compare the total notional amounts of designated hedging instruments and undesignated hedges.

Data from the Context:

Undesignated Hedges:  
Total notional amount as of October 31, 2009: \$38 million.  
Designated Hedging Instruments:  
Interest rate swap agreements accounted as fair value hedges: \$375 million.  
Other hedging instruments (the text mentions "\$128.0 million of..."; assuming this is part of designated hedging instruments): \$128 million.  
Total Designated Hedging Instruments: \$375 million + \$128 million = \$503 million.

Calculation:

Net Difference = Total Designated Hedging Instruments − Total Undesignated Hedges  
= \$503 million − \$38 million  
= \$465 million  

Answer:  
The net difference between the amounts used as hedging instruments is \$465 million.
\\
        \\
        \rule{\textwidth}{0.4pt}
         \textbf{Reference Answer:} 247.0       
    \end{tcolorbox} 
\end{figure}

\begin{figure}[H] 
    \centering 
    \begin{tcolorbox}[colback=blue!5!white, colframe=blue!75!black, width=\textwidth, arc=5mm, auto outer arc, boxrule=0.5mm, title=Prompt and Answer, fonttitle=\small, fontupper=\tiny] 
        \textbf{Prompt:} \\
        \tiny Read the PROBLEM. Follow the INSTRUCTIONS to answer the PROBLEM.\\
        \\
        \textbf{PROBLEM:} \\
        Please answer the given financial question based on the context. Context: table of contents notes to consolidated financial statements of american airlines group inc. secured financings are collateralized by assets , primarily aircraft , engines , simulators , rotable aircraft parts , airport leasehold rights , route authorities and airport slots . at december 31 , 2015 , the company was operating 35 aircraft under capital leases . leases can generally be renewed at rates based on fair market value at the end of the lease term for a number of additional years . at december 31 , 2015 , the maturities of long-term debt and capital lease obligations are as follows ( in millions ) : . |2016|\$ 2266| |2017|1598| |2018|2134| |2019|3378| |2020|3587| |2021 and thereafter|7844| |total|\$ 20807| ( a ) 2013 credit facilities on june 27 , 2013 , american and aag entered into a credit and guaranty agreement ( as amended , restated , amended and restated or otherwise modified , the 2013 credit agreement ) with deutsche bank ag new york branch , as administrative agent , and certain lenders that originally provided for a \$ 1.9 billion term loan facility scheduled to mature on june 27 , 2019 ( the 2013 term loan facility ) and a \$ 1.0 billion revolving credit facility scheduled to mature on june 27 , 2018 ( the 2013 revolving facility ) . the maturity of the term loan facility was subsequently extended to june 2020 and the revolving credit facility commitments were subsequently increased to \$ 1.4 billion with an extended maturity date of october 10 , 2020 , all of which is further described below . on may 21 , 2015 , american amended and restated the 2013 credit agreement pursuant to which it refinanced the 2013 term loan facility ( the \$ 1.9 billion 2015 term loan facility and , together with the 2013 revolving facility , the 2013 credit facilities ) to extend the maturity date to june 2020 and reduce the libor margin from 3.00\% ( 3.00 \% ) to 2.75\% ( 2.75 \% ) . in addition , american entered into certain amendments to reflect the ability for american to make future modifications to the collateral pledged , subject to certain restrictions . the \$ 1.9 billion 2015 term loan facility is repayable in annual installments , with the first installment in an amount equal to 1.25\% ( 1.25 \% ) of the principal amount commencing on june 27 , 2016 and installments thereafter , in an amount equal to 1.0\% ( 1.0 \% ) of the principal amount , with any unpaid balance due on the maturity date . as of december 31 , 2015 , \$ 1.9 billion of principal was outstanding under the \$ 1.9 billion 2015 term loan facility . voluntary prepayments may be made by american at any time . on october 10 , 2014 , american and aag amended the 2013 credit agreement to extend the maturity date of the 2013 revolving facility to october 10 , 2019 and increased the commitments thereunder to an aggregate principal amount of \$ 1.4 billion while reducing the letter of credit commitments thereunder to \$ 300 million . on october 26 , 2015 , american , aag , us airways group and us airways amended the 2013 credit agreement to extend the maturity date of the 2013 revolving facility to october 10 , 2020 . the 2013 revolving facility provides that american may from time to time borrow , repay and reborrow loans thereunder and have letters of credit issued thereunder . as of december 31 , 2015 , there were no borrowings or letters of credit outstanding under the 2013 revolving facility . the 2013 credit facilities bear interest at an index rate plus an applicable index margin or , at american 2019s option , libor ( subject to a floor of 0.75\% ( 0.75 \% ) , with respect to the \$ 1.9 billion 2015 term loan facility ) plus a libor margin of 3.00\% ( 3.00 \% ) with respect to the 2013 revolving facility and 2.75\% ( 2.75 \% ) with respect to the \$ 1.9 billion 2015 term loan facility ; provided that american 2019s corporate credit rating is ba3 or higher from moody 2019s and bb- or higher from s\&p , the applicable libor margin would be 2.50\% ( 2.50 \% ) for the \$ 1.9 billion 2015 term loan . Question: what is the amount of the first installment of the 19 billion 2015 term loan facility payable on june 27 , 2016 in billions Answer: \\
        INSTRUCTIONS: what is the amount of the first installment of the 19 billion 2015 term loan facility payable on june 27 , 2016 in billions\\
        \rule{\textwidth}{0.4pt} 
        \textbf{Answer:} \\
      To determine the amount of the first installment of the \$1.9 billion 2015 term loan facility payable on June 27, 2016, we use the information provided:

First installment percentage: 1.25\% of the principal amount.
Calculation:

Principal Amount:  
\$1.9 billion

First Installment Percentage:  
1.25\% = 0.0125

First Installment Amount:  
First Installment = Principal Amount × First Installment Percentage  
First Installment = \$1.9 billion × 0.0125  
First Installment = \$0.02375 billion

Answer:  
The amount of the first installment of the \$1.9 billion 2015 term loan facility payable on June 27, 2016, is \$0.02375 billion.
\\
        \\
        \rule{\textwidth}{0.4pt}
         \textbf{Reference Answer:} 0.02375.       
    \end{tcolorbox} 
\end{figure}

\begin{figure}[H] 
    \centering 
    \begin{tcolorbox}[colback=blue!5!white, colframe=blue!75!black, width=\textwidth, arc=5mm, auto outer arc, boxrule=0.5mm, title=Prompt and Answer, fonttitle=\small, fontupper=\tiny] 
        \textbf{Prompt:} \\
        \tiny Read the PROBLEM. Follow the INSTRUCTIONS to answer the PROBLEM.\\
        \\
        \textbf{PROBLEM:} \\
        Please answer the given financial question based on the context. Context: in the ordinary course of business , based on our evaluations of certain geologic trends and prospective economics , we have allowed certain lease acreage to expire and may allow additional acreage to expire in the future . if production is not established or we take no other action to extend the terms of the leases , licenses , or concessions , undeveloped acreage listed in the table below will expire over the next three years . we plan to continue the terms of many of these licenses and concession areas or retain leases through operational or administrative actions . net undeveloped acres expiring year ended december 31 . |( in thousands )|net undeveloped acres expiring year ended december 31 , 2015|net undeveloped acres expiring year ended december 31 , 2016|net undeveloped acres expiring year ended december 31 , 2017| |u.s .|211|150|94| |e.g .|36|2014|2014| |other africa|1950|1502|1089| |total africa|1986|1502|1089| |other international|88|2014|2014| |total|2285|1652|1183| oil sands mining segment we hold a 20 percent non-operated interest in the aosp , an oil sands mining and upgrading joint venture located in alberta , canada . the joint venture produces bitumen from oil sands deposits in the athabasca region utilizing mining techniques and upgrades the bitumen to synthetic crude oils and vacuum gas oil . the aosp 2019s mining and extraction assets are located near fort mcmurray , alberta , and include the muskeg river and the jackpine mines . gross design capacity of the combined mines is 255000 ( 51000 net to our interest ) barrels of bitumen per day . the aosp operations use established processes to mine oil sands deposits from an open-pit mine , extract the bitumen and upgrade it into synthetic crude oils . ore is mined using traditional truck and shovel mining techniques . the mined ore passes through primary crushers to reduce the ore chunks in size and is then sent to rotary breakers where the ore chunks are further reduced to smaller particles . the particles are combined with hot water to create slurry . the slurry moves through the extraction process where it separates into sand , clay and bitumen-rich froth . a solvent is added to the bitumen froth to separate out the remaining solids , water and heavy asphaltenes . the solvent washes the sand and produces clean bitumen that is required for the upgrader to run efficiently . the process yields a mixture of solvent and bitumen which is then transported from the mine to the scotford upgrader via the approximately 300-mile corridor pipeline . the aosp's scotford upgrader is located at fort saskatchewan , northeast of edmonton , alberta . the bitumen is upgraded at scotford using both hydrotreating and hydroconversion processes to remove sulfur and break the heavy bitumen molecules into lighter products . blendstocks acquired from outside sources are utilized in the production of our saleable products . the upgrader produces synthetic crude oils and vacuum gas oil . the vacuum gas oil is sold to an affiliate of the operator under a long-term contract at market-related prices , and the other products are sold in the marketplace . as of december 31 , 2014 , we own or have rights to participate in developed and undeveloped leases totaling approximately 163000 gross ( 33000 net ) acres . the underlying developed leases are held for the duration of the project , with royalties payable to the province of alberta . synthetic crude oil sales volumes for 2014 averaged 50 mbbld and net-of-royalty production was 41 mbbld . in december 2013 , a jackpine mine expansion project received conditional approval from the canadian government . the project includes additional mining areas , associated processing facilities and infrastructure . the government conditions relate to wildlife , the environment and aboriginal health issues . we will evaluate the potential expansion project and government conditions after infrastructure reliability initiatives are completed . the governments of alberta and canada have agreed to partially fund quest ccs for \$ 865 million canadian . in the third quarter of 2012 , the energy and resources conservation board ( "ercb" ) , alberta's primary energy regulator at that time , conditionally approved the project and the aosp partners approved proceeding to construct and operate quest ccs . government funding commenced in 2012 and continued as milestones were achieved during the development , construction and operating phases . failure of the aosp to meet certain timing , performance and operating objectives may result in repaying some of the government funding . construction and commissioning of quest ccs is expected to be completed by late 2015. . Question: what are total african net undeveloped acres in thousands for 12/31/15? Answer: \\
        INSTRUCTIONS: what are total african net undeveloped acres in thousands for 12/31/15?\\
        \rule{\textwidth}{0.4pt} 
        \textbf{Answer:} \\
        To determine the total African net undeveloped acres in thousands as of December 31, 2015, we need to sum the net undeveloped acres expiring in 2015 and future years, as these represent the total undeveloped acreage in Africa at that time.

Data from the Table:

Total Africa Net Undeveloped Acres Expiring:  
Year Ended December 31, 2015: 1,986 thousand acres  
Year Ended December 31, 2016: 1,502 thousand acres  
Year Ended December 31, 2017: 1,089 thousand acres  

Calculation:

Total African Net Undeveloped Acres = Acres expiring in 2015 + Acres expiring in 2016 + Acres expiring in 2017  
= 1,986 thousand acres + 1,502 thousand acres + 1,089 thousand acres  
= 4,577 thousand acres  

Answer:  
The total African net undeveloped acres as of December 31, 2015, is 4,577 thousand acres.
\\
        \\
        \rule{\textwidth}{0.4pt}
         \textbf{Reference Answer:} 3936.0      
    \end{tcolorbox} 
\end{figure}

\begin{figure}[H] 
    \centering 
    \begin{tcolorbox}[colback=blue!5!white, colframe=blue!75!black, width=\textwidth, arc=5mm, auto outer arc, boxrule=0.5mm, title=Prompt and Answer, fonttitle=\small, fontupper=\tiny] 
        \textbf{Prompt:} \\
        \tiny Read the PROBLEM. Follow the INSTRUCTIONS to answer the PROBLEM.\\
        \\
        \textbf{PROBLEM:} \\
        Please answer the given financial question based on the context. Context: asset category target allocation total quoted prices in active markets for identical assets ( level 1 ) significant observable inputs ( level 2 ) significant unobservable inputs . ||level 3| |balance as of january 1 2018|\$ 278| |actual return on assets|-23 ( 23 )| |purchases issuances and settlements net|-25 ( 25 )| |balance as of december 31 2018|\$ 230| balance as of january 1 , 2017 . . . . . . . . . . . . . . . . . . . . . . . . . . . . . . . . . . . . . . . . . . . . . . . . . . . . . . . . . \$ 140 actual return on assets . . . . . . . . . . . . . . . . . . . . . . . . . . . . . . . . . . . . . . . . . . . . . . . . . . . . . . . . . . 2 purchases , issuances and settlements , net . . . . . . . . . . . . . . . . . . . . . . . . . . . . . . . . . . . . . . . . . . . . 136 balance as of december 31 , 2017 . . . . . . . . . . . . . . . . . . . . . . . . . . . . . . . . . . . . . . . . . . . . . . . . . . . . . . \$ 278 the company 2019s postretirement benefit plans have different levels of funded status and the assets are held under various trusts . the investments and risk mitigation strategies for the plans are tailored specifically for each trust . in setting new strategic asset mixes , consideration is given to the likelihood that the selected asset allocation will effectively fund the projected plan liabilities and meet the risk tolerance criteria of the company . the company periodically updates the long-term , strategic asset allocations for these plans through asset liability studies and uses various analytics to determine the optimal asset allocation . considerations include plan liability characteristics , liquidity needs , funding requirements , expected rates of return and the distribution of returns . in 2012 , the company implemented a de-risking strategy for the american water pension plan after conducting an asset-liability study to reduce the volatility of the funded status of the plan . as part of the de-risking strategy , the company revised the asset allocations to increase the matching characteristics of fixed- income assets relative to liabilities . the fixed income portion of the portfolio was designed to match the bond- . Question: by what percentage level 3 balance decrease during 2018? Answer: \\
        INSTRUCTIONS: by what percentage level 3 balance decrease during 2018?\\
        \rule{\textwidth}{0.4pt} 
        \textbf{Answer:} \\
        To determine the percentage decrease in the Level 3 balance during 2018, we use the following data:

\textbf{Level 3 Balance:}

\begin{itemize}
    \item \textbf{Beginning Balance (January 1, 2018):} \$278 million
    \item \textbf{Ending Balance (December 31, 2018):} \$230 million
\end{itemize}

\textbf{Calculation:}

\begin{align*}
\text{Decrease Amount} &= \text{Beginning Balance} - \text{Ending Balance} \\
&= \$278\ \text{million} - \$230\ \text{million} \\
&= \$48\ \text{million}
\end{align*}

\textbf{Percentage Decrease}:

\[
\text{Percentage Decrease} = \left( \dfrac{\text{Decrease Amount}}{\text{Beginning Balance}} \right) \times 100\%
\]

Plugging in the numbers:

\[
\text{Percentage Decrease} = \left( \dfrac{\$48\ \text{million}}{\$278\ \text{million}} \right) \times 100\% \approx 17.266\%
\]

\textbf{Answer:}

The Level 3 balance decreased by approximately \textbf{17.27\%} during 2018.
\\
        \\
        \rule{\textwidth}{0.4pt}
         \textbf{Reference Answer:} -0.17      
    \end{tcolorbox} 
\end{figure}

\newpage
\subsection{Logical Reasoning}

\begin{figure}[H]
    \centering
    \begin{tcolorbox}[colback=blue!5!white, colframe=blue!75!black, width=\textwidth, arc=5mm, auto outer arc, boxrule=0.5mm, title=Prompt and Answer for Question 2, fonttitle=\small, fontupper=\tiny]
        \textbf{Prompt:} \\
        \tiny Read the PROBLEM. Follow the INSTRUCTIONS to answer the PROBLEM.\\
        \\
        \textbf{PROBLEM:} \\
        Zhang Ming, Li Ying, Wang Jia and Chen Rui work in a team. They come from four provinces of Jiangsu, Anhui, Fujian and Shandong. Each person can only speak a dialect of their origin. Speaking of southern Fujian dialects, Shandong people have the highest educational level and can speak Central Plains Mandarin. Wang Jia has a lower education level than Fujianese. Li Ying can speak Huizhou dialect and is a colleague with his classmates from Jiangsu. Chen Rui does not understand the dialect of southern Fujian. According to the above statement, you can get which of the following? \\
        A. Chen Rui can't speak the Central Plains Mandarin \\
        B. Zhang Ming speaks Minnan dialect \\
        C. Li Ying is from Shandong \\
        D. Wang Jia can speak Huizhou dialect \\
        INSTRUCTIONS: Study the PROBLEM above and select the answer option you believe is correct.\\
        \rule{\textwidth}{0.4pt} 
        \textbf{Answer:} \\
        \tiny Given Information: \\
        1. Civic Park is at the center. \\
        2. Four communities are built around the Civic Park, located in the southeast (SE), northwest (NW), southwest (SW), and northeast (NE) directions. \\
        3. The communities are designated as the cultural area, leisure area, commercial area, and administrative service area. \\
        4. Administrative service area is southwest of the cultural area. \\
        5. Cultural area is southeast of the leisure area. \\
        Step-by-Step Analysis: \\
        1. Assign Communities to Directions: We need to assign each community to one of the four quadrants (SE, NW, SW, NE) around the Civic Park such that the given positional relationships are satisfied. \\
        2. Determine Possible Positions: Let's consider the following assignment: \\
        - Leisure Area: Northwest (NW) quadrant \\
        - Cultural Area: Southeast (SE) quadrant \\
        - Administrative Service Area: Southwest (SW) quadrant \\
        - Commercial Area: Northeast (NE) quadrant \\
        3. Validate the Positional Relationships: \\
        a. Cultural Area is Southeast of the Leisure Area: \\
        - Leisure Area is in the NW quadrant. \\
        - Cultural Area is in the SE quadrant. \\
        - From the Leisure Area in the NW, moving southeast, you pass through the center and reach the Cultural Area in the SE. \\
        - This satisfies the condition that the Cultural Area is southeast of the Leisure Area. \\
        b. Administrative Service Area is Southwest of the Cultural Area: \\
        - Cultural Area is in the SE quadrant. \\
        - Administrative Service Area is in the SW quadrant. \\
        - From the Cultural Area in the SE, moving southwest, you reach the Administrative Service Area in the SW. \\
        - This satisfies the condition that the Administrative Service Area is southwest of the Cultural Area. \\
        4. Evaluate Each Option: \\
        Option A: Civic Park is north of the administrative service area. \\
        - Civic Park is at the center. \\
        - Administrative Service Area is in the SW quadrant. \\
        - The Civic Park (center) is north of the Administrative Service Area (which is south of the center). \\
        - Conclusion: True \\
        Option B: The leisure area is southwest of the cultural area. \\
        - Leisure Area is in the NW quadrant. \\
        - Cultural Area is in the SE quadrant. \\
        - From the Cultural Area, the Leisure Area is northwest, not southwest. \\
        - Conclusion: False \\
        Option C: The cultural area is in the northeast of the commercial area. \\
        - Cultural Area is in the SE quadrant. \\
        - Commercial Area is in the NE quadrant. \\
        - From the Commercial Area, the Cultural Area is south, not northeast. \\
        - Conclusion: False \\
        Option D: The commercial area is southeast of the leisure area. \\
        - Leisure Area is in the NW quadrant. \\
        - Commercial Area is in the NE quadrant. \\
        - From the Leisure Area, the Commercial Area is directly east, not southeast. \\
        - Conclusion: False \\
        Final Answer: Option A is the correct statement that can be derived from the given information. \\
        **Answer:** A
        \\
        \rule{\textwidth}{0.4pt}
         \textbf{Reference Answer:} \\
         a       
    \end{tcolorbox}
    \caption{Example of Prompt and Answer Analysis}
\end{figure}

\begin{figure}[H]
    \centering
    \begin{tcolorbox}[colback=blue!5!white, colframe=blue!75!black, width=\textwidth, arc=5mm, auto outer arc, boxrule=0.5mm, title=Prompt and Answer for Question 3, fonttitle=\small, fontupper=\tiny]
        \textbf{Prompt:} \\
        \tiny Read the PROBLEM. Follow the INSTRUCTIONS to answer the PROBLEM.\\
        \\
        \textbf{PROBLEM:} \\
        The person in charge of the relevant department said that pilots will be carried out in some areas this year to prepare for the system and policy preparations for the comprehensive clean-up of "small property houses", and require all localities to confirm the registration and certification of rural collective land. All small-property houses are not subject to registration and are not protected by law. Therefore, this newly built house in Hexi Village is not protected by law. Which of the following is true will most weaken the above argument? \\
        A. This newly built house in Hexi Village has been tacitly approved by relevant departments \\
        B. This newly built house in Hexi Village is a small property house \\
        C. This newly built house in Hexi Village is built on rural collective land \\
        D. Some of the workers’ children in the statistics may be former farmers’ children \\
        INSTRUCTIONS: Study the PROBLEM above and select the answer option you believe is correct.\\
        \rule{\textwidth}{0.4pt} 
        \textbf{Answer:} \\
        \tiny To solve this problem, we'll carefully analyze the given statements and deduce the correct answer. \\
        People and Provinces: \\
        We have four team members: \\
        1. Zhang Ming \\
        2. Li Ying \\
        3. Wang Jia \\
        4. Chen Rui \\
        They come from four provinces: \\
        • Jiangsu \\
        • Anhui \\
        • Fujian \\
        • Shandong \\
        Each person can only speak the dialect of their origin. \\
        Given Clues: \\
        1. The Shandong person has the highest educational level and can speak Central Plains Mandarin. \\
        2. Wang Jia has a lower education level than the Fujianese. \\
        3. Li Ying can speak Huizhou dialect and is a colleague with his classmate from Jiangsu. \\
        4. Chen Rui does not understand the dialect of southern Fujian (Minnan dialect). \\
        Step-by-Step Deduction: \\
        1. Determine Li Ying's Origin: \\
        • Li Ying can speak Huizhou dialect. \\
        • Huizhou dialect is spoken in Anhui province. \\
        • Therefore, Li Ying is from Anhui. \\
        • Li Ying is a colleague with his classmate from Jiangsu. \\
        2. Determine Chen Rui's Origin: \\
        • Chen Rui does not understand the Minnan dialect (southern Fujian dialect). \\
        • Remaining possible origins for Chen Rui are Jiangsu and Shandong. \\
        • The Shandong person has the highest educational level and can speak Central Plains Mandarin. \\
        3. Determine Wang Jia's Origin: \\
        • Remaining provinces for Wang Jia and Zhang Ming are Jiangsu and Fujian. \\
        • Wang Jia has a lower education level than the Fujianese. \\
        4. Determine Zhang Ming's Origin: \\
        • By elimination, Zhang Ming must be from Fujian. \\
        Final Answer: B. Zhang Ming speaks Minnan dialect \\
        **Answer:** B\\
        \rule{\textwidth}{0.4pt} 
        \textbf{Reference Answer:} \\
        \tiny b
    \end{tcolorbox}
    \caption{Example of Prompt and Answer Analysis}
\end{figure}

\begin{figure}[H]
    \centering
    \begin{tcolorbox}[colback=blue!5!white, colframe=blue!75!black, width=\textwidth, arc=5mm, auto outer arc, boxrule=0.5mm, title=Prompt and Answer for Question 4, fonttitle=\small, fontupper=\tiny]
        \textbf{Prompt:} \\
        \tiny Read the PROBLEM. Follow the INSTRUCTIONS to answer the PROBLEM.\\
        \\
        \textbf{PROBLEM:} \\
        There are five teams participating in the game. The audience had the following comments on the results? (1) The champion is either the Shannan team or the Jiangbei team. (2) The champion is neither Shanbei nor Jiangnan. (3) The champion is Jiangnan Team. (4) The champion is not the Shannan team. The result of the match showed that only one argument was correct, so who won the championship? \\
        A. Shannan \\
        B. Jiangnan \\
        C. Shanbei \\
        D. Jiangbei \\
        INSTRUCTIONS: Study the PROBLEM above and select the answer option you believe is correct.\\
        \rule{\textwidth}{0.4pt} 
        \textbf{Answer:} \\
        \tiny To determine which option most weakens the argument, let's first understand the argument's structure and then evaluate each option. \\
        Argument Summary: \\
        • Premise 1: All small-property houses are not subject to registration and are not protected by law. \\
        • Premise 2 (Implied): The newly built house in Hexi Village is a small-property house. \\
        • Conclusion: Therefore, this newly built house in Hexi Village is not protected by law. \\
        Our Goal: Find an option that, if true, would most weaken the conclusion that the newly built house in Hexi Village is not protected by law. \\
        Analyzing Each Option: \\
        Option A: This newly built house in Hexi Village has been tacitly approved by relevant departments. \\
        • If the relevant departments have tacitly approved the house, it implies a level of official acceptance or legality. \\
        • This approval could mean the house is an exception to the rule that small-property houses are not protected by law. \\
        • Conclusion: This option weakens the argument by suggesting that the house may, in fact, be protected by law due to the tacit approval. \\
        Option B: This newly built house in Hexi Village is a small property house. \\
        • Confirms the implied premise that the house is a small-property house. \\
        • Reinforces the conclusion that it is not protected by law. \\
        • Conclusion: This option strengthens the argument rather than weakens it. \\
        Option C: This newly built house in Hexi Village is built on rural collective land. \\
        • Aligns with the characteristics of small-property houses (often built on rural collective land). \\
        • Supports the premises leading to the conclusion. \\
        • Conclusion: This option does not weaken the argument; it may even strengthen it. \\
        Option D: Some of the newly built houses in Hexi Village are not small property houses. \\
        • Indicates that not all houses in Hexi Village are small-property houses. \\
        • However, it doesn't specify that this particular house is not a small-property house. \\
        • Conclusion: This option does not directly weaken the conclusion about the specific house in question. \\
        Final Evaluation: \\
        Option A introduces information that directly challenges the conclusion by implying the house may be legally protected. \\
        Options B and C either support or do not affect the argument. \\
        Option D is too general and does not specifically address the house in question. \\
        Answer: This newly built house in Hexi Village has been tacitly approved by relevant departments. \\
        **Answer:** A\\
        \rule{\textwidth}{0.4pt} 
        \textbf{Reference Answer:} \\
        \tiny a
    \end{tcolorbox}
    \caption{Example of Prompt and Answer Analysis}
\end{figure}

\begin{figure}[H]
    \centering
    \begin{tcolorbox}[colback=blue!5!white, colframe=blue!75!black, width=\textwidth, arc=5mm, auto outer arc, boxrule=0.5mm, title=Prompt and Answer for Question 5, fonttitle=\small, fontupper=\tiny]
        \textbf{Prompt:} \\
        \tiny Read the PROBLEM. Follow the INSTRUCTIONS to answer the PROBLEM.\\
        \\
        \textbf{PROBLEM:} \\
        Compared with small and medium-sized cities, especially small cities and towns, large cities have higher living costs, which inevitably limits the entry of rural population. Therefore, the development of large cities alone cannot actually achieve urbanization. Which of the following is the conclusion must be assumed? \\
        A. Urbanization is the only way for China's development \\
        B. Simple development of large cities is not conducive to the promotion of urbanization \\
        C. To achieve urbanization, the city must fully absorb the rural population \\
        D. The attractiveness of large cities to the rural population in the outside world is significantly lower than that of small and medium-sized cities \\
        INSTRUCTIONS: Study the PROBLEM above and select the answer option you believe is correct.\\
        \rule{\textwidth}{0.4pt} 
        \textbf{Answer:} \\
        \tiny To determine who won the championship, we'll analyze each of the audience's comments and consider each possible team as the champion. \\
        Teams Participating: \\
        1. Shannan \\
        2. Jiangbei \\
        3. Shanbei \\
        4. Jiangnan \\
        Audience Comments: \\
        1. Statement 1: The champion is either the Shannan team or the Jiangbei team. \\
        2. Statement 2: The champion is neither Shanbei nor Jiangnan. \\
        3. Statement 3: The champion is Jiangnan Team. \\
        4. Statement 4: The champion is not the Shannan team. \\
        Testing Each Team as Champion: \\
        Option A: Champion is Shannan \\
        • Two statements are true (Statements 1 and 2). This contradicts the condition. \\
        Option B: Champion is Jiangnan \\
        • Two statements are true (Statements 3 and 4). This contradicts the condition. \\
        Option C: Champion is Shanbei \\
        • Only one statement is true (Statement 4). This satisfies the condition. \\
        Option D: Champion is Jiangbei \\
        • Three statements are true (Statements 1, 2, and 4). This contradicts the condition. \\
        Conclusion: The only scenario where exactly one statement is true is when the Shanbei team is the champion. \\
        Answer: C. Shanbei \\
        **Answer:** C\\
        \rule{\textwidth}{0.4pt} 
        \textbf{Reference Answer:} \\
        \tiny c
    \end{tcolorbox}
    \caption{Example of Prompt and Answer Analysis}
\end{figure}

\begin{figure}[H]
    \centering
    \begin{tcolorbox}[colback=blue!5!white, colframe=blue!75!black, width=\textwidth, arc=5mm, auto outer arc, boxrule=0.5mm, title=Prompt and Answer for Question 6, fonttitle=\small, fontupper=\tiny]
        \textbf{Prompt:} \\
        \tiny Read the PROBLEM. Follow the INSTRUCTIONS to answer the PROBLEM.\\
        \\
        \textbf{PROBLEM:} \\
        Researchers recently discovered that there is an area called the thalamus pillow deep in the human brain, which is like an information desk operator, responsible for sorting out the external stimulus information and focusing human attention on the most important information for behavior and survival. The researchers pointed out that this discovery is expected to bring new treatments for disorders such as attention deficit hyperactivity disorder and schizophrenia caused by lack of attention. Which of the following is assumed by the above argument? \\
        A. Some schizophrenia is not caused by lack of attention \\
        B. Visual information is only transmitted through the visual cortex to obtain a neural network to transmit \\
        C. Researchers have developed a new technology that can directly track the communication between the nerve cortex in the visual cortex and thalamus occipital region \\
        D. The brain cannot process too much information in detail at the same time, and will only selectively focus on the things most relevant to behavior \\
        INSTRUCTIONS: Study the PROBLEM above and select the answer option you believe is correct.\\
        \rule{\textwidth}{0.4pt} 
        \textbf{Answer:} \\
        \tiny To determine which option is a necessary assumption for the argument, let's first analyze the argument carefully.\\
        \\
        **Argument Summary:** \\
        The argument discusses the thalamus pillow and its role in filtering external information, with the assumption that the brain focuses selectively on important information related to behavior and survival.\\
        \\
        **Analyzing Each Option:** \\
        - **Option A:** Some schizophrenia is not caused by lack of attention. \\
            - Conclusion: This does not directly support or negate the assumption about how the brain processes attention.\\
        - **Option B:** Visual information is only transmitted through the visual cortex to obtain a neural network to transmit.\\
            - Conclusion: This statement is unrelated to the assumption about selective brain function.\\
        - **Option C:** Researchers have developed a new technology to track communication between the visual cortex and thalamus occipital region.\\
            - Conclusion: This option does not align with the assumption that the brain processes selective information.\\
        - **Option D:** The brain cannot process too much information in detail at the same time, and will only selectively focus on the things most relevant to behavior.\\
            - **Negation Test:** If this assumption is not true, the brain would process all information equally, contradicting the argument's premise.\\
            - Therefore, **Option D** is a necessary assumption for the argument to hold.\\
        \\
        **Conclusion:** Option D is the correct answer.\\
        \rule{\textwidth}{0.4pt}
        \textbf{Reference Answer:} D
    \end{tcolorbox}
    \caption{Example of Prompt and Answer Analysis}
\end{figure}

\begin{figure}[H]
    \centering
    \begin{tcolorbox}[colback=blue!5!white, colframe=blue!75!black, width=\textwidth, arc=5mm, auto outer arc, boxrule=0.5mm, title=Prompt and Answer for Question 7, fonttitle=\small, fontupper=\tiny]
        \textbf{Prompt:} \\
        \tiny Read the PROBLEM. Follow the INSTRUCTIONS to answer the PROBLEM.\\
        \\
        \textbf{PROBLEM:} \\
        According to a report, in the year to March, mainland Chinese buyers became the second largest group of foreign buyers buying US real estate, with a transaction volume of US \$ 9 billion, second only to Canada. This is 23\% higher than the previous year’s USD 7.3 billion transaction volume and 88\% higher than the previous year’s USD 4.8 billion transaction volume. Some people think that there are more and more rich people in China who are transferring their property abroad. Which of the following is true is the best argument against the above argument? \\
        A. Many Chinese people buy homes for their children to study in the US \\
        B. Although the turnover has increased by 23\%, the volume of Chinese buyers has not increased this year \\
        C. There are groups of real estate speculators among the wealthy Chinese who have speculated in Beijing and Shanghai \\
        D. The risk of the US real estate market has been small in recent years and has a certain function of value preservation and appreciation \\
        INSTRUCTIONS: Study the PROBLEM above and select the answer option you believe is correct.\\
        \rule{\textwidth}{0.4pt} 
        \textbf{Answer:} \\
        \tiny To determine which option is the best argument against the idea that Chinese rich people are transferring property abroad, we need to consider the reasoning and the facts provided.\\
        \\
        **Analyzing Each Option:** \\
        - **Option A:** Many Chinese people buy homes for their children to study in the US.\\
            - This option doesn't directly argue against the idea of transferring property abroad, as it could be seen as a reason for such transfers.\\
        - **Option B:** Although the turnover has increased by 23\%, the volume of Chinese buyers has not increased this year.\\
            - This statement contradicts the idea that more rich Chinese people are transferring property abroad, as it suggests that the number of buyers has remained stable.\\
            - This is the best argument against the premise.\\
        - **Option C:** There are groups of real estate speculators among the wealthy Chinese who have speculated in Beijing and Shanghai.\\
            - This option is unrelated to the premise about transferring property abroad, focusing instead on domestic speculation.\\
        - **Option D:** The risk of the US real estate market has been small in recent years and has a certain function of value preservation and appreciation.\\
            - This option supports the idea of transferring property abroad, as it highlights the stability of the US market.\\
        \\
        **Conclusion:** Option B provides the strongest argument against the idea that Chinese rich people are increasingly transferring property abroad.\\
        \rule{\textwidth}{0.4pt}
        \textbf{Reference Answer:} B
    \end{tcolorbox}
    \caption{Example of Prompt and Answer Analysis}
\end{figure}

\begin{figure}[H]
    \centering
    \begin{tcolorbox}[colback=blue!5!white, colframe=blue!75!black, width=\textwidth, arc=5mm, auto outer arc, boxrule=0.5mm, title=Prompt and Answer for Question 8, fonttitle=\small, fontupper=\tiny]
        \textbf{Prompt:} \\
        \tiny Read the PROBLEM. Follow the INSTRUCTIONS to answer the PROBLEM.\\
        \\
        \textbf{PROBLEM:} \\
        People often use microwave ovens to heat food. Some people think that microwave oven heating changes the molecular structure of food, producing molecules that the human body cannot recognize. These strange molecular formulas are unacceptable to the human body, and some are toxic and may even cause cancer. Therefore, people or animals who often eat microwave foods will have serious physiological changes in their bodies, which will cause serious health problems. Which of the following can best question the above view? \\
        A. Microwave heating will not cause more nutrient loss than other cooking methods \\
        B. China's microwave oven production standards are consistent with international standards and EU standards \\
        C. It is also common for developed countries to use microwave ovens \\
        D. The microwave only heats the water molecules in the food, the food does not change chemically \\
        INSTRUCTIONS: Study the PROBLEM above and select the answer option you believe is correct.\\
        \rule{\textwidth}{0.4pt} 
        \textbf{Answer:} \\
        \tiny To determine which option best questions the idea that microwave heating alters the molecular structure of food and causes health problems, we need to evaluate each option carefully.\\
        \\
        **Analyzing Each Option:** \\
        - **Option A:** Microwave heating will not cause more nutrient loss than other cooking methods.\\
            - This statement addresses nutrient loss but does not challenge the idea that molecular structures are altered.\\
        - **Option B:** China's microwave oven production standards are consistent with international standards and EU standards.\\
            - This statement concerns production standards and does not address the claim about molecular changes.\\
        - **Option C:** It is also common for developed countries to use microwave ovens.\\
            - This statement indicates widespread usage but does not challenge the claim that microwaves alter molecular structures.\\
        - **Option D:** The microwave only heats the water molecules in the food, the food does not change chemically.\\
            - **Negation Test:** If this statement is false, the argument holds. This directly contradicts the claim that microwave heating causes molecular changes.\\
            - Therefore, **Option D** is the strongest challenge to the argument.\\
        \\
        **Conclusion:** Option D provides the best counter to the claim that microwave heating changes the molecular structure of food.\\
        \rule{\textwidth}{0.4pt}
        \textbf{Reference Answer:} D
    \end{tcolorbox}
    \caption{Example of Prompt and Answer Analysis}
\end{figure}

\begin{figure}[H]
    \centering
    \begin{tcolorbox}[colback=blue!5!white, colframe=blue!75!black, width=\textwidth, arc=5mm, auto outer arc, boxrule=0.5mm, title=Prompt and Answer for Question 9, fonttitle=\small, fontupper=\tiny]
        \textbf{Prompt:} \\
        \tiny Read the PROBLEM. Follow the INSTRUCTIONS to answer the PROBLEM.\\
        \\
        \textbf{PROBLEM:} \\
        A research report shows that the proportion of children of cadres in Peking University increased from 20
        A. Many of the cadres counted in the past 20 years are enterprise cadres, which used to include only government officials \\
        B. Compared with foreign countries, Chinese education provides workers and peasants with more opportunities for education and social mobility \\
        C. After the founding of New China, more and more children of workers and peasants entered the university \\
        D. Some of the workers’ children in the statistics may be former farmers’ children \\
        INSTRUCTIONS: Study the PROBLEM above and select the answer option you believe is correct.\\
        \rule{\textwidth}{0.4pt} 
        \textbf{Answer:} \\
        \tiny To question the media’s view that the proportion of cadres' children has been rising, we need to find an option that addresses potential flaws in the data or conclusions drawn from it.\\
        \\
        **Analyzing Each Option:** \\
        - **Option A:** Many of the cadres counted in the past 20 years are enterprise cadres, which used to include only government officials.\\
            - This suggests that changes in classification, not actual demographic changes, explain the increase, casting doubt on the media's view.\\
        - **Option B:** Compared with foreign countries, Chinese education provides workers and peasants with more opportunities for education and social mobility.\\
            - This does not directly challenge the specific increase in cadres' children at Peking University.\\
        - **Option C:** After the founding of New China, more and more children of workers and peasants entered the university.\\
            - This focuses on broader educational trends rather than the specific group of cadres’ children.\\
        - **Option D:** Some of the workers’ children in the statistics may be former farmers’ children.\\
            - This option is irrelevant to the argument about cadres' children.\\
        \\
        **Conclusion:** Option A provides the strongest argument against the media’s conclusion, questioning the data classification.\\
        \rule{\textwidth}{0.4pt}
        \textbf{Reference Answer:} A
    \end{tcolorbox}
    \caption{Example of Prompt and Answer Analysis}
\end{figure}

\begin{figure}[H]
    \centering
    \begin{tcolorbox}[colback=blue!5!white, colframe=blue!75!black, width=\textwidth, arc=5mm, auto outer arc, boxrule=0.5mm, title=Prompt and Answer Question 10, fonttitle=\small, fontupper=\tiny]
        \textbf{Prompt:} \\
        \tiny Read the PROBLEM. Follow the INSTRUCTIONS to answer the PROBLEM.\\
        \\
        \textbf{PROBLEM:} \\
        A research report shows that the proportion of children of cadres in Peking University increased from 20\% in the 1980s to nearly 40\% in 1977, surpassing the children of workers, farmers and professional technicians, and becoming the largest source of students. According to this, some media believe that the proportion of students and children of Peking University students has been rising for 20 years, surpassing other classes. Which one, if true, most doubts the views of the above media? \\
        A. Many of the cadres counted in the past 20 years are enterprise cadres, which used to include only government officials \\
        B. Compared with foreign countries, Chinese education provides workers and peasants with more opportunities for education and social mobility \\
        C. After the founding of New China, more and more children of workers and peasants entered the university \\
        D. Some of the workers’ children in the statistics may be former farmers’ children \\
        INSTRUCTIONS: Study the PROBLEM above and select the answer option you believe is correct.\\
        \rule{\textwidth}{0.4pt} 
        \textbf{Answer:} \\
        \tiny To challenge the view that the proportion of cadres' children has been increasing, we need to identify any changes in the classification system or misinterpretation of data.\\
        \\
        **Analyzing Each Option:** \\
        - **Option A:** Many of the cadres counted in the past 20 years are enterprise cadres, which used to include only government officials.\\
            - This suggests that the increase in the proportion of cadres' children is due to a broader classification rather than an actual demographic shift.\\
        - **Option B:** Compared with foreign countries, Chinese education provides workers and peasants with more opportunities for education and social mobility.\\
            - This does not directly address the proportion of cadres' children at Peking University.\\
        - **Option C:** After the founding of New China, more and more children of workers and peasants entered the university.\\
            - This option focuses on general trends and does not question the increase in cadres' children.\\
        - **Option D:** Some of the workers’ children in the statistics may be former farmers’ children.\\
            - This option does not effectively question the media’s view about cadres’ children.\\
        \\
        **Conclusion:** Option A directly questions the classification of cadres, providing the strongest doubt against the media’s conclusion.\\
        \rule{\textwidth}{0.4pt}
        \textbf{Reference Answer:} A
    \end{tcolorbox}
    \caption{Example of Prompt and Answer Analysis}
\end{figure}

\newpage

\subsection{Analogical Reasoning}
\label{appen_ana_reason}
In this appendix section, we provide the rest examples for the experiment. 

\begin{figure}[H]
    \centering
    \begin{tcolorbox}[colback=blue!5!white, colframe=blue!75!black, width=\textwidth, arc=5mm, auto outer arc, boxrule=0.5mm, title=Prompt and Answer for Question 2, fonttitle=\small, fontupper=\tiny]
        \textbf{Prompt:} \\
        \tiny Read the PROBLEM. Follow the INSTRUCTIONS to answer the PROBLEM.\\
        \\
        \textbf{PROBLEM:} \\
        Question:\\
        lime => U\\
        king => C C C\\
        queen => c c c\\
        man => C C\\
        woman =>\\
        INSTRUCTIONS: We are conducting an experiment on general reasoning ability. You will be presented with a series of words and drawings. Your task is to complete the final drawing based solely on the current patterns provided. Please respond concisely with only the final drawing. \\
        \rule{\textwidth}{0.4pt} 
        \textbf{Answer:} \\
        \tiny 
            c c\\
        \rule{\textwidth}{0.4pt}
        \textbf{Reference Answer:}
            c c
    \end{tcolorbox}
\end{figure}

\begin{figure}[H]
    \centering
    \begin{tcolorbox}[colback=blue!5!white, colframe=blue!75!black, width=\textwidth, arc=5mm, auto outer arc, boxrule=0.5mm, title=Prompt and Answer for Question 3, fonttitle=\small, fontupper=\tiny]
        \textbf{Prompt:} \\
        \tiny Read the PROBLEM. Follow the INSTRUCTIONS to answer the PROBLEM.\\
        \\
        \textbf{PROBLEM:} \\
        Question:\\
        puppy => Q Q Z Z I I\\
        kitten => Q Q Z Z I I Q Q Z Z I I\\
        cat => Q Z I Q Z I\\
        dog =>\\
        INSTRUCTIONS: We are conducting an experiment on general reasoning ability. You will be presented with a series of words and drawings. Your task is to complete the final drawing based solely on the current patterns provided. Please respond concisely with only the final drawing. \\
        \rule{\textwidth}{0.4pt} 
        \textbf{Answer:} \\
        \tiny 
            Q Z I\\
        \rule{\textwidth}{0.4pt}
        \textbf{Reference Answer:}
            Q Z I
    \end{tcolorbox}
\end{figure}

\begin{figure}[H]
    \centering
    \begin{tcolorbox}[colback=blue!5!white, colframe=blue!75!black, width=\textwidth, arc=5mm, auto outer arc, boxrule=0.5mm, title=Prompt and Answer for Question 4, fonttitle=\small, fontupper=\tiny]
        \textbf{Prompt:} \\
        \tiny Read the PROBLEM. Follow the INSTRUCTIONS to answer the PROBLEM.\\
        \\
        \textbf{PROBLEM:} \\
        Question:\\
        feet => *\\
        inches => * * * * * * * * * * * *\\
        quarter => * * * *\\
        dollar => *\\
        dime => * * * * * * * * * *\\
        week =>\\
        INSTRUCTIONS: We are conducting an experiment on general reasoning ability. You will be presented with a series of words and drawings. Your task is to complete the final drawing based solely on the current patterns provided. Please respond concisely with only the final drawing. \\
        \rule{\textwidth}{0.4pt} 
        \textbf{Answer:} \\
        \tiny 
            * * * * * * *\\
        \rule{\textwidth}{0.4pt}
        \textbf{Reference Answer:}
            * * * * * * *\\
    \end{tcolorbox}
\end{figure}

\begin{figure}[H]
    \centering
    \begin{tcolorbox}[colback=blue!5!white, colframe=blue!75!black, width=\textwidth, arc=5mm, auto outer arc, boxrule=0.5mm, title=Prompt and Answer for Question 5, fonttitle=\small, fontupper=\tiny]
        \textbf{Prompt:} \\
        \tiny Read the PROBLEM. Follow the INSTRUCTIONS to answer the PROBLEM.\\
        \\
        \textbf{PROBLEM:} \\
        Question:\\
        pants => H\# H\\
        glove => X \# X\\
        torso => V\\
        foot => Z\\
        head => M\\
        shirt => V \# V\\
        hat =>\\
        INSTRUCTIONS: We are conducting an experiment on general reasoning ability. You will be presented with a series of words and drawings. Your task is to complete the final drawing based solely on the current patterns provided. Please respond concisely with only the final drawing. \\
        \rule{\textwidth}{0.4pt} 
        \textbf{Answer:} \\
        \tiny 
            M \# M\\
        \rule{\textwidth}{0.4pt}
        \textbf{Reference Answer:}
            M \# M\\
    \end{tcolorbox}
\end{figure}

\begin{figure}[H]
    \centering
    \begin{tcolorbox}[colback=blue!5!white, colframe=blue!75!black, width=\textwidth, arc=5mm, auto outer arc, boxrule=0.5mm, title=Prompt and Answer for Question 6, fonttitle=\small, fontupper=\tiny]
        \textbf{Prompt:} \\
        \tiny Read the PROBLEM. Follow the INSTRUCTIONS to answer the PROBLEM.\\
        \\
        \textbf{PROBLEM:} \\
        Question:\\
        \} \^ \{\\
        \^\\
        \} * \{\\
        INSTRUCTIONS: We are conducting an experiment on general reasoning ability. You will be presented with a series of words and drawings. Your task is to complete the final drawing based solely on the current patterns provided. Please respond concisely with only the final drawing. \\
        \rule{\textwidth}{0.4pt} 
        \textbf{Answer:} \\
        \tiny 
            *\\
        \rule{\textwidth}{0.4pt}
        \textbf{Reference Answer:}
            *\\
    \end{tcolorbox}
\end{figure}

\begin{figure}[H]
    \centering
    \begin{tcolorbox}[colback=blue!5!white, colframe=blue!75!black, width=\textwidth, arc=5mm, auto outer arc, boxrule=0.5mm, title=Prompt and Answer for Question 7, fonttitle=\small, fontupper=\tiny]
        \textbf{Prompt:} \\
        \tiny Read the PROBLEM. Follow the INSTRUCTIONS to answer the PROBLEM.\\
        \\
        \textbf{PROBLEM:} \\
        Question:\\
        king => C C C\\
        queen => c c c\\
        man => C C\\
        woman =>\\
        INSTRUCTIONS: We are conducting an experiment on general reasoning ability. You will be presented with a series of words and drawings. Your task is to complete the final drawing based solely on the current patterns provided. Please respond concisely with only the final drawing. \\
        \rule{\textwidth}{0.4pt} 
        \textbf{Answer:} \\
        \tiny 
            c c\\
        \rule{\textwidth}{0.4pt}
        \textbf{Reference Answer:}
            c c\\
    \end{tcolorbox}
\end{figure}

\begin{figure}[H]
    \centering
    \begin{tcolorbox}[colback=blue!5!white, colframe=blue!75!black, width=\textwidth, arc=5mm, auto outer arc, boxrule=0.5mm, title=Prompt and Answer for Question 8, fonttitle=\small, fontupper=\tiny]
        \textbf{Prompt:} \\
        \tiny Read the PROBLEM. Follow the INSTRUCTIONS to answer the PROBLEM.\\
        \\
        \textbf{PROBLEM:} \\
        Question:\\
        black => \} \^ \{\\
        gray => \^\\
        red => \} * \{\\
        door =>\\
        INSTRUCTIONS: We are conducting an experiment on general reasoning ability. You will be presented with a series of words and drawings. Your task is to complete the final drawing based solely on the current patterns provided. Please respond concisely with only the final drawing. \\
        \rule{\textwidth}{0.4pt} 
        \textbf{Answer:} \\
        \tiny 
            *\\
        \rule{\textwidth}{0.4pt}
        \textbf{Reference Answer:}
            *\\
    \end{tcolorbox}
\end{figure}

\begin{figure}[H]
    \centering
    \begin{tcolorbox}[colback=blue!5!white, colframe=blue!75!black, width=\textwidth, arc=5mm, auto outer arc, boxrule=0.5mm, title=Prompt and Answer for Question 9, fonttitle=\small, fontupper=\tiny]
        \textbf{Prompt:} \\
        \tiny Read the PROBLEM. Follow the INSTRUCTIONS to answer the PROBLEM.\\
        \\
        \textbf{PROBLEM:} \\
        Question:\\
        motorcycle => * *\\
        unicycle => !\\
        tricycle => ! ! !\\
        skateboard => ! ! ! !\\
        car => * * * *\\
        bicycle =>\\
        INSTRUCTIONS: We are conducting an experiment on general reasoning ability. You will be presented with a series of words and drawings. Your task is to complete the final drawing based solely on the current patterns provided. Please respond concisely with only the final drawing. \\
        \rule{\textwidth}{0.4pt} 
        \textbf{Answer:} \\
        \tiny 
            ! !\\
        \rule{\textwidth}{0.4pt}
        \textbf{Reference Answer:}
            ! !\\
    \end{tcolorbox}
\end{figure}

\begin{figure}[H]
    \centering
    \begin{tcolorbox}[colback=blue!5!white, colframe=blue!75!black, width=\textwidth, arc=5mm, auto outer arc, boxrule=0.5mm, title=Prompt and Answer for Question 10, fonttitle=\small, fontupper=\tiny]
        \textbf{Prompt:} \\
        \tiny Read the PROBLEM. Follow the INSTRUCTIONS to answer the PROBLEM.\\
        \\
        \textbf{PROBLEM:} \\
        Question:\\
        grandfather => !\\
        grandmother => *\\
        mother => * *\\
        father => ! !\\
        brother => ! ! !\\
        sister =>\\
        INSTRUCTIONS: We are conducting an experiment on general reasoning ability. You will be presented with a series of words and drawings. Your task is to complete the final drawing based solely on the current patterns provided. Please respond concisely with only the final drawing. \\
        \rule{\textwidth}{0.4pt} 
        \textbf{Answer:} \\
        \tiny 
            * * *\\
        \rule{\textwidth}{0.4pt}
        \textbf{Reference Answer:}
            * * *\\
    \end{tcolorbox}
\end{figure}

\subsection{Scientific Reasoning}

\begin{figure}[H]
    \centering
    \begin{tcolorbox}[colback=blue!5!white, colframe=blue!75!black, width=\textwidth, arc=5mm, auto outer arc, boxrule=0.5mm, title=Exemplary prompt to test Scientific Reasoning]
    \begin{small}
        \textbf{Prompt:} \\
        Read the \textbf{PROBLEM} carefully and provide a solution using given the \textbf{INSTRUCTIONS}.\\
        \\
        \textbf{PROBLEM}: \\
         Two studies estimate the mean caffeine content of an energy drink. Each study uses the same test on a random sample of the energy drink. Study 1 uses 25 bottles, and study 2 uses 100 bottles. Which statement is true? 
         
        a. The estimate of the actual mean caffeine content from each study will be equally uncertain.
        
        b. The uncertainty in the estimate of the actual mean caffeine content will be smaller in study 1 than in study 2. 
        
        c. The uncertainty in the estimate of the actual mean caffeine content will be larger in study 1 than in study 2. 
        
        d. None of the above 
        
        \textbf{INSTRUCTIONS}: \\
        Select the one correct answer from options a, b, c, and d.
    \end{small}
    \end{tcolorbox}
    \captionof{figure}{Example prompt to assess the Scientific Reasoning of the o1-preview model. Item used from the TOSLS test.}\label{fig:scr_reason_prompt2}
\end{figure}

\begin{figure}[H]
    \centering
    \begin{tcolorbox}[colback=blue!5!white, colframe=blue!75!black, width=\textwidth, arc=5mm, auto outer arc, boxrule=0.5mm, title=Answer]
    \begin{small}
            
    \textbf{Answer:} c. The uncertainty in the estimate of the actual mean caffeine content will be larger in Study 1 than in Study 2.

    \textbf{Explanation:}
    
    The uncertainty in estimating a population mean decreases as the sample size increases. This is because the standard error of the mean (SEM), which measures this uncertainty, is inversely proportional to the square root of the sample size:
    \[
    \text{SEM} = \frac{s}{\sqrt{n}},
    \]
    where \( s \) is the sample standard deviation and \( n \) is the sample size.
    
    Study 1 uses 25 bottles (\( n = 25 \)).
    
    Study 2 uses 100 bottles (\( n = 100 \)).
    
    Comparing the two studies:
    
    The standard error in Study 1 is
    \[
    \text{SEM}_1 = \frac{s}{\sqrt{25}} = \frac{s}{5}.
    \]
    The standard error in Study 2 is
    \[
    \text{SEM}_2 = \frac{s}{\sqrt{100}} = \frac{s}{10}.
    \]
    Since 
    \[
    \frac{s}{10} < \frac{s}{5},
    \]
    the uncertainty (standard error) is smaller in Study 2. Therefore, the uncertainty in the estimate from Study 1 is larger than that from Study 2.
    
    \end{small}
    \end{tcolorbox}
    \captionof{figure}{Answer the o1-preview model given to the prompt shown in  Figure \ref{fig:scr_reason_prompt}.}\label{fig:sc_reason_answer2}
\end{figure}

\end{CJK}
\end{document}